\documentclass[11 pt,a4paper]{article}

\usepackage{graphics,graphicx}
\usepackage{bm}
\usepackage{amsmath}
\usepackage{amssymb}
\usepackage{natbib}
\usepackage{color}

\begin{document}
\pagestyle{plain}
\pagenumbering{arabic}

\title{}
\author{}

\date{}

\maketitle

\begin{center}

{\bf \Large White Paper on East Asian Vision for mm/submm VLBI:}

\vspace{0.5cm}

{\bf \Large Toward Black Hole Astrophysics down to Angular Resolution of 1~R$_{S}$}

\vspace{2cm}
{\bf \large Editors} 

\vspace{0.2cm}
{\large Asada,~K.$^{1}$, Kino,~M.$^{2,3}$, Honma,~M.$^{3}$, Hirota,~T.$^{3}$, Lu,~R.-S.$^{4,5}$,\\
Inoue,~M.$^{1}$, Sohn,~B.-W.$^{2,6}$, Shen,~Z.-Q.$^{4}$, and Ho,~P.~T.~P.$^{1,7}$ }

\vspace{0.5cm}
{\bf \large Authors} 

\vspace{0.2cm}
{\large 
Akiyama,~K.$^{3,8}$,
Algaba,~J-C.$^{2}$,
An,~T.$^{4}$,
Bower,~G.$^{1}$,
Byun,~D-Y.$^{2}$,
Dodson,~R.$^{9}$,
Doi,~A.$^{10}$,
Edwards,~P.G.$^{11}$,
Fujisawa,~K.$^{12}$, 
Gu,~M-F.$^{4}$,
Hada,~K.$^{3}$, 
Hagiwara,~Y.$^{13}$,
Jaroenjittichai,~P.$^{15}$,
Jung,~T.$^{2,6}$,
Kawashima,~T.$^{3}$,
Koyama,~S.$^{1,5}$,
Lee,~S-S.$^{2}$,
Matsushita,~S.$^{1}$,
Nagai,~H.$^{3}$,
Nakamura,~M.$^{1}$,
Niinuma,~K.$^{12}$,
Phillips,~C.$^{11}$, 
Park,~J-H.$^{15}$,
Pu,~H-Y.$^{1}$,
Ro,~H-W.$^{2,6}$,
Stevens,~J.$^{11}$,
Trippe,~S.$^{15}$,
Wajima,~K.$^{2}$,
Zhao,~G-Y.$^{2}$
}

\vspace{3cm}
\begin{minipage}{170mm}
{\footnotesize
$^{1}$
Institute of Astronomy and Astrophysics, Academia Sinica, P.O. Box 23-141, Taipei 10617, Taiwan \\
$^{2}$
Korea Astronomy and Space Science Institute, Daedukudae-ro 776, Yuseong-gu, Daejeon 34055, Republic of Korea \\
$^{3}$
National Astronomical Observatory of Japan, 2-21-1 Osawa, Mitaka, Tokyo, 181-8588, Japan\\
$^{4}$
Shanghai Astronomical Observatory, Chinese Academy of Sciences, 80 Nandan Road, Shanghai 200030, China\\
$^{5}$
Max-Planck-Institut f\"{u}r Radioastronomie, Auf dem H\"{u}gel 69, D-53121 Bonn, Germany \\
$^{6}$
University of Science and Technology, 217 Gajeong-ro, Yuseong-gu, Daejeon 34113, Republic of Korea\\
$^{7}$
East Asian Observatory, 660 N. A’ohoku Place University Park, Hilo, Hawaii 96720, USA \\
$^{8}$
Massachusetts Institute of Technology, Haystack Observatory, Route 40, Westford, MA 01886, USA\\
$^{9}$
The International Centre for Radio Astronomy Research,
The University of Western Australia, M468, 35 Stirling Highway, Crawley, Perth, WA 6009, Australia\\ 
$^{10}$
Institute of Space and Astronautical Science, Japan Aerospace Exploration Agency, 
3-1-1 Yoshinodai, Chuou-ku, Sagamihara, Kanagawa 252-5210, Japan\\ 
$^{11}$
Commonwealth Scientific and Industrial Research Organisation,
Astronomy and Space Science, P.O. Box 76, Epping NSW 1710, Australia\\ 
$^{12}$
Yamaguchi University, Yoshida 1677-1, Yamaguchi, Yamaguchi 753-8512, Japan\\
$^{13}$
Toyo University, 5-28-20 Hakusan, Bunkyo-ku, Tokyo 112-8606, Japan\\
$^{14}$
 National Astronomical Research Institute of Thailand, 
 Siripanich Building 191 Huay Kaew Road, Muang District, Chiangmai 50200, Thailand \\ 
$^{15}$
Seoul National University, 1 Gwanak-ro, Gwanak-gu, 08826, Seoul, Republic of Korea\\}
\end{minipage}

\end{center}

\maketitle\thispagestyle{empty}

\newpage

\begin{center}
\begin{minipage}{170mm}

\vspace{20mm}

\begin{center}
{\bf \Large Preamble}
\end{center}

This White Paper details the intentions and plans of the East Asian Very Long Baseline Interferometry (VLBI) community for pushing the frontiers of millimeter/submillimeter VLBI. To this end, we shall endeavor to actively promote coordinated efforts in the East Asia region.
Our goal is to establish firm collaborations among the East Asia VLBI community in partnership with related institutes in North America and Europe and to expand existing global mm/submm VLBI arrays for (a) exploring the vicinity of black holes with an ultimate angular resolution down to 1~R$_{S}$ (Schwarzschild radius) and (b) investigating the dynamics of circumstellar gas in star-forming regions and late-type stars, and circumnuclear gas around active galactic nuclei (AGNs).

In the first half of this White Paper, we highlight scientific accomplishments of the East Asia (EA) VLBI community. Various VLBI research results on M87, Sgr A*, blazars, narrow-line Seyfert~1 galaxies, and compact symmetric objects are described, and future visions of our VLBI science are briefly presented. Maser science of star formation, stellar evolution, and physics of accretion disks around AGNs are also discussed. A new vision for conducting multi-transition maser studies using mm/submm VLBI together with the Atacama Large Millimeter/Submillimeter Array (ALMA) is described. In the second half of this White Paper, we describe the EA community's vision for using mm/submm VLBI arrays in the framework or extended version of the Event Horizon Telescope (EHT) and the Global Millimeter VLBI Array (GMVA). In 2014, the East Asia Observatory (EAO) was established with the adoption of the James Clerk Maxwell Telescope (JCMT), and the first VLBI fringes at 230\,GHz were detected in Japan. In Korea, the Seoul Radio Astronomy Observatory (SRAO) plans to acquire VLBI capabilities, and a 230\,GHz receiver has been installed on one of the Korean VLBI Network (KVN) antennas. 
Furthermore, at submillimeter wavelengths, the capability of phasing-up ALMA is available from 2017 spring, 
and the Greenland Telescope (GLT) is expected to be commissioned in 2017/2018 in Greenland. 
In the next few years, activities of EA telescopes in the submillimeter waveband (JCMT, GLT, SRAO, Nobeyama 45\,m and the Solar Planetary Atmosphere Research Telescope (SPART)) can develop their own capabilities. For instance, EA stations such as SRAO/KVN stations, GLT, and Nobeyama/SPART will provide unique baselines for the mm/submm VLBI array, while JCMT together with the Submillimeter Array (SMA) will provide zero-spacing and redundant baselines, simplifying the calibration process. The phase frequency transfer technique, which was mainly developed by the Korea Astronomy and Space Science Institute, is a powerful tool for compensating for the phase fluctuations that can dramatically increase coherence time; the compensation is especially crucial for high-frequency observations. Furthermore, the National Astronomical Observatory of Japan has recently developed and demonstrated a sparse-modeling--based imaging technique that can help to overcome the diffraction limit on a synthesized beam. All these advances are expected to facilitate the imaging of the black hole (BH) shadows in Sgr A* and M87, and help in revealing the nature of the accretion flows and improving our understanding of the jet. We stress the importance of the baselines between the phased ALMA, JCMT/SMA, and GLT at 350\,GHz, which is crucial for attaining the ultimate angular resolution of $1~R_{s}$ at the base of the M87 jet. In the millimeter waveband, recent activities of the EA VLBI Network (EAVN), including the VLBI Exploration of Radio Astrometry array and the KVN and the TianMa 65\,m telescope at 43\,GHz will be further extended to 86\,GHz in collaboration with the GMVA and Australia Telescope Compact Array. 
In 2017 spring,  an intensive campaign observation for Sgr A* and M87 with EAVN at 43\,GHz
was conducted coherently scheduled with the EHT campaign 
observation  at 230\,GHz.
High dynamic range EAVN images in the millimeter wavebands are valuable 
for understanding the jet launching mechanism, and jet dynamics in great detail.

The accomplishment of the aforementioned goal will maximize the overall scientific outcomes of mm/submm VLBI in the world.


\end{minipage}
\end{center}

\newpage
\tableofcontents


\newpage
\newpage

\section{Introduction}

\begin{flushleft}
{\bf What happens at the vicinity of a black hole?}
\end{flushleft}

Black holes provide a unique environment for studying the physics of strong gravity. Despite its great success, Einstein's general relativity (GR) had been tested only in the perturbative regime (including double pulsar tests), where it makes small corrections to Newtonian gravity (e.g., Will, 2014, and references therein). Nevertheless, it has played a critical role in helping us acquire an understanding of astronomical objects, such as compact stars, and the cosmos, often invoking the strongly nonlinear regime. The Event Horizon Telescope (EHT) project that includes the phased the Atacama Large Millimeter/submillimeter Array (ALMA) opens a new window to probe the strong-gravity regime around black holes via synchrotron emission from magnetized plasma. 

\bigskip

\begin{flushleft}
{\bf How do black holes power and form jets?}
\end{flushleft}

This is a longstanding problem in astrophysics. Black holes are also known as the ultimate drivers of a variety of energetic phenomena, ranging from $\gamma$-ray bursts to active galactic nuclei (AGNs). To determine the relativistic jet formation mechanism, we should clarify the structure and strength of magnetic fields, the radio-$\gamma$ connection, and velocity fields at the jet source in AGNs.

\bigskip

We believe that millimeter/submillimeter (mm/submm)-wavelength very long baseline interferometry  (VLBI) has the potential to provide information on these  fundamental aspects. This technique has the unique capability to provide the highest angular resolution images of supermassive black hole systems and associated relativistic jets. For the nearest black holes, Sgr A* and M87, millimeter (mm) VLBI is capable of providing a linear resolution of a few Rs (Schwarzschild radius), thereby facilitating unique GR tests in the strong-field regime. In addition, observations of Sgr A*, M87, and other systems can provide information on the nature of accretion flows and the formation of relativistic jets. We are operating and shall continue to operate the James Clerk Maxwell Telescope (JCMT), which is part of the EHT project  and a network of East Asian (EA) VLBI stations. Three EA VLBI stations --- the Greenland Telescope (GLT), the 10m Solar Planetary Atmosphere Research Telescope (SPART), and the 6m Seoul Radio Astronomy Observatory (SRAO) telescope --- will play a key role because they offer unique combinations of mm/submm VLBI subarrays. In particular, we emphasize the importance of observations at 345\,GHz made using the GLT, JCMT/Submillimeter Array (SMA), and phased ALMA. They provide the longest baselines among the mm/submm VLBI networks facing the northern sky, and therefore, they play a critical role in imaging the shadow of M87 with an angular resolution as high as 1\,Rs. Another key project is the EA VLBI Network (EAVN), which is advantageous for exploring the physics of relativistic jets. Currently, the combination of the Korean VLBI Network (KVN) and the VLBI Exploration of Radio Astrometry (VERA) array, part of the EAVN and referred to as KaVA, has been used for intensive monitoring of M87 with the aim of clarifying the velocity field of the M87 jet. KaVA is also used for the monthly monitoring of Sgr A* by exploiting the availability of many short baselines. The EAVN is expected to considerably improve our understanding of AGN jet physics.

\bigskip

\begin{flushleft}
{\bf Nature of astrophysical masers}
\end{flushleft}

Masers are powerful probes when studied at VLBI angular resolutions. {\it What do maser lines tell us?} Because of their high brightness, masers enable us to measure proper motions at better than milliarcseconds per year. Therefore, VLBI observations of masers provide information on the dynamic motion of circumstellar gas in star-forming regions and late-type stars, and circumnuclear gas around AGNs. However, because of nonlinear amplification processes, it is difficult to infer the physical properties of masing gas clumps. In the near future, ALMA will provide new mm/submm data at comparable spatial resolution at less than 10\,mas. Multi-transition maser studies that combine ALMA and mm/submm VLBI observations can determine the dynamic and physical properties of maser sources. Such information can appreciably help in acquiring a complete understanding of star formation, stellar evolution, and the physics of accretion disks around AGNs.

\bigskip

This White Paper is organized as follows. In Section~2, we highlight scientific accomplishments of the EA VLBI community. The section is divided into four parts: Imaging the Shadows of Supermassive Black Holes, Determining the Nature of Accretion Flows, Understanding Jet Formation Mechanism, and Advancing Maser Science. In Section~3, the current status of EA contributions to mm/submm VLBI is summarized, and finally, in Section~4, the vision of the EA VLBI community for the field of mm/submm VLBI is described.

\newpage

\newpage
\section{Scientific Objectives}

\subsection{Shadow of Super Massive Black Holes}
\subsubsection{General Objectives}\label{sec:2.1.1}

\begin{figure*}
	\centering
	\begin{tabular}{cc}
		\begin{minipage}[b]{0.45\textwidth}
			\centering
			\includegraphics[angle=-90,width=1.0\textwidth]{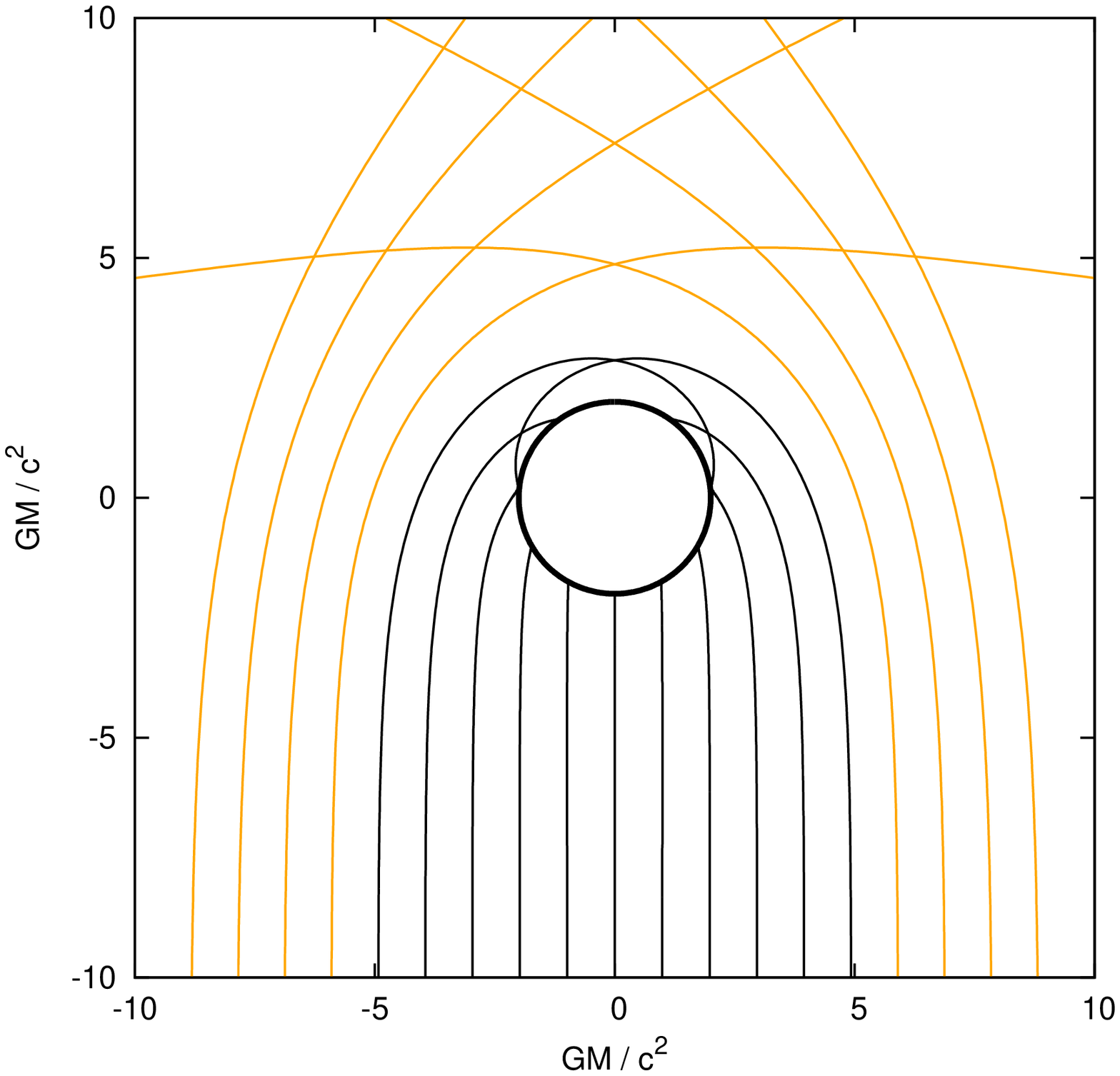}
		\end{minipage} &
		\begin{minipage}[b]{0.45\textwidth}
			\centering
			\includegraphics[angle=-90,width=1.0\textwidth]{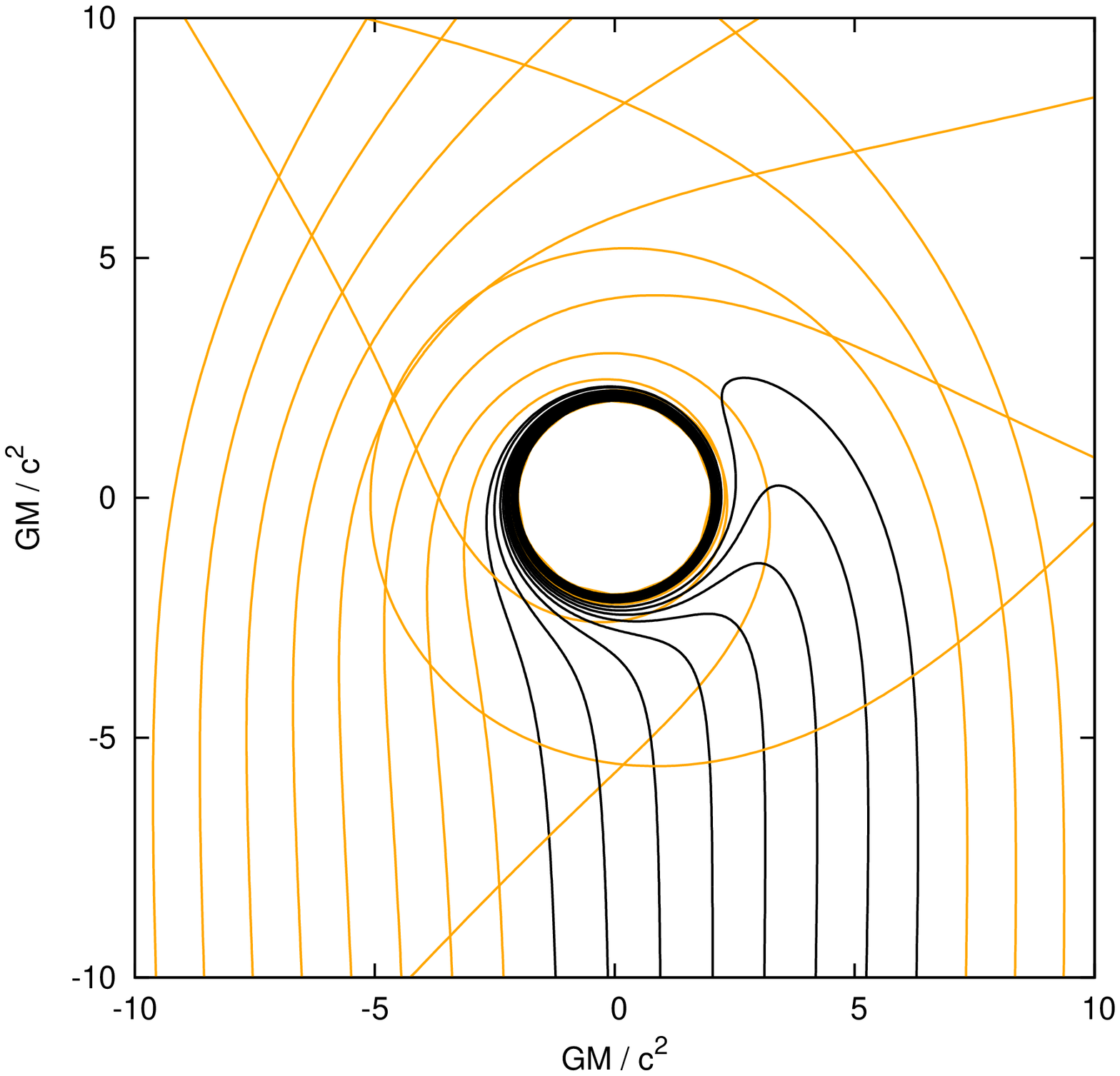}
		\end{minipage}\\
		\begin{minipage}[b]{0.35\textwidth}
			\centering
			\includegraphics[width=0.8\textwidth]{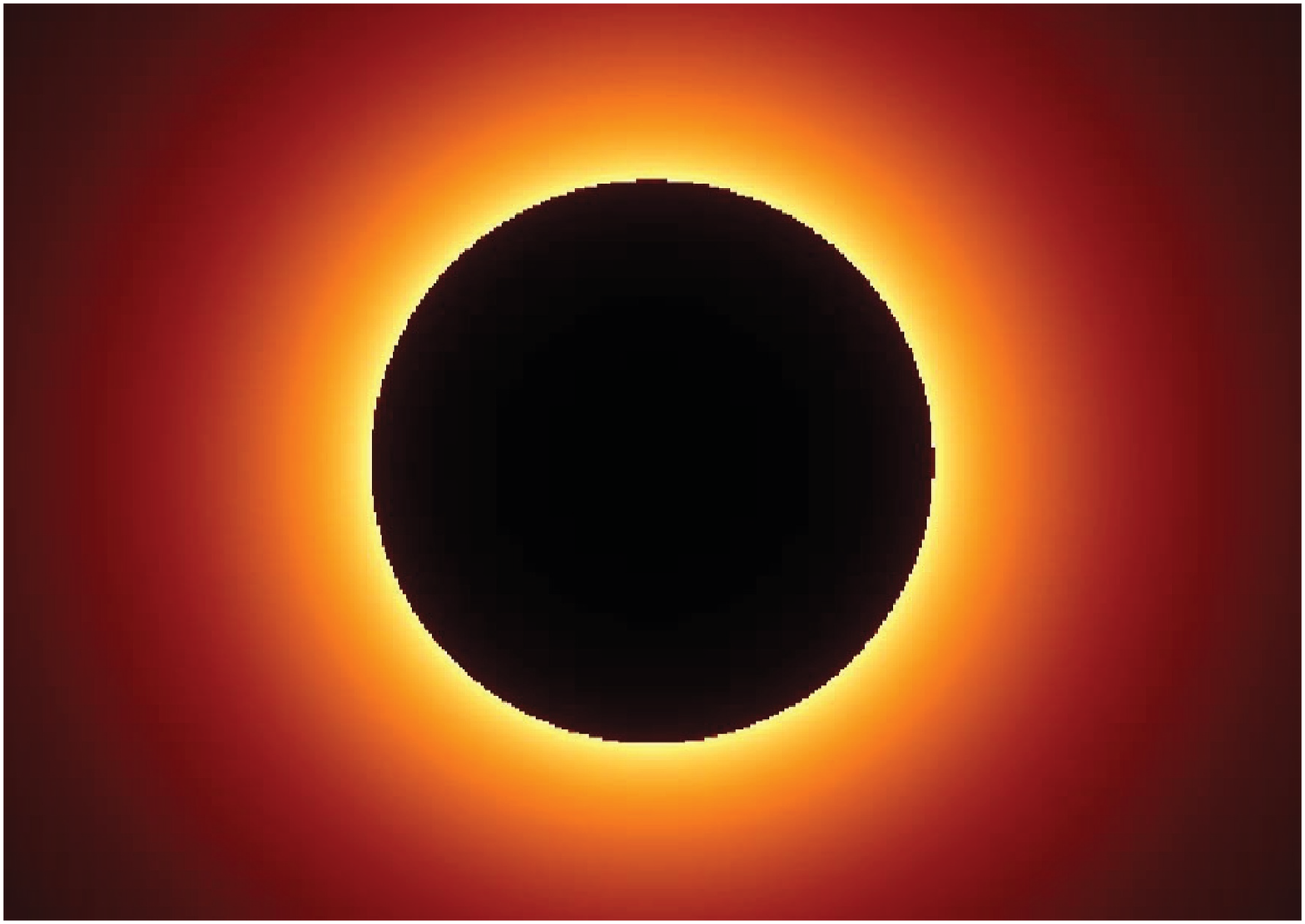}
		\end{minipage} &
		\begin{minipage}[b]{0.35\textwidth}
			\centering
			\includegraphics[width=0.8\textwidth]{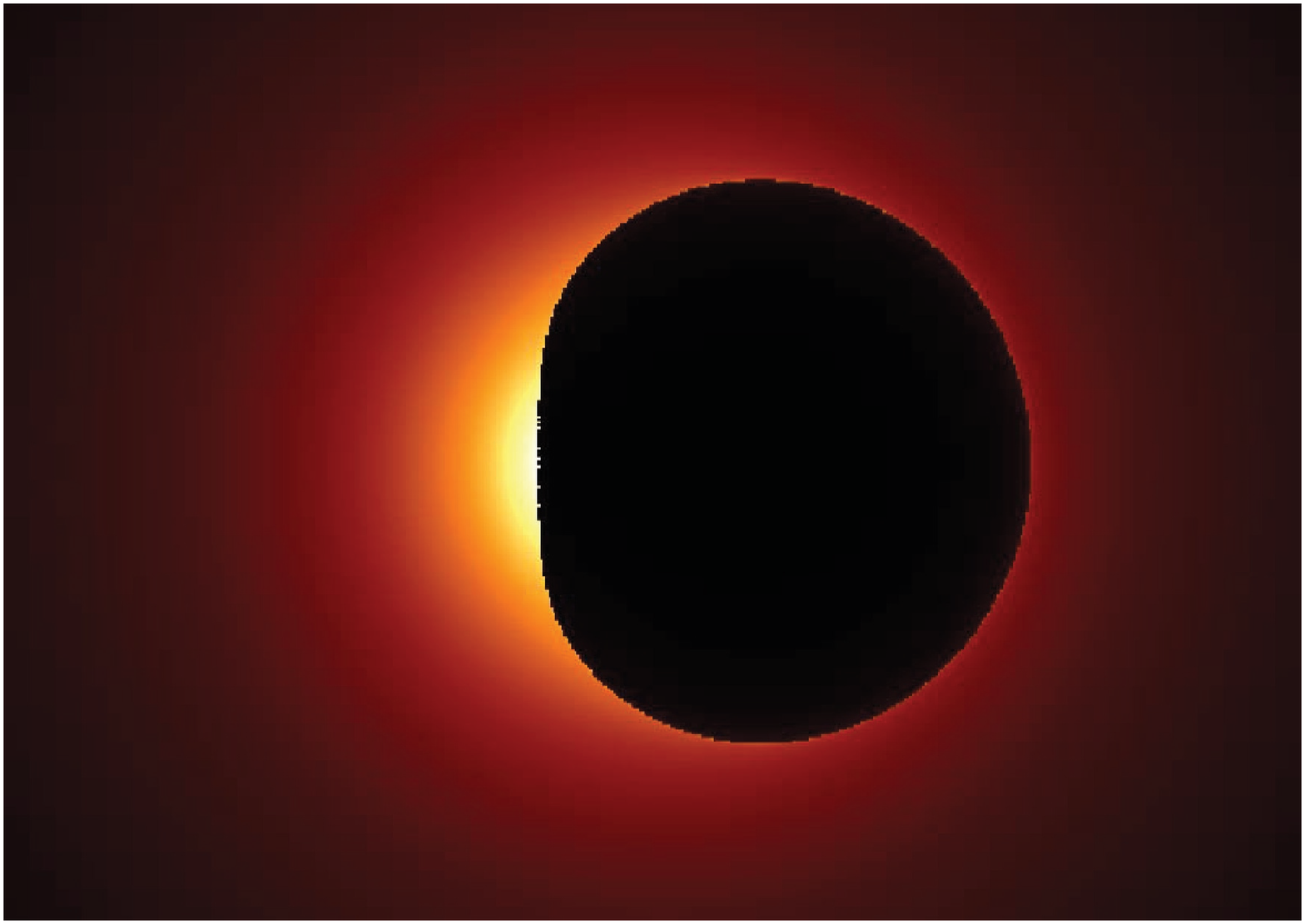}
		\end{minipage} \\
	\end{tabular}
	\caption{
	(Top panel): Frontal view of null geodesics (backward in time) on the equatorial plane for a non-rotating (left;  $a/M=0$) and a rapidly rotating (right; $a/M=0.998$) black hole. Geodesics that contribute to the shadow region are indicated in black. Geodesics are plotted in Cartesian coordinates. In the right panel, the black hole is rotating counter-clockwise and geodesics are ``dragged'' toward the same direction. (Bottom panel): Corresponding black hole image with optically thin, free-fall materials uniformly surrounding the black holes shown in the top panel. The observer is assumed to be edge-on. The shadow cast by the black hole photon capture region is clearly seen.
	\label{fig:hypu1}}
\end{figure*}

The photon capture region around a black hole casts a ``shadow'' or ``silhouette'' when it is surrounded by luminous, optically thin materials such as accretion flows or jets \citep[][]{1973blho.conf..241B}. The size and shape of the black hole shadow are determined by null geodesics. Because the geodesics are determined by a background spacetime metric, observations of the black hole shadow provide a direct test of Einstein's general relativity and other gravity theories in a strong gravity regime. In particular, the shadow size and shape can constrain the mass and spin of the black hole by tracing trajectories of the received photon backward in time. Mm/submm VLBI observations with microscale resolution (the angular resolution is $\sim$ 27\,$\mu$as for a 10,000\,km baseline at 1.3\,mm/230\,GHz) will enable the shadows of the supermassive black holes at the centers of Sgr A* and M87 to be resolved. Figure~\ref{fig:hypu1} illustrates the null geodesics for non-rotating and rapidly rotating black holes, assuming the background spacetime is described by the Kerr metric. In contrast to the non-rotating black hole, the frame-dragging effect around the rotating black hole forces all light rays to corotate in the same direction as the black hole when the rays are inside the ergosphere. Therefore, the shadow is not symmetric.

In the case of astronomical environments, the observed image of the black hole shadow is also related to background emission sources in the vicinity of the black hole, such as accretions or jets. The black hole shadow is revealed only when the surrounding materials are sufficiently optically thin at the observation frequency. In the vicinity of the black hole, general relativistic effects on radiative transfer, such as the energy shift between the fluid rest frame around the black hole and the distant observer, are essential \citep[e.g.,][]{2012A&A...545A..13Y,2016ApJ...820..105P}.  In the following we discuss the cases for Sgr A* and M87.

\subsubsection{Sgr A$^{*}$}\label{sec:2.1.2}

The event horizon of the supermassive black hole at the Galactic center, associated with the compact radio source Sgr A*, subtends the largest angular scale of all such objects in the sky. The angular size of the shadow is $\sim$50\,$\mu$as given by the distance of $\sim$8\,kpc (determined by VLBI astrometry and NIR observations) and the mass of $\sim 4\times 10^6$~$M_\odot$ (European and American groups), and is comparable to the spatial resolution of the global mm/submm arrays. In addition, the extremely low bolometric luminosity ($L_{\rm bol}\sim10^{-8} L_{\rm Edd}$) and the presence of a submillimeter bump in its broadband spectrum (ref) suggest that Sgr A* is powered by radiatively inefficient accretion flows (RIAFs) \citep[e.g.][]{2014ARA&A..52..529Y} that are optically thin and therefore illuminate the black hole shadow at mm/submm wavelengths. Thus, it is a suitable laboratory with which to study fundamental physics in a strong gravity field. Hence, force imaging and characterization of the black hole shadow cast by the event horizon, as a fundamental prediction of general relativity (GR), permits experimental testing of the existence of black holes and probing of a strong gravity field ~\citep{2000ApJ...528L..13F}. The detection of horizon-scale structures in Sgr A* \citep{2008Natur.455...78D} showed that direct imaging of emission structures on the event-horizon scale is achievable.

To identify the source structure at mm/submm wavelengths, numerous theories have predicted different images depending on various physical conditions and assumptions such as dynamics, steadiness of solutions, microphysics of plasma, inclination angles, and black hole spin. Current theoretical models commonly predict the existence of a clear signature for the black hole shadow illuminated by an accretion flow at short-millimeter or submillimeter wavelengths, regardless of the use of non-steady GRMHD  \citep[e.g.][]{2009ApJ...703L.142D,2012ApJ...752L...1M,2015ApJ...798...15P,2015ApJ...799....1C} or steady semi-analytic models \citep[e.g.][]{2000ApJ...528L..13F, 2006JPhCS..54..448B,2011ApJ...729...86T,2016ApJ...831....4P} or the presence of radio-emitting jets \citep{2014A&A...570A...7M}.

\begin{figure*}
	\centering{}
	\includegraphics[width=1.0\textwidth]{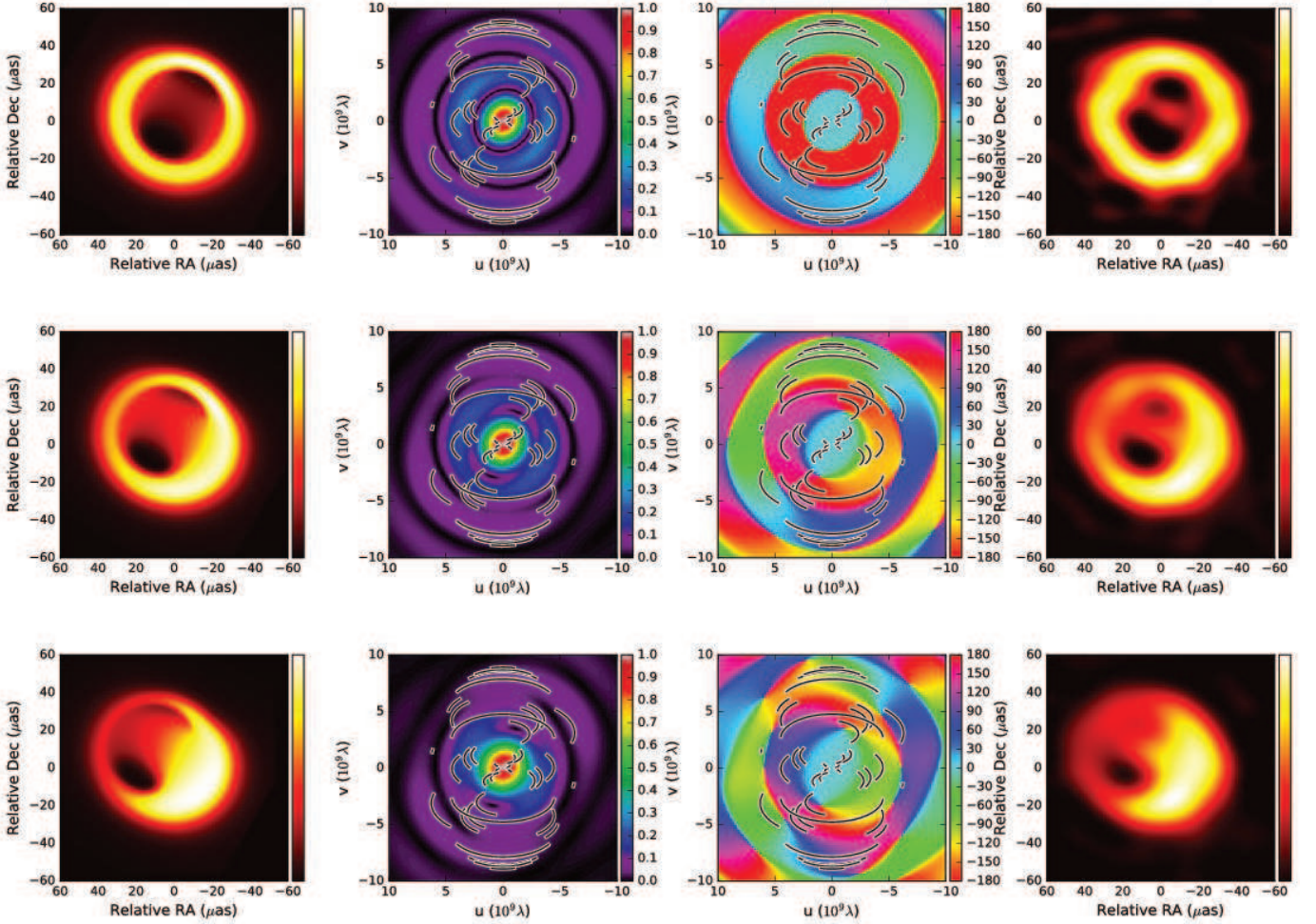}
	\caption{
Left-most panels: model image; second and third panels from left: visibility ampligtude and phase distributions; fourth panels from left: scattered images; and right-most panels: reconstructed images, of semianalytic RIAF models \citep{2016ApJ...831....4P} for Sgr A* with various free-fall dynamics. Top panels: no angular momentum at infinity; middle panels: sub-Keplerian rotation; and bottom panels: Keplerian rotation. According to the suggestion of \citet{2011ApJ...735..110B}, the black hole spin parameter $a = 0$ and inclination angle $i = 68^{\circ}$ are adopted. The distribution of flow materials is prescribed similarly to  \citep[][]{2006JPhCS..54..448B}, in which the electron population decreases rapidly when approaching the funnel region. The black lines show the $uv$-coverage of observations at 1.3\,mm using existing US stations, the LMT in Mexico, the IRAM 30 m telescope in Spain, the NOEMA in France, the SPT at the South Pole, and ALMA/APEX in Chile. The images were reconstructed by the latest version of our imaging softwares utilizing sparse modeling from visibility amplitudes and closure phases \citep{Akiyama2016a} with the mitigating process described in \citet{2014ApJ...795..134F} (Kuramochi et al. in prep.).
\label{fig:akiyama_sgra_1}}
\end{figure*}

At mm/submm wavelengths, the accretion environment around the black hole becomes optically thin, and a considerable energy shift caused by flow dynamics can be distinguished. The Doppler effect caused by the accretion flow rotation (and also the spacetime rotation caused by a rotating black hole) can cause an intensity contraction on the left and right sides of the black hole spin axis. Figure~\ref{fig:akiyama_sgra_1} depicts sample images of a semi-analytic RIAF (sub-Keplerian rotating) model, as well as those of a Keplerian rotating model and free-fall model . The signature of the black hole shadow appears in the Fourier domain of the image (i.e., the visibilities) as null regions of the visibility amplitude, where the sign of the visibility phase rapidly changes (Figure \ref{fig:akiyama_sgra_1}). Inside the first null regions, the visibility amplitude distribution is Gaussian and broadly consistent with observational data regardless of the model, whereas significant differences between models appear over the regions at the baseline length of $\gtrsim 3-4$ G$\lambda$, which can be investigated with future global mm/submm arrays.

\begin{figure}
	\begin{center}
		\includegraphics[width=0.25\textwidth,angle=-90]{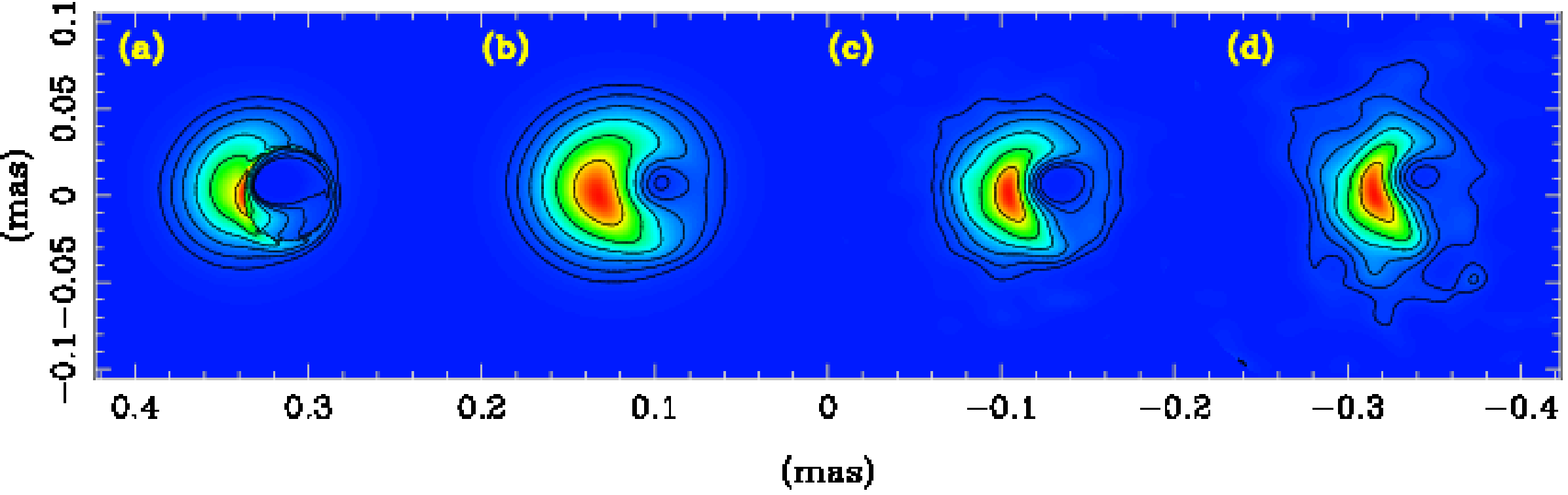}
		\caption{
		Horizon-scale structure reconstruction of Sgr A*. Time average of a variable source structure (a) is convolved with a scattering kernel (b). Reconstruction of synthetic visibilities after dividing by the Fourier transform of the scattering kernel (c) is very close to the original unscattered static image. Image reconstructed from the scattered variable structure (d) using corrected visibilities enables recovering the black hole shadow and photon ring \citep[see][for details]{2015arXiv151208543L}.}
		\label{fig:figure6}
	\end{center}
\end{figure}

The detectability of the black hole shadow also depends on the array performance. Imaging the black hole shadow in Sgr A* is challenged by two effects: (1) Scattering caused by interstellar electrons blurs the strong GR features near the black hole \citep[see][for a review]{2015ApJ...805..180J,2016ApJ...826..170J}; and (2) the emission structure surrounding the black hole is intrinsically variable on a time scale of minutes, whereas a typical VLBI imaging experiment takes several hours. Examination based on a ``full'' EHT array configuration supports the possibility of directly imaging the black hole shadow in Sgr A* despite the scattering effects and time variability. \citet{2014ApJ...795..134F} showed that angular broadening, the predominant effect of scattering \citep[e.g.][]{2015ApJ...805..180J}, can be mitigated by subsequent processing of the observed visibilities. Figure~\ref{fig:akiyama_sgra_1} also presents reconstructed images of models obtained by combining our sparse imaging techniques \citep[see \S\ref{sec:2.1.4};][]{Akiyama2016a} with the mitigating procedure proposed in \citet{2014ApJ...795..134F} (Kuramochi et al. in prep.).  Furthermore, we have shown that the effects of time variation can be significantly mitigated by sampling sufficient visibilities to obtain a time-averaged image from multi-epoch observations \citep[Figure~\ref{fig:figure6},][]{2015arXiv151208543L}.

\subsubsection{M87}\label{sec:2.1.3}

M87, second in apparent size to Sgr A*, is another ideal target for imaging the shadow of a supermassive black hole. Its proximity \citep[$D=16.7\pm 0.6$~Mpc][]{2009ApJ...694..556B} and the remarkably massive black hole \citep[$M_{\rm BH} \sim 3-6\times10^{9}$~$M_\odot$;][]{1997ApJ...489..579M,2011ApJ...729..119G,2013ApJ...770...86W} subtend $1R_s\sim 3-7$~$\mu$as, thus enabling the resolution of the event-horizon scale structure through mm/submm VLBI, in a similar manner to Sgr A*. Recent observations have revealed that the emission at the jet base is optically thin at $\lambda \lesssim 1.3$\,mm, as required for imaging black hole shadows (see \S\ref{sec:2.1.1}); this finding was based on measurements of the frequency-dependent position of the radio core \citep[][]{2011Natur.477..185H} and the discovery of a break in its radio spectrum at $\sim$1.3\,mm, indicating the opacity transition \citep{2013EPJWC..6108008D}. Previous 1.3-mm VLBI observations have resolved a compact structure with a size comparable to the innermost stable circular orbit of the central black hole \citep{2012Sci...338..355D,2015ApJ...807..150A}, as predicted by the jet collimation profile \citep{2012ApJ...745L..28A,2012Sci...338..355D,2013ApJ...775..118N,2013ApJ...775...70H,2016ApJ...817..131H}. Results have particularly demonstrated that M87 is another excellent target for accessing black hole shadows at mm/submm wavelengths. Notably, M87 does not suffer from two of the issues relevant to imaging Sgr A*. First, the scattering effects are negligable for M87, because it is located outside the Galactic plane ($b\sim+74^\circ$). Second, the large black hole mass of M87 causes variability timescales substantially longer than those of Sgr A*, providing a stable structure during a single interferometric observation. Therefore, future mm/submm VLBI observations can take sharp snapshots of the intrinsic source structure of M87 on event-horizon scales.

\begin{figure}
	\centering{}
	\includegraphics[width=0.8\textwidth]{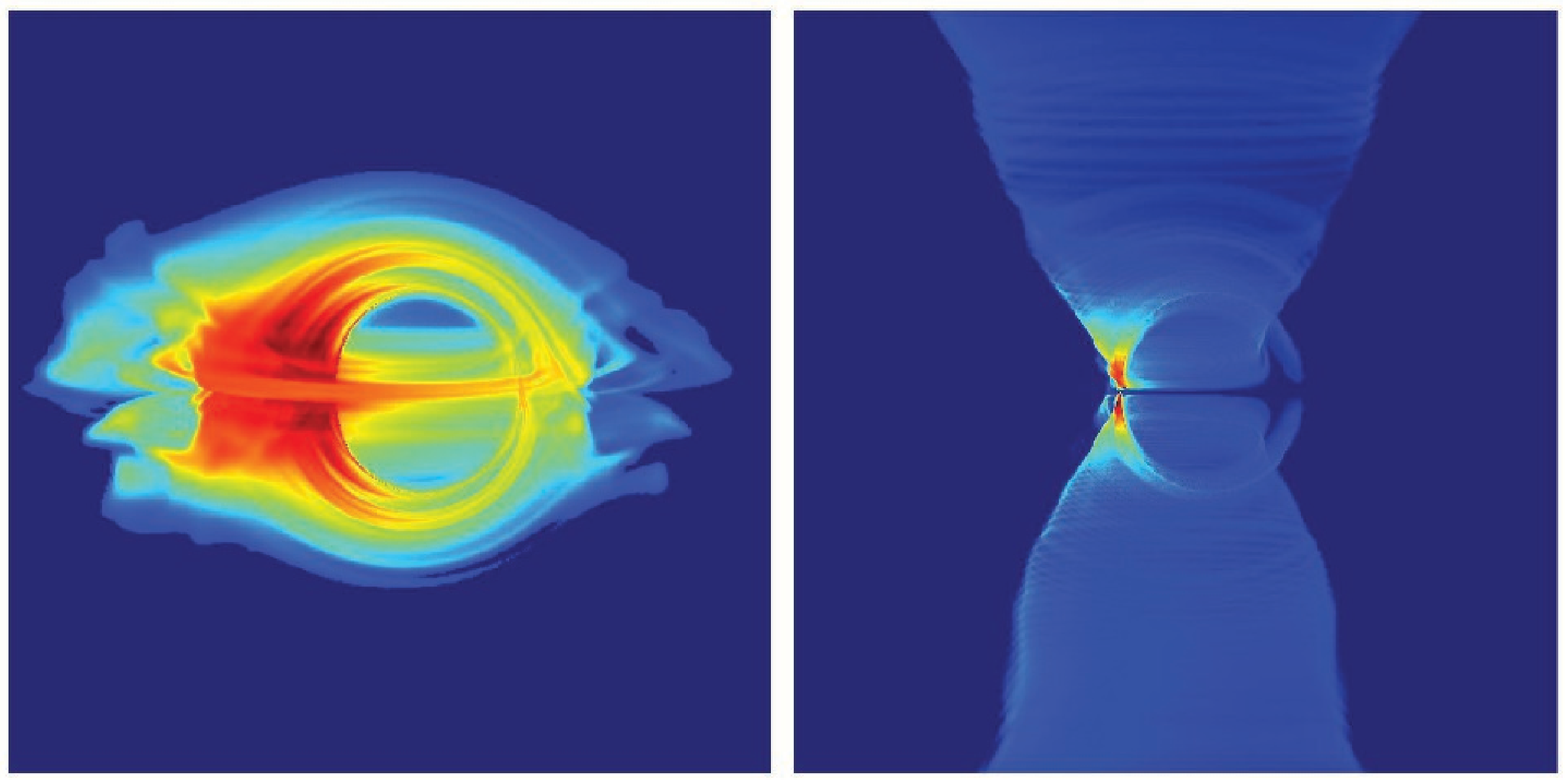}
	\caption{
	Example thermal synchrotron emission images obtained by post-processing 2D GRMHD simulation data of HARM code \citep[][]{2003ApJ...589..444G,2006ApJ...641..626N}, assuming  $M_{\rm BH}=4.3\times 10^{6} M_{\odot}$ and $\dot{M}\sim 10^{-9}\dot{M}_{\rm Edd}$. The ratio between ion temperature $T_{\rm i}$ and electron temperature $T_{\rm e}$ is assumed to be constant ($T_{\rm i}/T_{\rm e}=3$; left) and follows a function of $\beta=P_{\rm gas}/P_{\rm mag}$ (following the relation described in \citet[][]{2016A&A...586A..38M}; right). The observation frequency is assumed to be 230\,GHz, and the inclination angle is 90$^{\circ}$. The black hole spin $a/M=0.9$. Fast-light approximation is adopted for the data with $t=3000 GM/c^{3}$.
\label{fig:hypu_m87_1}}
\end{figure}

In the context of shadow imaging, the presence of a bright jet is one of the major differences with Sgr A*. The mm/submm emission could be dominated by synchrotron emission from either the jet \citep[e.g.][]{2008ApJ...679..990Z,2009ApJ...695..503G,2009ApJ...697.1164B,2012MNRAS.421.1517D,2016A&A...586A..38M} or the accretion disk \citep[e.g.][]{1996MNRAS.283L.111R,2003ApJ...582..133D,2010ApJ...711..222N,2011ApJ...729...86T,2012MNRAS.421.1517D} in the RIAF regime \citep[e.g.][]{2014ARA&A..52..529Y} with a low mass accretion rate of $<9.2\times10^{-4}$~M$_\odot$~y$^{-1}$ \citep{2014ApJ...783L..33K}. A black hole shadow is formed by the last photon orbit illuminated by a counter jet or accretion disk in the vicinity of the black hole. Thus, the features of the black hole shadow become obvious when the accretion disk is predominant or when the bright emission region in the jet is sufficiently close to the central black hole that the counter jet can illuminate the photon ring \citep[see discussion in][]{2015ApJ...807..150A}.  Modeling jet emissions on the horizon scale requires detailed knowledge of the energy and spatial populations of electrons as well as of the jet launching sites and mechanisms. In GRMHD schemes, the absence of information regarding the non-thermal electron distribution and the ratio of the ion temperature to the thermal electron temperature introduces major difficulties for predicting jet images. For example, Figure~\ref{fig:hypu_m87_1} presents two thermal synchrotron emission images on the horizon scale, computed by postprocessing the same GRMHD simulation data, but with different assumptions regarding the ratio of the ion temperature to the electron temperature. The investigation and prediction of the characteristic features of different jet models is important for future VLBI observations of M87.

Future mm/submm VLBI observations can effectively constrain the existence of the black hole shadow in M87. Regardless of the presence of a clear signature for the black hole shadow, these physically motivated models are broadly consistent with each other and with the current observational data at short American baselines (baseline length within 3--4\,G$\lambda$). However, they predict dramatically different measurements at intercontinental baselines. In particular, models in which the image is dominated by contributions close to the horizon (right three images; counter-jet-dominated and accretion-disk-dominated models) exhibit nulls in the visibility amplitude where the visibility phase is flipped, in contrast to those dominated by more distant emissions \citep[e.g. extreme cases of approaching-jet-dominated models; see][]{2014ApJ...788..120L}. These signatures of the black hole shadow will be effectively probed by longer baselines including ALMA \citep[e.g., ][]{2015ApJ...807..150A}.

\begin{figure}
	\begin{center}
		\includegraphics[width=0.3\textwidth,angle=-90]{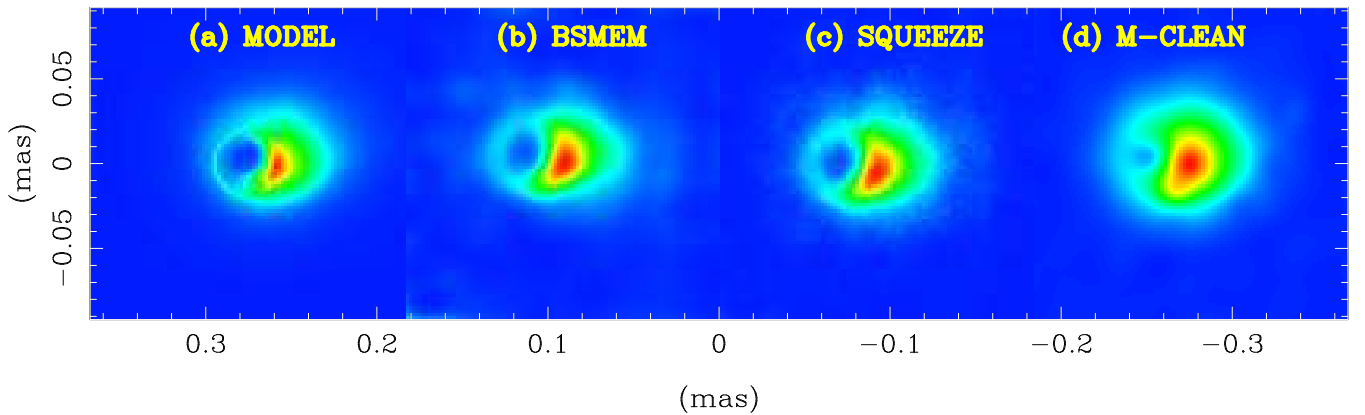}
		\caption{
		M87 model image at 230\,GHz and corresponding reconstructions with widely used imaging algorithms assuming a global 1.3\,mm VLBI array \citep[see][for details]{2014ApJ...788..120L}.}
		\label{fig:m87}
	\end{center}
\end{figure}

\begin{figure*}
	\centering{}
	\includegraphics[width=0.8\textwidth]{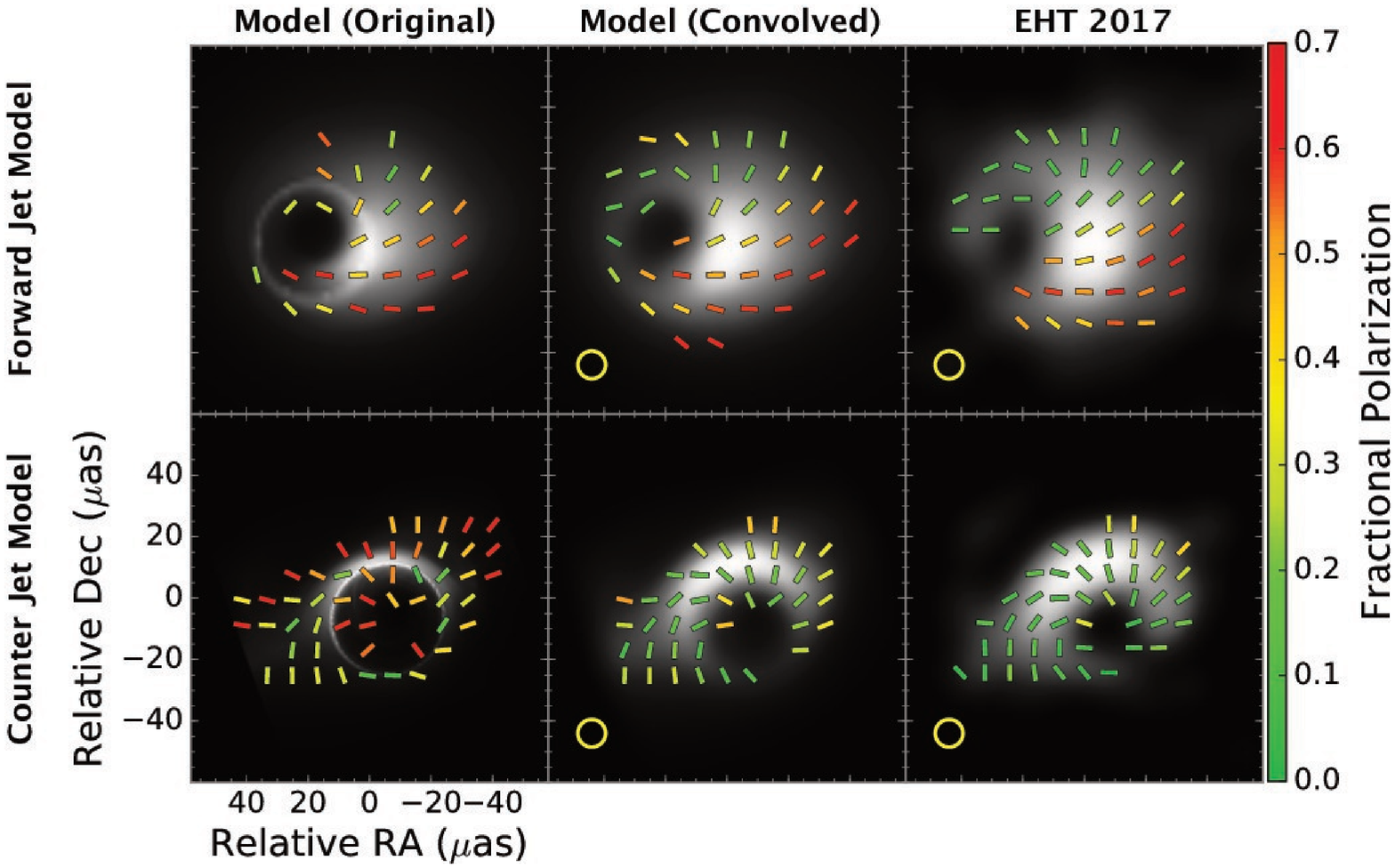}
	\caption{
		Stokes $I$ images (gray-scale contour) and linear polarization maps (colored bars) of M87 at 1.3~mm. Left panels show the original model images, while middle and right panels show the model images and reconstructed images, respectively, convolved to an optimal resolution (yellow circles) achievable with the latest version of our imaging software \citep{Akiyama2016a,Akiyama2016b}. The magnetic field has a coherent, helical structure in both cases. In the forward (approaching) jet model \citep[top;][]{2009ApJ...697.1164B}, this leads to coherent polarization on small scales (long baselines). In the counter-jet model \citep[bottom;][]{2012MNRAS.421.1517D}, the emission is de-polarized at typical spatial resolutions of EHT by general relativistic parallel transport due to strong light bending and also Faraday Rotation.
		\label{fig:akiyama_m87_1}}
\end{figure*}

Our simulations showed that the arrays of future mm/submm VLBI can sample sufficient data in the Fourier domain with sufficient sensitivity to image the black hole shadow of M87.  \citet{2014ApJ...788..120L} examined the detectability of the black shadow of M87 with an EHT by performing realistic VLBI simulations (Figure \ref{fig:m87}) for the jet models of \citet{2009ApJ...697.1164B} by applying the well-established Bi-spectrum Maximum Entropy Method \citep{1994IAUS..158...91B} implemented in the BSMEM and SQUEEZE \citep[e.g., see][]{2010SPIE.7734E..2IB,2012SPIE.8445E..1EB} packages, as well as Multi-resolution CLEAN \citep[M-CLEAN; e.g.][]{2008ISTSP...2..793C,2009AJ....137.4718G} implemented in the CASA package.
We also examined the imaging capability of the physical models by using sparse modeling techniques (see \S\ref{sec:2.1.4}). Figure~\ref{fig:akiyama_m87_1} shows samples of the images reconstructed from the simulated visibility amplitude and closure phase, and also linear polarization data by using sparse imaging \citep{Akiyama2016a,Akiyama2016b}. These imaging techniques can reconstruct the source images regardless of the presence of the black hole shadow or the location of the brightest emission region at full polarizations. This also demonstrates that future mm/submm VLBI observations can trace real-time structural changes of the magnetized plasma on some Schwarzschild radius scales at full polarizations, such as those implicated by the VHE flaring activity of M87.

\citet{2014ApJ...788..120L}  identified the minimum necessary requirements for clearly detecting black hole shadows and studying event-horizon--scale jet launching. For example, the emission from the counter jet must be sufficiently bright for the black hole to cast a jet against, and the array must have sufficient resolution and sensitivity. Although the inclusion of the phased ALMA in the array is critical for black hole shadow imaging, in general, the array is robust against the loss of a station. With the full array, future observations can provide the best evidence for the presence of a black hole and assist with the understanding of the jet-launching processes near the black hole in M87.

\subsubsection{Imaging with Sparse Modeling}\label{sec:2.1.4}
Since the size of a BH shadow is extremely small, high resolution is essential for BH imaging with mm/submm VLBI. Basically, the resolution of any interferometric array is given by $\theta\sim \lambda/D$, where $\theta$ is the angular resolution, $\lambda$ is the observing wavelength, and $D$ is the diameter of the array (i.e., the maximum baseline length). 
However, improvement of the angular resolution by a factor of a few is possible by introducing a novel interferometric imaging technique, and recently, imaging with sparse modeling was proposed as an approach to obtain super-resolution images \citep[see \S 3.3.1 and ][for a review]{2016JPhCS.699a2006H}.

\citet{2014PASJ...66...95H} demonstrated that a super-resolution image can be obtained by directly solving an underdetermined Fourier equation for interferometric imaging by using a class of sparse modeling techniques called least absolute shrinkage and selection operator (LASSO). They showed that sparse modeling can trace the BH shadow of M87 even in the smaller mass case of $M=3\times 10^9 M_\odot$, whereas conventional imaging with CLEAN cannot. 

Since then, further researches have been conducted to make sparse modeling applicable to more realistic data. For instance, in the early phase of mm/submm VLBI observations, direct observables are likely to be bi-spectra and/or visibilty amplitudes + closure phases \citep[e.g.][]{2015ApJ...807..150A}.
Recently, \citet{2016PASJ...68...45I} found that imaging with closure phases is possible based on a newly-proposed technique called phase retrieval from closure phase (PRECL). 
Another approach of imaging using visibility amplitudes and closure phases with sparse modeling is proposed by \citet[][]{Akiyama2016a}, in which the regularization term is extended to $l_1$-term (LASSO) plus total variation (TV). In \citet{Akiyama2016a}, we also propose a practical technique to determine imaging parameters from data themselves with Cross Validation (CV) techniques, enabling to obtain images without special fine tuning. Also developed are tools for better imaging with self-calibration, scattering mitigation (particularly for Sgr A*), and full-stokes imaging for polarimetry  \citep[e.g.][]{Akiyama2016b}.

To highlight the results of the recent developments, in Figure \ref{fig:akiyama_sgra_1} and \ref{fig:akiyama_m87_1}, we present simulated images for the EHT observations of Sgr A* and M87 in 2017.
Figures \ref{fig:akiyama_sgra_1} are the image reconstruction results for Sgr A*, where angular broadening from the diffractive scattering effect is also taken into account. The initial models are from \citet{2016ApJ...831....4P} with different inclination angles, and regardless of the model geometry, sparse modeling can reproduce the detailed structure of Sgr A*. Figures \ref{fig:akiyama_m87_1} show the image reconstruction for model images by \citet{2009ApJ...697.1164B} or \citet{2012MNRAS.421.1517D} at full polarization. As seen in the figures, the full-stokes imaging is now available using sparse modeling, and will be expected to provide detailed structures of magnetic field around the super-massive black holes \citep{Akiyama2016b}. All results shown in our works demonstrate that sparse modeling is an attractive choice for the next generation mm-/sub-mm VLBI observations in addition to other techniques developed recently (e.g. CHIRP: \citealt{2015arXiv151201413B}; Bi-spectral/Polarization MEM: \citealt{2016ApJ...829...11C}; and also see \citealt{2016Galax...4...54F} for an overview of state-of-art imaging techniques).





\subsection{Revealing Nature of Accretion Flow}
\subsubsection{General Objectives}

Mass accretion onto supermassive black holes (SMBHs) is one of the most fundamental processes characterizing AGNs, because it supports the nuclear luminosity of AGNs and presumably powers their outflow. Radiatively inefficient accretion flow (RIAF) is a subclass of accretion flows that are generally quasi-spherical accretion flows in which the viscous timescale is considerably longer than the free-fall time, rather than an optically thick and geometrically thin accretion disk (the well-known standard Shakura--Sunyaev disk); the longer viscous timescale leads to the decoupling of the electron and ion temperatures \citep[e.g., ][]{2014ARA&A..52..529Y}. In particular, the electron temperature can be considerably lower than the ion temperature, leading to a considerable amount of the potential energy of accretion being transported across the event horizon rather than being radiated. This is referred to as advection-dominated accretion flow (ADAF) \citep[e.g.,][]{2011MNRAS.415.3721N}. Both convection and magnetic fields may play a role in suppressing the accretion flow \citep[convection-dominated accretion flow (CDAF); e.g.,][]{2002ApJ...566..137I}, and a considerable outflow is expected for an adiabatic inflow-outflow solution \citep[ADIOS; e.g.,][]{1999MNRAS.303L...1B}. 

Approximately one-third of nearby galaxies show some level of weak activity in their nucleus, with these activities being quite distinct from typical stellar processes~\citep{1997ApJ...487..568H}.  These AGNs are the so-called low-luminosity AGNs (LLAGNs), and they form a subclass of AGNs. They are characterized by sub-Eddington luminosity and they account for the majority of AGNs in the local universe~\citep{2008ARA&A..46..475H}. Their low luminosity is thought to be caused by an optically thin and geometrically thick {\it hot}  accretion flow with a very low accretion rate and very low radiative eﬃciency. Observational understanding of accretion flows, especially in the context of the spatial structure, is very limited, and in this section, we present scientific objectives related to accretion flows.\\

\textbf{\textsl{Test for the RIAF paradigm}}\\

The accretion flow theory predicts that a RIAF would have considerably high brightness temperature in
its innermost region and that bright emission is expected at mm/submm wavelengths; on the other hand, the emission is expected to be very dim at mm/submm wavelength for a standard disk because of its low black-body temperature, which is approximately 10$^{5-6}$\,K (e.g., Kato, Fukue \& Mineshige 1998 for review).  Therefore, the primary objective of studies on accretion flows is to resolve the innermost region of accretion flows and image its structure. Detection of the expected very hot plasma would facilitate the development of a test for directly detecting the presence of RIAF, and will pioneer the new parameter space: observational constraints on spatial structure of the accretion flow. \\

\textbf{\textsl{Geometrical structure}}\\

In the context of the aforementioned objective, it is not known if RIAFs tend to have disk-like structures on small scales ($R \sim 10 R_S$). In general, a RIAF is characterized by quasi-spherical accretion flows onto the central SMBH, and a thick disk-like structure (H/R\,$\sim$\,1)  is expected in the innermost region of the accretion flows on the basis of numerical simulations (e.g., Sadowski et al., 2016). Direct imaging of the accretion flow will reveal its structure.\\

\textbf{\textsl{Magnetic fields}}\\

Transfer of angular momentum is essential for mass accretion onto central objects. The angular momentum is presumably transferred outward by the flow through magneto-rotational instability, and strong and turbulent magnetic fields are generated in the accretion flow, with the azimuthal component of the magnetic fields being dominant in the equatorial plane of the flow (e.g., Sadowski et al., 2015). VLBI polarimetric observations at mm/submm wavelengths can reveal the magnetic field topology in the accretion flow and enable us to probe it in comparison with the various models.\\

\textbf{\textsl{Jet--accretion flow connection}}\\

The magnetized jet from a rotating black hole (BH) has been widely discussed by many studies involving state-of-the-art general relativistic magnetohydrodynamic simulations. As one of the scenario, the ``magnetically arrested'' accretion disc has been suggested to have a strong outflow (Narayan et al. 2003). In this process, together with the spin of the rotating BH, the rate of accretion of the strong coherent magnetic field onto the spinning BH, which is related to the mass accretion rate, is supposed to be a crucial parameter for determining the jet power (e.g., Tchekovskoy et al. 2011). Therefore, future mm/submm VLBI observations are expected to facilitate, for the first time, the imaging of the time series of the structural evolution for the purpose of investigating the mass accretion process together with jet ejection.

\subsubsection{Sgr A* as the Nearest Accretion Flow}

Sgr A* is undoubtedly the best laboratory for understanding BH accretion flows (and possibly outflows) because it harbors the nearest and best constrained SMBH candidate. Despite rigorous studies having been conducted on the accretion flow of Sgr A*, many problems remain unsolved. \\

{\bf {\em Sgr A* on 10 Rs scale}}\\

Observations of Sgr A* have indicated that its bolometric luminosity ($L_{\rm bol}$) is approximately $10^{36}$\,erg\,s$^{-1}$, which is considerably lower than the Eddington luminosity $L_{\rm Edd}~\,(5{\times}10^{44}~\,{\rm erg}~\,{\rm s}^{-1}$) for the SMBH at the Galactic center. Because of such a low Eddington ratio, the plasma accreting onto Sgr A* should be a hot accretion flow; in other words, the plasma is an advection-dominated and radiatively inefficient accretion flow \citep{1977ApJ...214..840I, 1994ApJ...428L..13N, 1995ApJ...452..710N, 2014ARA&A..52..529Y}.

Several lines of evidence support the hypothesis of a RIAF being present in Sgr A*. First, the accretion rate $\dot{M}_{\rm Bondi}$ at the Bondi radius ($R_{\rm Bondi}$)  of approximately $10^5 R_S$ is determined by the stellar mass loss and is observed to be approximately  $10^{-4}$ M$_\odot$~y$^{-1}$ \citep{1999ApJ...517L.101Q}. If the accretion rate near the BH were to be comparable to $\dot{M}_{\rm Bondi}$, the luminosity would be orders of magnitude higher.

Second, linear polarization measurements definitively show that much of the accreting material does not actually reach the BH. The polarized source that is produced in the inner  $10 R_S$  at mm wavelengths can be viewed through the dense, magnetized plasma of the accretion flow. The position angle of the linear polarization undergoes Faraday rotation with a rotation measure RM of $-5 \times 10^5$\,rad\,m$^{-2}$ \citep[e.g.,][]{2003ApJ...588..331B}.  
Under the assumption of a power-law distribution of electron density with radius, and energy equipartition between the magnetic and thermal energy densities, $\dot{M}$ can be constrained to approximately $10^{-8 \pm 1}$ M$_\odot$~y$^{-1}$,
with the majority of RM originating between $10^2$ and $10^4 R_S$. Both the drop in $\dot{M}$ and a two-temperature solution for electrons and ions lead to the low luminosity of Sgr A*.

Third, high-resolution imaging provides constraints on the nature of accretion flow. ADAF models require a steep variation of electron density and temperature with radius, and such a variation should leads to a bright, compact source. A more gradual RIAF profile is consistent with the observed brightness temperature.

Linear polarization that originates at approximately 10\,$R_S$ supports the transition of a spherical accretion flow
into a thin disk. Such flows require an ordered magnetic field. In particular, submillimeter VLBI imaging shows that the linear
polarization originates from a few Rs and that the accretion disk may have a magnetized structure on even smaller linear scales \citep{2015Sci...350.1242J}.

A major limitation of accretion models for Sgr A* is that they do not reproduce the observed radio spectrum. Accretion models produce only thermal particles and under-predict the spectrum. Nonthermal particles are primarily introduced through  {\em ad hoc} mechanisms, such as a fraction of the total particle energy being assigned to nonthermal particles with a power-law distribution. \\

The frontiers of mm/submm VLBI probes of Sgr A* are in several domains.

\begin{itemize}
\item{
Constraints on imaging of the accretion flow: Current observations show that the submm emission region can be modeled by a Gaussian with a characteristic size of approximately $4 R_S$ \citep{2008Natur.455...78D} and a small degree of asymmetry \citep{2016arXiv160205527F}. However, all theoretical models indicate that more complex structures are likely to be present \citep[e.g.,][]{2014A&A...570A...7M}. Increasing the density of ($u,v$)-plane coverage through the introduction of more submm stations is critical for improving the sensitivity and fidelity of VLBI images. High-quality images (or models constructed directly from visibilities) can help verify the presence of large-scale structures and clarify whether the expected crescent shape is present; moreover, the images can also help determine fundamental geometric parameters such as the inclination and orientation angle of the disc.
}

\item{
Polarimetric imaging of magnetic field structures: The first polarimetric submm VLBI images of Sgr A* indicate a magnetic field geometry that is more complex than the total intensity distribution \citep{2015Sci...350.1242J}. The magnetic field structure can provide vital clues to the nature of accretion flows and the origin of synchrotron emission. Similar to total intensity imaging, higher  
fidelity and sensitivity can be achieved through the addition of new stations. The combination of long-term polarization stability, stable persistent asymmetry, and tangled magnetic fields is a challenging theoretical puzzle.
}

\item{
Short-timescale measurements of total intensity and polarized intensity for material orbiting the BH: The period of the last stable orbit for the non-spinning BH case is approximately 30\,min. Time-domain studies have shown that Sgr A* has a characteristic variability timescale of $8^{+3}_{-2}$\,h, corresponding to the Keplerian period for material at a radius of approximately 10\,$R_S$ \citep{2014MNRAS.442.2797D}. Furthermore, polarized emission varies substantially on a timescale of hours, and sometimes, it is in the form of circles in the $Q-U$ plane \citep{2006JPhCS..54..354M}. Both these timescales are smaller than or comparable to a VLBI imaging observation of Sgr A*. Thus, either the generation of short-timescale images or the use of non-imaging techniques is necessary to search for effects in the region of accretion flow. Periodic imaging effects are especially a strong indicator of accretion flows. Such observations will strongly depend on the sensitivity of the phased ALMA, which will be used in conjunction with other VLBI stations.
}

\item{
Astrometry of the total intensity signal to the celestial reference frame or a background source is unlikely to be possible, given the very short coherence times of the array. However, relative astrometry of the total and polarized signals is possible. These signals have been shown to have different structures, and therefore, they may not vary together. In particular, on short timescales, polarized structures tracing the accretion flow may be formed relative to the average total intensity background.
}

\end{itemize}

Handling the effects of time-variable scattering may pose a major challenge. Long-wavelength observations have characterized the ensemble average scattering properties very well \citep{2006ApJ...648L.127B}.  However, the minimum variability timescale of the scattering properties at a wavelength of 1.3\,mm could be as short as hours, which is comparable to the timescale of an observation or of intrinsic variability. Observations that can characterize the scattering medium at long and short wavelengths more thoroughly are necessary to completely characterize the scattering medium.

A detailed physical understanding of accretion flow physics is necessary to obtain convincing evidence of general relativistic effects. \\

{\bf {\em Sgr A* on 100 Rs scale}}\\

For Sgr A*, 1\,mas corresponds to 100 Rs, and therefore, the plasma properties accreting onto the Galactic center (GC) can be investigated in great detail by mm/submm VLBI. \\

\textsl{Kawashima prediction: Imprint of the G2-encounter}

\begin{figure}[h]
 \centering
 \includegraphics[width=17cm, angle=0]{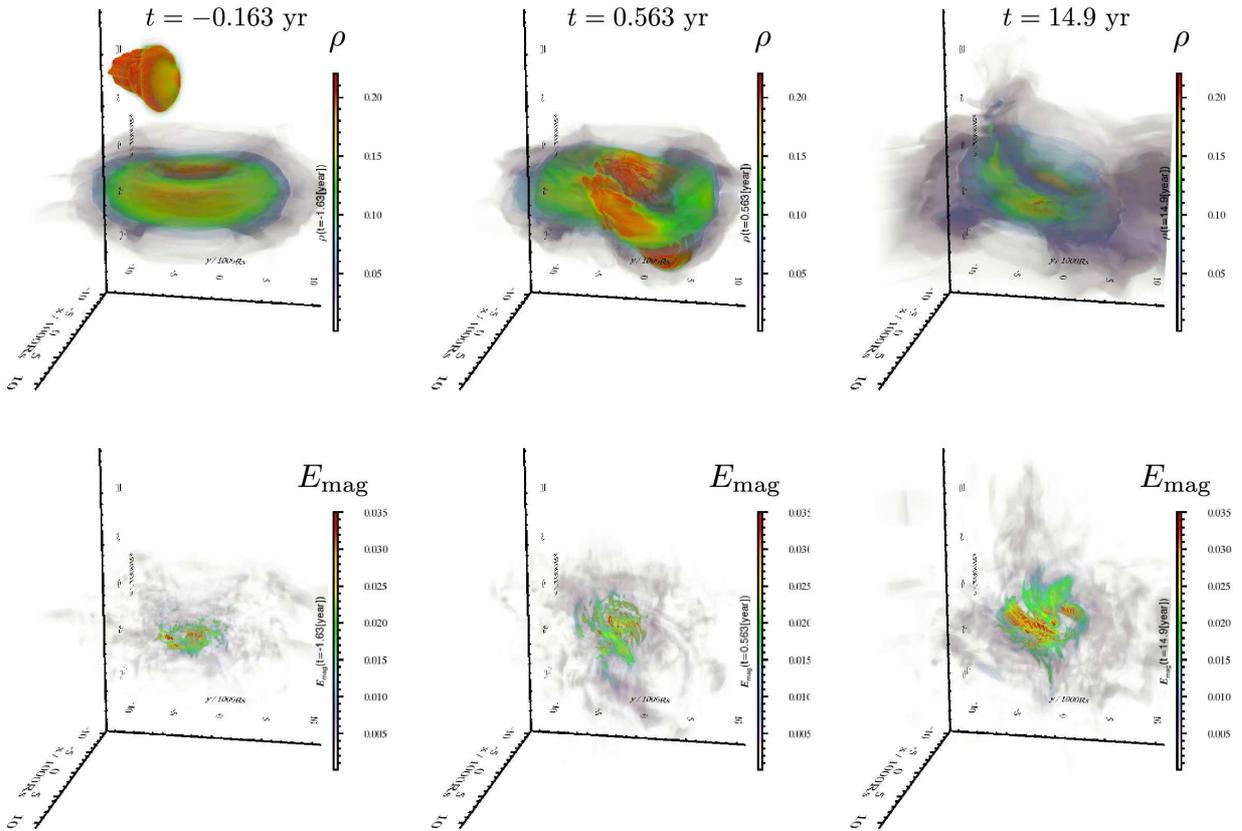}
 \caption{
Time evolution of three-dimensional MHD simulations of the accretion flow of Sgr A* interacting with the G2 cloud \citep{2017arXiv170207903K}. The G2 cloud passed the pericenter at time t = 0 yr.
 \label{fig:222-1}
 }
 \end{figure}


G2, which is a composite of gas and dust passing through the Sgr A* region \citep{2012Natur.481...51G}, may strongly affect the dynamics of the accretion flow of Sgr A*. The location of the pericenter of G2 is close to the SMBH; the distance of the pericenter from the GC is only approximately $2{\times}10^3$\,Rs. The size of the gas cloud is as large as the distance of the pericenter from the SMBH, and Br-$\gamma$ observations indicate the size to be approximately 15\,mas, (i.e., $ {\sim}2{\times}10^3$\,Rs.). The mass of the gas component of G2 is estimated to be approximately  3\,$M_{\oplus}$, and it is comparable to the expected mass of the accretion flow of Sgr A*. Therefore, G2 is expected to affect the dynamics of the accretion flow and trigger a series of flares. Following the discovery of G2, many numerical simulations of G2 have been performed to predict the brightening of Sgr A* in the future \citep{2012ApJ...750...58B,2012ApJ...755..155S,2013MNRAS.433.2165S,2013ApJ...776...13B,2014MNRAS.440.1125A,2014PASJ...66....1S}.
However, no flaring event appearing to be related to the approach of G2 has been observed in Sgr A* so far \citep{2015ApJ...798L...6T,2015ApJ...802...69B}.

\begin{figure}[h]
 \centering
 \includegraphics[width=16cm, angle=0]{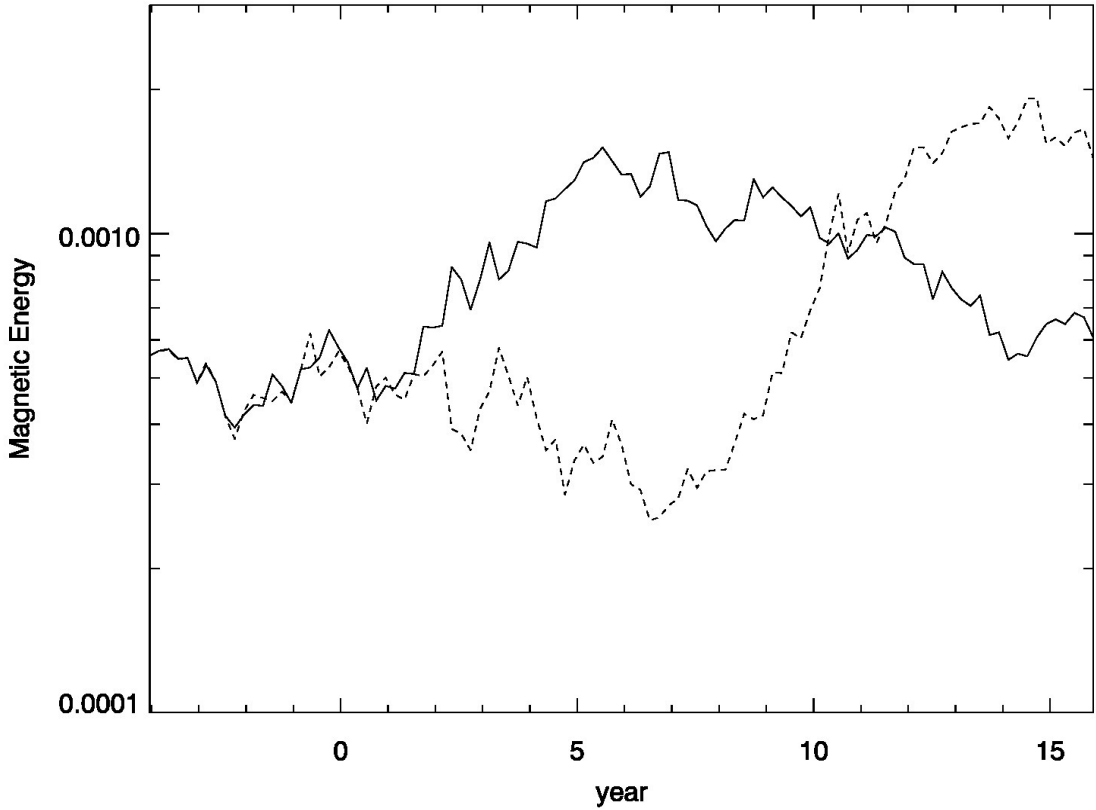}
 \caption{
 Time evolution of the magnetic energy within  $750 R_{\rm s}$ \citep{2017arXiv170207903K}. The solid and dashed curves represent results for models with inclination angles of $i=0^{\circ}$ and $i=60^{\circ}$, respectively.
 \label{fig:222-2}
 }
 \end{figure}

Recently, \citet{2017arXiv170207903K} conducted long-term three-dimensional magnetohydrodynamic (MHD) simulations of the accretion flow of Sgr A* interacting with a gas cloud, by considering the effects of radiative cooling (Figure~\ref{fig:222-1}). Hereafter, it is referred to as the {\it Kawashima prediction}. To predict the radio light curve of Sgr A*, it is necessary to perform long-term MHD simulations because synchrotron emission is the main radiative process in the radio band emission of Sgr A*. A time lag is necessary between the instant of magnetic field amplification at approximately $10^3 R_{\rm s}$ and that of magnetic field increase in the vicinity of the BH. If we assume that the amplified magnetic field is transported to the innermost region of the accretion flow of Sgr A* on the timescale of the $\alpha$-viscosity for $\alpha= 0.1$, the time lag is ${\lesssim} ~2$, which is sufficiently less than the timescale of magnetic field amplification by a G2-induced disk dynamo. According to the Kawashima prediction, the magnetic energy gradually increases by three to four times because of the disk dynamo after 5--10 years of the passage of the G2 cloud through the pericenter (Figure~\ref{fig:222-2}). The increase in the radio luminosity, which follows the synchrotron emissions through the amplified magnetic field, should be detectable. The inclination angle $i$, which is the angle between the orbital plane of G2 and the equatorial plane of the accretion flow, is expected to change because of the dynamic impact of the G2 encounter. The change in the emission region after the collision of G2 with the accretion flow, which is expected for $i=60^{\circ}$ (Figure~\ref{fig:222-1}), can be spatially resolved in mm/submm VLBI observations. Therefore, because we know the orbital plane of G2, it should be possible to constrain the orientation of the rotation axis of the accretion flow of Sgr A* by comparing the simulation results with the timing of the brightening, which is expected to be obtained through radio observations. \\


\textsl{Comprehensive monitoring of Sgr A*: testing the Kawashima prediction} \\

\begin{figure}[htbp]
   \centering
	\includegraphics[width=6cm, angle=0]{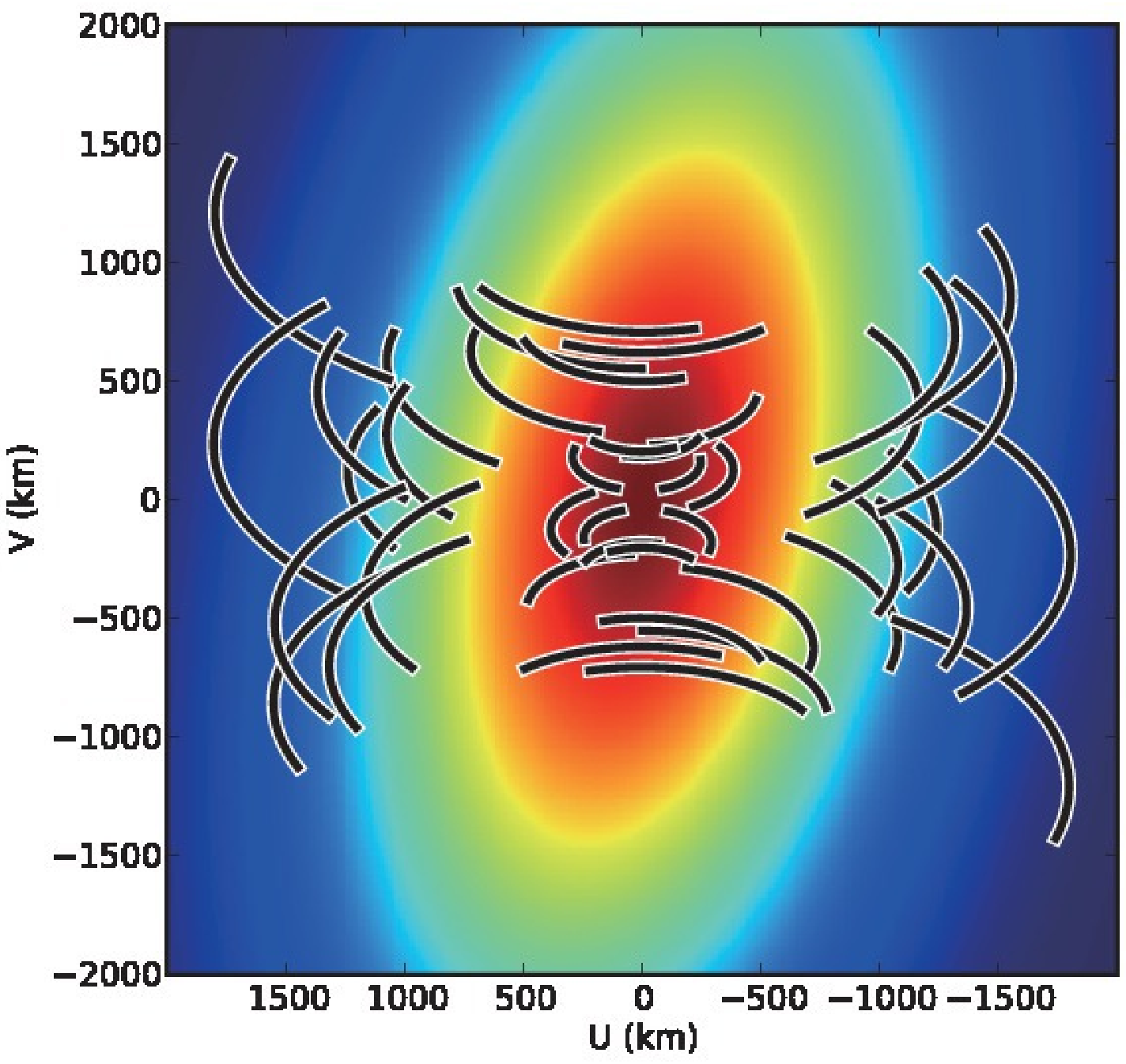}
	\includegraphics[width=6cm, angle=0]{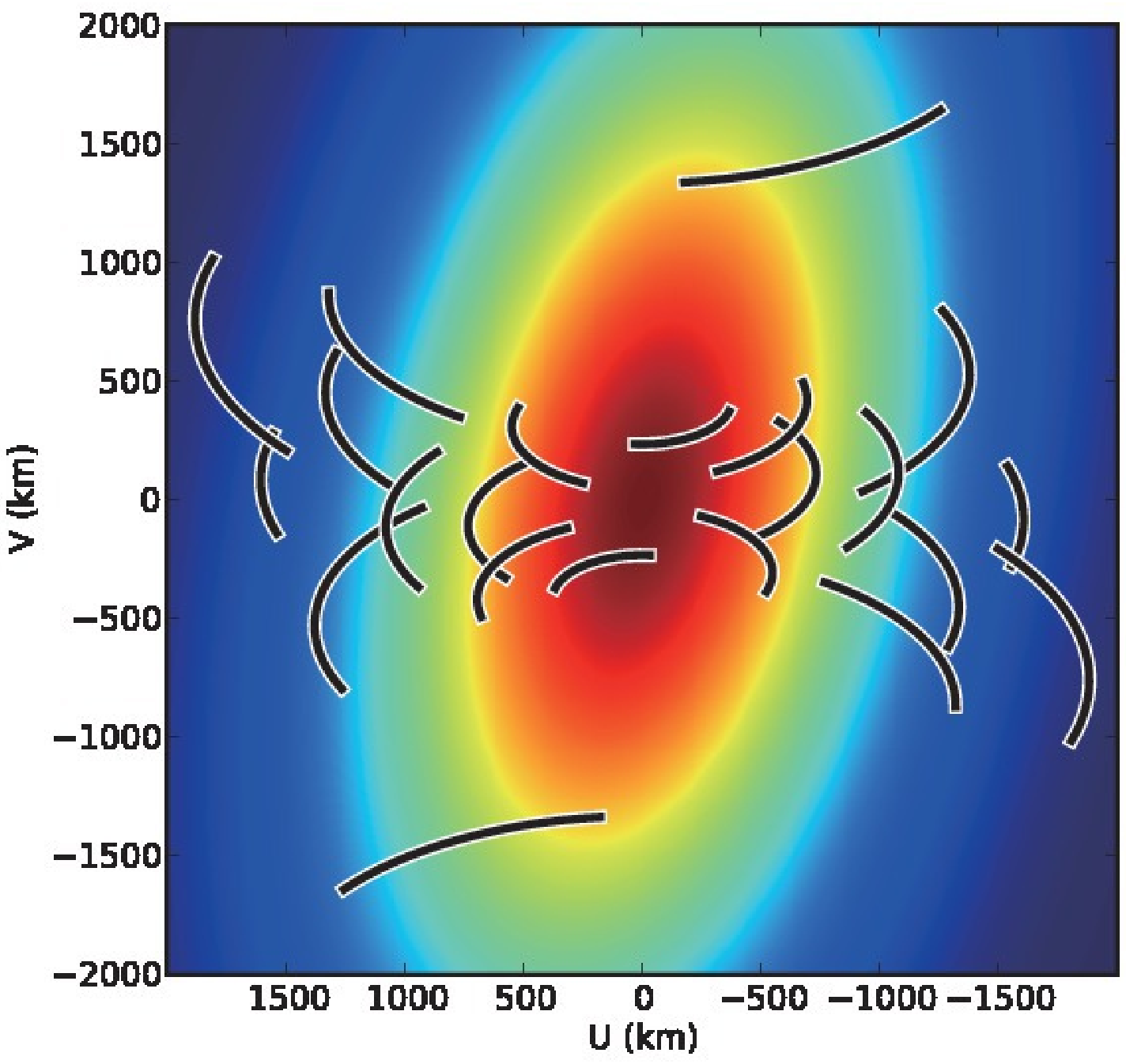}
   \caption{The ($u,v$)-coverage of (left) KaVA and (right) the VLBA for Sgr A*. The curves represent the ($u,v$)-coverage of the array, and the colored contours represent the visibility amplitude distribution obtained from the best-fit elliptical Gaussian model by  \citet{2004Sci...304..704B}. This figure is taken from \citet{2014IAUS..303..288A}
   \label{fig:uvplot}
         }
   \end{figure}

To test the Kawashima prediction, monthly monitoring of Sgr A* at 43\,GHz (7\,mm) using KaVA was started in September 2014. Notably, this is the only regular long-term VLBI monitoring project for Sgr A*. The main advantage of using KaVA is the superior ($u,v$)-coverage. KaVA has more baselines in the short and intermediate ($u,v$)-range ($50-300 M\lambda$) than other arrays. At longer baselines, the source is resolved out due to the extended structure as a result of interstellar scattering at this frequency \citep[e.g.][for a review]{2008AIPC..968..340S}. Figure~\ref{fig:uvplot} shows the ($u,v$)-coverage of KaVA in comparison to the Very Long Baseline Array (VLBA). For Sgr A*, KaVA can offer more closure quantities (amplitude and phase), which are crucial for obtaining high-quality maps and size measurements \citep[e.g.][]{2011A&A...525A..76L, 2004Sci...304..704B, 2014ApJ...790....1B}.


    \begin{figure}[htbp]
   \centering
	\includegraphics[width=6cm, angle=270]{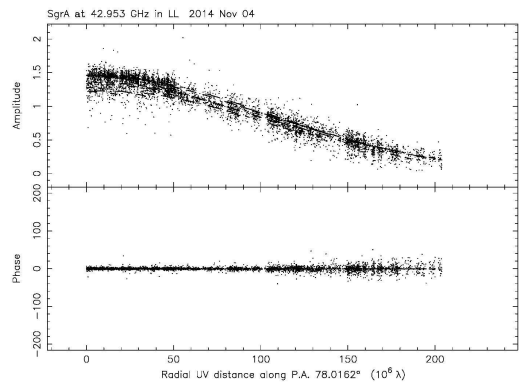}
   \caption{Visibility amplitude and phase of Sgr A* as functions of the projected ($u,v$)-distance at the epoch of 2014 November 4. All visibilities are plotted for a position angle of $78.1^{\circ}$ east of north, which corresponds to the major axis of the model fit. The solid lines represent the model obtained through elliptical Gaussian fitting; the model is also shown in Figure~\ref{fig:mapplot}.
   \label{fig:radplot}
         }
   \end{figure}
 
\begin{figure}[htbp]
   \centering
	\includegraphics[width=8cm, angle=0]{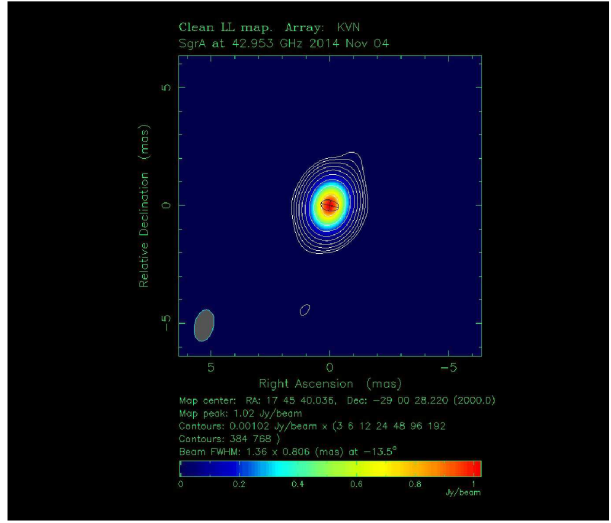}
   \caption{
    Natural weighted contour map of Sgr A* at the epoch of 2014 November 4. The contours start from 3 times the image rms and increase by factors of two. The beam is plotted at the bottom left corner. The elliptical contours represent the Gaussian model fits.
   \label{fig:mapplot}
         }
   \end{figure}

The observation frequency was selected by considering the effects of scattering and antenna performance. At 43\,GHz, the amount of scatter broadening decreased and the measured size differed from the size determined by extrapolating the scattering law. We present preliminary results from an observation made on 2014 November 4, one of the epochs with good weather conditions. Figure~\ref{fig:radplot}  shows the visibility amplitude and phase as a function of ($u,v$)-distance. Fringe detections range up to 250\,M$\lambda$, and the minimum detected flux is approximately 200\,mJy. Figure~\ref{fig:mapplot} shows a natural weighted contour map superposed with an elliptical Gaussian model fit. The model fit shows an elongated structure with an almost east-west alignment, which is consistent with previous results \citep[e.g.][]{2013PASJ...65...91A, 2014ApJ...790....1B}. Because of the superior ($u,v$)-coverage and good weather conditions, we achieved a very high dynamic range in this image ( $\sim$ 1000).

Our preliminary multi-epoch results showed no appreciable change in the flux density and intrinsic size of Sgr A*. This result is consistent with the single-epoch Global Millimeter VLBI Array (GMVA) observation of October 2013 \citep{2015A&A...576L..16P} and multi-epoch Very Large Array (VLA), SMA, and ALMA observations during 2012--2014 \citep{2015ApJ...802...69B}.  If no brightening is observed over the next 10 years, say, there could be two possible reasons: the radius of the outer edge of the accretion flow is smaller than the distance of the pericenter of G2 from the GC, or the gas component of G2 is less massive than expected and too small to affect the dynamics of the accretion flow of Sgr A*. Studies with EA mm/submm VLBI are expected to open new avenues for constraining the global structure of the accretion flow of Sgr A*.




\subsubsection{Accretion flow of M87}



\begin{figure}[tp]
\center \scalebox{0.28}{\includegraphics{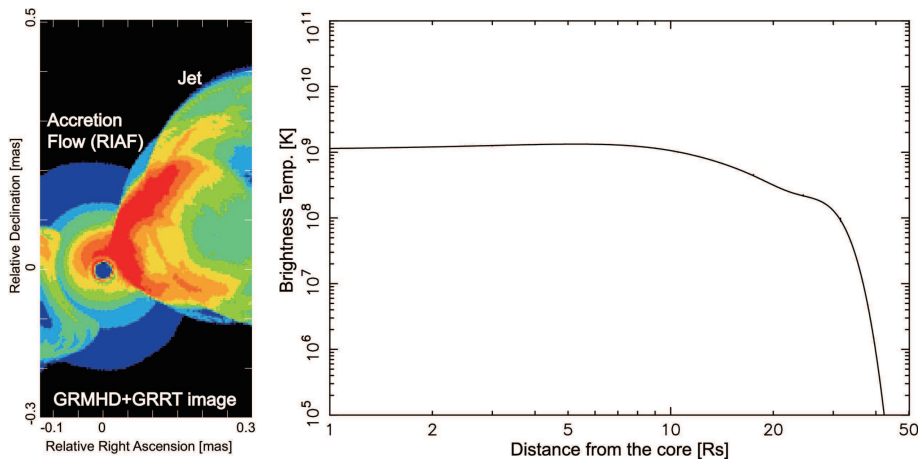}} \caption{
(Left) General relativistic MHD (GRMHD) numerical simulation model of M87 \citep{2009ApJ...692..411N} with general relativistic radiative transfer (Pu et al.\ 2016). The hole at the center of the accretion flow is due to a funnel at the jet base, and it is not associated with the BH. (Right) Expected profile of the brightness temperature in a direction transverse to the jet at 86\,GHz.
\label{fig:AF}}
\end{figure}

The accretion flow theory predicts that the electron temperature of the RIAF has a radial profile with a high-temperature plateau at the center (see Figure~\ref{fig:AF}). The electron temperature in the central region is expected to reach approximately 10$^{9}$\,K, and the radius of the (partially) optically thick photosphere is expected to be 10\,R$_{s}$ (0.08\,mas) at 86\,GHz for a BH mass of 6 $\times$ 10$^{9}$\,M$_{\cdot}$. The innermost region of the jet is also expected to have an optically thick core (e.g., the so-called VLBI core), and therefore, distinguishing the photosphere of an accretion flow from the innermost jet is difficult with a single VLBI observation. However, the accretion flow component is expected to have diffuse and extended emission, which is associated with the optically thin and relatively cold component toward the periphery of the accretion flow. The detectable radius of the accretion flow is estimated to be 20\,R$_{s}$ (0.16\,mas) at 86\,GHz. Since the emission from the accretion flow component should extend in a direction perpendicular to the jet axis, we expect to easily distinguish this emission from the jet emission. At 230\,GHz and above, the jet and accretion flow become more transparent, and hence, the relative locations of their emissions with respect to the BH shadow are considerably clearer. In some cases, it is expected that both the accretion flow and counter-jet emission illuminate the central BH, resulting in a ring-like or crescent structure that corresponds to the photon ring of the SMBH. Although this phenomenon is crucial and significant for studies of GR, it hinders the determination of the origin of the emission (a counter-jet or an accretion flow) by causing confusion. This dichotomy can be resolved by observing the spectral index distributions because the jet and accretion flow would have different spectral indices. Because the M87 jet is elongated in a direction almost parallel to the east–west direction, the main direction for the emission from the accretion flow component is the north–south direction. Therefore, by considering north–south baselines, we can directly image the photosphere and the associated optically thin and diffuse extension of the accretion flow in the direction perpendicular to the jet axis; such direct imaging has not been performed to date. Direct imaging of the accretion flow has never been achieved in any jet-accretion flow system so far, and therefore, it provides an opportunity to verify the RIAF paradigm. The detected radial structure of the accretion will provide constraints on submodels (ADAF, ADIOS, and CDAF). Subsequently, the rate of mass accretion on to the SMBH can be estimated, and the spin parameter of the SMBH can be determined by comparing the accreting power with the kinematic power of the jet.

Polarimetric and mm/submm VLBI observations toward the accretion flow are crucial. Recent EHT observations of Sgr A* have revealed very high fractional polarization on the event horizon scale \citep{2015Sci...350.1242J}.  Since emission from Sgr A* is expected to originate from the accretion flow onto the SMBH, the observed increase in the fractional polarization toward longer VLBI baselines is interpreted to reflect the presence of ordered magnetic fields on different scales. Finally, if the observed fringe spacing attains a value that is a few times the event horizon, a highly coherent/high-order magnetic field is observed. This suggests a scale of the turbulence in the accretion flow. \citet{2014ApJ...783L..33K} performed SMA observations at 230\,GHz and found that the M87 nucleus is weakly polarized ($\sim$1.5\%). It is expected that fractional polarization will increase if observations are performed with longer baselines, similar to the case of Sgr A* \citep{2015Sci...350.1242J}.

Accretion flows can be probed on the basis of observations of the Faraday rotation measure (RM). Performing RM observations at mm/submm wavelengths is one of the established methods for obtaining the mass accretion rate; this method was developed mainly for our GC, Sgr A* (also see Section 2.2.2). RM observations have revealed the presence of very dense plasma, probably associated with accretion flows (e.g., \citealt{2006ApJ...640..308M}; \citealt{2006ApJ...646L.111M} references therein).   The RM is related to the electron
density $n_{\rm e}$ and the magnetic field component parallel to the line of sight, $B_{||}$, as $RM \sim \int_{\rm LOS} n_{\rm e} B_{||} dr$, where
$\int_{\rm LOS} dr$ represents integration along the line of sight. According to \citet{2006ApJ...640..308M}, the RM is related to $\dot{M}_{in}$ as follows:

\vspace{-0.4cm}
{\small
\begin{equation*}
\dot M_{in} =1.1\times10^{-8}\left[1-(r_{out}/r_{in})^{-(3\beta-1)/2}\right]^{-2/3} \times \left( \frac{M_{BH}}{6.6\times 10^9 M_{\odot}}\right)^{4/3} \left( \frac{2}{3\beta-1}\right)^{-2/3} r_{in}^{7/6} RM^{2/3} [M_{\odot} \, {\rm yr}^{-1}],
\end{equation*}
}

\begin{figure}[tp]
\center \scalebox{0.4}{\includegraphics{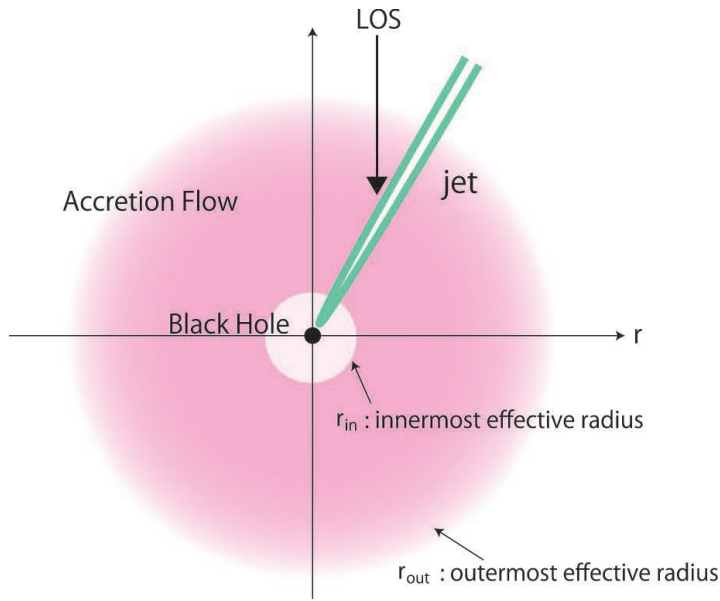}} 
\caption{
The envisioned accretion flow in M87 for obtaining the RM. It is assumed that the M87 jet offers the background polarized emission for Faraday rotation and that the dominant contribution comes from the innermost core of the jet. In the figure, $r_{in}$ and $r_{out}$ are the inner and outer edges of the Faraday screen in the accretion flow. In the accretion flow, the electrons are sub-relativistic and the magnetic field is coherent.
\label{fig:AF2}}
\end{figure}

\vspace{-0.4cm}
where $\beta$ is a parameter that depends on the subclass of the RIAF model (3/2 for an ADAF, 1/2 for a CDAF, and values between these two values can be considered as the ADIOS). Although this method has been successful for Sgr A*, it has been tested only for a very limited number of other LLAGNs. Very recently, we conducted observations by using the SMA and Combined Array for Research in Millimeter-wave Astronomy (CARMA) for applying the method to M87 \citep[][; see also Figure~\ref{fig:AF2}]{2014ApJ...783L..33K} and Per~A \citep{2014ApJ...797...66P}. 
The method was also tested by applying it to a blazar jet by using ALMA Science Verification data \citep{2015Sci...348..311M}.    In particular, \citet{2014ApJ...783L..33K}  applied this method to M87 and obtained an RM of ($-$2.1\,$\pm$\,1.8) $\times$ 10$^{5}$\,rad\,m$^{-2}$; an upper limit of $|$RM$| <$7.5$\times$10$^{5}$\,rad\,m$^{-2}$ was obtained at the 3$\sigma$ confidence level. From the RM value, the range of $\dot{M}_{in}$ was calculated to be between 0 and 9.2 $\times$10$^{-4}$ M$_{\odot}$~yr$^{-1}$.

Although, the RM value obtained was an upper limit because of the wide range of RM, the estimated $\dot{M}_{in}$ is two orders of magnitude smaller than the accretion rate in the outer part of the accretion flow\footnote{It corresponds to 3.8 $\times$ 10$^{5}$ $R_{s}$ \citep{2003ApJ...582..133D}}  (0.1\,$M_{\odot}$\,yr$^{-1}$), which was estimated through X-ray observations \citep{2003ApJ...582..133D}.   This indicates that $\dot{M}$ is strongly suppressed in the inner region of the accretion flow, and it disfavors an ADAF and favors an ADIOS or a CDAF (for the ADAF, it is predicted that $\dot{M}_{\rm ADAF}$ $\sim$ 0.1 $M_{\odot}$ yr$^{-1}$, whereas for the ADIOS and CDAF, the predictions are  6 $\times$ 10$^{-5}$ $<$ $\dot{M}_{\rm ADIOS}$ $<$ 0.1  and $\dot{M}_{\rm CDAF}$ $\sim$ 6 $\times$ 10$^{-5}$ $M_{\odot}$ yr$^{-1}$). Although this method was validated for connected array observations, VLBI observations can enable us to provide a stronger constraint on the mass accretion rate. Moreover, VLBI observations are expected to reveal the radial profile of the RM, from which the type of accretion model (i.e., ADAF, CDAF, or ADIOS) can be determined.

\subsubsection{Low Luminosity AGNs}


The majority of AGNs in the local universe are LLAGNs ~\citep{2008ARA&A..46..475H}. In particular, the accretion state of an AGN near a BH is thought to be an ADAF or RIAF~\citep[][]{1995ApJ...452..710N}. Many past observations suggest that at centimeter wavelengths, the ADAF/RIAF alone is not sufficient to reproduce the observed radio power and spectra~\citep{2001ApJ...562L.133U, 2004ApJ...603...42A}, and an additional radio-emitting component is required. The presence of a compact radio jet, which is a natural emission source, has been suggested, in analogy with bright radio-loud AGNs with powerful relativistic jets~\citep{1999A&A...342...49F}. This suggestion is supported by recent sophisticated theoretical studies based on MHD simulations~\citep[e.g.,][]{2012ApJ...761..130Y}. These studies revealed that the creation of an outflow or wind suggestive of a jet is a major effect of the gas dynamics of ADAF-/RIAF-type flows.

At mm-to-submm wavelengths, the observational appearance of an LLAGN may change considerably~\citep{2005MNRAS.363..692D}. It is expected that the ADAF/RIAF releases accretion energy through emissions, mostly in these bands (the so-called submm bump), and the extended jet emission becomes dim at such high frequencies because of its steep spectral nature. Moreover, the decrease in the opacity of the extended jet emission to the opacity of synchrotron emission facilitates a close view of the vicinity of the central BH. Therefore, by performing high-resolution, high-sensitivity mm VLBI observation of a nearby LLAGN, we may have a rare chance to directly image the accretion flow (possibly together with the jet base) in the vicinity of the BH.

\bigskip
\noindent
{\bf The Sombrero galaxy}

The Sombrero galaxy (M104, NGC\,4594) is associated with one of the nearest LLAGNs. Because of its proximity \citep[9.0\,Mpc;][]{2006AJ....132.1593S} and the large mass of the central BH~\citep[$1\times10^{9}\,M_{\odot}$;][]{1996ApJ...473L..91K},  we can access a physical scale as small as 1\,mas = 0.04\,pc = 454\,R$_{\rm s}$. In terms of Rs, such a spatial scale is approximately 5 or 20 times finer than that of other famous LLAGNs, such as M81 or NGC\,4258; moreover, the spatial scale is comparable to the gravitational scale accessible for Sgr A* or M87. Despite these merits, there have been few VLBI studies on the nucleus of this source.

\begin{figure}[htbp]
\centering \includegraphics[width=0.9\columnwidth]{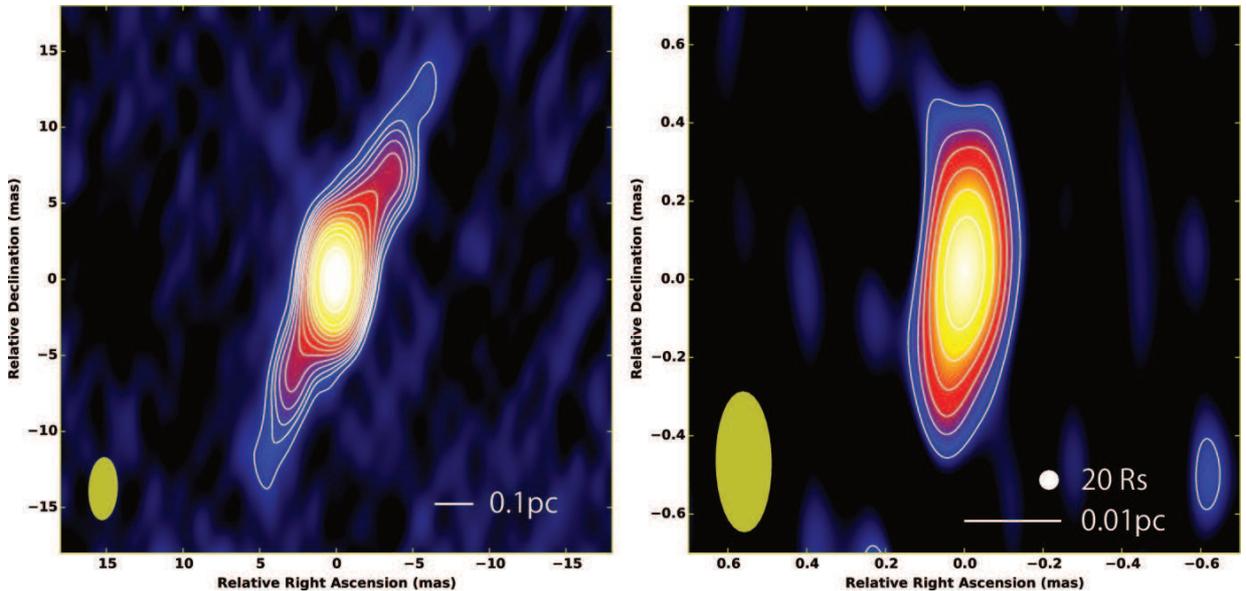} \caption{
 VLBA images of the Sombrero galaxy nucleus: (left) a 5-GHz image and (right) a 43-GHz image. These images are taken from Hada et al. (2013b). The yellow ellipse indicates the beam size of each image.
 }  \label{fig:kazuhiro.hada.1}
\end{figure}

Hada et al.\ (2013) recently performed a dedicated VLBA observation of the Sombrero galaxy, successfully obtaining high-quality VLBI images of the nucleus at seven frequencies between 1.4 and 43\,GHz.
Two images are shown in Figure~\ref{fig:kazuhiro.hada.1}.
At 43\,GHz, the nuclear structure was imaged on a linear scale under 0.01\,pc or 100\,$R_{\rm s}$, and a compact radio core with a high brightness temperature ($\gtrsim$$3\times 10^9$\,K) was observed.
In contrast, at lower frequencies, an extended structure originating from two sides of the core and extending along the northwest–southeast direction was detected.
The nuclear radio spectra show a clear spatial gradient, which is similar to that seen in more luminous AGNs with powerful relativistic jets.
These findings indicate the central engine of M104 powers radio jets which overwhelm the emission from the underlying RIAF at the observed cm frequencies.

\begin{figure}[htbp]
\centering \includegraphics[width=1.0\columnwidth]{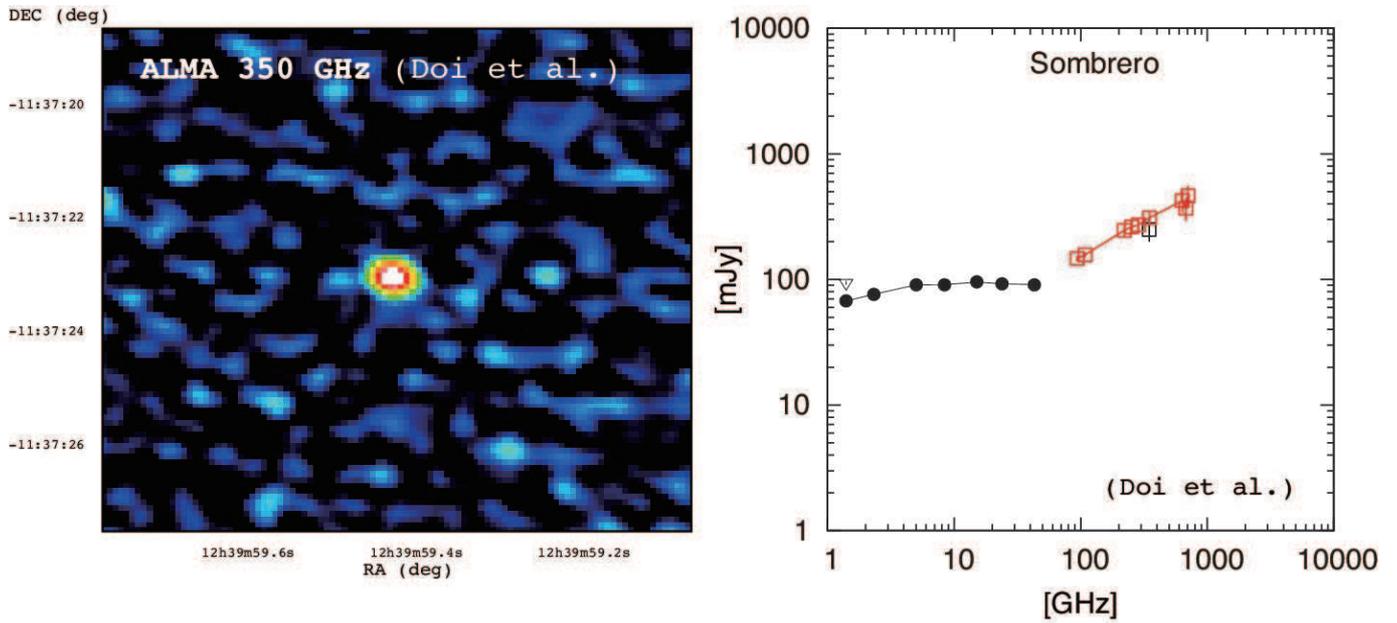}
\caption{
ALMA continuum spectrum of the Sombrero galaxy nucleus (Doi et al., 2016a, in preparation), and the VLBI core spectrum \citep{2013ApJ...779....6H}}
\label{fig:submmSombrero_whitepaper}
\end{figure}

After this study, we performed a mm-to-submm multifrequency observation with ALMA (Doi et~al.~2016a in preparation). Figure~\ref{fig:submmSombrero_whitepaper} shows the observed ALMA continuum spectrum of M104's nucleus and the VLBI core spectrum at cm wavelengths \citep{2013ApJ...779....6H}.  The mm-to-submm flux densities clearly increase at higher frequencies. By contrast, the cm spectrum is relatively flat. Therefore, future mm VLBI observations toward the BH vicinity of the Sombrero galaxy nucleus hold promise for detecting fringes at a good signal-to-noise ratio  (SNR) if the phased ALMA is added to the existing mm/submm VLBI network.

\begin{figure}[htbp]
\centering \includegraphics[width=0.5\columnwidth]{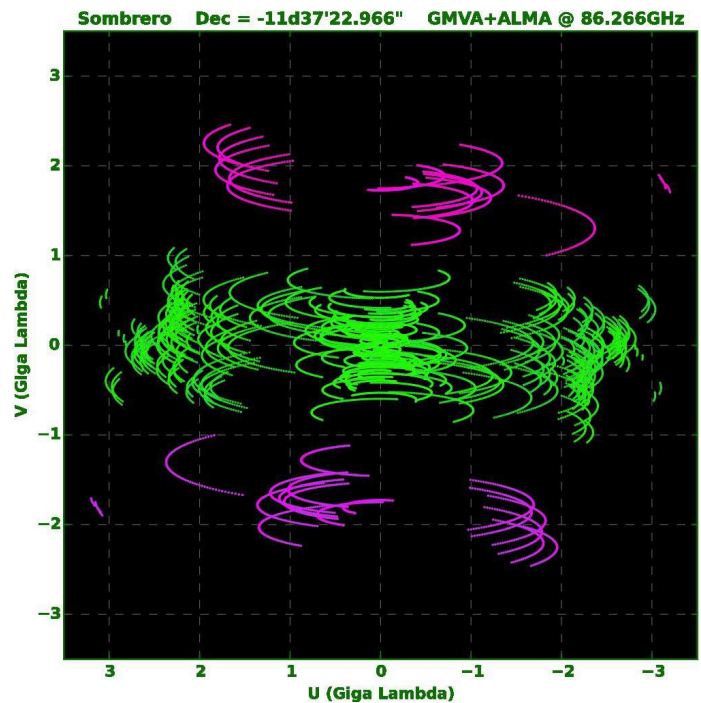}
\caption{
 Simulated ($u,v$)-coverage of the Sombrero galaxy and a full-track 86-GHz GMVA observation for the case where ALMA is added to the GMVA. The magenta lines indicate baselines to ALMA.}
\label{fig:kazuhiro.hada.2}
\end{figure}

\bigskip
\noindent
{\bf Perspective for future ALMA-VLBI}

As described, the Sombrero galaxy is an ideal target for obtaining a general picture of the nuclear activity of LLAGNs. VLBI observations in cm bands and multifrequency ALMA spectral measurements in the mm-to-submm bands have begun to provide interesting insights into the nuclear structure of this source. Naturally, the next crucial step is to directly image the nucleus using mm-to-submm VLBI. A problem is that this nucleus is dimmer (typically $\lesssim 100$\,mJy)  than other luminous radio sources and the current GMVA or EHT may not be capable of securely detecting signals. The inclusion of ALMA at bands 3 and 6 is crucial for improving the array sensitivity. Furthermore, as noted, the spectral shape of these bands appears to become highly inverted, further increasing the chances of signal detection.

The benefits of including ALMA are actually twofold for this source. As an example, in Figure~\ref{fig:kazuhiro.hada.2}  we show the expected ($u,v$)-coverage of the Sombrero galaxy and a full-track 86-GHz GMVA observation for the case where ALMA is included. Since the Sombrero galaxy is located in the southern hemisphere, the addition of baselines to ALMA dramatically improves the angular resolution in the north–south direction. Compared to the ($u,v$)-coverage without ALMA, the ($u,v$)-coverage with ALMA provides an angular resolution that is 2.5 times higher in this direction (and probably even better depending on weighting scheme of the data). Because the jet is elongated in the north–south
direction, resolving the nuclear structure in this direction is crucial to separate the emission from the jet base and accretion flow. The expected angular resolution at 86\,GHz is approximately 50\,$\mu$as, corresponding to approximately 23\,$R_{\rm s}$. With the EHT at 230\,GHz, the resolution is approximately three times greater. Therefore, mm-to-submm VLBI observations of the M104 nucleus are useful for probing the disk-jet connection site of the LLAGN on a scale of the event horizon.

\begin{figure}[htbp]
\centering \includegraphics[width=1.0\columnwidth]{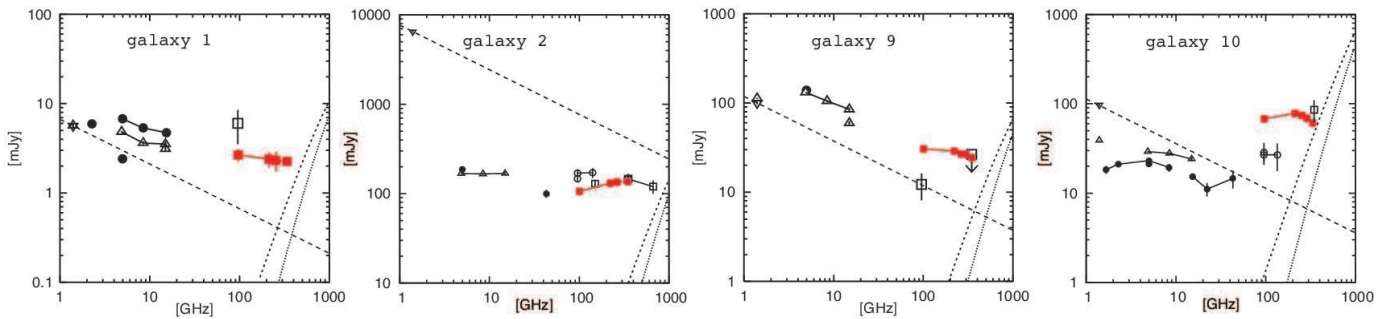} \caption{Examples of results from our ALMA observations toward a dozen of nearby LLAGNs (Doi et al., 2016b, in preparation).\label{fig:four_Virgo_galaxies.eps} }
\end{figure}

Finally, we note that the inclusion of ALMA for mm VLBI is also beneficial for studies of other nearby LLAGNs. Figure ~\ref{fig:four_Virgo_galaxies.eps} shows four examples of the results obtained from ALMA observations toward a dozen nearby LLAGNs (Doi et al., 2016b, in prep.). We found that the fraction of LLAGNs showing flat or inverted spectra in the mm-to-submm bands is not small. This ALMA result indicates the potential of VLBI for detecting other LLAGNs in the mm-to-submm regime.

\bigskip
\noindent
{\bf Time Domain Constraints}

All compact radio sources are variable. Time-domain observations provide several crucial constraints that provide an important context for submm VLBI imaging. The simplest use of such observations is for the detection and characterization of flares, which may be coupled with ejection of jet components or structural changes in accretion disk flows. Multiwavelength characterization of flares and their variability can provide strong constraints on the physical properties of jets and accretion flows, including those during events such as the G2 passage \citep[e.g.,][]{2015ApJ...802...69B}.  Flares can also make previously undetectable sources suﬃciently bright, facilitating the study of the sources using submm VLBI. For this purpose, monitoring campaigns that study a large number of interesting objects on a monthly basis would be valuable.

\begin{figure}[tbh]
\includegraphics{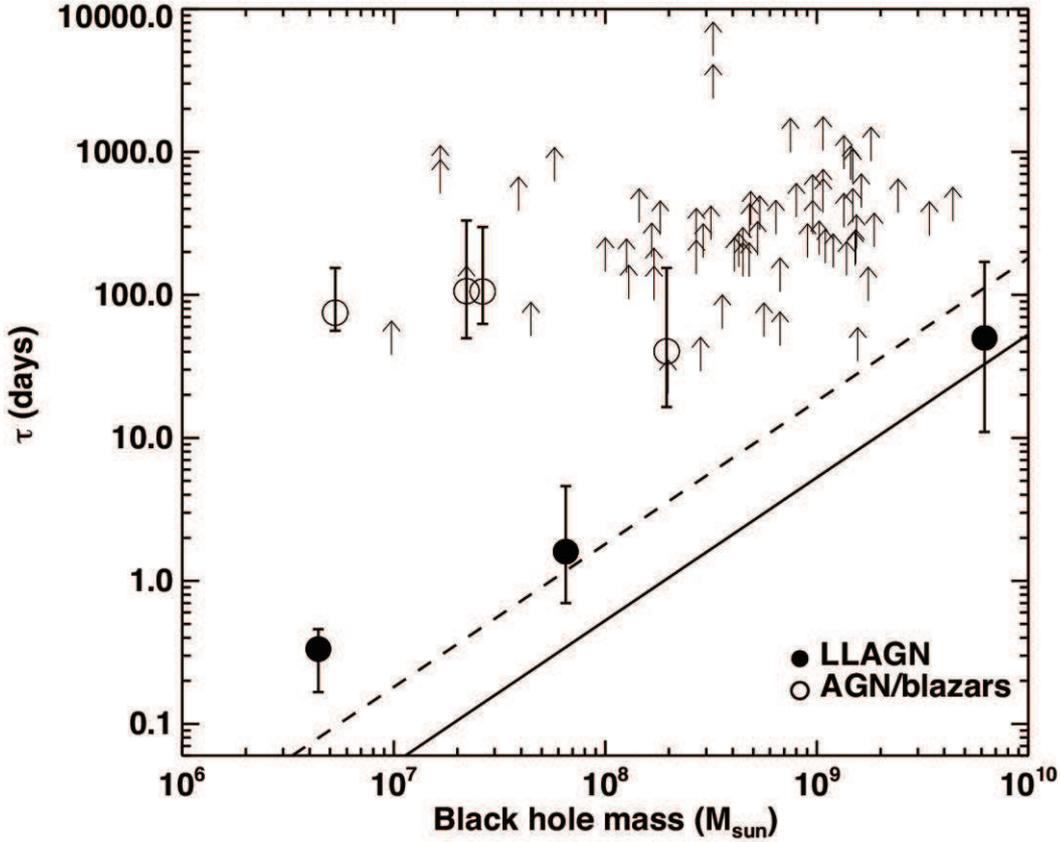}
\caption{BH mass versus damped random walk timescale for the SMA calibrator source sample \citep{2015ApJ...811L...6B}. The LLAGN Sgr A*, M81, and M87 are shown by filled circles. Other AGNs from the SMA calibrator database with BH masses are shown as open circles for detections and upward arrows for lower limits on $\tau$. For a given BH mass, the timescales associated with AGNs/blazars are considerably longer than those associated with the LLAGN. The inferred timescales for LLAGN variability increase with the BH mass and are consistent with the linear relationship predicted by GR. The solid line gives the period of the innermost stable circular orbit (ISCO), and the dashed line gives the infall time for a disk with radius $5 R_S$, a viscosity ($\alpha=1$) of 1, and a height ($H/R$) of 1.
\label{fig:taumbh}
}
\end{figure}

The second use of time-domain information is in the statistical analysis of light curves. \citet{2015ApJ...811L...6B} used SMA archival data to analyze the light curves of blazars and an LLAGN. The two sources clearly belong to different classes, with the LLAGN showing characteristic variability timescales comparable to the period of the last stable orbit (Figure~\ref{fig:taumbh}). This observation strongly supports the inference that the emission originates very close the BH. Further, the statistical constraints indicated that LLAGNs have remarkably stable flux densities over long periods of time. Evidence for blazars, which are strongly variable, having a detectable characteristic time was limited: they appeared to vary on timescales as long as or longer than 10 years. This is characteristic of emissions originating in a jet far from the BH. This type of statistical analysis permits us to clearly separate sources belonging to the two classes.

\clearpage


\subsection{Understanding Jet Formation Mechanism}
\subsubsection{General Objectives}

The formation mechanism of relativistic jets in AGNs remains a longstanding, unresolved problem in astrophysics. Emissions from radio-bright AGNs, such as quasars, blazars, and luminous radio galaxies, are dominated by nonthermal radiation from their outflows. In particular, through synchrotron emissions from jets comprising magnetized hot plasma, the violent ejection process has been intensively investigated, and it is well known that jets are accelerated up to relativistic speeds. However, despite decades-long research efforts and recent major advances in high-resolution VLBI observations, the launching mechanism of jets in AGNs is not well understood. The key factors that can help us in understanding the jet formation mechanism are as follows.

\bigskip

\textbf{\textsl{Magnetic fields in jets}}\\

The role of magnetic fields should be clarified. For clarifying the field topology, polarimetric observation is considered to be crucial (e.g., Asada et al., 2002).

Although magnetically driven jet models (e.g., the Blandford-Znajek mechanism) have been widely discussed, 
the actual value of the strength of the magnetic field and the field geometry 
at the base of the jet are still poorly constrained by observations. 
To test the magnetic jet paradigm, it is necessary to clarify 
the energy density of the magnetic fields and that of the particles, 
and the field geometry in great detail at the upstream end of the jet \citep[Kino et al. 2014, ][and references therein]{2015ApJ...803...30K}. 
Rotation measure (RM) at mm/submm wavelengths can provide tight constraints on the physical properties at the jet base (Kuo et al. 2014).


\bigskip

\textbf{\textsl{Velocity fields of jets}}\\

Where and when do jets in AGNs attain their relativistic speed? Understanding the acceleration mechanism of relativistic jets is a fundamental requirement to probe the jet formation mechanism \citep[][and references therein]{2014ApJ...781L...2A}. The velocity field of jets basically appears to reflect the fundamental properties of the central engine composed of SMBHs and accreting flows onto the engine. High-resolution VLBI monitoring of jet components is the most promising way to explore it. The KaVA science team have started a densely-sampled M87 monitoring at 22 and 43\,GHz in order to clarify the velocity field in M87.

\bigskip

\textbf{\textsl{Internal structures of jets}}\\



Streamline and  internal structure of jets remains an open question. While a limb-brightened structure, the so-called sheath,  is seen in many jets in AGNs (e.g., Asada \& Nakamura 2012; Hada et al.\ 2013; Nagai et al.\ 2014; Koyama 2016), the spine structure has not been seen so far. Detailed structures and relationships among various possible dissipation regions such as recollimation shocks (Marscher et al.\ 2008), multiple turbulent zones (Marscher 2014), internal shocks (e.g., Kino et al. 2008, and references therein), and reconnection zone (Sikora et al. 2005) are not well understood.

\bigskip

\textbf{\textsl{Radio $\gamma$ connection in jets}}\\

There may be a relationship between variations in the radio flux and outbursts of high-energy $\gamma$-ray radiation. Statistical analyses of selected samples of AGNs have found a correlation between the radio flux and the $\gamma$-ray emission, with the $\gamma$-ray emission leading the radio emission, typically by approximately one month in the source frame \citep{2010ApJ...722L...7P}. This can be understood if the emission is caused by a single event in an AGN core and the delay between the radio flux and the $\gamma$-ray emission is caused by the opacity of the jet plasma. However, it appears that such correlations can be found only for a few per cent of all $\gamma$-ray bright AGNs \citep{2009AJ....138.1874L}; this raises the question of whether a general radio-$\gamma$ relationship indeed exists. Confirmation of the presence or absence of a radio-$\gamma$ connection can provide new insights into the emission mechanisms of high-energy radiation in blazar jets (i.e., leptonic vs. hadronic origin).


\bigskip

\textbf{\textsl{AGN Feedback}}\\

Although AGN feedback (i.e., the energy released by the AGN into the interstellar medium) is invoked to solve many astrophysical problems, it is not clear how the feedback actually works. For mechanical feedback, multiwavelength observations of radio galaxies have shown a tight correlation between accretion and jet formation: the kinetic power of jets is approximately 1\% of the Bondi accretion power (i.e., Bondi accretion rate $\times$ c$^{2}$) of the AGN \citep{2006MNRAS.372...21A, 2013MNRAS.432..530R}. Variations of this efficiency factor appear to be related to the optical AGN type, implying an interplay between radio jets and the spatial distribution of circumnuclear matter \citep{2014JKAS...47..159T}.  Understanding this relationship can provide crucial insights into the question of how jets are regulated (i.e, turned ``on'' and ``off''), which is closely linked to AGN feedback.

\subsubsection{On a possible jet in Sgr A* }

As described in Section~2.2.2, numerous lines of evidence support the accretion flow onto Sgr A* being an RIAF. Evidence for a jet is more indirect and inconclusive. Scaled models of disk-jet systems produce a jet that is consistent with constraints obtained from VLBI imaging \citep{2007MNRAS.379.1519M}. Importantly, jet models can produce the radio spectrum of Sgr A*, which thermal accretion models cannot naturally do.

There have been several claims of arcsecond-scale jet structures in the GC, but none of these are unambiguously associated with Sgr A* \citep[e.g.,][]{2013ApJ...779..154L}. In addition, high-resolution VLBI imaging at 7\,mm shows an intrinsic source that is extended and oriented roughly in the east–west direction \citep{2014ApJ...790....1B}. The result of an analysis of submm VLBI closure phases is also consistent with the east-west elongation \citep{2016arXiv160205527F}. The elongation may be interpreted as indicating either an accretion flow or a jet.


Submm VLBI has the potential to detect and characterize the jet in Sgr A*, if it exists
\citep[e.g.,][]{2014A&A...570A...7M}.
Both the accretion flow and jet are likely to be present in any image, complicating the analysis
required to robustly detect and characterize each.  Nevertheless, high resolution imaging in
total and polarized intensity has the potential to reveal in unprecedented detail the coupling
between an accretion flow and jet.  Polarized imaging is particularly important since  standing
shocks in the jet nozzle could have highly polarized features.

High-time-resolution imaging is also important for characterizing jet features. A jet will launch features that are visible for tens of minutes to hours before escaping from the system. Current jet models suggest that jet flow velocities $(\beta)$ are approximately 0.5 \citep{2015A&A...576A..41B}, implying angular velocities of approximately 0.5\,mas/h. Jet proper motions are likely to exhibit very different signatures in the closure phase or imaging products other than accretion orbits around the BH.

\subsubsection{M 87 Jet}

The radio galaxy M87 is one of the closest example of an AGN with relativistic jets, 
and it has been investigated across the entire electromagnetic spectrum for a several decades. 
Its proximity and the large estimated mass of its central BH provide an unprecedented opportunity to study jet formation processes at its base. 
It is widely believed that jets are powered electromagnetically in accreting BHs  (Blandford \& Znajek 1977 [BZ77]; Blandford \& Payne 1982 [BP82]). 
They are magnetically accelerated as a consequence of the conversion of Poynting flux to kinetic flux along the poloidal stream. 
Thus, elucidating magnetic fields is crucial for understanding AGN jets (Zamaninasab et al. 2014; Ghiselini et al. 2014). 
VLBA observations suggest that AGN jets are highly relativistic; this is well known from superluminal motions up to approximately 50 c (Lister et al. 2009). 
The theory of relativistic MHD jets (Li et al. 1992; Vlahakis \& Konigl 2003) outlines the process of bulk acceleration up to a few tens of the Lorentz factor ($\Gamma$). 
Over the last decade, numerical simulations have investigated the behavior of acceleration and 
collimation of relativistic jets in detail (McKinney, 2006; Komissarov et al., 2007; Tchekhovskoy et al., 2009), 
but the launching process of AGN jets is poorly constrained by observations.

\bigskip

{\bf {\em Origin of relativistic jets}}\\  

Submm VLBI observations toward nearby radio galaxies, especially M87, can provide an answer to this fundamental quest. The behaviors of GRMHD outflows from a rotating BH has been examined (e.g., McKinney, 2006; Pu et al., 2015). In the cold limit, the position of the inflow/outflow stagnation (along the large-scale, hole-threading poloidal field) depends on the largeness of the BH spin. This property could be considered in combination with the VLBI core shift measurements for M87 (Hada et al., 2011). \\

{\bf {\em Parabolic jet hypothesis} } \\

Over the past decades, it was widely believed that conical jets were the norm in AGNs (Blandford \& Konigl, 1979). Furthermore, the MHD jet theory suggests that a parabolic jet may be feasible to have an effr¥Êent bulk acceleration (Blandford \& Payne, 1982; Li et al., 1992). Asada \& Nakamura (2012) indeed discovered the parabolic stream in the M87 jet within the Bondi radius ($\sim 10^{5}$ Rs), where the bulk acceleration presumably occurs (Asada et al., 2014). It is imperative to test the universality of the parabolic jet by performing direct measurements in nearby radio galaxies (Tseng et al., 2016)
 and  a notable quasar 3C273 (Akiyama et al. in prep.)
 and/or core shift measurements in distant quasars (Asada et al., in prep.; Algaba et al. submitted). \\

{\bf {\em Origin of the jet}} \\

\begin{figure}[htbp]
\centering
 \includegraphics[scale=0.7, angle=0]{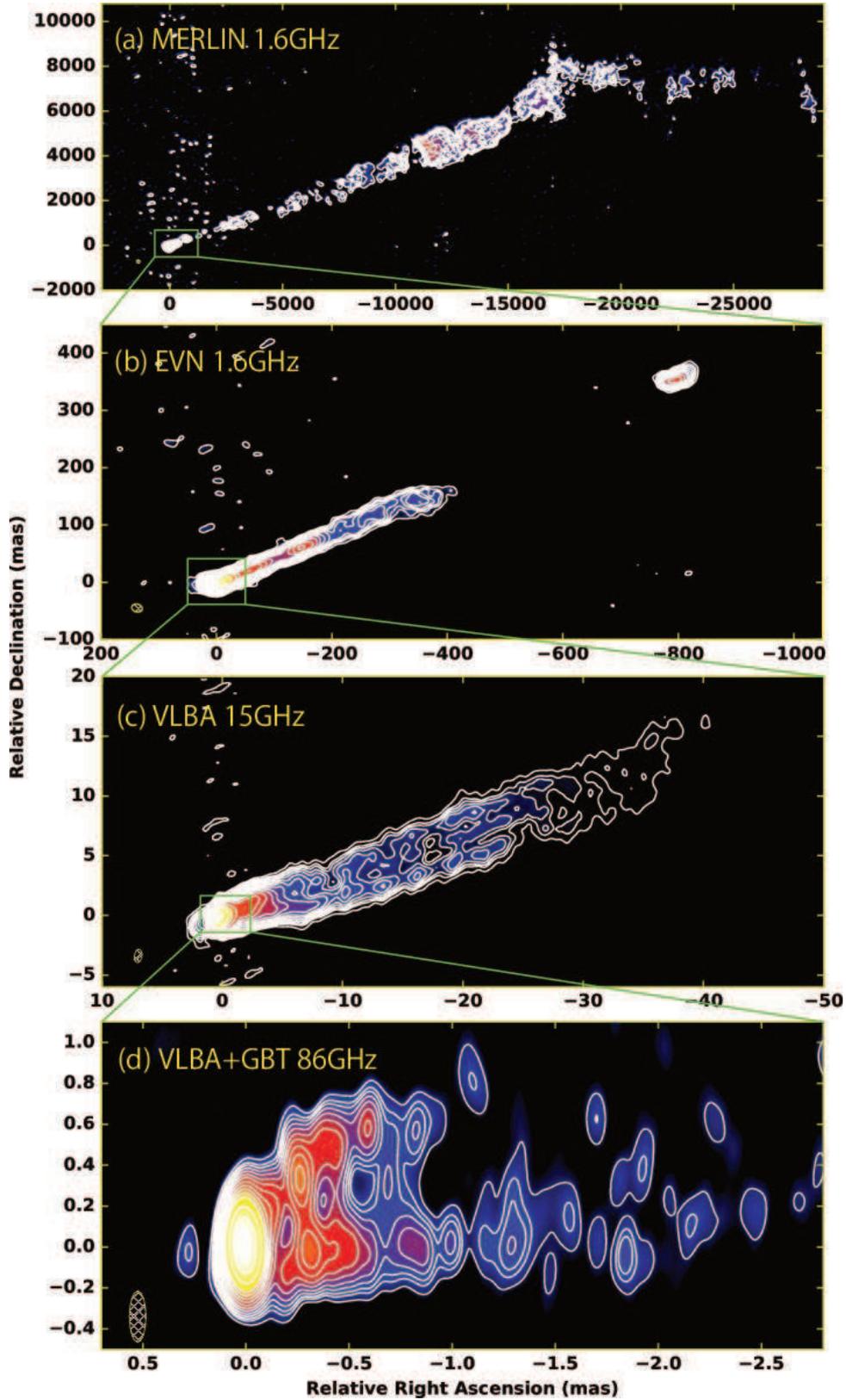}
 \caption{Radio images of the M87 jet. (a) MERLIN 1.6GHz image (taken from Asada \& Nakamura 2012). (b) EVN 1.6GHz image (taken from Asada \& Nakamura 2012). (c) VLBA 15GHz image (taken from MOJAVE archive). (d) HSA (VLBA+GBT) 86GHz image (taken from Hada et al. 2016). }
 \label{fig:hada87.1}
\end{figure}

One of the central goals in high-energy astrophysics is to ascertain the origin of relativistic jets. 
We believe that we can investigate this exciting problem with the highest angular resolution through submm VLBI observations toward M87. 
Recent GRMHD simulations have substantially improved our understanding of BH-ergosphere-driven jets from RIAFs (McKinney \& Gammie, 2004; McKinney, 2006; McKinney \& Narayan, 2007). 
An analysis of streamlines by using various VLBI observations toward M87 provides clues for constraining existing theoretical models of relativistic jets (Asada \& Nakamura, 2012; Nakamura \& Asada, 2013; Hada et al., 2013). 
We reveal that the M87 jet exhibits a collimated streamline with a semiparabolic index $(\alpha)$ of $\simeq 1.7$ ($z \propto r^{\alpha}$) in the sphere of gravitational influence (SGI) of the BH (i.e., Bondi radius $\sim 7 \times 10^5\, r_{\rm g}$), 
whereas the jet transforms into an uncollimated conical streamline ($\alpha \sim 1$) beyond the SGI. 
The semiparabolic streamline inside the SGI in M87 appears to be consistent with GRMHD simulations (Nakamura et al., in prep.), as shown in Figure 21. 
A jet with a wider opening angle is more realistic than the parabolic solution $z \propto r^2$ for genuine BZ77; 
this solution can be described by the approximate force-free, steady jet solution for the thin disk-like configuration in the ergosphere (McKinney \& Narayan, 2007; Tchekhovskoy, 2015).

\begin{figure}[htbp]
\centering
 \includegraphics[scale=0.8, angle=0]{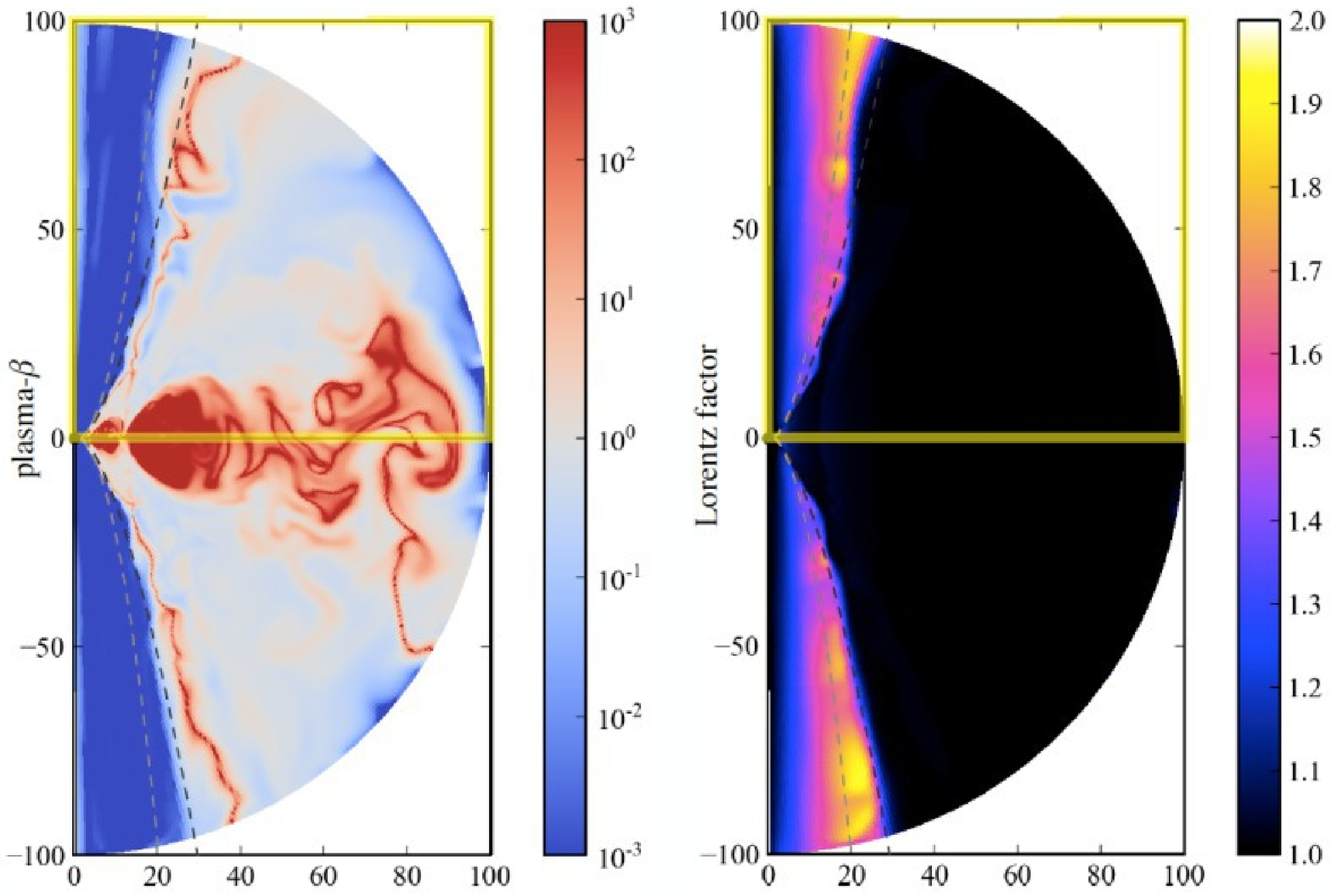}
 \caption{Parabolic jet in GRMHD simulations by the HARM code version
 1.0 (Gammie et al. 2003; Noble et al. 2006). Computational domain
 covers up to $r = 100\, r_{\rm g}$. The BH spin parameter $a=0.9$ is
 adopted. {\em Left} panel shows plasma-$\beta$ (gas pressure/magnetic
 pressure); highly magnetized parabolic jet is confined by the gas
 pressure-dominated coronal wind $\beta=1-10$. {\em Right} panel shows
 the Lorentz factor. The BH jet in the middle to lower latitude is
 accelerated up to $\sim 0.85\,c$ at $r \gtrsim 50\, r_{\rm g}$. Further
 acceleration would be expected further downstream. Two break lines
 denote the parabolic streamline (BZ77; {\em right gray}) and
 semi-parabolic streamline ({\em dark gray}). {\em Yellow} box shows
 $10^2\, r_{\rm g} \times 10^{2}\, r_{\rm g}$, corresponding to the
 region of interest in mm/submm VLBI observations (Figure \ref{fig:nakamura.2}).
 \label{fig:nakamura.1}}
\end{figure}

In the RIAF environment, a powerful relativistic jet emerges from the BH ergosphere (McKinney, 2006) in the BZ77 process, and uncollimated wind is ejected from the RIAF coronal region (e.g., Sadowski et al., 2013). It is also known that a funnel-wall jet exists at the interface between the BH jet and the coronal wind (e.g., Hawley \& Krolik, 2006). What is the counterpart of the outflow observed in M87? An investigation of the streamline of M87 is a promising means to determine the origin of the outflow. The current observations can resolve the M87 parabolic jet up to $\sim$ a few of $10^{2}\, r_{\rm g}$ (at VLBA 43 GHz) (Asada \& Nakamura, 2012; Hada et al. 2013) and approximately $\sim 10^{2}\, r_{\rm
g}$ (at High-Sensitivity Array (HSA)/GMVA 86 GHz) (Hada et al. 2016; Asada et al., in prep.). Further upstream can be examined by considering either the VLBI core (Hada et al., 2011, 2013; Nakamura \& Asada, 2013) or through direct imaging using submm VLBI. In other words, the region of interest obtained using submm VLBI can be directly compared with GRMHD simulations (Figure \ref{fig:nakamura.1}). This is a unique opportunity to investigate the origin of relativistic jets. \\

{\bf {\em Spine-sheath structure and its dynamics}}\\

\begin{figure}[htbp]
\centering 
\includegraphics[scale=0.8, angle=0]{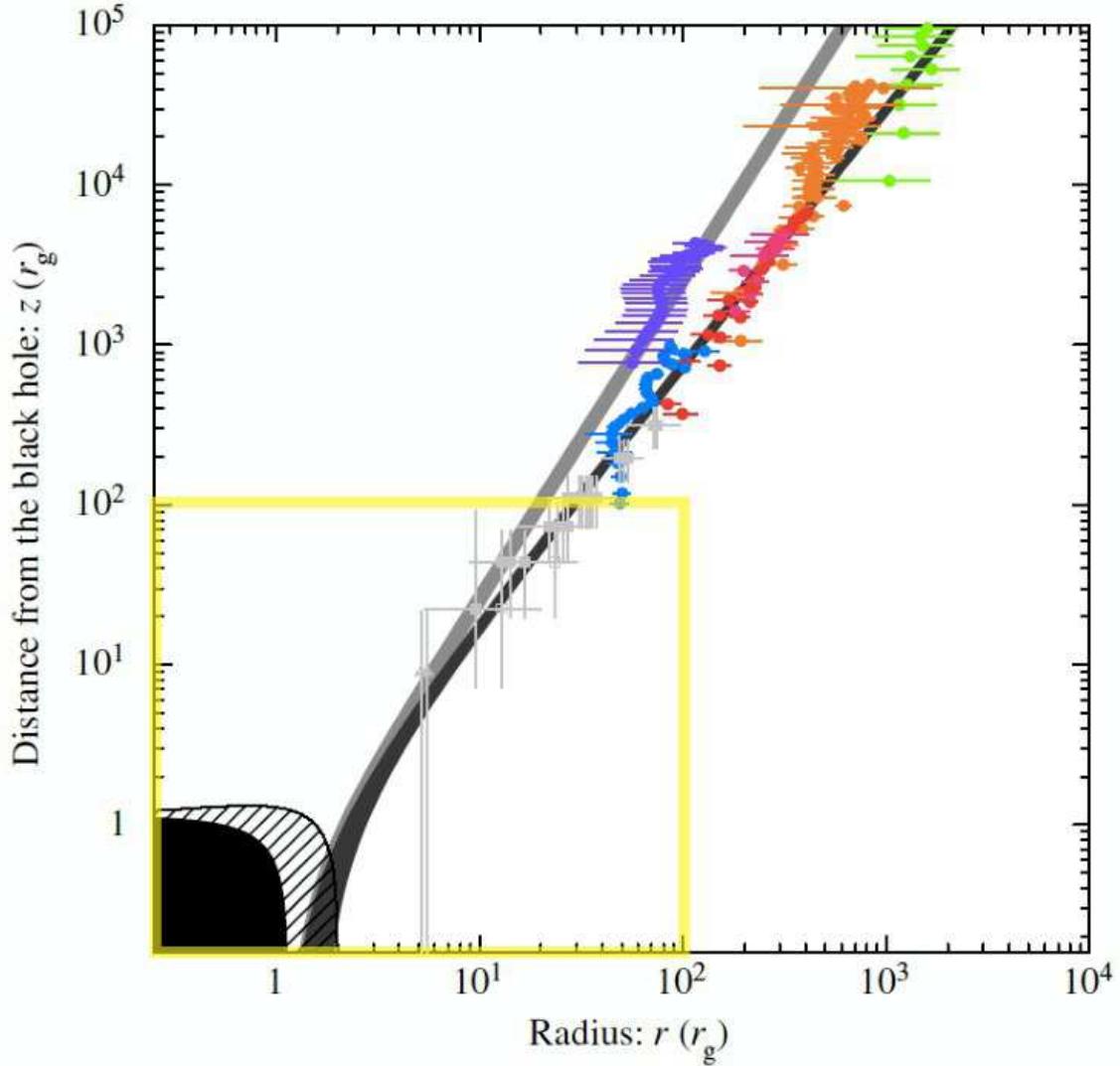}
 \caption{
Streamlines of the jet in M87 (Nakamura et al., in prep.). The data points were obtained from 1.6-GHz EVN (green; Asada \& Nakamura, 2012), 15-GHz VLBA (orange; Asada \& Nakamura, 2012; Hada, et al., 2013), 22-GHz VLBA (pink; Hada et al., 2013), 43-GHz VLBA (red; Asada \& Nakamura, 2012; Hada et al., 2013), 86-GHz GMVA (blue; Asada et al., in prep.), and 1.6-GHz VSOP observations (purple; Asada et al., submitted). VLBI cores (5 - 230 GHz; Nakamura \& Asada, 2013; Hada et al., 2013; Doeleman et al., 2012; Akiyama et al., 2015; Asada et al., in prep.) were used to determine the innermost jet emissions at given frequencies (Hada et al., 2011). The yellow box shows $10^2\, r_{\rm g} \times 10^{2}\, r_{\rm g}$, corresponding to the region of interest in the GRMHD simulations (Figure \ref{fig:nakamura.1}). The shaded area (light gray) denotes the outermost parabolic streamline (BZ77), whereas the outer shaded area (dark gray) represents the outermost semiparabolic streamline.
 \label{fig:nakamura.2}}
\end{figure}

In Figure  \ref{fig:nakamura.2}, we summarize the current status of the streamline analysis of the parabolic jet of M87. The outer sheath (semiparabolic streamline) is revealed by ground-based VLBI observations (Asada \& Nakamura, 2012; Hada et al., 2013; Nakamura \& Asada, 2013), whereas the inner spine (genuine parabolic streamline) is determined through space VLBI observations (VLBI Space Observatory Programme (VSOP); Asada et al. in press). From a comparison of Figure \ref{fig:nakamura.2} with Figure \ref{fig:nakamura.1}, we speculate that the spine jet (relativistic BH jet) may be covered by the sheath jet (nonrelativistic funnel-wall jet). It is noteworthy that the funnel-wall jet can be accelerated up to the relativistic regime further downstream, as discussed by Nakamura \& Asada (2013) and Asada et al. (2014). Currently, we can access the innermost region ($\lesssim 10^{2}\, r_{\rm g}$) of the M87 jet only by using VLBI core shift (Hada et al., 2011, 2013; Nakamura \& Asada, 2013). Therefore, imaging the jet on this scale may provide information on the multilayer system of the M87 jet. Furthermore, GRMHD simulations indicate that the initial acceleration of the BH jet depends on the BH spin. Therefore, the observed image is useful for understanding the properties of the BH jet and constraining the BH spin. \\

\begin{figure}[htbp]
\centering
 \includegraphics[scale=0.8, angle=0]{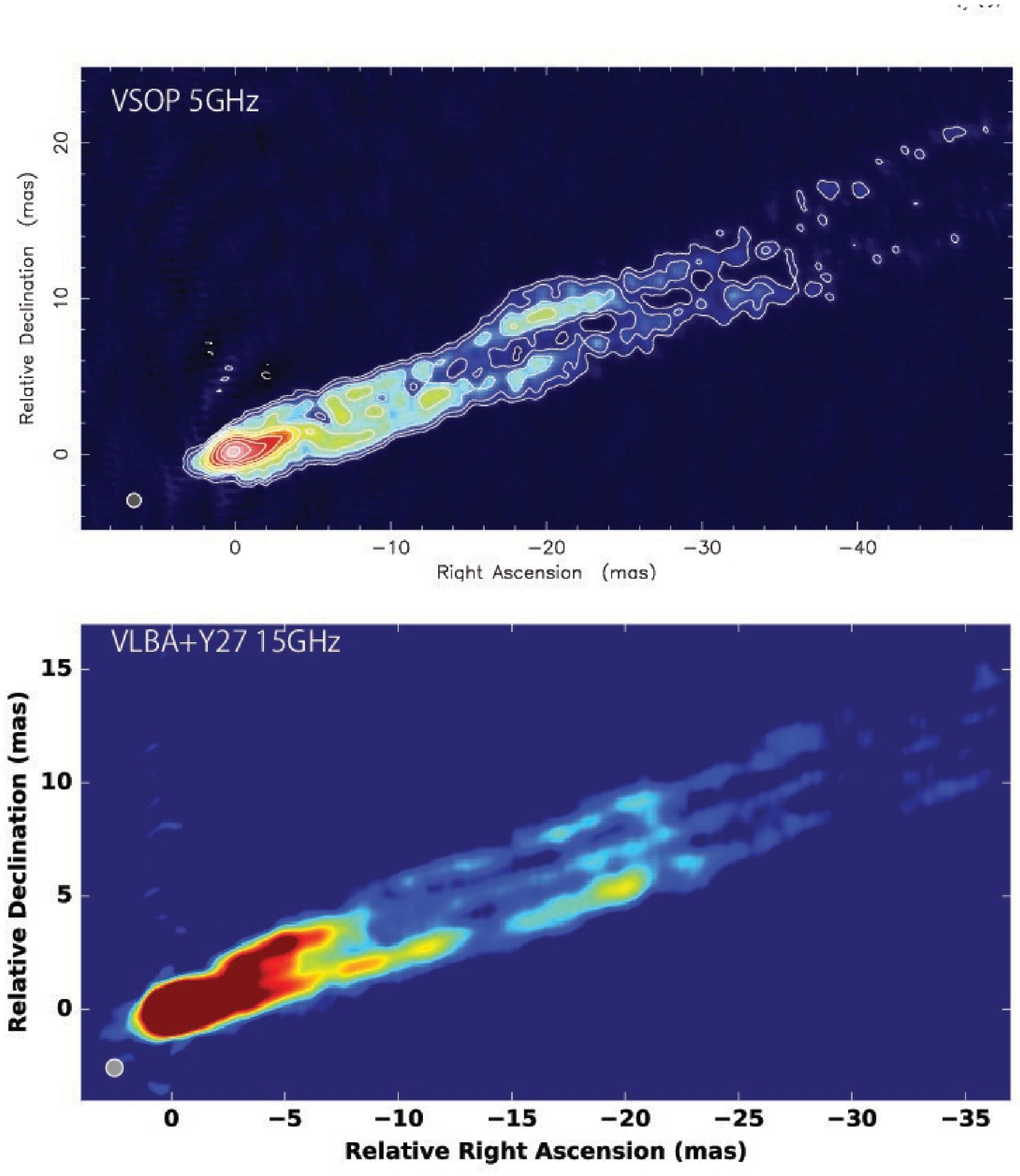}
 \caption{
 Centimeter-VLBI images of the M87 jet that have revealed the central spine component as well as the sheath. The top panel shows a VSOP 5GHz image (Asada et al., accepted) that has discovered the central ridge near the core, while the bottom panel indicates a recent high-sensitivity VLBA+Y27 15GHz image that have revealed a continuing triple-ridge profile further down the jet (Hada et al., in prep.).
 \label{fig:m87spine}}
\end{figure}

Following the discovery of the spine component in the VSOP image, we have recently made a new full-track, high-sensitivity VLBA observation of M87 at 15GHz in concert with the phased-VLA. The aim of this observation is to search for further evidence of the spine component in the M87 jet. Thanks to the excellent quality of the dataset, we have obtained a pc-scale jet image of M87 at an unprecedented dynamic range (Figure~\ref{fig:m87spine}). The most remarkable discovery in this image is that we clearly detected a triple-ridge structure that is continuing further down the jet (for a distance of ~10-30mas). The central ridge keeps a remarkably narrow shape, suggesting the presence of differential collimation between the exterior sheath and the interior spine. Therefore, future coordinated observations of EHT/GMVA jointly with high-dynamic-range lower-frequency VLBI will allow us to reveal the detailed collimation, flow acceleration (via brightness analysis or multi-epoch monitor) and particle energetics (via spectral analysis) of the multi-layered jet all the way from the launching to large scales.

{\bf {\em Counter-jet}}\\

To better constrain fundamental parameters of a relativistically beamed jet, information obtained from the counter-jet is also crucial because the apparent contrast between the main jet and the counter-jet directly provides the Doppler factor $\delta$, viewing angle $\theta$, and speed $v$ of the jets (Urry $\&$ Padovani, 1995). For M87, the exact values of these parameters near the jet base are still controversial, and there are large uncertainties in the likely values (e.g., Acciari et al., 2009). Previous low-frequency (15 GHz/43 GHz) VLBI imaging studies have revealed the presence of a weak counter-jet near the core (Kovalev et al., 2007; Ly et al., 2007), and recently, the counter-jet was detected at 86 GHz also with the VLBA + GBT (Hada et al., 2016), although the data are insufficient for obtaining the entire structure of the counter emission. Therefore, higher-sensitivity/higher-resolution imaging of the counter-jet using mm/submm VLBI and ALMA is eagerly being awaited. Such imaging has of the potential to constrain $\theta$, $v$ and (therefore)  $\delta$ of the jet(s) closer to the BH. The determination of these parameters should also help in constraining geometrical parameters of the surrounding accretion flow (see Section 2.2.4) and the central BH shadow (see Section 2.1.3).

Once the counter-jet images are obtained using mm/submm VLBI, we emphasize that it is imperative to perform joint analyses of this structure and low-frequency VLBI images. Because VLBI observations at different frequencies image the counter-jet at different radial distances from the BH, we can investigate the bulk acceleration profile of the M87 jet by determining the evolution of the jet-to-counter-jet brightness ratio as a function of distance from the central BH (from several\,$R_{\rm s}$ to thousands\,$R_{\rm s}$) on the basis of cm-to-submm images. It is expected that the M87 jet becomes superfast magnetosonic at approximately 100- 1000\,$R_{\rm s}$ from the BH in the framework of the magnetically-driven jet (Nakamura \& Asada, 2013; Pu et al., 2015). If this is true, we should be able to see a gradual increase in the brightness ratio with an increase in distance from the BH. \\

{\bf {\em Polarization}}\\

High-resolution VLBI polarimetry can be directly used for probing magnetic field structures associated with relativistic jets. In particular, M87 offers an exceptional opportunity to compare polarimetric observations with GRMHD jet simulations at a matched spatial resolution (e.g., Broderick \& Loeb, 2009). Nevertheless, although polarimetric properties of the M87 jet have been mainly studied on the kiloparsec scale in the radio and optical bands 
(e.g., Owen et al., 1989; Perlman et al., 1999; Chen et al., 2011; Algaba et al. 2016), the polarization structure below parsec-to-subparsec scales is highly uncertain. This is because polarization signals from the M87 inner jet are quite weak (typically a few percent of the fractional polarization degree or lower), presumably because of the strong Faraday depolarization effects of the dense foreground medium at these scales (Zavala \& Taylor, 2002). Therefore, using the higher-frequency regime is crucial for determining the intrinsic polarimetric structure of the jet.

\begin{figure}[htbp]
\centering
 \includegraphics[scale=0.5, angle=0]{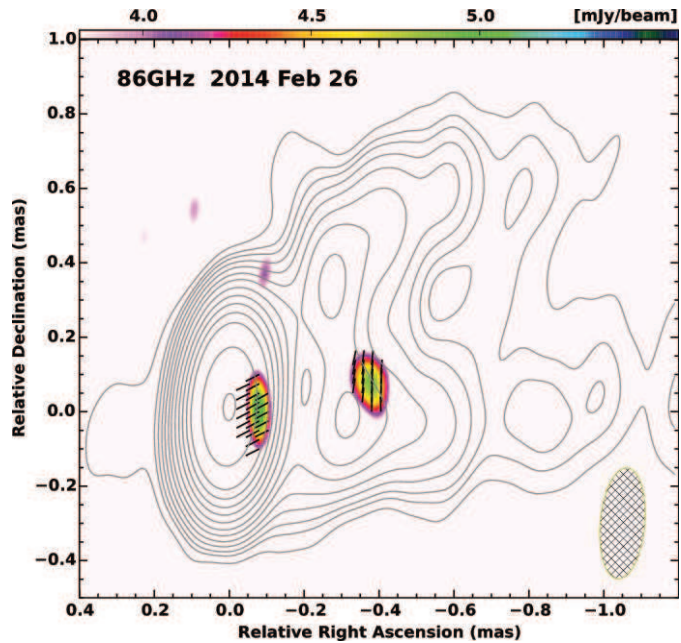}
 \caption{
 Polarimetric result obtained for M87 on February 26, 2014, by using VLBA + GBT at 86 GHz. The color map, vectors, and contours indicate the observed polarized intensity, observed electric vector polarization angle, and total intensity distribution, respectively. The convolving beam is shown at the bottom right corner.
 \label{fig:hada87.2}}
\end{figure}

In this context, we recently performed the first 86-GHz VLBI polarimetric experiment involving the M87 jet by using VLBA + GBT (Hada et al., 2016; see Figure~\ref{fig:hada87.2}). We discovered a highly polarized ($\sim$20\%) feature near the jet base; this is the highest fractional polarization ever seen for this jet below the parsec scale. This result
demonstrates that mm VLBI observations are in fact suitable for overcoming the depolarization problem. Moreover, such a highly polarized signal suggests the presence of a well-ordered magnetic field near the jet base, which is indicative of a magnetically dominated jet. Therefore, the use of an even higher frequency such as 230\,GHz for probing the magnetic field properties is very promising, as already demonstrated for Sgr A* (Johnson et al., 2015).

Ultimately, increasing the array sensitivity is crucial for more fruitful polarimetry studies. The inclusion of ALMA in the EHT, which is expected in the near future, will lead to an enormously powerful array. This array will facilitate the direct testing of the magnetically driven jet paradigm by imaging the entire spatial distribution of the polarimetric structure at the jet-launching site.
\\

{\bf {\em Mapping velocity field}}\\

According to current understanding, AGN jets are driven by a magnetic ($B$) force and subsequently accelerated by the progressive conversion of magnetic energy into kinetic energy. Relativistic MHD models have indicated that this conversion is gradual in relativistic jets. M87 is known as the best target to test this understanding because of its proximity to and the largeness of the BH. One of the objectives of this study was to map the velocity field of the M87 jet for testing the $B$-driven jet paradigm. Indeed, mapping the velocity field of M87 has been intensively explored in previous studies, and it was discussed by Asada \& Nakamura (2014). Surprisingly, the apparent velocities in the literature differ considerably, rendering the mapping controversial
(see also Mertens et all. 2015, Mertens et al. 2016). 
One of the possible reasons is the misidentification of components due to sparse interval of  monitoring and/or mismatching of components measured at different frequencies. Another possible reason is that different velocity fields were obtained in different periods. 

\begin{figure*}[htbp]
\centering
\includegraphics[trim=62mm 50mm 40mm 0mm, clip, width = 75mm]{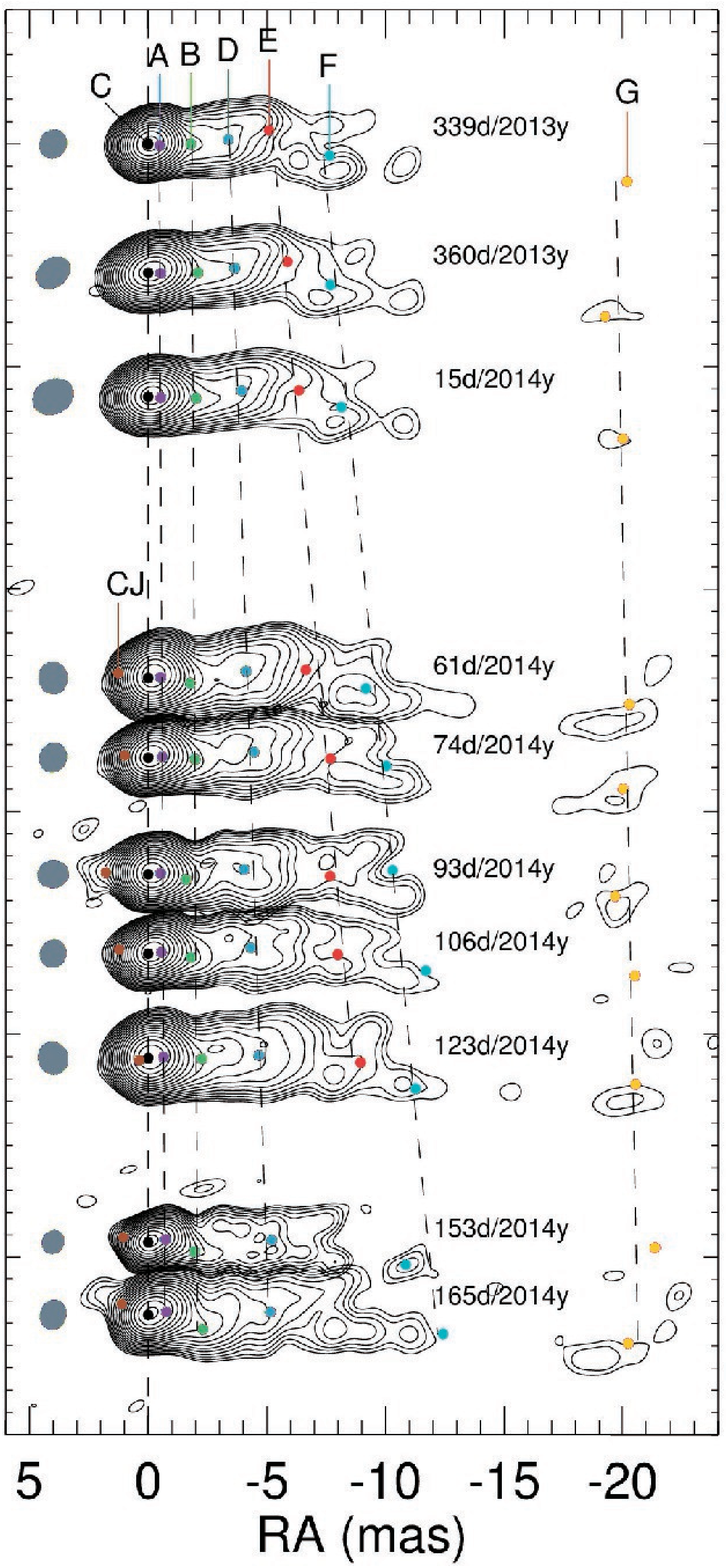}
\includegraphics[trim=62mm 50mm 40mm 0mm, clip, width = 75mm]{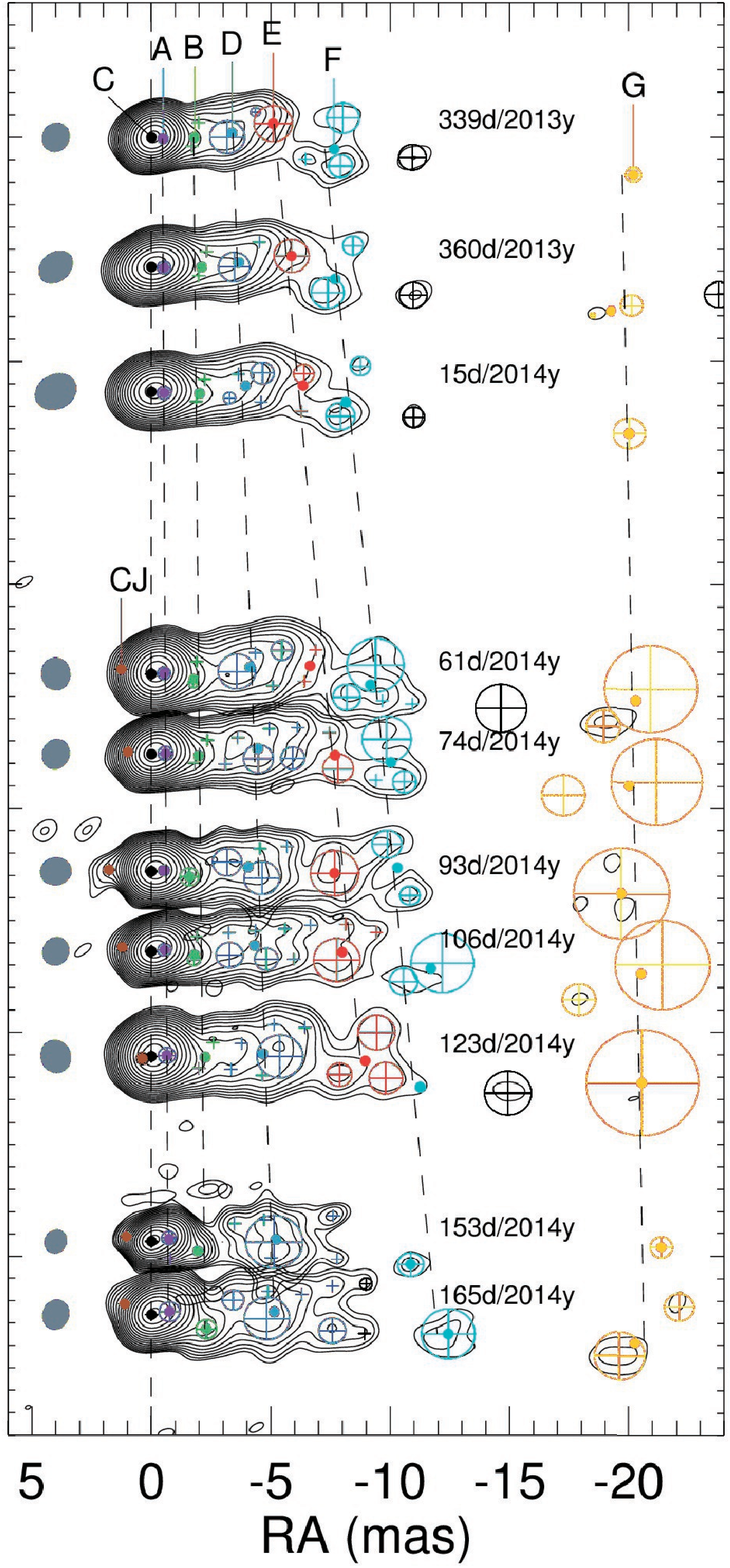}
\caption{
Images obtained in the KaVA pilot  monitoring of M87 at 22~GHz in 2014.
\emph{Left} : CLEAN images for all epochs as function of time from top to bottom. The model components, labelled A...G, obtained from the modelfit and the grouping (see text for details) are overlaid with coloured small circles. The black dashed lines are the best fit lines to the separation from core as function of time for each group component. The gray shaded ellipses at RA = 4 mas show the FWHM beam size. The observation date and year are noted for each image. The distance between adjacent contours is proportional to the separation in the observation dates. \emph{Right} : Images constructed by the modelfit. The crosses surrounded by circles are the fitted circular Gaussian components, while those without circles denote the fitted point source components. The components with the same color are grouped and their average positions weighted by their flux densities are shown with the coloured small circles (same as in the left panel). The components with black color were not used for grouping. All the images are aligned with respect to the position of component C and are rotated by 20 degree in a clockwise direction. Contours start from 4.07 and 6.10 mJy / beam and increase by factors of $\sqrt{2}$ for the left and right figures, respectively. \label{evol}}
\end{figure*}

To make a robust mapping  of the jet velocity field in M87, 
a promising means is a densely sampled multi-frequency monitoring. 
To this end, the KaVA large program  has been started from 2016 March. 
It is a dual-frequency (22 and 43 GHz) biweekly monitoring 
during the same period aiming for quasi-simultaneous
detection of the same components at both frequencies 
($http://radio.kasi.re.kr/kava/large_programs.php\#sh2$).
As brightness distributions of the M87 jet are quite smooth
and characterized by a limb-brightening, describing the
jet as several Gaussian components obtained from the
model fit procedure in \emph{Difmap} is not enough to model
the complex jet structure.
To overcome this problem, a technique which uses many more components and groups
several neighboring components presumably in the same
jet region into a single component has been developed
and applied to the KaVA data observed in 2014 at 22~GHz.
In  Figure~\ref{evol}, 
we present the observed CLEAN images and the model
fit images obtained in the pilot  monitoring of M87 at 22~GHz in 2014 
in the left and the right panel, respectively.
From this, we obtain the velocity map of the M87 jet 
down to 20mas scale from the radio core
(Hada et al., 2016, in preparation).

\begin{figure}[h]
\centering
\includegraphics[width=160mm]{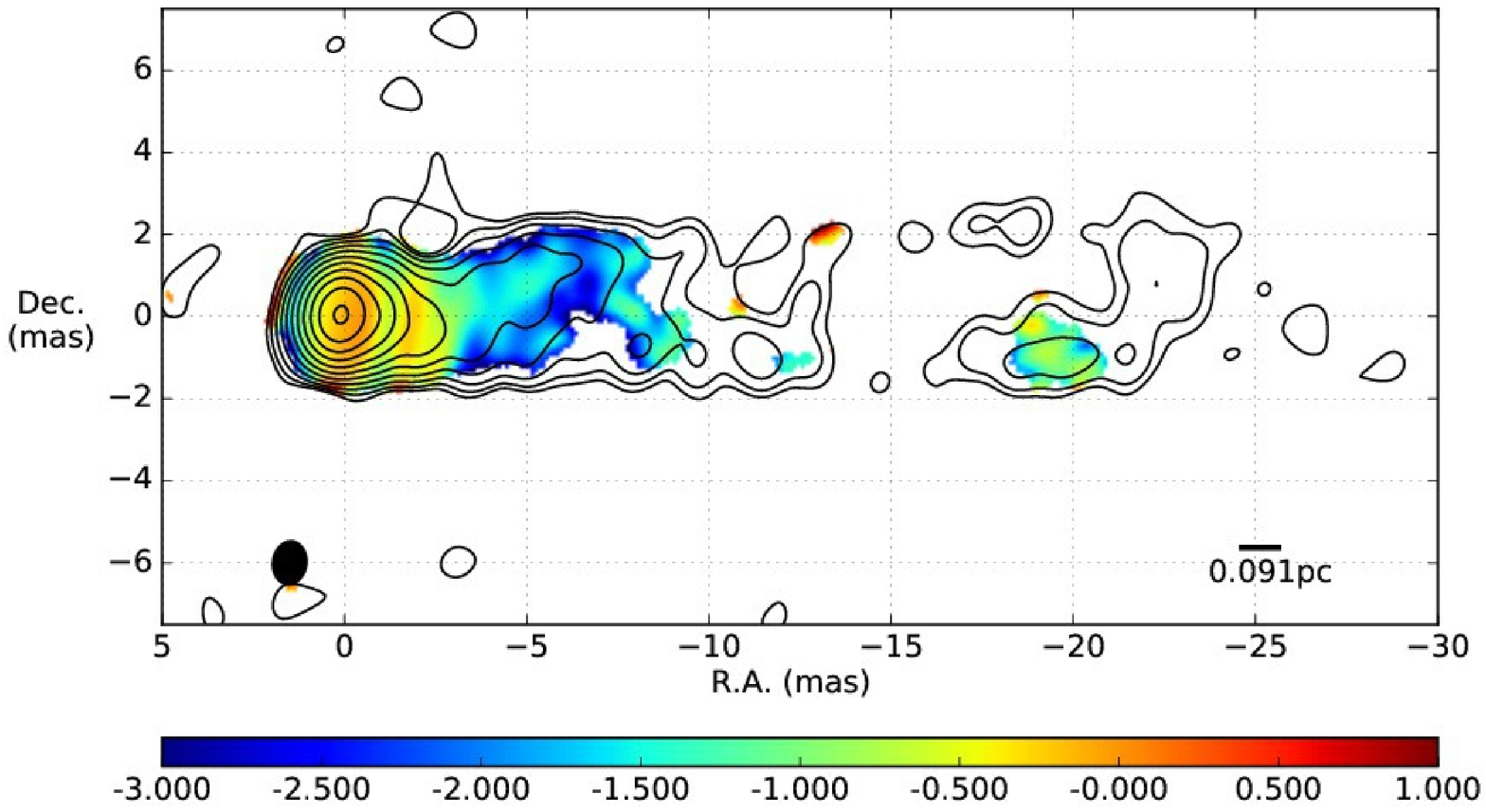}
\vspace*{10mm}
\caption{
Spectral index map of M87 between 22 and 43GHz on one epoch of KaVA Large Program (March 09, 2016). The spectral index is shown in color, overlaid on the 22GHz total intensity contours. For the spectral index calculation, region where the total intensity is less than 3 sigma level in 22 or 43GHz is excluded. The beam size and the scale bar are displayed on the bottom left and bottom right respectively. 
\label{fig:M87indexmap}}
\end{figure}

Further detailed investigations using the new data obtained in the KaVA large program 
are now actively ongoing  (Park et al., in preparation; Ro et al., in preparation).
 As one of the early results,  we present a spectral index map  of the M87 jet  
 in Figure~\ref{fig:M87indexmap}.

\textbf{\textsl{Radio $\gamma$ connection at the jet base}} \\

M87 is known to show $\gamma$-ray emissions up to the very high energy (VHE; $>$100 GeV) regime, and in this regime, it often exhibits active flaring episodes (Abramowski et al., 2012). 
Here, we choose a VHE flare event of M87 that occurred in the spring of 2012, because the event was intensively monitored using VERA (Hada et al. 2014). 
Furthermore, the combination of observational results of VERA and the EHT (Akiyama et al., 2015) provided new constraints on the properties of the $\gamma$-ray emitting region. 
During the event, VERA detected a remarkable increase in the radio flux density from the unresolved jet base (radio core) at 22 and 43 GHz, coincident with the VHE activity. 
Furthermore, EVN observations at 5 GHz confirmed that the peculiar knot HST-1, which is an alternative favored $\gamma$-ray production site located 120 pc from the nucleus, 
remained quiescent in terms of its flux density and structure. These results in the radio bands strongly suggest that the VHE $\gamma$-ray activity in 2012 originated at the jet base within 56 $r_{\rm s}$ from the central SMBH.

Akiyama et al. (2015) showed that no obvious flux change was seen in the EHT data at 230GHz between 2009 and 2012, despite an increase in the core flux on the arcsecond scale. One plausible explanation is that the VHE flare emitting region has an extended structure that is resolved out by the current EHT. Therefore, the minimum full width at half-maximum (FWHM) size is estimated to be $\mu$as at 20 $r_{\rm s}$, which has a half-width at half-maximum size of 600M$\lambda$ in the visibility plane of the EHT data. This limitation is consistent with at least two aspects of VHE flares. First, the two-month duration of the 2012 VHE event
implies that the minimum size of the VHE emission region on the basis of causality considerations. The second constraint,on the maximum size of the VHE emission region $<0.44$~mas $\sim60$~$r_{\rm s}$, was obtained by Hada et al. (2014) by using VERA at 43 GHz  because the flare component was not spatially resolved in the observations. Combining them, we can constrain the size of the VHE emission region ($R_{\gamma}$) as 20 $r_{\rm s}$ $\le$ $R_{\gamma}$ $\le$ 60 $r_{\rm s}$.

\begin{figure}[h]
\centering
\includegraphics[width=70mm]{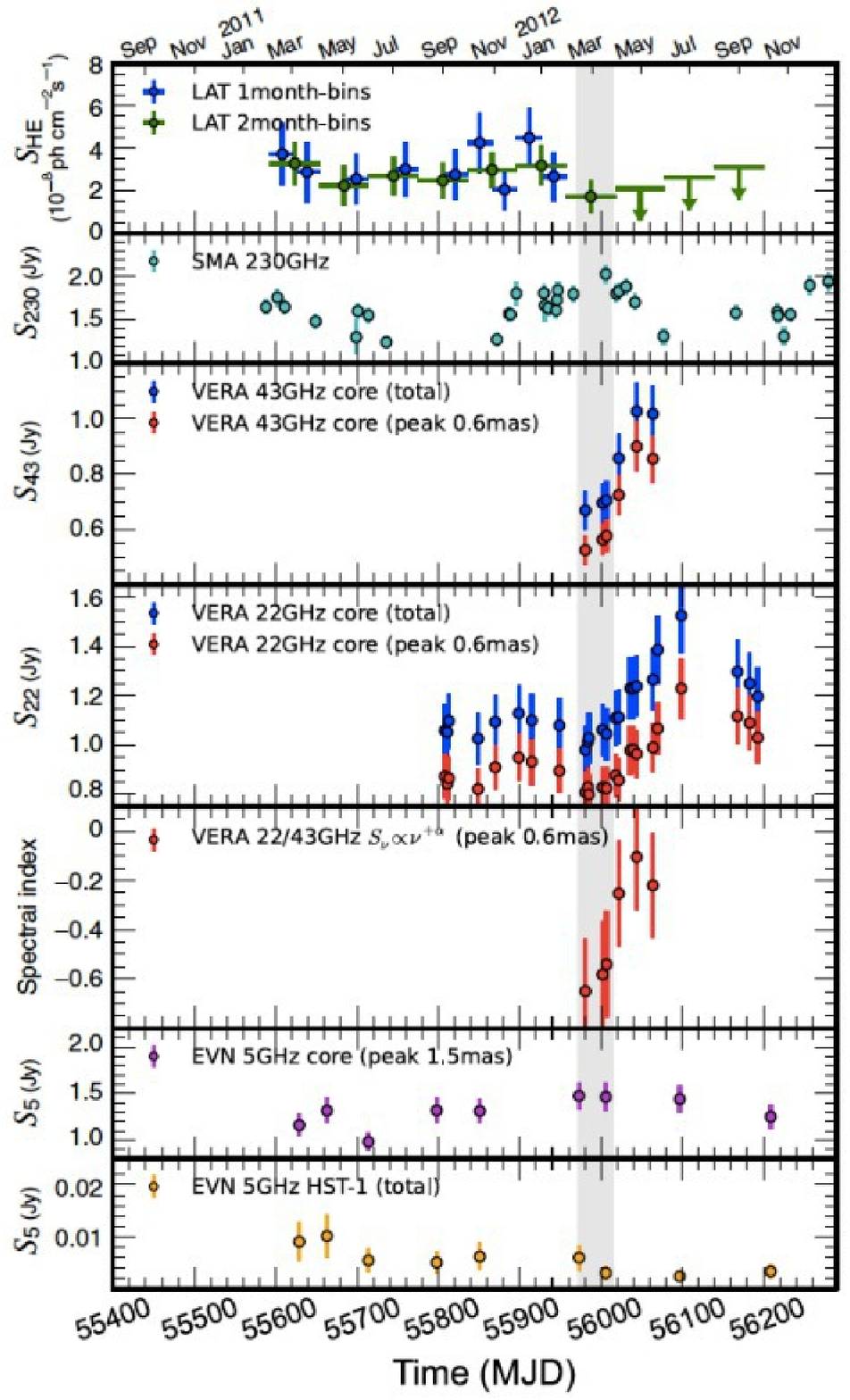}
\includegraphics[width=90mm]{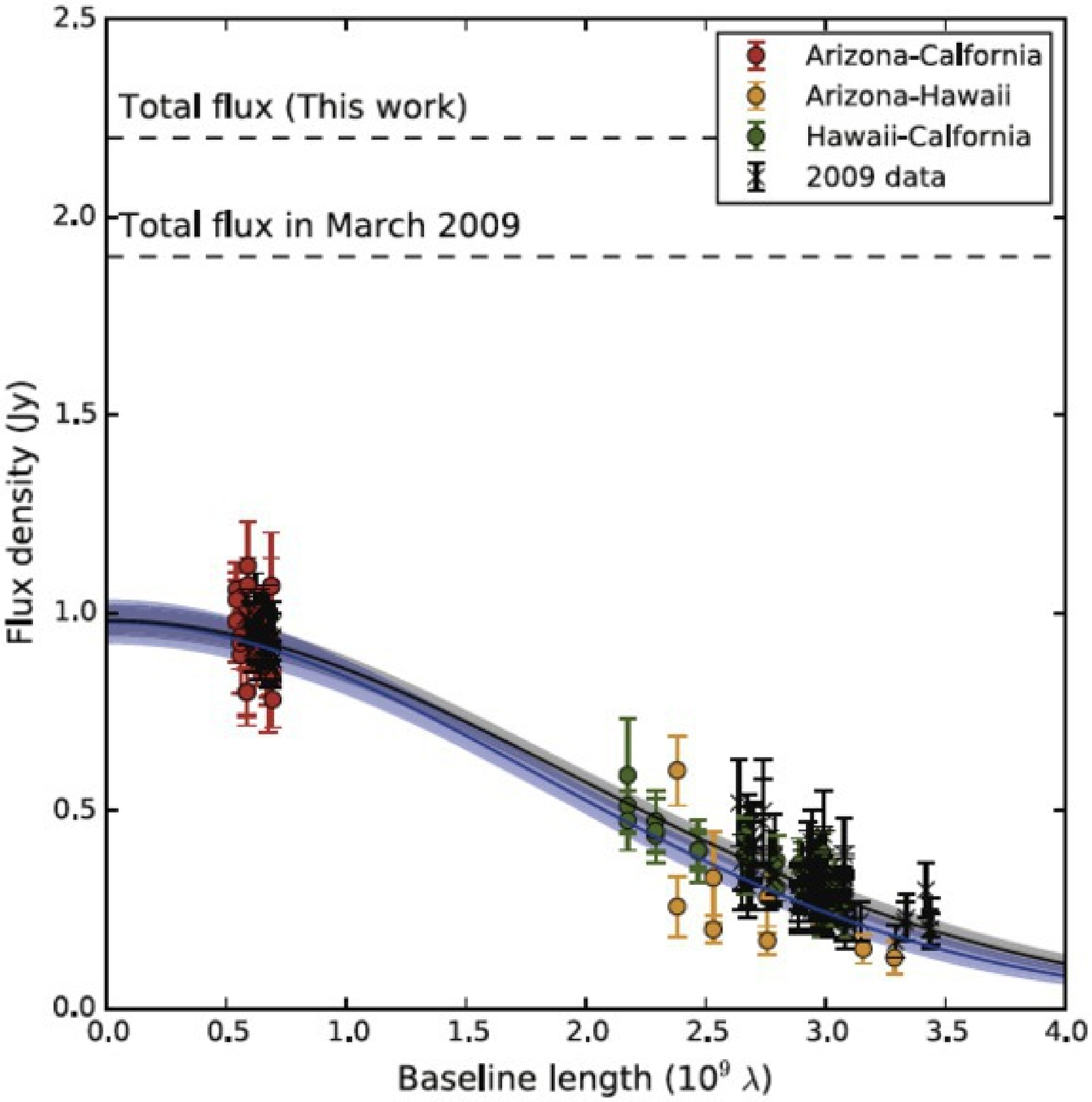}
\caption{
(Left) Multiwavelength light curves of M87 for the period between February 2011 and December 2012. The vertical shaded area over the plots indicates the period of elevated VHE emission reported by Beilicke et al. (2012). See the text for more detailed descriptions of each panel of the light curves. 
(Right) Measured correlated flux density of M87. Circles and crosses indicate the correlated flux densities observed in 2012 (this work) and 2009 (Doeleman et al., 2012), respectively. Errors are 1$\sigma$. The blue line and light-blue region are the best-fit models for the 2012 data and the three uncertainties on it, respectively, and the black line and gray region are the best fits for the 2009 data and the uncertainties on it, respectively.
\label{fig:M87radiogamma}}
\end{figure}

In addition, Hada et al. (2014) also conducted VERA astrometry for the M87 core at six epochs during the VHE flaring period, and they detected core shifts between 22 and 43 GHz, the mean value of which is similar to that obtained by previous astrometric measurement (Hada et al., 2011). A clear frequency-dependent evolution of the radio core flare at 43, 22, and 5 GHz was also observed (left panel of Figure ~\ref{fig:M87radiogamma}); the radio flux density increased more rapidly at higher frequencies and larger amplitudes, and the light curves clearly showed a time lag between the peaks at 22 and 43 GHz in VLBA observations, indicating that a new radio-emitting component was created near the BH in the period of the VHE event and that the component subsequently propagated outward with a progressive decrease in the synchrotron opacity. By combining the obtained core shift and time lag, we can estimate the apparent speed of the component propagating through the opaque region between the cores at 22 and 43 GHz. A subluminal speed (less than 0.2c) was obtained for this component. This value is considerably lower than the superluminal ($\sim 1.1c$) features of the core during
the prominent VHE flaring event of 2008, suggesting that stronger VHE activity can be associated with the production of a higher Lorentz factor jet in M87. \\



\textbf{\textsl{Magnetic fields strength at the jet base }}\\

\begin{figure*}[htbp]
\centering
\includegraphics[width=80mm]{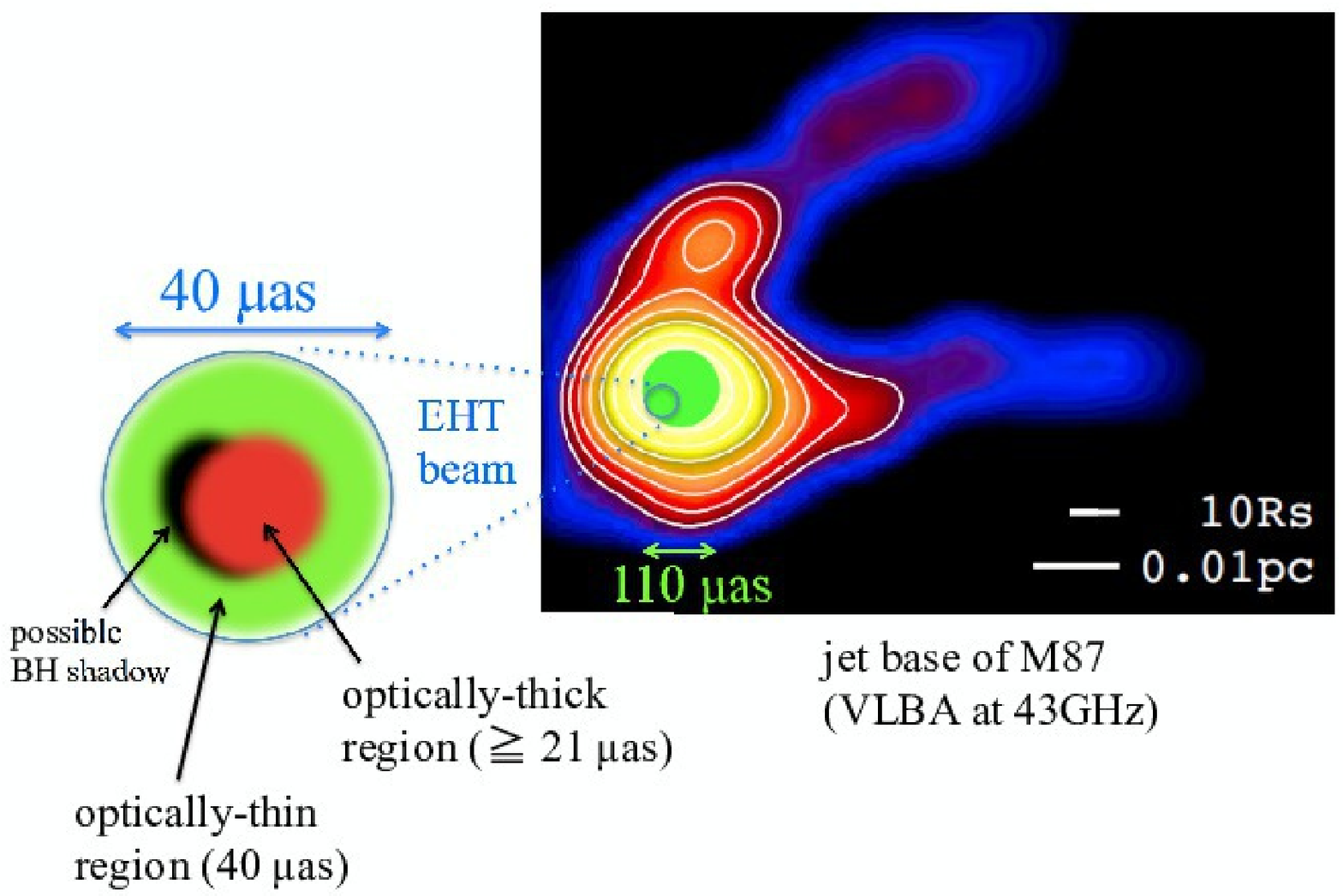}
\includegraphics[width=80mm]{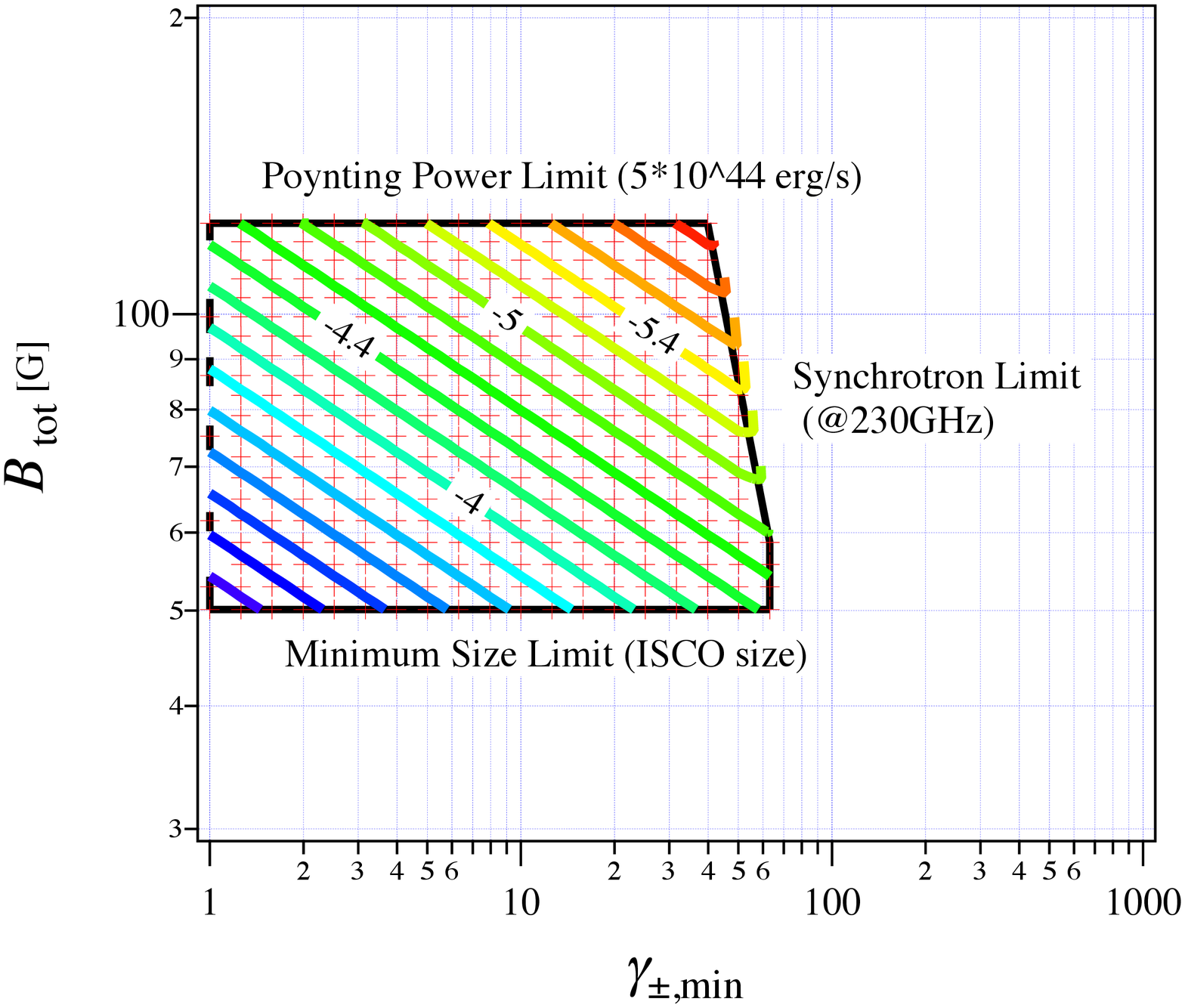}
\caption{
(Left) Illustration of the jet base of M87 on the EHT region scale. The right panel shows an actual VLBA image of M87 at 43 GHz (adopted from Hada et al. (2013)). The yellow-green circle shows the one-zone region, which has a diameter of 110 $\mu$as, investigated by Kino et al. (2014). The EHT region detected by Doeleman et al. (2012) is shown as a blue circle. The left-side figure shows an illustration of the internal structure of the EHT region. The red-colored region represents an SSA-thick compact region in the SSA-thin region. The black-colored region conceptually shows a possible BH shadow image. On the basis of the smallness of the closure phase reported by Akiyama et al. (2015), a certain level of symmetry is maintained in this figure. 
(Right) Allowed region of minimum Lorentz factor of nonthermal electrons/positrons ($\gamma_{\rm min}$) and the energy density of magnetic fields ($B_{\rm tot}$) (red crosses in the black trapezoid). The colored contour lines show the allowed values of log $U_{\pm}/U_{B}$. The tags log ($U_{\pm}/U_{B}$) = -4, -4.4, -5, and -5.4 denote reference values.
\label{fig:M87-kino15}}
\end{figure*}

Magnetic field ($B$-field) strength at the jet base of M87 is an important physical quantity that is to be determined. Higher-resolution VLBI measurements can facilitate the determination of the absolute strength of the B-field at the jet-launching site on the basis of the standard theory of synchrotron self-absorption (SSA) with less uncertainty (Marscher, 1983; Kino et al., 2014). This will provide another test for the Blandford-Znajek model, which predicts a very strong $B$-field. Kino et al. (2015) investigated the degree of magnetization at the jet base of M87 by using observational data obtained by Doeleman et al. (2012) with the EHT at 230 GHz. If the overall EHT region is fully optically thick against SSA, then the predicted field strength is approximately 300 G. The corresponding Poynting power at the EHT region is then considerably greater than the mean kinetic power of the jet, inferred from the large-scale dynamics of the jet. The difficulity associated with a very large Poynting power can be overcome if the EHT region is considered to be composed of both SSA-thick and SSA-thin regions (left panel of Figure~\ref{fig:M87-kino15}). In this two-zone model, the magnetic energy dominates in the SSA-thick part of the EHT-detected region. The magnetic energy dominates even in the case of the fully SSA-thin EHT region, unless protons in the region are relativistic.

To test the two-zone model proposed by Kino et al (2015), inclusion of baselines longer than 3 G$\lambda$ is necessary. For example, the visibility amplitude of the SSA-thick component is detactable above 3 G$\lambda$ at a wavelength of 1.3 mm. Obtaining polarimetric observations with ALMA is clearly one of the promising first steps for improving the blending of multiple polarized substructures, and for examining the magnetic field geometry in detail. Obviously, in the final stage, submm VLBI polarimetric observations are indispensable for preventing contamination from the extended region.

\subsubsection{Blazars \label{sec:agn}}

\textbf{\textsl{Radio-$\gamma$ connection in blazars}}\\

The location and the physical properties of VHE $\gamma$-ray emission from relativistic jets in AGNs are crucial for investigating the jet-launching mechanism. Monitoring programs for $\gamma$-ray loud blazars have been conducted with VERA and the KVN. Blazars are extremely bright AGNs with variable jets. Here, we
review and highlight the findings of
two programs in which the radio-$\gamma$ connection has been intensively investigated: the Gamma-Ray Emitting Notable AGN Monitoring by Japanese VLBI (GENJI) and Interferometric Monitoring of Gamma–Ray Bright AGNs (iMOGABA) programs.


\bigskip

{\bf {\em 1. GENJI programme}}\\

Since October 2010, we have been conducting intensive VLBI monitoring of $\gamma$-ray AGNs, primarily at
1.3\,cm and typically biweekly. We termed this monitoring program ``\textit{\textbf{G}amma-Ray \textbf{E}mitting \textbf{N}otable AGN 
Monitoring by \textbf{J}apanese VLB\textbf{I} (\textbf{GENJI})}" \citep{2013PASJ...65...24N}.  
This program is aimed at specifying the radio counterpart of $\gamma$-ray emissions using VERA, which is operated by the Mizusawa VLBI Observatory of the NAOJ. For this program, we selected ten objects: DA\,55, 3C\,84, OJ\,287, M87, PKS\,1510$-$089, DA\,406, NRAO\,530, BL\,Lac, CTA\,102, and 3C\,454.3. For VERA's Galactic maser astrometry programs, observations are performed throughout the year. For each VERA observation, bright AGNs are observed as calibrators every 80 min. By using these calibrator slots, we can obtain a snapshot of each object. We have therefore been able to perform intensive VLBI observations for each object throughout the year.

The GENJI program, which focuses on the time resolution, provides data that complement other VLBI monitoring projects, such as MOJAVE\footnote{http://www.physics.purdue.edu/MOJAVE/}
\citep[e.g.,][]{2009AJ....138.1874L} and the Boston University Blazar Projects\footnote{http://www.bu.edu/blazars/VLBAproject.html}
\citep[e.g.,][]{2010ApJ...710L.126M}  in the Northern Hemisphere, and Tracking Active Galactic Nuclei with Austral Milliarcsecond Interferometry (TANAMI)\footnote{http://pulsar.sternwarte.uni-erlangen.de/tanami/}
\citep[e.g.,][]{2010A&A...519A..45O}  in the Southern Hemisphere. Additionally, GENJI observations do not negatively impact the VERA project observations. Therefore, the GENJI program is a type of commensal observation program \citep[e.g., see][]{2015aska.confE..51F}, and the commensality is a key reason for continuing the program.

%

\begin{figure}[htbp]
	\begin{center}
	\begin{minipage}{0.46\hsize}
		\begin{center}
			\includegraphics[width=0.87\linewidth]{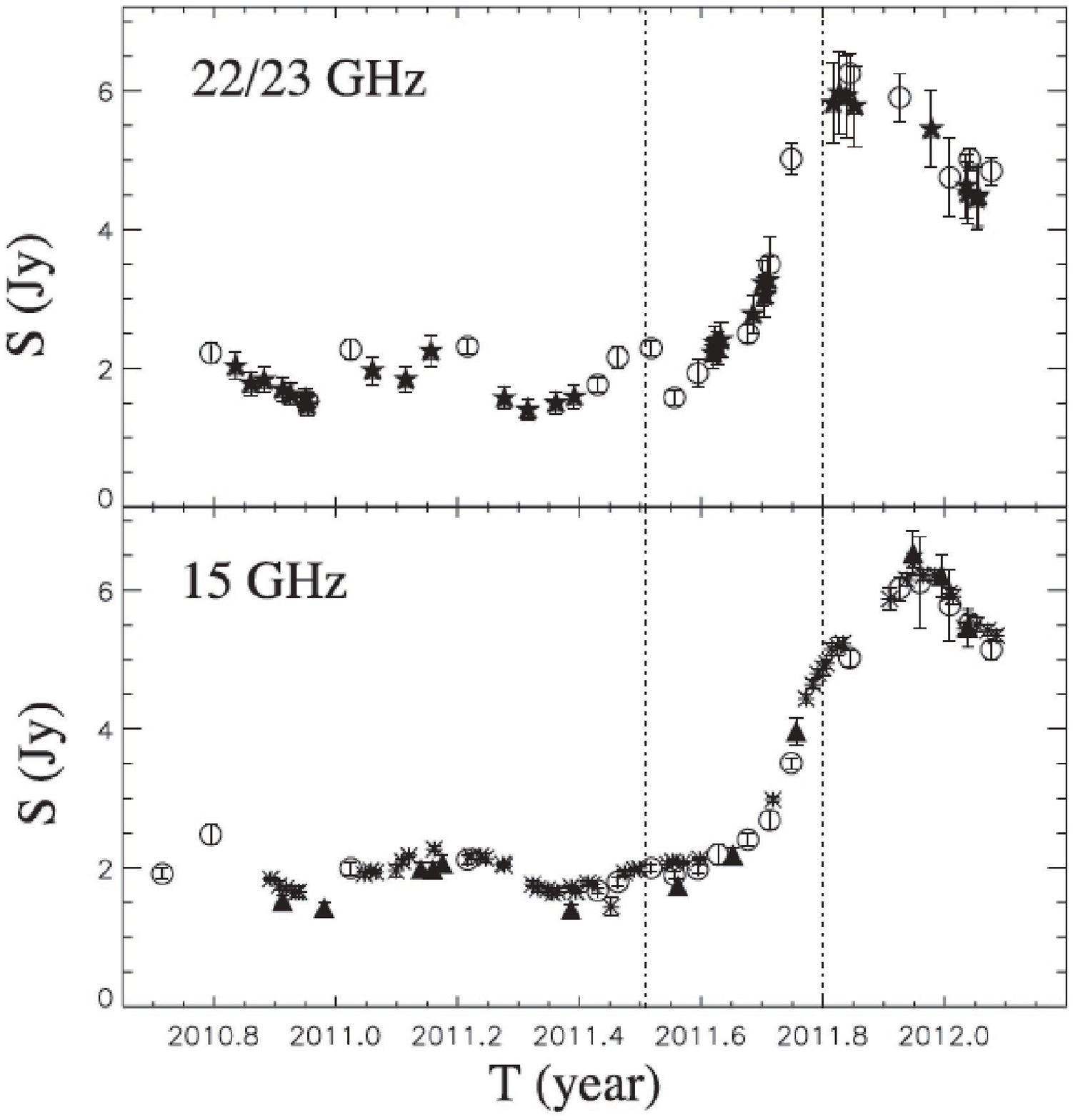}
		\end{center}
	\end{minipage}
	\begin{minipage}{0.46\hsize}
		\begin{center}
			\includegraphics[width=0.9\linewidth]{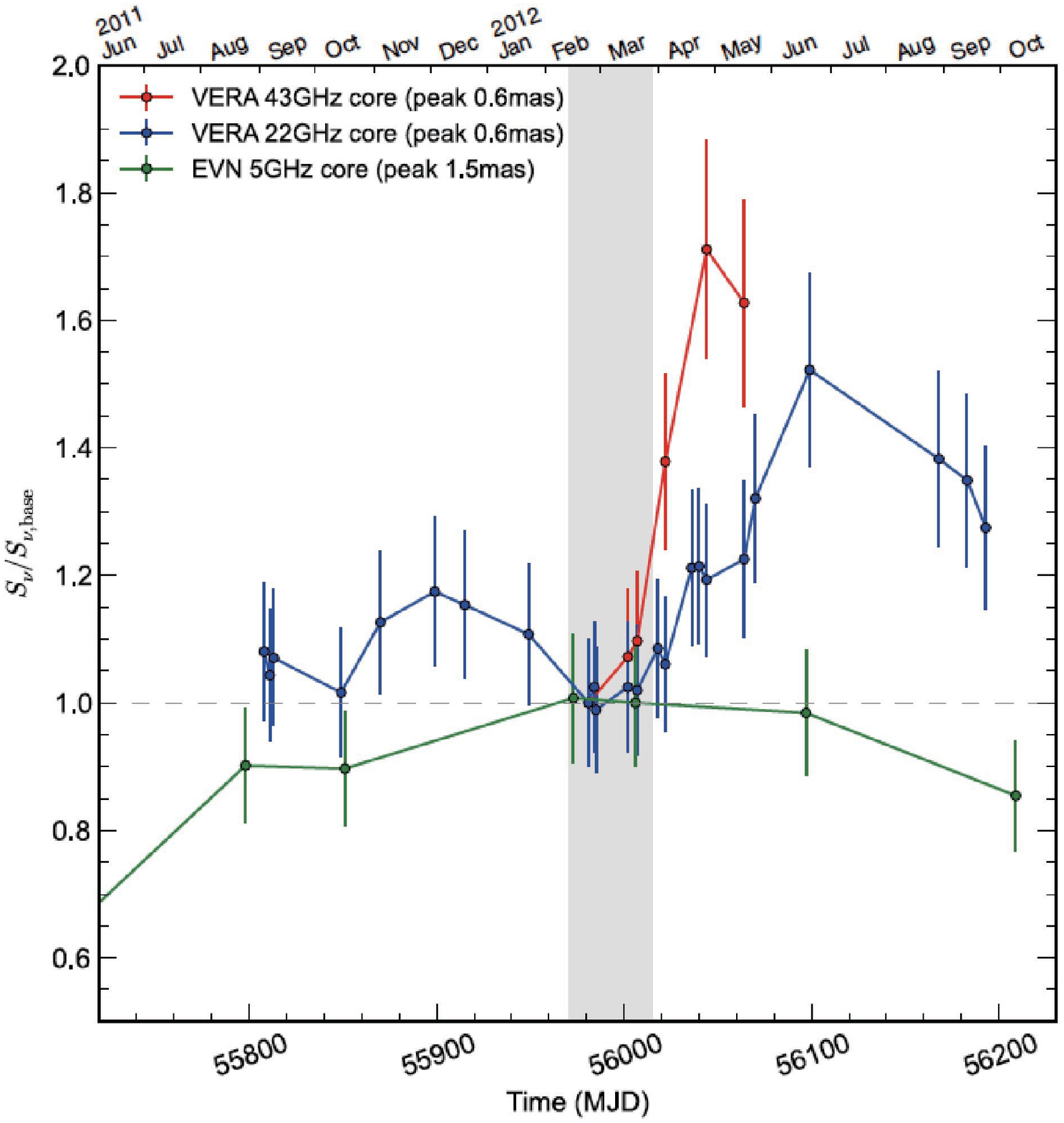}
		\end{center}
	\end{minipage}
	\caption{
	Recent results of GENJI. (Left) Radio light curves at 22 and 23\,GHz (upper panel) and 15\,GHz (lower panel) of PKS 1510$-$089. The empty circles represent Fermi Gamma-ray Space Telescope AGN Multi-frequency Monitoring Alliance (F-GAMMA) data at 15 and 23 GHz, the filled stars are VERA data at 22 GHz, and the asterisks and filled triangles represent OVRO and MOJAVE data at 15\,GHz, respectively. The vertical lines indicate the time of $\gamma$-ray flares \citep[from][]{2013MNRAS.428.2418O}. (Right) Normalized radio light curves of the M87 core at 43, 22, and 5\,GHz. The light curves at 43, 22 (VERA), and 5\,GHz (EVN) are normalized to the flux values of MJD 55981, MJD 55981, and MJD 55936, respectively. The vertical shaded area over the plots indicates the period of the elevated VHE state \citep[from][]{2014ApJ...788..165H}.
	\label{fig:genji}}
	\end{center}
\end{figure}

The continuous intensive monitoring by VERA facilitated VLBI follow-up observations immediately after high-energy flares occurred in several GENJI objects. 
Follow-up observations showed a delayed increase in the radio core flux for GeV- or VHE $\gamma$-ray flares \citep[e.g.,][see Figure~\ref{fig:genji}]{2013MNRAS.428.2418O, 2014ApJ...788..165H}.

In East Asia, a new mm VLBI facility, KaVA, consisting of seven radio telescopes and offering both long and short baselines, will provide comprehensive ($u,v$)-coverage; moreover, KaVA is sufficiently powerful to reveal complex extended structures of AGNs \citep{2014PASJ...66..103N}. Recently, in the GENJI program, observations of 3C\,84, M87, and PKS\,1510$-$089 have commenced; the observations are being made in coordination with similar observations with KaVA on the basis of a collaborative key science project, which is one of three key science projects (on stellar evolution, star formation, and AGNs). We term these observations KaVA--GENJI observations. Because of the improvement of the image dynamic range of KaVA, we expect to obtain information on not only the relationship between flux variations at radio energies and high energies but also the kinematics of jet components possibly associated with high-energy flaring phenomena for identifying the radio counterpart of the $\gamma$-ray emission.\\

{\bf \em{2. iMOGABA}}\\

The iMOGABA project (Lee et al., in preparation) was launched in 2015 as a key science program of the KVN. This project uses the KVN for monthly interferometric monitoring of more than 30 $\gamma$-ray bright AGNs at 22, 43, 86, and 129\,GHz simultaneously (see http://sslee.kasi.re.kr for preliminary results). The project is especially aimed at determining the potential connection between $\gamma$-ray outbursts and the formation of new jet components through the investigation of the potential correlation of $\gamma$-ray light curves with the brightness and milliarcsecond-scale structures of the inner jets. The monitoring cadence of a month and the observing frequencies of 22--129\,GHz make this project unique and appropriate for studying $\gamma$-ray flaring AGNs.

The iMOGABA observations consist of ten 24-h VLBI observations per year for the period 2015--2017. In every observation, all target sources are observed, yielding snapshot images in four frequency bands for an on-source integration time of 30\,min to 1\,h. The first priority of this project is to measure changes in jet structures (and hence the formation of new jet components) of approximately 30 $\gamma$-ray bright AGNs on the milliarcsecond scale when they are flaring in the $\gamma$-ray band. The target sources are chosen to have flares in the $\gamma$-ray band, and they are detected with the Fermi Large Area Telescope (LAT) on the Fermi $\gamma$-ray Space Telescope. Since most of them are very bright at the operating frequencies of the KVN, they can be detected on KVN baselines for coherence times in the range 20--100\,s at the corresponding frequencies. However, we have confirmed that it is possible to improve the fringe detection sensitivity with an integration time longer than canonical coherence times by using a frequency phase transfer (FPT) technique \citep[][; see also \S 3.4.2]{2015JKAS...48..237A}. Therefore, we observe each source with several 5-min-long scans distributed over its local sidereal time range to ensure uniform ($u,v$)-coverage. With these snapshot mode observations, we can detect the sizes and flux densities of compact jets. However, some target sources with complicated jet structures at KVN's resolution should be carefully imaged. This problem may be resolved by performing full-track observations for each source. The second priority is to measure flux density changes of AGN jets on the milliarcsecond scale. Since atmospheric fluctuations are very high at mm wavelengths, careful amplitude calibration should be conducted for individual observations. For typical VLBI imaging observations, the amplitude can be self-calibrated with closure amplitudes. However, because there are only three stations in the KVN, amplitude self-calibration is not possible. Therefore, we conduct very frequent measurements of atmospheric opacity (every hour) and the antenna system noise temperature (in every scan). More importantly, constant monitoring and adjustment for antenna pointing (and/or focus) are performed.

In addition to the iMOGABA project, single-dish rapid response observations (SD RROs) for flaring AGNs are also conducted. The SD RROs may consist of twelve 7-h observations per $\gamma$-ray flaring source (every week for three months after triggering) with two KVN SD telescopes at 22, 43, and 86\,GHz (and/or 129\,GHz) in dual polarization. We expect to study one or two sources per year through SD RROs. First results of SD RROs have been reported for one of the brightest quasars, 3C~279 \citep{2015JKAS...48..257K}.\\


\begin{figure}[h]
\centering
\includegraphics[width=55mm]{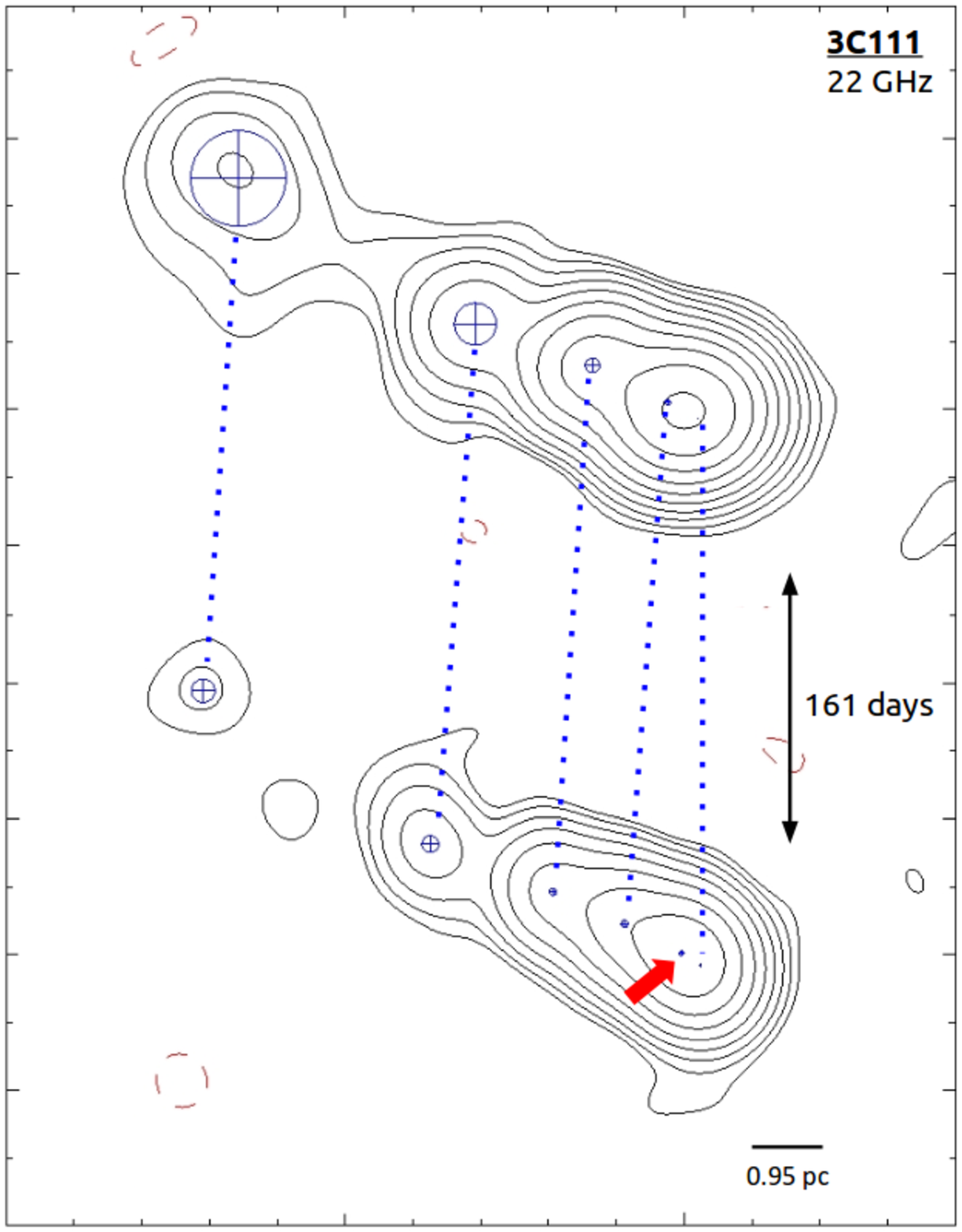}
\includegraphics[width=55mm]{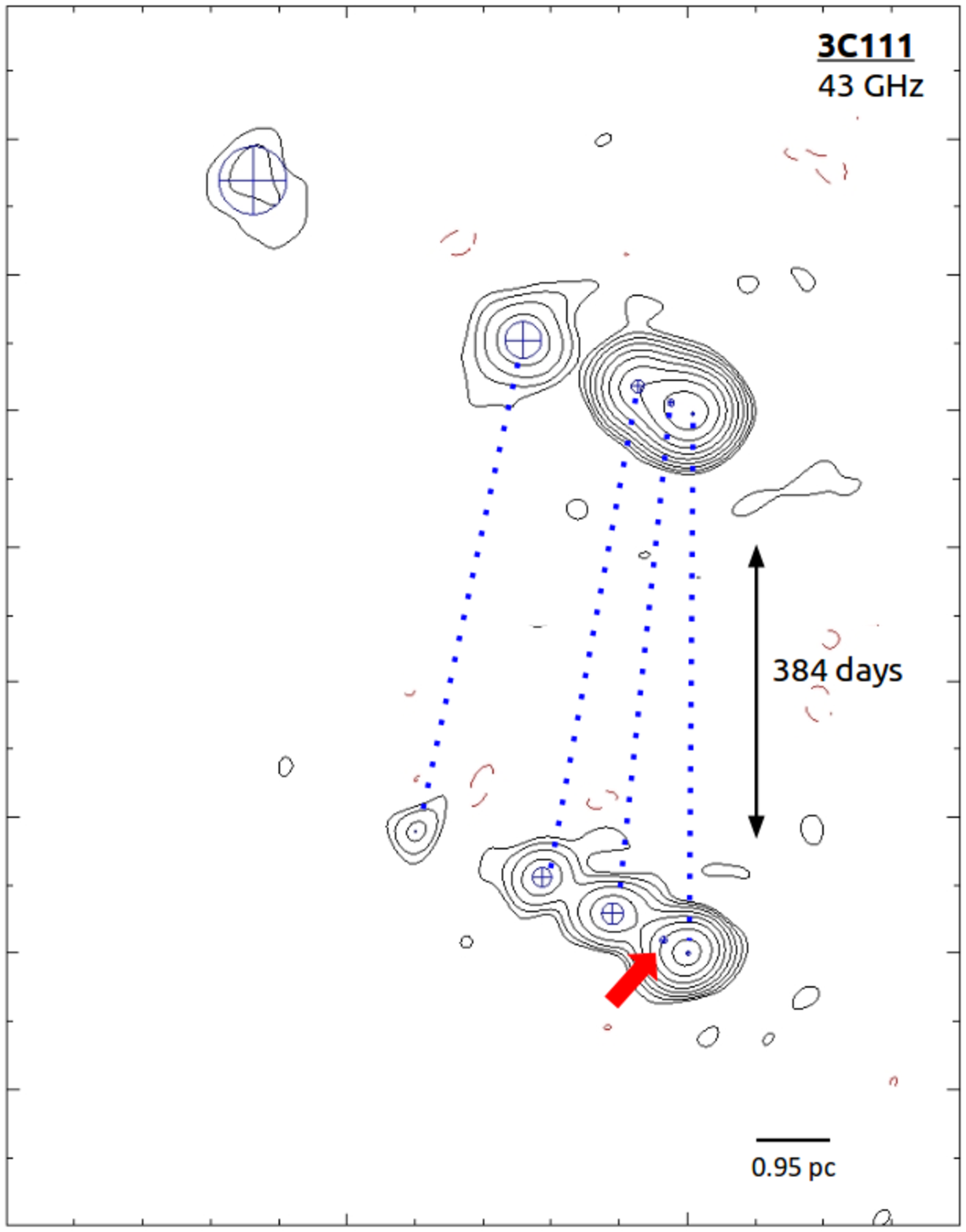}
\caption{
Jet kinematics of 3C 111 observed at 22\,GHz (left) and 43\,GHz (right) with KaVA; maps obtained in 2014 and 2015 and separated by 161 and 384 days, respectively, were used. Cross-identified components are connected by dashed lines, and red arrows mark a new component. The fast proper motion, which is visible by eye, reaches approximately 5$c$. (Figures from \citealt{2015JKAS...48..299O}.) \label{fig:pagan}}
\end{figure}

{\bf \em{Polarimetry: Probing magnetic field geometry}}\\

In the current jet theory, on the basis of GRMHD studies, the interaction between the magnetic field and strong gravity is considered as the source of jet formation \citep[e.g.,][]{2006MNRAS.368.1561M},  and observations of the magnetic fields are crucial to understanding how jets are accelerated and collimated. VLBI polarimetry is an established method to probe the magnetic field structure of relativistic jets on the parsec scale. Observations of the polarization position angle in the radio (and optical) band have been used to probe the projected component of the magnetic field, because the emission from the jet is due to synchrotron radiation; the polarization position angle is perpendicular to the direction of the magnetic field in the optically thin case. The RM has been used to investigate the line-of-sight component of the magnetic field, and in particular, the RM gradient transverse to the jet direction has been considered as an indicator of the presence of a toroidal/helical magnetic field structure, which is frequently invoked by many MHD jet models. Existence of such an RM gradient was first discovered toward 3C\,273 (\citealt{2002PASJ...54L..39A}) and it has subsequently been detected for parsec-scale jets of several blazar sources  (e.g., \citealt{2008ApJ...682..798A}, \citealt{2005BaltA..14..363G}, \citealt{2015MNRAS.450.2441G}, \citealt{2008ApJ...681L..69G}, \citealt{2012AJ....144..105H}). 

Furthermore, observations of circular polarization have also been used to probe the magnetic field structure. Generation of circular polarization is most likely through Faraday conversion of linear polarization to circular polarization, and the systematic change in the sign of circular polarization is believed to be directly associated with the presence of helical jet B-fields \citep[e.g., ][]{2008MNRAS.384.1003G}. Furthermore, because comparison between the amount of Faraday rotation and conversion provides limited information on the plasma contents (normal plasma and/or pair plasma) of the relativistic jet \citep[e.g.,][]{2009ApJ...706.1253H}, which is another fundamental factor that contributes to our understanding of the relativistic jet, polarization observations are very useful for investigating jet physics. Future submm VLBI polarimetric observations can push these efforts forwards to the higher angular resolution. The recent discovery of very high fractional polarization toward Sgr A* from EHT observations \citep{2015ApJ...813..132J} is encouraging, despite Sgr A* being very weakly polarized at cm wavelengths. Such a crucial submm VLBI polarimetric observation has been conducted only for Sgr A*. In particular, submm VLBI polarimetric observations can provide a unique opportunity for probing the magnetic field structure in the innermost/jet base region for the first time.
\\

\begin{figure}
\begin{center}
\includegraphics[width=15cm]{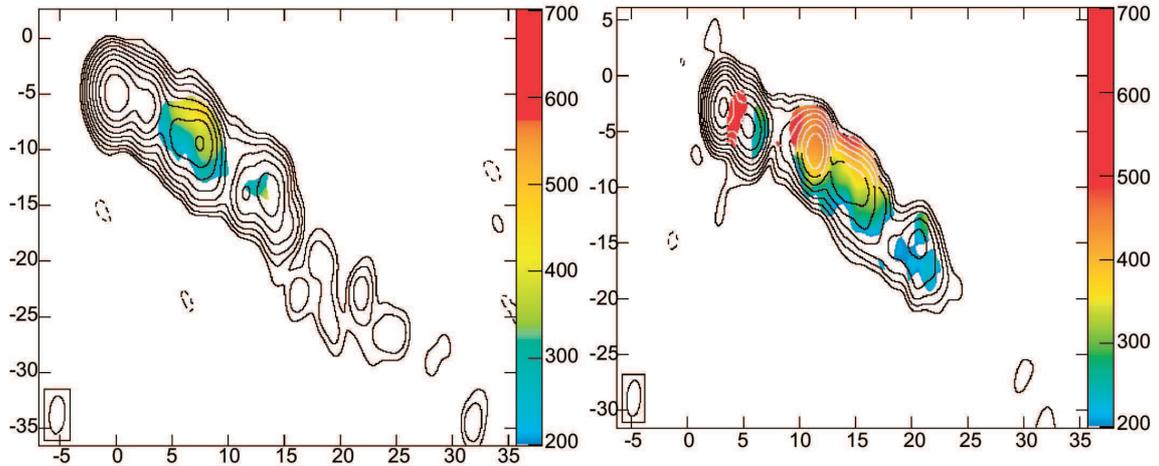}
\end{center}
\caption{(Left) RM distribution of 3C 273 (color) superposed on the total intensity (contour) in 1995 \citep{2002PASJ...54L..39A}. (Right) Same as the left panel, but in 2003 \citealt{2008ApJ...675...79A}. RM gradients across the jet are clearly visible. It is expected that submm VLBI observations can facilitate the probing of the upper stream of the jet.
}
\label{fig:3C273RM}
\end{figure}

The  \emph{P}lasma-physics of \emph{A}ctive \emph{Ga}lactic \emph{N}uclei (PAGaN)  project uses the KVN and KaVA jointly for observing selected radio AGNs. This project is aimed at determining the internal plasma physical conditions of AGN outflows. Because AGN jets are generally optically thin synchrotron emitters, deep insights can be obtained through (a) systematic time-resolved VLBI mapping of outflow structures and their temporal evolution and (b) systematic multi-frequency observations (which provide spectral index and optical depth information), and (c) systematic polarimetry (for probing magnetic fields of AGN jets). The PAGaN project started in 2013, and the following main results have been obtained.

\citet{2015JKAS...48..285K} observed seven nearby radio-bright AGNs at frequencies of 22, 43, 86, and 129\,GHz in dual-polarization mode with the KVN. Their observations constrained the apparent brightness temperatures of jet components and radio cores of the AGN samples to values $>10^{8.01}$\,K and $>10^{9.86}$\,K, respectively. The degree of linear polarization  $m_{L}$ was relatively low overall ($<$10\%), indicating suppression of polarization by strong turbulence in the jets. An exceptionally high degree of polarization was found in a jet component of BL Lac at 43\,GHz, with  $m_{L}$ being approximately 40\%. If a transverse shock front propagating downstream along the jet is assumed, the shock front being almost parallel to the line of sight can explain the high degree of polarization.


\citet{2015JKAS...48..299O} observed eight selected AGN at 22 and 43\,GHz in single polarization (LCP) between March 2014 and April 2015 with KaVA. Each source was observed for 6 to 8 hours per observing run to obtains a good ($u,v$) coverage. They obtained a total of 15 deep high-resolution images permitting the identification of individual circular Gaussian jet components and three spectral index maps of BL Lac, 3C~111 and 3C~345 from simultaneous dual-frequency observations. The spectral index maps show trends in agreement with general expectations --- flat core and steep jets --- while the actual value of the spectral index for jets shows indications for a dependence on AGN type. They analyzed the kinematics of jet components of BL Lac and 3C~111, detecting superluminal proper motions with maximum apparent speeds of about $5c$ (see also Figure~\ref{fig:pagan}). This constrains the lower limits of the intrinsic component velocities to $\sim0.98c$ and the upper limits of the angle between jet and line of sight to $\sim$ 20$^{\circ}$. In agreement with global jet expansion, jet components show systematically larger diameters $d$ at larger core distances $r$, following the global \emph{linear} relation $d\approx0.2r$ (which is consistent with conical jets), albeit within substantial scatter.

In addition, complementary KVN single-dish observations of \citet{2015ApJ...808L..26L} provided constraints on the
minute-to-hour scale variability of four radio-bright AGNs. The PAGaN project is continuing, and papers reporting new observations obtained using the KVN and KaVA are in preparation. Major improvements are expected in the near future from (a) the partial addition of a geodetic antenna at Sejeong to the KVN and (b) the commissioning of dual-polarization receivers at Iriki and Mizusawa antennas; these developments will increase the number of antennas that can be used for polarimetry from three to five.
\\

{\bf \em{Astrometry: Shedding light on radio core location}}\\


The radio jet bases (radio cores) located at the upstream ends of AGN jets in radio images have been considered to be opaque surfaces against SSA or free-free absorption, and this is supported by frequency-dependent offsets of radio core positions detected in VLBI observations \citep[e.g.,][]{1983AJ.....88.1133M}. Furthermore, the radio core of M87 observed at a mm wavelength was found to be located only approximately 0.01\,pc from the upstream end of the jet through VLBI astrometry \citep{2011Natur.477..185H}. On the basis of light curve studies of radio cores at radio-to-GeV $\gamma$-ray wavelengths, one of the leading hypotheses suggests that the radio cores of blazars correspond to stationary standing shocks located from several parsecs to approximately 10\,pc downstream of the central engines.

For understanding AGN jet formations and propagation, it is crucial to investigate the origin of such discrepancies in the locations of radio cores between radio galaxies and blazars and their stationarity. For this purpose, we started performing blazar radio core astrometry for the most nearby high-synchrotron peaked blazars in 2011 using VERA, which is dedicated to radio astrometry; Mrk~421 ($z$=0.031) and Mrk~501 ($z$=0.034) were the first targets.

\begin{figure}[htbp]
	\centering
	\includegraphics[width=0.8\linewidth]{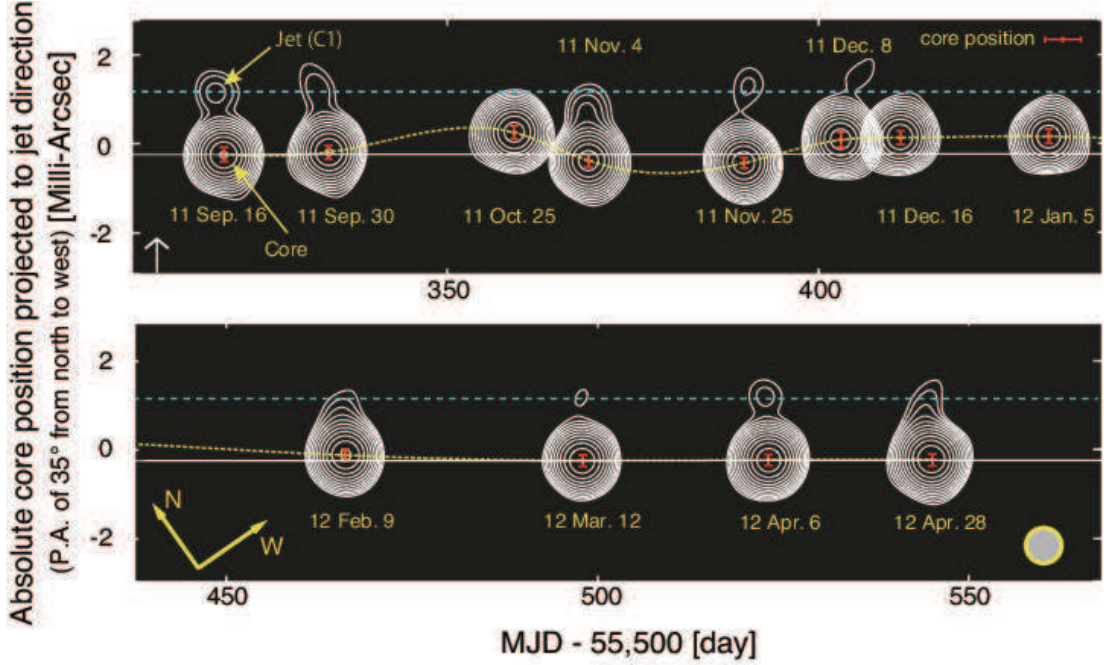}
	\caption{
	Radio core positions projected onto the jet axis, and radio images of Mrk 421. These images are overlaid on their corresponding absolute core positions. The white arrow for September 2011 indicates the date of the large X-ray flare. The white solid and blue dashed lines indicate the positions of the radio core and the jet component (C1) as measured on 2011 September 16, respectively. The yellow dashed line represents the cubic spline interpolation of each core position \citep[from][]{2015ApJ...807L..14N}.\label{fig:mrk421_wander}}
\end{figure}

During our observations of Mrk 501, which was relatively quiescent at high energiies, the radio core was stationary to within an astrometric accuracy of approximately 0.2\,mas \citep{2015PASJ...67...67K}. By contrast, the radio core of Mrk 421 showed noticeable wandering behavior within the seven months following
a large X-ray flare \citep[][see figure \ref{fig:mrk421_wander}]{2015ApJ...807L..14N}. A few similar observations of VHE $\gamma$-rays have also been performed for the radio galaxy M87 during its active state \citep{2009Sci...325..444A, 2014ApJ...788..165H}.  However, the radio core was stationary (moving less than  $40~R_\mathrm{s}$). The structured jet model offers a possible explanation for the discrepancy between M87 and Mrk 421.

Although we consider that the wandering behavior of the radio core of Mrk 421 can be naturally explained on the basis of the internal shock scenario, which is a leading model for explaining the blazar phenomenon, this new finding is expected to be a new manifestation associated with high-energy flares, and to offer crucial information about the innermost region of the jets for resolving the key question of jet origin regardless of its mechanism. Furthermore, the new finding also suggests that such a flare-associated radio core must be located
at least 3.5\,pc or $10^5$ $R_\mathrm{s}$ downstream of the persistent radio core. This is consistent
with the location of the radio core of the flat-spectrum radio quasar CTA 102, determined by the image alignment method at frequencies up to 86\,GHz  \citep{2015A&A...576A..43F}.

We now intend to perform multi-epoch VLBI astrometry for these objects at multiple frequencies to understand the behavior and location of the radio core of blazars. Along with the development of new techniques on phase referencing  \citep[e.g.,][]{2011PASJ...63..375J},  short-mm VLBI astrometry is expected to provide new insights into AGN jets.

\subsubsection{Young radio Sources}

In this subsection, we focus on the radio source 3C\,84 ($z$=0.0176), which is the best example of a young radio source. Because of its brightness and proximity, 3C\,84 is one of the best-studied radio sources. From 2005, increased activity has been detected, and VLBI observations have revealed that the flux density increase originated in the central parsec-scale core and accompanied the ejection of a long-lived new jet component (Nagai et al. 2010).
\\

{\bf \em{Jet collimation profile}}\\

\begin{figure}
\begin{center}
\includegraphics[width=15cm]{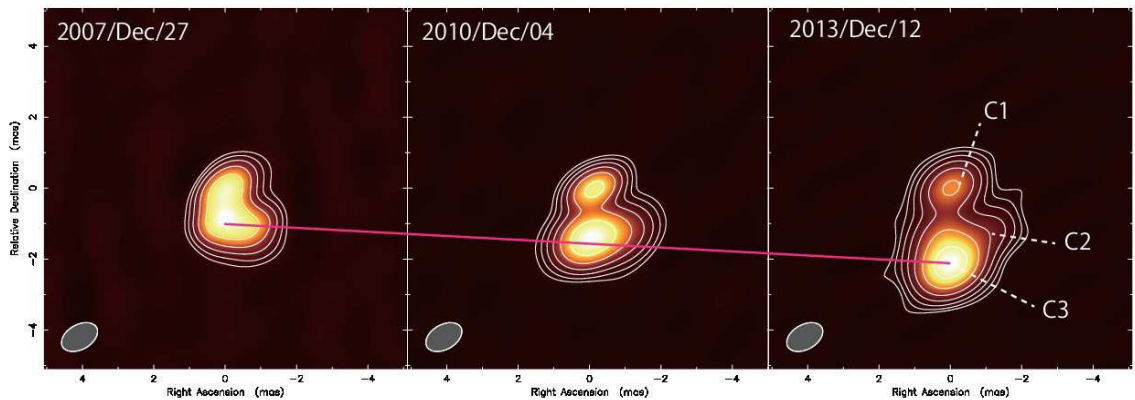}
\end{center}
\caption{VERA images of 3C\,84, obtained on 2007 December 27, 2004 December 04, and 2013 October 12. The synthesized beam had a size of $1.1\times0.7$\,mas at a position angle of $-60^{\circ}$. The contours are plotted at the level of 111\,mJy $\times (1, 2, 4, 8, 16, 32, 64)$. The peak intensities are 3.9, 6.4, and 10.3\,Jy~beam$^{-1}$ for 2007, 2010, and 2013 images, respectively.
}
\label{fig:3C84VERA}
\end{figure}

\begin{figure}
\begin{center}
\includegraphics[width=10cm]{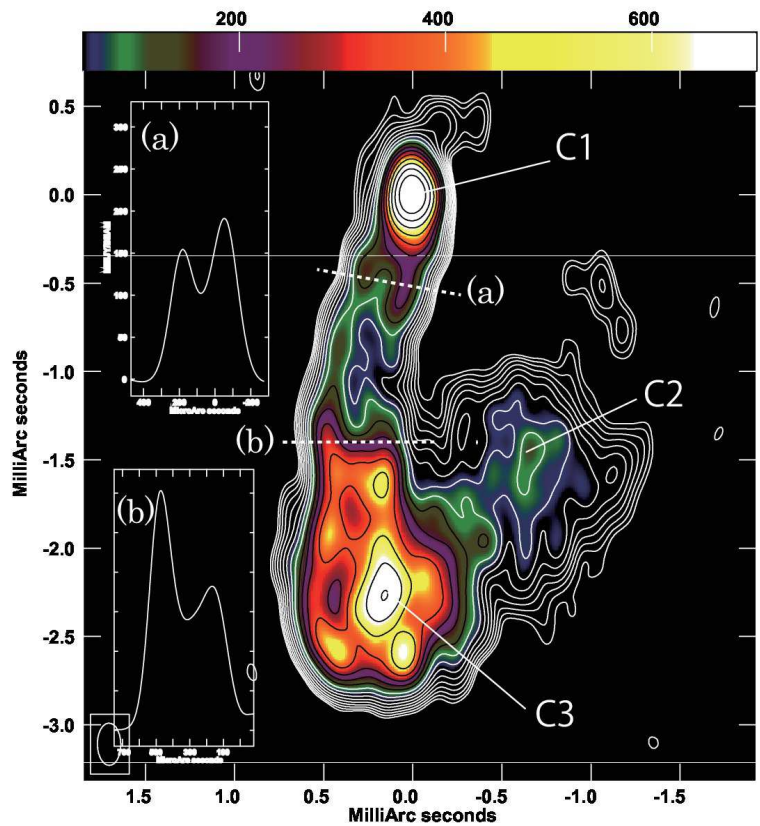}
\end{center}
\caption{A 43\,GHz VLBA image of 3C\,84. The contours are plotted at the level of 5.43 $\times$ ($-$1.41, 1, 1.41, 2.83, 4, 5.66, 8, 11.3, 16, 22.6, 32, 45.3, 64, 90.5, 128, 181, 256) mJy beam$^{−1}$.  The insets (a) and (b) show sliced profiles across the jet at the positions indicated by the broken lines in the figure.
}
\label{fig:3C84VLBA}
\end{figure}

VLBI observations have revealed that the flux increase of 3C\,84 originated in the central parsec-scale core and accompanied the ejection of a new jet component, namely C3 \citep[][ Figure \ref{fig:3C84VERA}]{2010PASJ...62L..11N, 2012MNRAS.423L.122N}. The morphology of C3 resembles that of a hot spot in the radio lobe, with a size smaller than approximately 1\,pc. Notably, the jet between C1 and C3 was resolved in the transverse direction and a clear limb-brightened structure was detected by performing VLBA 43-GHz observations \citep[][, Figure \ref{fig:3C84VLBA}]{2014ApJ...785...53N}. This facilitates the study of the jet width ($W_{\rm j}$)  as a function of distance, i.e., the jet collimation profile. This profile is a key factor that differentiates theoretical models of jet formation. The observed profile of 3C\,84 is  $W_{\rm j} \propto r^{0.25\pm0.03}$ \citep{2014ApJ...785...53N}. This is a rather collimated profile, unlike that observed for M87, which has a parabolic shape at 10$^{2-5}$\,Rs \citep{2012ApJ...745L..28A, 2013ApJ...775...70H}. Because a deeper inner region was resolved for M87, it is crucial to study the inner-jet collimation properties of 3C\,84 with higher angular resolution for performing a comparison. Accordingly, mm/submm VLBI is a powerful tool to probe the inner region of 3C\,84 and other nearby radio galaxies. \\

{\bf \em{Magnetic fields}}\\

Another important test for the jet production model is the study of the magnetic field structure. The use of mm/submm VLBI can help resolve the jet collimation region ($\sim100$--1000\,R$_{s}$), and the magnetic field morphology of the region can be compared with numerical simulations of the morphology, thereby enabling us to distinguish between models of jet production. A diﬃculty in probing magnetic fields in radio galaxies is that they show no or very little polarization in parsec-scale jets at cm wavelengths \citep{2005A&A...433..897M}. The absence of polarized flux in radio galaxies at cm wavelengths is probably because of depolarization. An observer viewing an AGN nearly edge-on is likely to see a dense ionized gas such as narrow emission line clouds, broad emission line clouds, and accretion flow. The gas acts as a Faraday screen, leading to a large RM \citep*{2002ApJ...566L...9Z}. If the Faraday screen is suﬃciently tangled within the observing beam, the polarized emission from different source segments will be cancelled out because of the spatially varying RM (beam depolarization effect). Indeed, recent mm and submm polarimetric observations performed using connected interferometry have successfully detected polarization from M87 \citep{2014ApJ...783L..33K} and 3C~84 \citep{2014ApJ...797...66P}, and the obtained
high RM ($\sim10^{5}$ rad m$^{-2}$) accounts for the absence of polarization at cm wavelengths because of depolarization\footnote{The same study is ongoing to measure RM of Centaurus A with ALMA (PI: H.\ Nagai).}.
Therefore, mm/submm VLBI will allow the magnetic field morphology to be probed in the jet collimation region.

As described, the distribution of the RM across the jet width (i.e., its gradient across the jet) has been reported for more than a dozen AGN jets \citep[e.g., ][]{2002PASJ...54L..39A, 2008ApJ...675...79A, 2015MNRAS.450.2441G}, and it has been interpreted as reflecting the presence of a helical magnetic field. Stronger evidence can be provided if sign changes in the RM gradient can be observed \citep[see][for more detail]{2008ApJ...675...79A}. Radio galaxies at moderate viewing angles are suitable for observing sign changes, because the contribution from the toroidal component would be considerably higher. By contrast, only one sign can be seen for blazars because of its small viewing angle. Radio galaxy jets, including 3C\,84, are the best sources for detailed studies of the RM gradient.\\

{\bf \em{ Radio-$\gamma$ connection}}\\

We note that and submm VLBI observations of nearby radio galaxies are intriguing in terms of the jet physics in connection with high energy emission. 3C\,84 and Centaurus~A were detected in high energy (HE) and VHE $\gamma$-ray bands despite the absence a strongly beamed jet \citep{2009ApJ...707...55A, 2014Sci...346.1080A, 2009ApJ...695L..40A}. The emission mechanism of $\gamma$-ray photonsin radio galaxies is also a subject of debate and not well understood. Many authors have tried to explain the HE emissions on the basis of one-zone synchrotron self-Compton models, but the derived Doppler factor is relatively low \citep[$\delta\simeq2$--4, e.g.,][]{2008MNRAS.386..945T}, which makes it difficult to explain the emissions in the framework of the misaligned blazar scenario. Furthermore, there is compelling evidence that many $\gamma$-ray--detected radio galaxies show a limb-brightened jet \citep{2007ApJ...668L..27K, 2006ApJ...641..158K, 2014ApJ...785...53N}. This is a strong indication of a structured jet, and the structure of the jet should be considered for spectral energy distribution modeling, as discussed by  \citet{2014MNRAS.443.1224T}. An interesting finding is that a sub-parsec jet of 3C\,84 showed single-ridge/central ridge brightening \citep{1998ApJ...498L.111D} in the 1990s, when the $\gamma$-ray luminosity was more than 10 times the current value; the jet currently shows double-ridge/limb brightening. As discussed by  \citet{2014ApJ...785...53N}, a change in the transverse jet structure might be related to a change in the $\gamma$-ray luminosity. In the coming ALMA-VLBI era, to acquire a better understanding of the HE emission mechanism in radio galaxies, it is crucial to study the time evolution of the transverse jet structure and its relationship with changes in the $\gamma$-ray flux. \\

{\bf \em{Radio jet feedback in young radio lobes}}\\

AGN feedback is a key ingredient in cosmological simulations of galaxy formation. In radio-quiet AGNs, the powerful AGN wind is a crucial feedback agent. The winds can expel the gas at the GC and quench star formation. AGN feedback thus regulates the correlation between the BH mass and velocity dispersions and masses of galaxy bulges. In radio-loud AGNs, the kinetic power of the jets is comparable to the energy release from the AGN winds, and therefore, the jet-induced outflows including young radio lobes also make a crucial contribution to the feedback process.

  \begin{figure}[htbp]
   \centering
   \includegraphics[width=0.7\textwidth, angle=0]{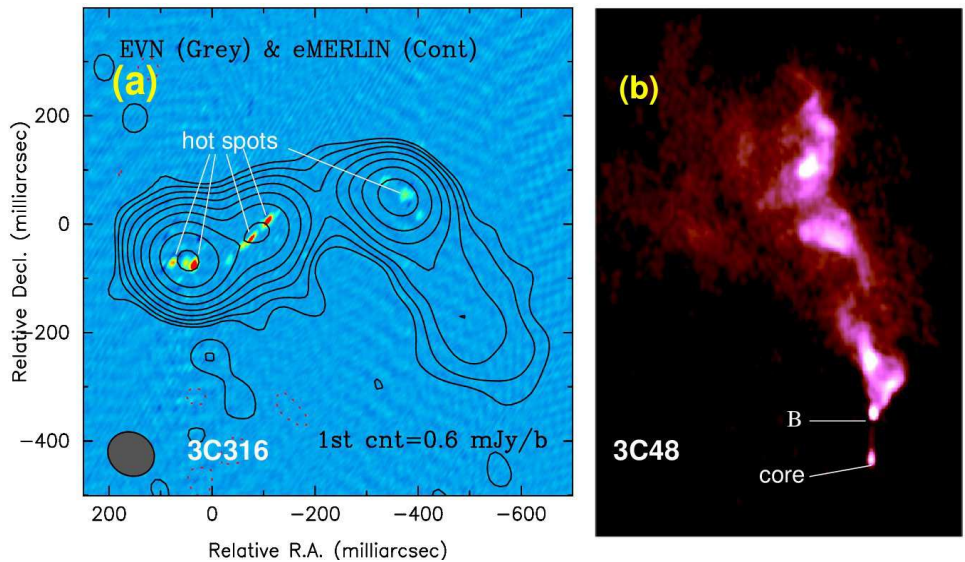} 
   \caption{\small
(a) Radio structure of the double-peaked narrow-line galaxy 3C\,316. The contours are the upgraded MERLIN (e-MERLIN) image, and the colored scale shows the EVN map \citep{2013MNRAS.433.1161A}. (b) Disrupted jet in the radio quasar 3C\,48 \citep{2010MNRAS.402...87A}.
  \label{fig:tao.an.2} }
   \end{figure}

Fast gas outflows in both warm ionized and cold neutral phases are ubiquitous in young radio sources \citep{2007NewAR..51..149B,2008MNRAS.387..639H,2015arXiv151101613M}; examples are the Giga-Hertz Peaked Spectrum Sources and compact symmetric objects (CSOs), which are often found in gas-rich host galaxies. Recent observations of double-peaked narrow-line (DPNL) AGNs, believed to be dual AGN candidates, indicate that the majority of these DPNL objects may not be genuine dual AGNs because the DPNL signature could result from narrow line region (NLR) kinematics. Figure~\ref{fig:tao.an.2}  displays the radio structure of a DPNL radio galaxy, 3C\,316. The VLBI image of the radio galaxy (on a color scale) shows the signature of strong interactions between the jets and the NLR clouds; in other words, it shows the knotty jet structure, in which the bright discrete jet knots are evidence of shock-brightening events. The broad-wing [O III] component in this source may be driven by the radiation pressure of the central AGN, and the narrow-velocity components may be ionized locally and accelerated by the bow shocks of the radio jets. A similar high-velocity [O III] line component was observed in the well-known quasar 3C\,48 \citep{2007ApJ...659..195S}, and it is spatially associated with the radio jet \citep[Figure~\ref{fig:tao.an.2}-{\it Right};][]{2010MNRAS.402...87A}. The initial section of the jet remains collimated. After the hot spot B, the radio structure is disrupted, and the main body shows a helically wiggling morphology, indicating strong interactions with the NLR interstellar medium. The jet-driven feedback in 3C\,48 is complicated. A submm component observed at the northeast end of the jet could be a star forming region, with the star formation being triggered by the impact of the jet on the local molecular clouds.

\subsubsection{Narrow Line Seyfert 1 Galaxies}

Narrow-line Seyfert 1 galaxies (NLS1s) are a subclass of AGNs, with strong permitted optical/UV Fe~II emission lines, relatively weak forbidden-line emission (i.e., $\rm [O III] 5007/H\beta < 3$). Their broad lines are narrower than those of normal broad-line Seyfert~1 galaxies with FWHM\,($\rm H\beta$) less than 2000\,$\rm km~s^{-1}$ \citep{1987ApJ...323..108O,1989ApJ...342..908G}. 
NLS1s are often thought to be young and still
evolving AGNs, and they host relatively small BH masses with high accretion rates \cite[see review by][]{2008RMxAC..32...86K}. The probability of NLS1s being radio loud is low, approximately 7\%, unlike normal broad-line AGNs \cite[e.g.,][]{2006AJ....132..531K,2006ApJS..166..128Z}. Interestingly, radio-loud NLS1s (RLNLS1s) are
inhomogeneous in their radio properties. As shown by \cite{2006AJ....132..531K}, most RLNLS1s in their sample are compact, steep spectrum sources, and hence, they are likely to be associated with compact steep-spectrum (CSS) radio sources; moreover, only a few NLS1s showed blazar-like behavior. Observational evidence has recently shown that a major fraction of RLNLS1s at the highest radio-loudnesses do display the characteristics of blazars, including large-amplitude radio flux and spectral variability, compact radio cores, very high variability brightness temperatures, enhanced optical continuum emission, flat X-ray spectra, and blazar-like spectral energy distributions  \cite[e.g.,][]{2008ApJ...685..801Y}. In the AGN correlation space, because of their small BH masses, high accretion rate, and very strong Fe II emission \citep{2006AJ....132..531K}, RLNLS1s are at the end opposite to classical radio-loud AGNs \cite[e.g.,][]{2008RMxAC..32...51S}. Therefore, RLNLS1s enable us to address some of the key questions, such as the physical conditions under which a jet can be launched, regarding the physics of jet formation at high mass accretion rates.
\\

{\bf \em{Radio $\gamma$ connection in RLNLS1s}}\\

Relativistic jets were postulated to exist in a few RLNLS1s by using VLBI observations and on the basis of the high brightness temperatures and inverted radio spectra if the RLNLS1s \citep{2007PASJ...59..703D}, and the jets were later confirmed through the detection of flaring $\gamma$-ray emission \citep{2009ApJ...699..976A,2009ApJ...707L.142A}. Previous studies based on VLBA data for a few sources have argued that RLNLS1s can be either intrinsically radio loud or apparently radio loud because of jet beaming effects \citep{2010AJ....139.2612G,2011ApJ...738..126D}. These observational results may be contrary to the well-known paradigm of jets being generally associated with elliptical host galaxies in typical radio-loud AGNs, and there is tentative evidence suggesting that at least a few low-redshift RLNLS1s are hosted by spiral galaxies  \cite[e.g.,][]{2007ApJ...658L..13Z,2008A&A...490..583A,2014ApJ...795...58L}.   Although a number of RLNLS1s have been imaged using VLBI \citep{2010AJ....139.2612G,2011ApJ...738..126D,2011A&A...528L..11G,2013MNRAS.433..952D,
2014ApJ...781...75W,2015ApJ...800L...8R},  comprehensive large-sample studies of their jet properties on the parsec scale have just commenced.

Eight NLS1s have been detected in the $\gamma$-ray band by the Fermi LAT with high significance\footnote{Three more tentative candidates, 
FBQS J1102+2239, SDSS J1246+0238, and RX J2314.9+2243 were mentioned by \cite{2011nlsg.confE....F} and \cite{2015A&A...574A.121K}.}  (PMN J0948+0022, Abdo et al.\ 2009a; PKS 1502+036, 1H 0323+342 and PKS 2004$-$447, Abdo et al.\ 2009b; SBS 0846+513, 
D'Ammando et al.\ 2012; FBQS J1644+2619, D'Ammando et al.\ 2015; B3 1441+476, Liao et al.\ 2015; SDSS J122222.55+041315.7, Yao et al.\  2015).  All these $\gamma$-ray NLS1s possess a one-sided jet, and they strongly favor the presence of the beaming effect in these $\gamma$-ray objects \cite[e.g.,][]{2013MNRAS.433..952D,2014ApJ...781...75W}. However, so far, SBS 0846+513 is the only source showing apparent superluminal motion, with $\beta_{\rm app}=(9.3\pm0.6)c$ \citep{2013MNRAS.436..191D}.   A maximum apparent speed of 0.92$c$ was reported for SDSS J122222.55+041315.7\footnote{http://www.physics.purdue.edu/astro/MOJAVE/sourcepages/1219+044.shtml}, the $\gamma$-ray NLS1 with the largest redshift ($z$=0.966) \citep{2015AJ....150...23Y}. B3 1441+476, the only NLS1 galaxy detected in the $\gamma$-ray band with high significance and with radio properties similar to those of CSS sources\footnote{Although the CSS-like NLS1 galaxy, PKS 2004$-$447, was also detected by $Fermi$/LAT, however,  the optical spectral classification as a NLS1 galaxy continues to remain uncertain.} \citep{2015arXiv151005584L}, shows a core-jet structure, with the jet extending to approximately 30\,mas \citep{2015ApJS..221....3G}. The core brightness temperature is approximately $10^{10.3}$\,K at 5\,GHz\footnote{However, the core identification is uncertain \citep{2015ApJS..221....3G}.}, and this value is at least one order of magnitude lower than those of other flat-spectrum $\gamma$-ray NLS1 galaxies, indicating intrinsic differences in the Doppler boosting of the jet emission.

\begin{figure*}
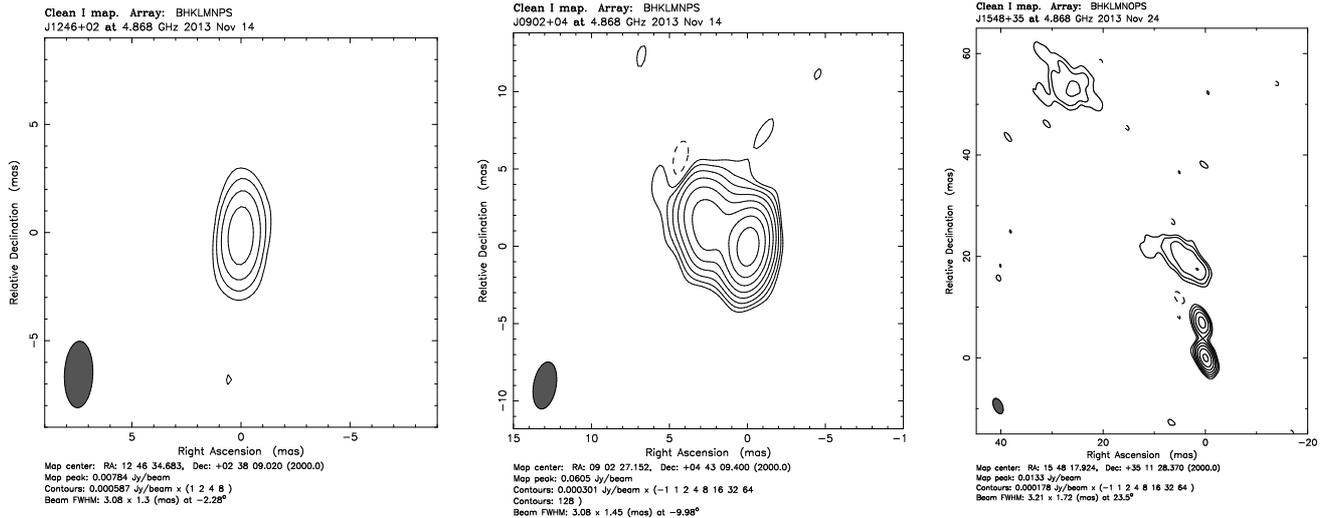

  \begin{center}
    \mbox{
     {\scalebox{0.292}{\includegraphics{j1246+0238a1i.eps}}}
      \quad

      {\scalebox{0.29}{\includegraphics{j0902+0443a1i.eps}}}
      \quad

      {\scalebox{0.255}{\includegraphics{j1548+3511b1i.eps}}}
      \quad}

  \end{center}
  \caption{Example VLBA 5 GHz images. Contours increase by factors of two. $Left$: SDSS J124634.65+023809.0 with a
 compact core only. $Middle$: SDSS J090227.16+044309.6 with a core-jet structure. $Right$: SDSS J154817.92+351128.0 
 with emission extending to about 60\,mas.  \label{fig1}}
\end{figure*}

To understand jet formation in RLNLS1s, \citep{2015ApJS..221....3G} performed the first systematic investigation of jet properties on the basis of VLBA observations of 14 sources. All fourteen NLS1 galaxies were detected as single unresolved sources within 1\,arcmin of the Sloan Digital Sky Survey position in both Faint Images of the Radio Sky at Twenty Centimeters (FIRST) and the NRAO VLA Sky Survey (NVSS), indicating a compact radio structure at the FIRST resolution (5\,arcsec). Although all these sources are very radio-loud with $R > 100$, two of their jets properties --- milliarcsecond-scale (parsec scale) morphology and overall radio spectral shape --- differ. From 5\,GHz VLBA images, it was found that seven sources had only a compact core, and the remaining seven objects had a core-jet structure (see example images in Figure~\ref{fig1}). The majority of these sources are compact on the VLBI milliarcsecond scale, with all components being within $\sim$5\,mas ($\sim$24--38\,pc, depending on the source redshift). Combined examination of the 5\,GHz VLBA images and archival multifrequency data revealed that seven sources showed flat or even inverted radio spectra ($\alpha\le0.5$, $f_{\nu}\propto\nu^{-\alpha}$), whereas the remaining seven objects exhibited steep spectra ($\alpha>0.5$).

From high-resolution VLBA images, the brightness temperature of the radio core in the rest frame can be estimated \citep[e.g., ][]{1993ApJ...407...65G}; the estimated brightness temperature is often used in combination with the assumed intrinsic brightness temperature to constrain the jet Doppler factor, either as the equipartition brightness
temperature of $5\times10^{10}$\,K \citep{1994ApJ...426...51R}, or the inverse Compton catastrophe brightness temperature of $\sim10^{12}$\,K \citep{1969ApJ...155L..71K}. The core brightness temperature ranges from $10^{8.4}$ to $10^{11.4}$\,K with a median value of $10^{10.1}$\,K, indicating that the radio emission is from nonthermal jets
and confirming that powerful jets can be formed in accretion systems that host relatively small black hole masses with high accretion rates \citep{2010AJ....139.2612G, 2011ApJ...738..126D}. However, the core brightness temperatures of these sources are considerably lower than those of blazars. VLBA core brightness temperatures of
blazars typically range between  $10^{11}$ and $10^{13}$\,K, with a median value near $10^{12}$\,K, and they can even extend to $5\times10^{13}$\,K \citep{2005AJ....130.2473K, 2009ApJ...696L..17K}.  The core brightness temperatures of the sources are also well below those of flat-spectrum $\gamma$-ray NLS1s. Therefore, the beaming effect is generally less significant in these sources, implying that the
bulk jet speed is likely to be low. The low jet speeds (i.e., less relativistic jets), in combination with the low kinetic/radio power, may offer an explanation for the compact VLBA radio structure of most sources. The mildly relativistic jets in these high accretion rate systems are consistent with a scenario where jets are accelerated from the hot corona above the disk by the magnetic field and the radiation force of the accretion disk \citep{2014ApJ...783...51C}.  Alternatively, the low jet bulk velocity can be explained by the low spin in the Blandford–Znajek mechanism. \\

{\bf \em{Future prospects with EAVN}}\\

There are few observations of NLS1s in mm/submm bands. Emissions in these bands can differ according to the nature of the source. In extreme cases, considerable flux variations are observed in the mm band for several $\gamma$-ray NLS1s; examples are flux variations at 142\,GHz for 1H 0323+342 and PMN J0948+0022  \cite{2015A&A...575A..55A}. Prominent spectral evolution from an inverted to a steep spectrum was clearly detected between 43 and 142\,GHz, indicating the transition from optically thick to thin, although the radio spectrum is mostly flat and even inverted in these objects with beamed jets. Because of its high resolution, mm/submm VLBI observations will be crucial for decomposing the core-jet structure and studying jet activity, such as, the transition between optically thick and thin, likely because as an event reveals its optically thin part, a new one emerges, that quickly hardens the otherwise softening spectrum  \cite{2015A&A...575A..55A}. By contrast, the mm/submm spectrum of another $\gamma$-ray NLS1, PKS 1502+036, is most likely to have an optically thin part  \cite{2015A&A...575A..55A}. VLBI mm/submm observations are expected to be vital for studying the structure of the most inner region in jets.

A systematical study of jet properties on the basis of VLBI observations has just been started. Although tentative evidence of low jet bulk speed has been found, fundamental questions, such as those on the jet formation mechanism and jet–accretion relationship, still remain unanswered. To study the jet properties of NLS1s with the EAVN, we will need the following information:
(a) More VLBI observations on fainter sources: Existing VLBI studies have focused on and are biased toward bright sources. However, most RLNLS1s have a faint, with FIRST\,1.4 GHz flux densities less than several milli-jansky. At such low flux densities, the phase reference technique should be used. The construction of a large sample of VLBI-observed NLS1s can facilitate detailed studies of jet properties. 
(b) Observations of radio-quiet objects toward the 1\,mJy level: These are required for studying the radio emission mechanism (e.g., from jets or during star formation). 
(c) Multi-epoch VLBI observations: These are necessary for detecting jet proper motion to directly constrain the jet bulk speed. They can also be used to constrain jet formation scenarios \cite[e.g.,][]{2015ApJS..221....3G}.\\

\subsubsection{Compact Symmetric Objects}

CSOs are a subclass of extragalactic radio sources characterized by compact symmetric double components and with a total size less than 1\,kpc \citep{1982A&A...106...21P,1994ApJ...432L..87W}. The compact double morphology is reminiscent of classical large-sized symmetric objects (LSOs), which are the so-called Fanaroff–Riley~II radio galaxies \citep{1974MNRAS.167P..31F}. CSOs, LSOs, and medium-sized symmetric objects (MSOs; with a size between 1 and 20\,kpc) are structurally similar, and on the basis of their size, they can be ordered in the morphological sequence CSO--MSO---LSO, in increasing order of size.

The kinematic age of CSOs, which can be determined from the ratio of the source size to the hot spot expansion speed, ranges from 20 to 3000~years \citep[e.g., ][]{2005ApJ...622..136G,2012ApJS..198....5A}. Some other methods, such as that involving the use of the spectral break (resulting from synchrotron aging of the relativistic electrons) of CSOs for determining their radiative age \citep[][Nagai et al. 2006]{1999A&A...345..769M}, indicate the consistent conclusion that CSOs are the youngest radio galaxies. CSO studies can shed light on the dynamic properties of extragalactic radio sources in early evolutionary stages.
\\

{\bf  \em{Dynamic evolution}}\\

\citet*{2012ApJ...760...77A} analyzed a sample of 24 CSOs with known redshifts and hotspot advance velocities and with a wide range of radio powers. They found that observables such as the radio power, projected linear source size, hotspot advance velocity, and kinematic age of the source were generally consistent with the predictions of a self-similar evolution model \citep[e.g.][]{1996cyga.book..209B}, which describes the evolution of kinematic structures and energetics of a shocked jet-based cocoon expanding into an ambient medium, applicable to FR~II AGNs. The radiative properties of the radio sources were controlled by the balance between the particle energy density, and energy losses associated with adiabatic expansion and radiation (Kaiser \& Alexander, 1997; Kaiser \& Best, 2007; An \& Baan, 2012).

High-frequency EAVN observations (e.g., at $\ge$8\,GHz) can be used for the detection, morphological classification, and study of the dynamic evolution of CSOs (An \& Baan, 2012). Principal objectives of such studies include understanding the kinematic and spectral flux evolution of the core, jet components, and lobes in the context of being early type AGN with similarities; determining possible evolution paths to FR~II and FR~I galaxies (O'Dea, 1998; Fanti, 2009; An \& Baan, 2012; Perucho, 2016); and examining their contribution to the extragalactic $\gamma$-ray background and their relationship to other $\gamma$-ray bright AGNs, including blazars and NLS1s (An et al., 2016). Gathering a large sample across cosmological epochs by using millijansky-level detection sensitivity with the EAVN (Hagiwara et al., 2015; Wajima et al., 2016), performing milliarcsecond scale resolution studies of CSO components (core, jet, and lobes) and their evolution in individual sources, and conducting multi-frequency multi-epoch radio observations can yield the following information: (a) broadband spectra that can be used to identify flat spectrum cores and steep spectrum lobes, (b) high-resolution images that can be employed to study compact cores, bright hotspots, and jet--interstellar medium (ISM) interaction, (c) derivable kinematic parameters, including mildly relativistic jet components and wide jet opening angles, and (d) light curves that can facilitate the study of core and jet flux variations, shock-based emission, and ejection of jet components. This information can help meet the aforementioned objectives and as a general methodology for studying radio-loud AGNs.

   \begin{figure}[htbp]
  \centering
	\includegraphics[width=7cm, angle=0]{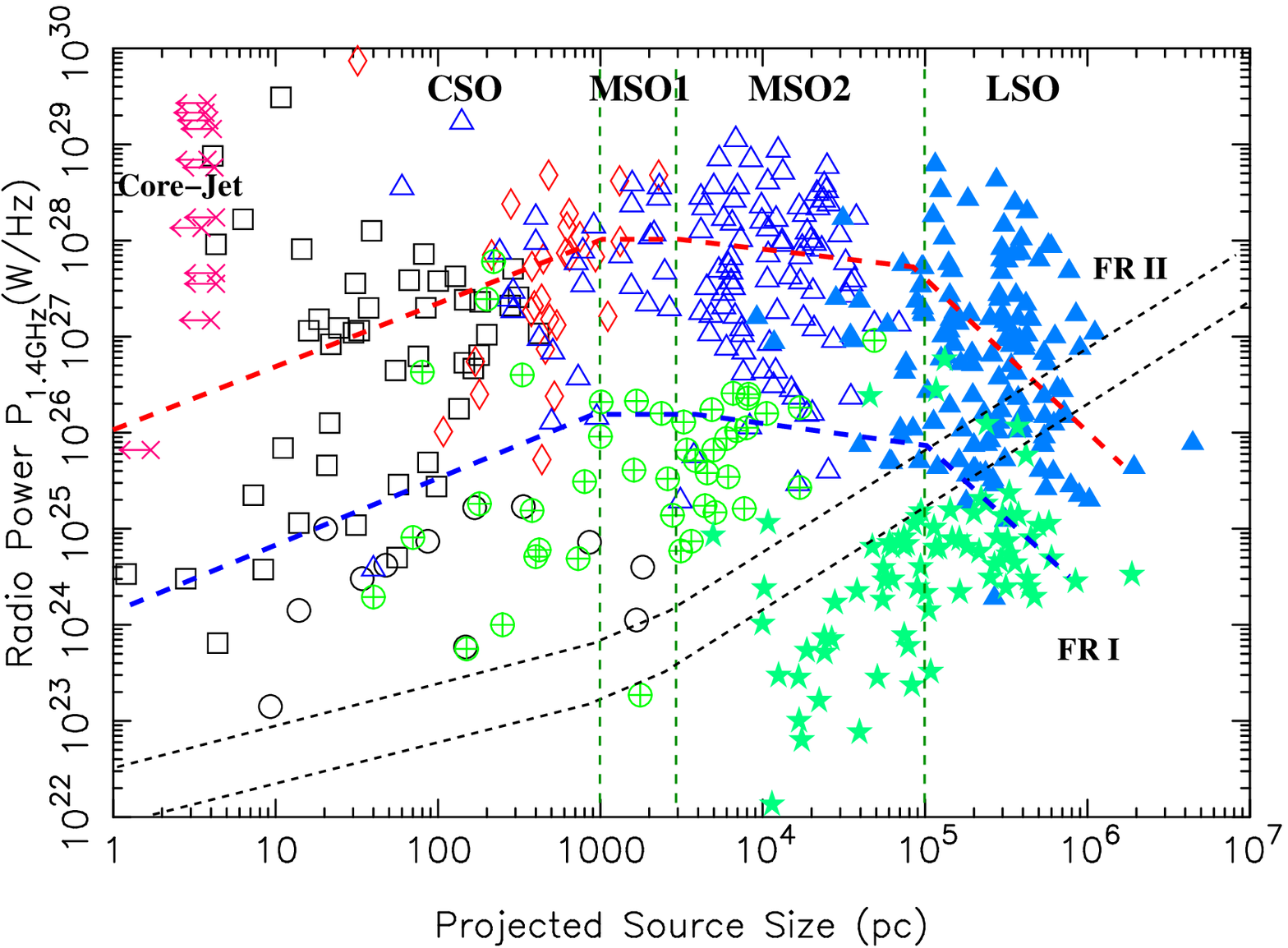} \hspace{5mm}
	\includegraphics[width=7cm, angle=0]{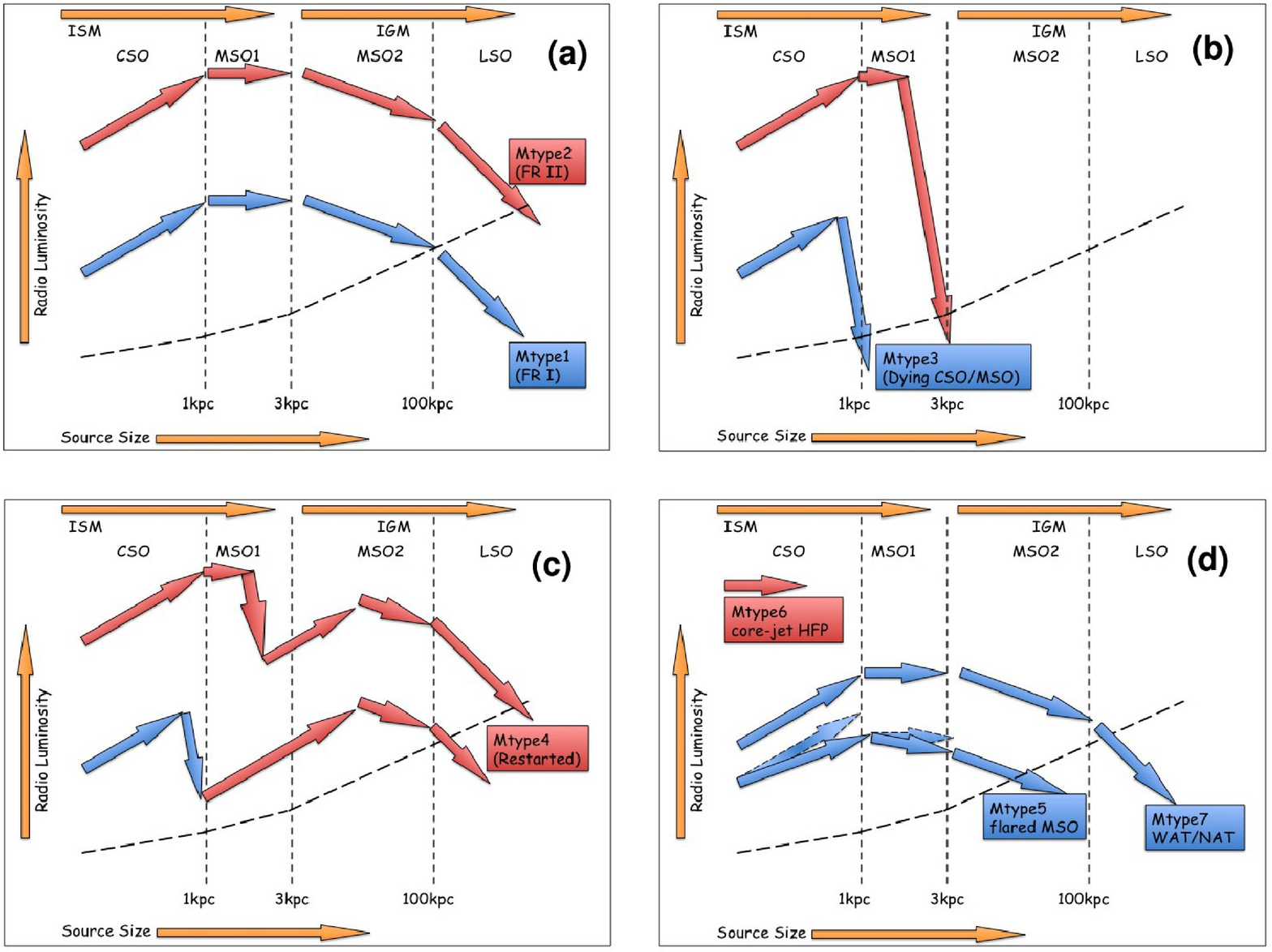}
   \caption{\small 
   (Left) Radio power as a function of the linear size of extragalactic radio sources (so-called P--D diagram). The red and blue dashed lines represent exemplary evolution tracks of high- and low-radio-power sources, respectively. The black dashed lines denote the boundary between stable laminar jets (above the lines) and unstable turbulent flows (below the lines). Symbols represent different classes of radio sources \citep[see details in][]{2012ApJ...760...77A}. (Right) Complex evolutionary pathways and the corresponding morphological types.
   \label{fig:tao.an.1}}
   \end{figure}

The left panel of Figure~\ref{fig:tao.an.1} depicts the evolution of radio power with an increase in the size of extragalactic radio sources. In general, all radio sources are thought to be born as CSOs and to subsequently follow the primary evolutionary tracks (shown by the red and blue dashed lines) as long as their jet power remains at an appropriate level for a suﬃciently long time. The actual evolution tracks are rather complex, and they are shown for various pathways and morphological types in the right panel of Figure~\ref{fig:tao.an.1}; they depend on the kinetic power of the jet, the duration of nuclear activity, the density gradients in the ambient medium in the host galaxy, and jet--ISM interactions.

In reality, most CSOs may not successfully evolve into FR~I and FR~II galaxies because of their activity ceasing midway. This can be inferred from their kinematic age distribution, which shows an overabundance of compact young CSOs less than 500 years old. In this sense, CSOs represent a transient phase during the evolution of radio AGNs. It is also inferred that a major fraction of CSOs may be radio-quiet and have low-power jets. They occupy the bottom-left corner of the P--D diagram in the left panel of Figure~\ref{fig:tao.an.1}. Fluid instabilities easily develop in such low-power jets, which then become turbulent and are disrupted, preventing the radio structure from expanding out of the host galaxy.
\\

{\bf \em{Gamma-ray emitting CSOs and CSSs}}\\


The physical origin and composition of the extragalactic $\gamma$-ray background (EGB) at GeV energies remains one of the unsolved fundamental questions in astrophysics. {\it Fermi}-LAT observations have revealed that the majority of extragalactic $\gamma$-ray sources are blazars. Recently, theoretical studies have predicted that $\gamma$-ray emission originates from hot spots and lobes of non-blazar AGNs \citep[e.g.,][]{2009MNRAS.395L..43K}. Young radio lobes have high electron density and temperature, and therefore, $\gamma$-ray emission from the youngest and most compact radio galaxies should be easily detectable. Latest investigations of $\gamma$-ray emitting, compact, and young radio sources show that these jets are mildly relativistic and aligned at a moderate viewing angle. Although the sample size of observed $\gamma$-ray CSOs and CSSs is still small because of the insuﬃcient sensitivity of $\gamma$-ray telescopes, because the nonbeamed AGN population is greater and because the lobes and cocoons have larger effective areas than compact jets, CSO and CSS sources could contribute appreciably to the isotropic extragalactic $\gamma$-ray background.

\subsection{Advancing Maser Science}



In this section, we discuss science cases for mm/submm VLBI observations of maser sources. Maser lines at mm/submm wavelengths are expected to be complementary to those at cm wavelengths, which have been well studied with currently available VLBI networks. In the EA region, we have started a KaVA Large Program to study both star formation and late-type stars on the basis of H$_{2}$O/CH$_{3}$OH and H$_{2}$O/SiO maser lines, respectively. In particular, the high imaging capability of KaVA will be used for understanding the spatial and velocity structure of multiple masers; the use of the high-imaging capability is analogous to high-frequency VLBI imaging by including phased ALMA and ALMA baselines. Here, we stress that continuing the search for new maser sources and good probes is imperative for optimizing maser science in mm/submm VLBI. In addition, the development of theoretical model calculations for maser pumping and source structures is crucial for estimating the physical parameters, such as hydrogen density, temperature, and molecular abundances, of masing gas clumps from new multi-transition maser observations. Such multi-transition maser observations can shed light on three-dimensional dynamical structures and the physical properties of young stellar objects and late-type stars associated with maser lines.


\subsubsection{General Objective}

Microwave amplification by stimulated emission of radiation (maser) lines are known to be associated with Galactic and extragalactic objects, including low- and high-mass young stellar objects (YSOs); shocked molecular gas in the ISM, such as that in star-forming regions (SFRs) and SNRs; circumstellar envelopes (CSEs) of late-type stars such as asymptotic giant branch (AGB) stars and red supergiants (RSGs); and circumnuclear disks around AGNs \citep{1981ARA&A..19..231R, 1992ARA&A..30...75E}.  Maser emission requires a coherent velocity structure for efficient amplification, and therefore, maser features have a spatially compact structure with extraordinarily high brightness temperatures (up to 10$^{16}$~K).
Because of such a structure, maser sources are ideal targets for high-resolution VLBI studies. The OH and H$_{2}$O lines at
1.6 and 22.2\,GHz, respectively, have been used for high resolution VLBI studies. Other well-studied probes are the 43\,GHz SiO maser and class~II\footnotemark \footnotetext{
Methanol (CH$_{3}$OH) masers are classified into two groups\citep[I and II; ][]{1991ApJ...380L..75M} according to excitation conditions and/or transitions.} CH$_{3}$OH maser at 6.7\,GHz.

One of the key features of maser observations is that three-dimensional (3D) velocity information can be obtained through multi-epoch VLBI monitoring observations. Multi-epoch high-resolution maser images facilitate the measurement of positions and proper motions of maser features with high accuracy; the highest accuracies for these measurements are on the
order of sub-milliarcseconds and milliarcseconds per year, respectively, at cm
wavelengths. Shorter-wavelength (i.e., mm and submm) VLBI observations can achieve a resolution that is higher than those  at cm wavelengths 
by a factor of 10 for positional and proper motion measurements. Consequently, we can measure proper motions of maser features toward more distant sources. Even for nearby sources, monitoring 3D motions of maser features on timescales shorter than those of previous VLBI monitoring observations ($\sim$ a few months or longer) is expected to become feasible in the near future.

Maser lines have been detected in the mm/submm wavelengths, and they include maser lines of H$_{2}$O, CH$_{3}$OH, SiO, and some other rarer species \citep{2007IAUS..242..471H}.  Nevertheless, VLBI observations of frequency bands higher than 86\,GHz are limited. ALMA has provided high-sensitivity and high-resolution imaging capabilities in new mm/submm windows at 0.05--0.1$''$ resolution in Cycle~3. Consequently, new maser lines have been mapped with ALMA. These maser lines can be potential targets for future mm/submm VLBI observations. 

In the following sections, we review recent progress in mm/submm maser observations and discuss new science cases. We mainly focus on star formation, stellar evolution of late-type stars, AGNs, and galactic astrometry.

\subsubsection{Young Stellar Objects  and Interstellar Matters}

Various physical processes, including fragmentation/contraction/collapse of filaments and cores, mass accretion through circumstellar disks, ejection through jets/outflows, interaction of jets/ouflows with ambient clouds, and feedback to environments, should be investigated for acquiring a complete understanding of star-formation processes. One of the key requirements for revealing each dynamical process is 3D velocity structures, to distinguish inward, outward, and rotational motions caused by these processes. For obtaining 3D velocity structures, VLBI observations of maser sources
is the only means for achieving suﬃciently high spatial resolution to measure proper motions of maser features at a resolution on the order of approximately 1~mas~yr$^{-1}$ (5~km~s$^{-1}$ at 1~kpc) or higher.

In the mm and submm bands, various  H$_{2}$O masers are detected \citep{2007IAUS..242..471H}.  Some of
them, namely H$_{2}$O lines at 183, 232, 321, 325, and 658\,GHz, have been mapped in selected sources (Cepheus-A, Serpens SMM1, and Orion~KL Source I) using interferometers such as the SMA and ALMA at the highest resolution of 0.2$''$ \citep[][Hirota et al. in press]{2007ApJ...658L..55P,2009ApJ...706L..22V,2012IAUS..287..184N,2012ApJ...757L...1H, 2014ApJ...782L..28H}.  For the most extensively observed source, Orion~KL Source I, submm H$_{2}$O masers have been suggested to be good probes for bipolar outflow and/or disk wind
within approximately 100\,AU scale from the newly born YSO. Multiple H$_{2}$O maser transitions at different energy levels
from 200--450\,K (183\,GHz, 325\,GHz), 1800\,K (321\,GHz), and greater than 2300\,K, including those of the highly
excited vibrational state ($\nu_{2}$=1; 658~GHz, 232~GHz), can be used as a diagnostic tool to identify a wide range of density/temperature regions  \citep{2016MNRAS.456..374G}.

Although SiO masers in SFRs are detected only toward three high-mass YSOs \citep[i.e. Orion~KL, W51N, and Sgr B2(M);][]{1992PASJ...44..373M, 2009ApJ...691..332Z}, one of the target sources, Orion~KL Source~I, has been extensively observed with VLBA and VERA in the 43~GHz SiO masers \citep{1998Natur.396..650G, 1999ApJ...510L..55D, 2008PASJ...60..991K, 2010ApJ...708...80M}. 
In contrast, the only VLBI observation of the 86~GHz SiO maser has been reported by \citet{2004ApJ...607..361D} with a single short baseline. 
The 43~GHz SiO masers around Source~I are found to be excited in the magneto-centrifugal disk wind emanating from a circumstellar disk \citep{2010ApJ...708...80M, 2013ApJ...770L..32G}. 
Although the 43~GHz SiO masers disappear in the midplane of the edge-on disk \citep{2008PASJ...60..991K, 2010ApJ...708...80M}, it might be due to optically thick continuum emission at lower frequency bands \citep{2014ApJ...782L..28H, 2015ApJ...801...82H}. 
If this is the case, higher frequency SiO masers \citep[e.g.][ detected in the ALMA Science Verification data]{2012A&A...548A..69N} will be able to reveal dynamical structures in the closer vicinity of massive YSOs deeply embedded in the circumstellar matters than that traced by the 43~GHz maser lines. 
Dynamical structures of disk-outflow/jet system around high-mass YSOs will be resolved by combining submillimeter H$_{2}$O and SiO maser data at $\sim$100~AU resolution \citep{2014ApJ...782L..28H, 2016ApJ...817..168H}. 
In addition, higher sensitivity phased-ALMA will be able to detect more distant weaker SiO masers associated with W51N \citep{2002ApJ...569..334E}, Sgr B2(M), and other new YSOs associated with SiO masers at $\sim$1~mas resolution. 

\begin{figure}[htbp]
\centering
\includegraphics[width=6cm]{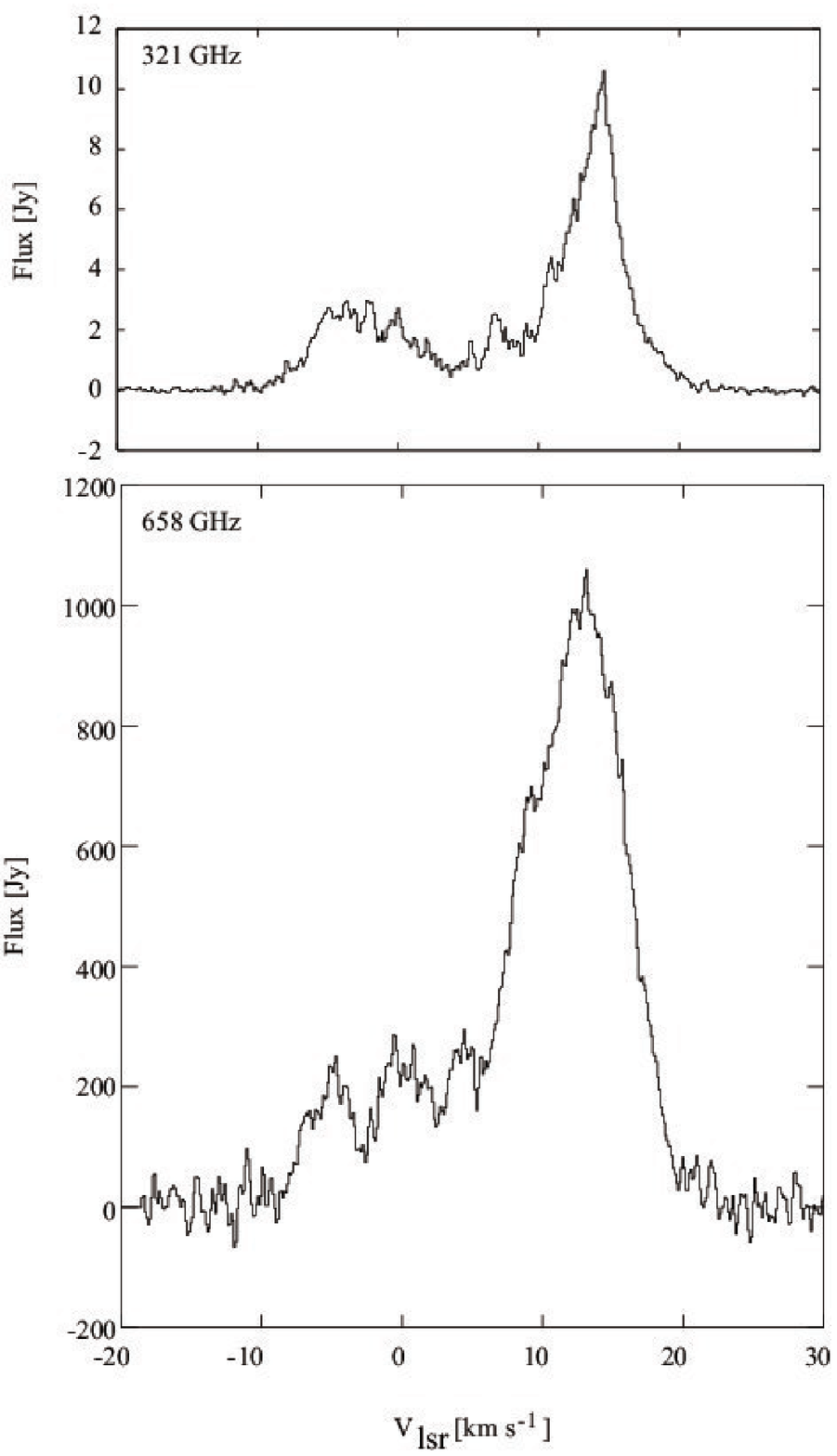}
\hspace{5mm}
\includegraphics[width=10cm]{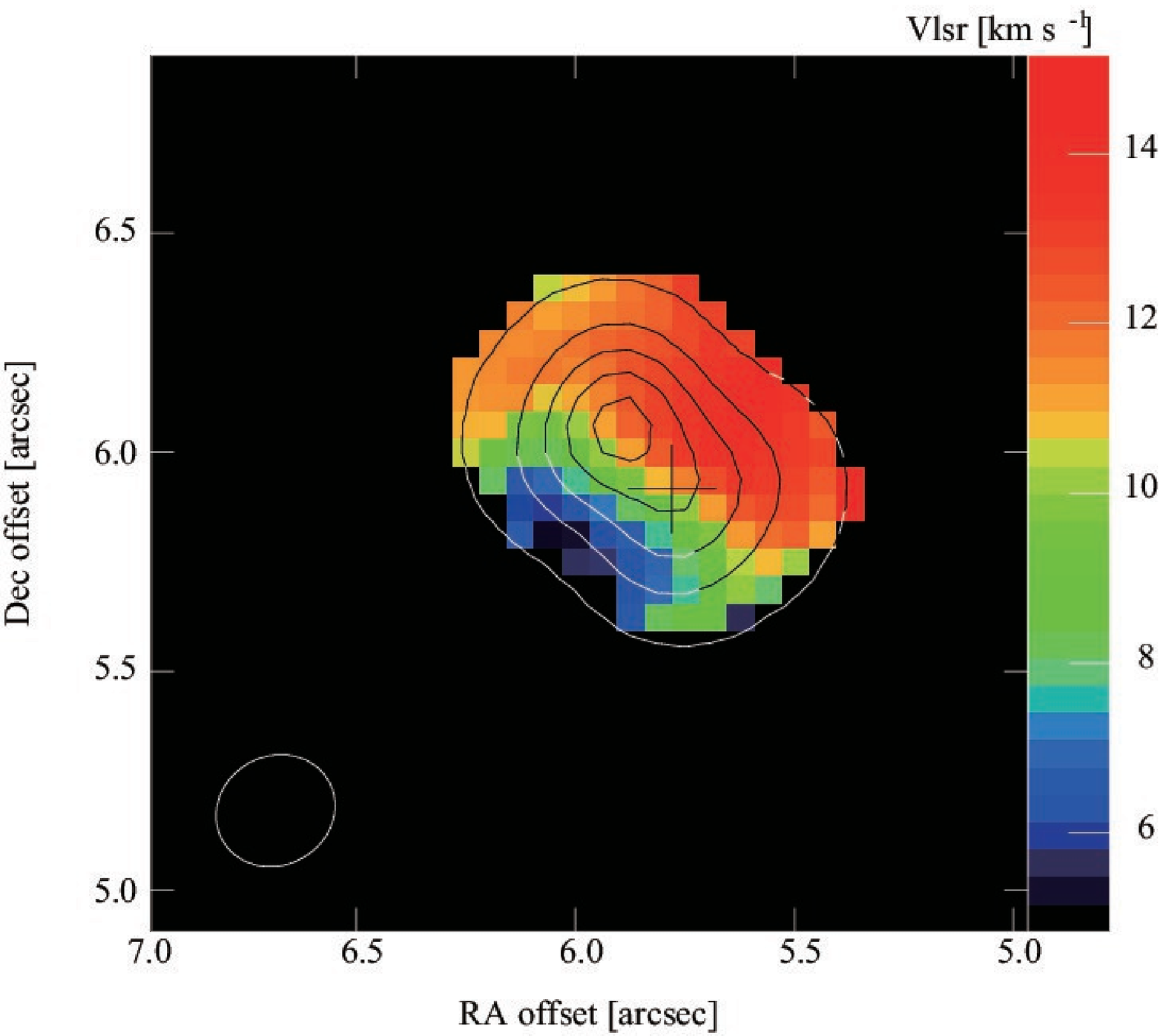}
\caption{(Left) Submillimeter H$_{2}$O maser spectra and (right) moment maps of the 658-GHz H$_{2}$O maser (right) toward Orion KL Source I observed in ALMA Cycle 0 \citep{2016ApJ...817..168H}. 
\label{fig:hirota.hirota2016}
}
\end{figure}

Several CH$_{3}$OH masers at mm/submm wavelengths have been detected and predicted \citep{2004A&A...428.1019M,2005MNRAS.360..533C, 2012IAUS..287..433V}.  However, none of these methanol maser lines has been observed with VLBI, and only a limited number of observational studies using interferometers have
been reported. According to observations with the Berkeley–Illinois–Maryland Array (BIMA), for investigating higher-density and/or higher-temperature regions  ($\sim$10$^{7}$~cm$^{-3}$ and $>$110~K),  the CH$_{3}$OH masers in the range of 86--107\,GHz are more suitable than those at 6.7, 12, and 20--40\,GHz \citep{2001ApJ...554..173S}. ALMA Cycle 0 observations detected the 278\,GHz Class~I CH$_{3}$OH maser in an infrared dark cloud (IRDC) \citep{2014ApJ...794L..10Y}; moreover, in the cloud, Class~I CH$_{3}$OH masers were found to be excited in weaker and/or older shocks than those traced by the thermal SiO line. Thus, these different tracers are complementary to each other for investigating small-scale shocked masing gas clumps, including their 3D velocity field, physical properties, and chemical composition. The first VLBI imaging observation of the 44\,GHz Class~I CH$_{3}$OH maser with KaVA was reported recently \citep{2014ApJ...789L...1M}, demonstrating the potential capability of compact VLBI networks with short (e.g. $<$100~M$\lambda$ for KaVA)  and dense ($u,v$)-coverage to provide Class~I CH$_{3}$OH maser observations in the future. With the aid of the long baseline of phased ALMA and shorter baselines in ALMA, VLBI observations of CH$_{3}$OH masers can be performed with both high resolution and high imaging capability.
The radio recombination line (RRL) maser of the hydrogen atom  \citep{1989A&A...215L..13M} is also a potential probe that can be used in star-formation studies usingmm/submm VLBI, and mm/submm RRL masers up to 662\,GHz have been detected  (H21$\alpha$) \citep{1994A&A...288L..25T}.  Although there are only a few known RRL maser objects, they are expected to play a crucial role in studies of circumstellar disks and radio jets at high temperatures ($T_{e}\sim$10000~K); such circumstellar disks and radio jets cannot be traced using other molecular masers \citep{1992ApJ...386L..23P,2008ApJ...677.1140W,2011ApJ...732L..27J,2013ApJ...764L...4J,2011A&A...530L..15M}.

Both linear and circular polarization measurements of maser lines are useful for investigating the magnetic field strength and structure at high resolutions \citep{2012IAUS..287...31V}.  Previous observational studies have revealed the magnetic field structure around newly born stars and their relationship to the outflows, and they have provided
 magnetic field strengths on the order of 10--100\,mG by performing Zeeman splitting measurements. Furthermore, mm/submm H$_{2}$O and CH$_{3}$OH  masers along with SiO masers associated with YSOs can help trace different volumes of gas from those traced by the lower-frequency transition, and hence, they are complementary to each other and can reveal magnetic field structures on different physical scales/structures. The RRL masers can also be used to trace the magnetic field structure in high-temperature ionized gas through Zeeman splitting measurements of the hydrogen atom \citep{1999A&A...344..923T}. The high sensitivity, high spectral resolution, and well calibrated polarization capability of ALMA is expected to facilitate the determination of the magnetic field strength and direction with unprecedented accuracy \citep{2013A&A...551A..15P}.

\subsubsection{Asymptotic Giant Branch Stars and Red Supergiants}

Similar to the star-formation studies, 3D velocity field measurements of AGB stars and RSGs associated with OH, SiO, and H$_{2}$O masers have been performed by using the masers. They indicate the physical properties of their CSEs and dynamics of mass-loss processes, which are crucial for understanding stellar evolutionary models, especially at the end of stellar lives. Spatially resolved images obtained using various masers are crucial for understanding pumping mechanisms of the target maser lines.

Multi-transition SiO maser observations, including observations of different rotational and vibrational states, are useful for studying the physical properties of CSEs. To date, most VLBI observations of SiO masers have been performed at 43\,GHz ($J$=1--0), and higher-frequency VLBI observations at 86\,GHz ($J$=2--1) and 129\,GHz ($J$=3--2) have also been tested \citep{1998ApJ...494..400D,2000AJ....119.3015P,2001AJ....122.2679P,2002evn..conf..223D,2004PASJ...56..475S,2003ApJ...588L.105P,2004A&A...426..131S,2005A&A...432L..39S,2007A&A...468L...1S,2013MNRAS.436.1708R}. The KVN is being used mainly for simultaneous multifrequency SiO maser observations, including  $J$=1--0, 2--1, and 3--2  with milliarcsecond resolutions \citep[e.g.][]{2015AJ....150..202R}.  Previous VLBI studies sometimes resolved out extended emission features of SiO masers, and hence, it was diﬃcult to compare spatial distributions traced with different SiO maser species, except for KaVA, which provides dense ($u,v$)-coverage at baseline lengths shorter than 500\,km. The high imaging capability of mm/submm VLBI observations obtained with phased ALMA and ALMA baselines is expected to overcome this problem. Furthermore, high-sensitivity VLBI with ALMA will facilitate the detection of higher vibrationally excited SiO masers
 \citep[e.g. $v$=3;][]{2010PASJ...62..431I} and rarer isotopic species, including $^{29}$SiO and $^{30}$SiO \citep{2005A&A...432L..39S, 2007A&A...468L...1S}. 
 
Submillimeter H$_{2}$O masers of late-type stars have been surveyed at frequencies of 183, 321, 325, and 658\,GHz, and selected target sources have been used to search for new rare maser lines at 293, 437, 439, 471, and 475\,GHz  \citep{2007IAUS..242..471H, 2009ASPC..417...83M}. Recent ALMA observations of VY-CMa during the course of long-baseline campaign observations revealed spatial distributions of  H$_{2}$O masers at 22, 321, 325, and 658\,GHz in addition to vibrationally excited SiO masers \citep{2014A&A...572L...9R}.  Furthermore, the submm H$_{2}$O maser line at 321\,GHz was detected toward water-fountain nebulae, which are thought to be stars ejecting high-velocity bipolar outflow in the post-AGB phase  \citep{2014A&A...562L...9T}; the detection was made with the Atacama Pathfinder Experiment (APEX). High-velocity ($>$100~km~s$^{-1}$) features suggest that the 321\,GHz H$_{2}$O masers co-exist with 22\,GHz H$_{2}$O masers excited in the bipolar jet \citep{2002Natur.417..829I, 2006Natur.440...58V}. Thus, high-resolution submm VLBI observations of the H$_{2}$O masers will be complementary to the 22\,GHz H$_{2}$O masers, and they can be used to determine the dynamical structure and physical properties of mass-losing AGB, post-AGB, and RSG stars.

\begin{figure}[htbp]
\centering
\includegraphics[width=8cm]{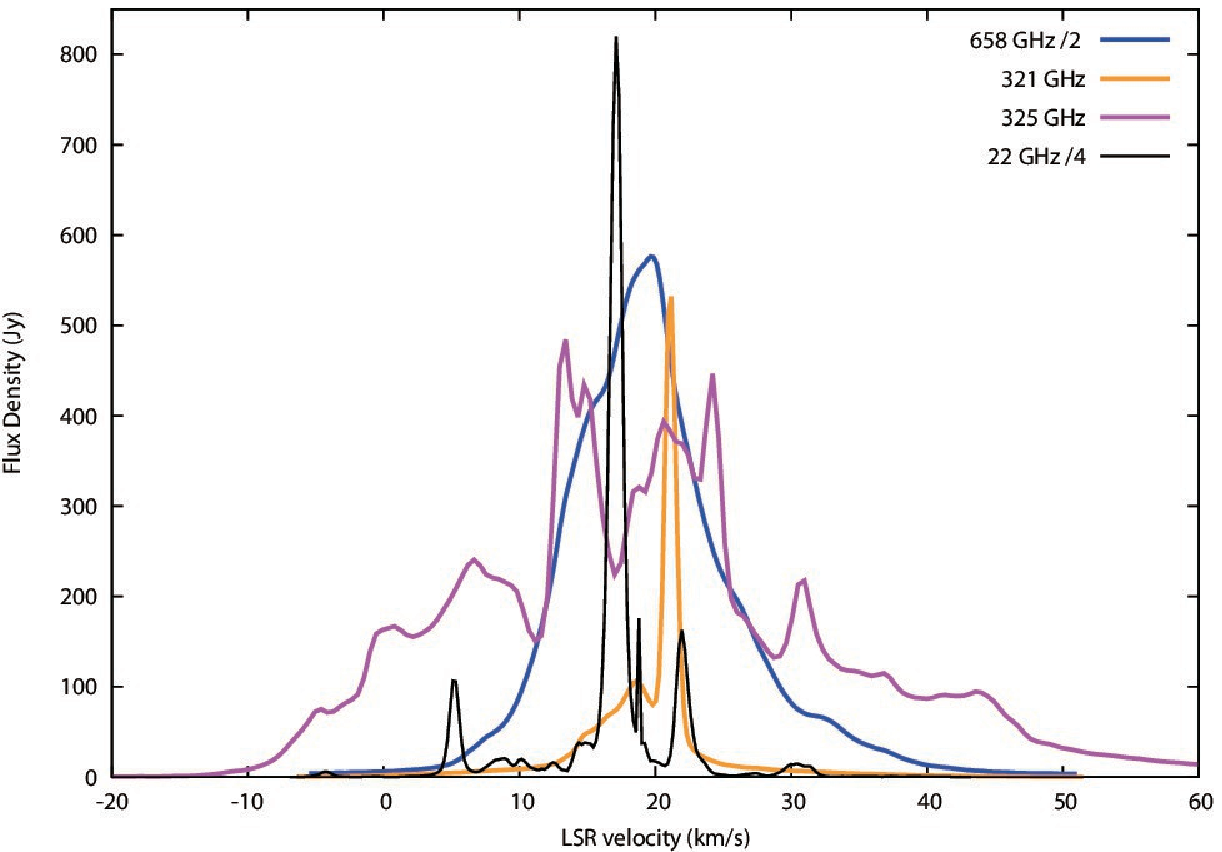}
\hspace{5mm}
\includegraphics[width=8cm]{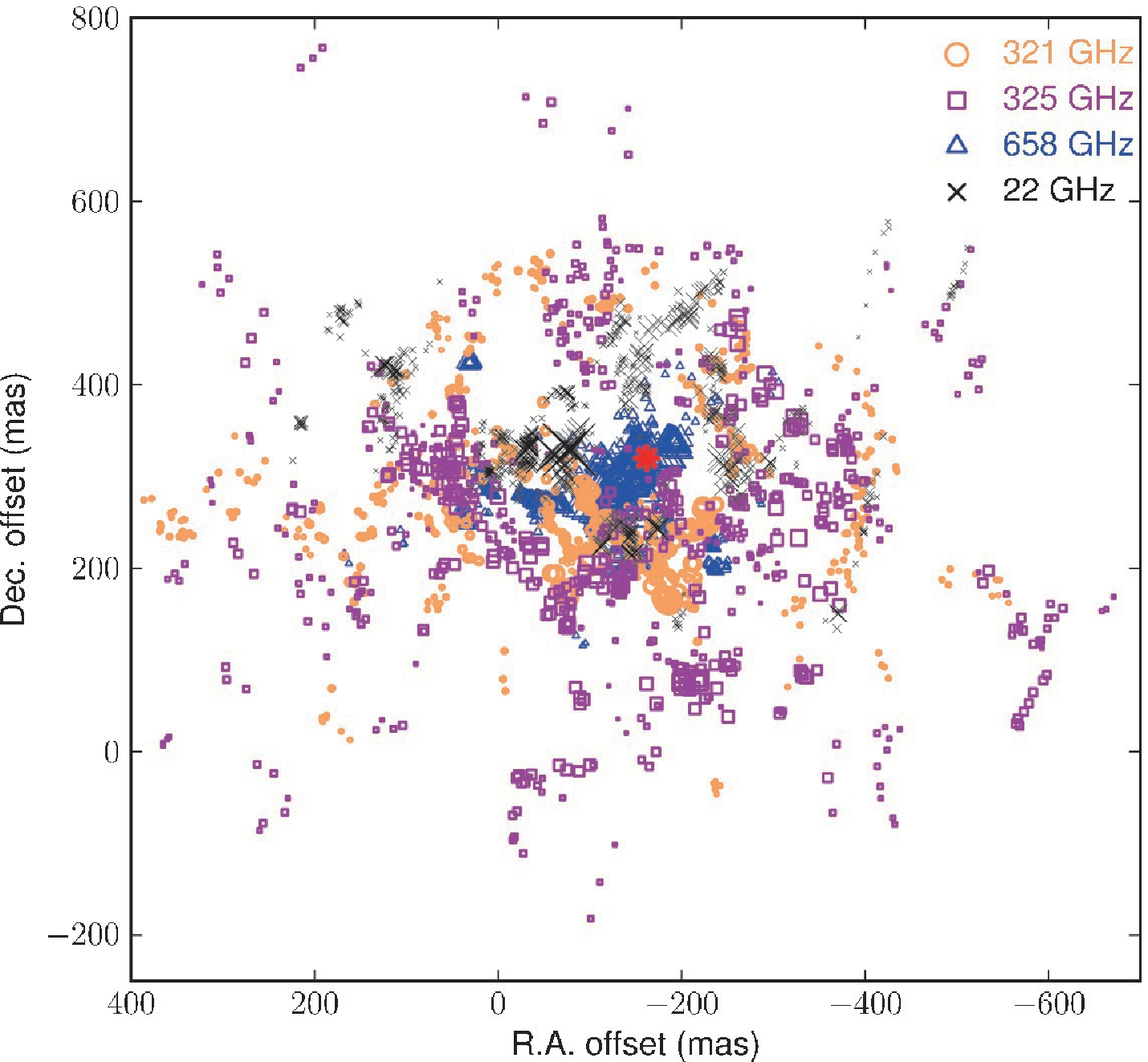}
\caption{
(Left) Submillimeter H$_{2}$O maser spectra and (right) their distribution toward VY-CMa observed in the ALMA long baseline campaign \citep{2014A&A...572L...9R}. 
\label{fig:hirota.2014A&A...572L...9R}
}
\end{figure}

Most of the previous maser studies have been conducted for oxygen-rich stars by considering oxygen-containing species, such as OH, SiO, and  H$_{2}$O.  At mm wavelengths, the vibrational ground and excited states of HCN masers from 88\,GHz ($J$=1--0) up to 890\,GHz \citep{1987A&A...176L..24G, 1987ApJ...323L..81I,1995ApJ...440..728I, 2003ApJ...583..446S} can be used to investigate carbon-rich stars. By using the extended SMA (eSMA) at 0.3$''$ resolution, the HCN maser emission region in a representative carbon star, IRC+10216, can be separated into the inner ($\sim$15~$R_{*}$)
accelerated gas and the outer CSE ($\sim$23~$R_{*}$) close to the terminal velocity  \citep{2009ApJ...698.1924S}.  These HCN maser results indicate that HCN masers can play a crucial role in determining the dynamical structure and physical properties of CSEs around carbon-rich stars and that they are equivalent to OH, H$_{2}$O, and SiO masers in oxygen-rich stars.

 Similar to YSOs in SFRs, polarization observations of SiO and H$_{2}$O masers around AGB stars and RSGs are useful for understanding the magnetic field structure and pumping mechanism of maser \citep{2011ApJ...728..149V, 2012IAUS..287...31V}. For instance, the SiO maser map with $v$=1 $J$=5--4 for VY-CMa, obtained using the SMA, shows a high degree of linear polarization (60\%) aligned with the bipolar outflow, suggesting radiative maser pumping \citep{2004ApJ...616L..47S}. As discussed in the previous section, mm/submm maser lines are complementary and can be used to probe higher-temperature/higher-density regions deep inside CSEs. Multitransition maser polarimetry can reveal the evolution of the 3D magnetic field structure of both oxygen- and carbon-rich AGBs  \citep{2013A&A...551A..15P}.

\subsubsection{Megamaser in AGNs}

The first discovery of extragalactic mm and submm masers was reported by  \citet{2005ApJ...634L.133H}. These researchers detected the 183\,GHz and 439\,GHz  H$_2$O masers toward the Seyfert 2/low-ionization nuclear emission-line region (LINER) galaxy NGC 3079 by using the SMA and JCMT, respectively, although the detection of the 439\,GHz maser was tentative. The 183\,GHz  H$_2$O maser emission features appear in the same velocity range as the 22\,GHz  H$_2$O line and coincide with the continuum peak. Although the spatial resolution is considerably lower than that of cm VLBI observations, the 183\,GHz  H$_2$O line is consistent with the maser emission originating from the nuclear region of the galaxy. The 183\,GHz  H$_2$O maser was also detected toward the nearest ultraluminous infrared galaxy (ULIRG), Arp 220, by the Institut de Radioastronomie Millim{\'e}trique (IRAM) 30\,m telescope \citep{2006ApJ...646L..49C}. 

\begin{figure}[hb]
\begin{center}
\includegraphics[width=6cm, angle=90]{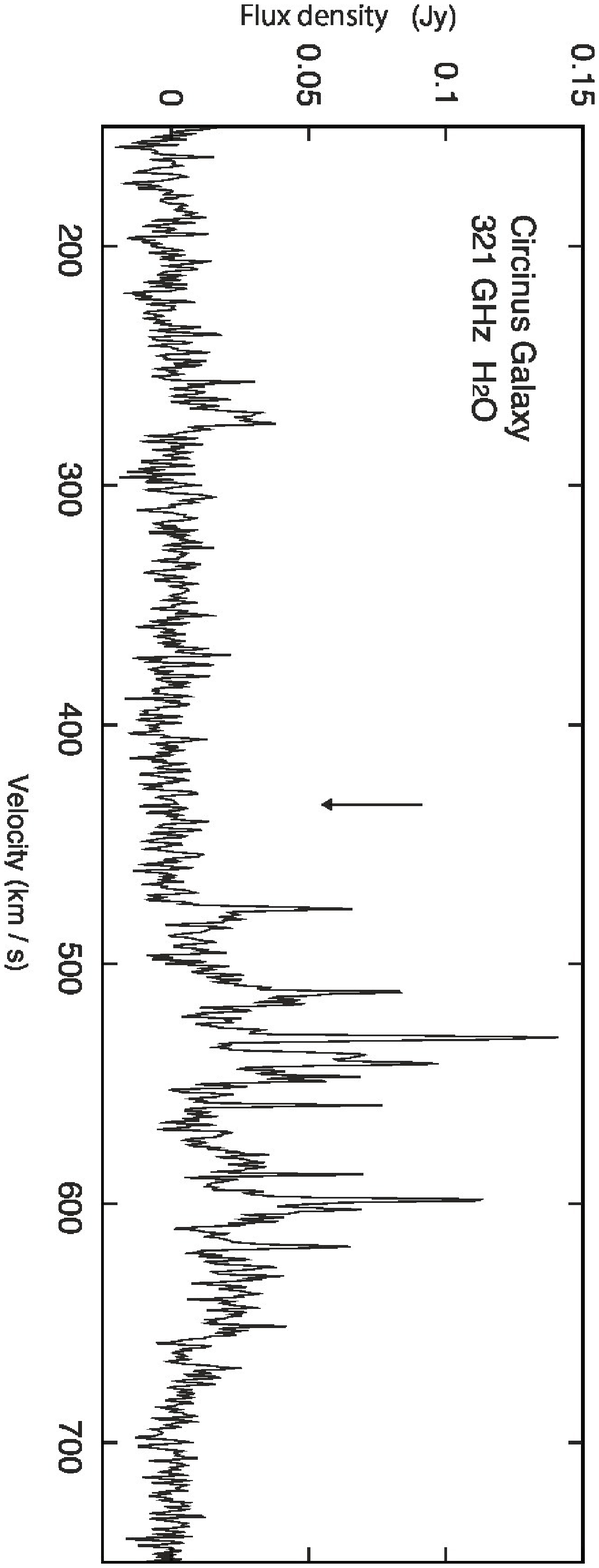}      
\caption{
Spectrum of the 321\,GHz H$_2$O maser in the Circinus galaxy, obtained on 2012 June 3 by  ALMA  \citep{2013ApJ...768L..38H}. The downward arrow marks the systemic velocity of the galaxy.
}
\end{center}
\begin{center}
\includegraphics[width=8cm]{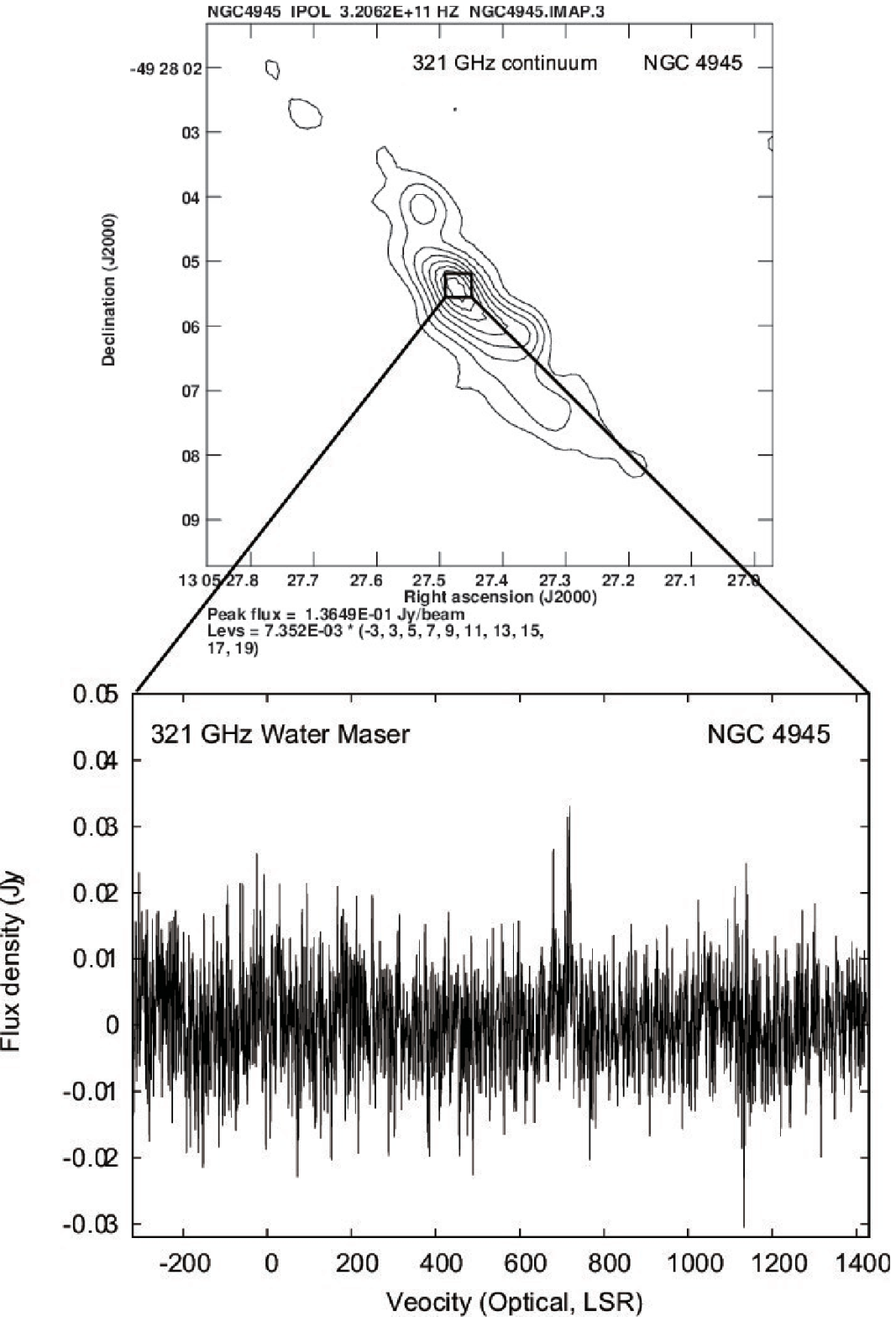}      
\caption{
The 321\,GHz continuum and H$_2$O maser in the nearby active galaxy NGC 4945, observed on 2012 June 3 by ALMA (Hagiwara et al., 2016). The 321\,GHz maser was detected toward the nuclear continuum, which is dominated by dust emission heated by an AGN in the galaxy.
\label{fig:hagi}}
\end{center}
\end{figure}

\citet{2013ApJ...768L..38H} first discovered the extragalactic submm H$_2$O maser in the 321\,GHz transition toward the Circinus galaxy (Figure  \ref{fig:hagi}), a nearby Seyfert~2 galaxy, using Cycle~0 ALMA data with a beam size of 0.66 arcsec. They found that the 321\,GHz H$_2$O maser occurs in a region similar to that  of the 22\,GHz H$_2$O maser. The ALMA data indicated the signature of a high-velocity redshifted component. Assuming the Keplerian rotation curve established by a previous VLBI study \citep{2003ApJ...590..162G}, they estimated the location of the high-velocity component to be at a radius of 0.018\,pc (or 1.2$\times$10$^5$\,Rs) from the central engine; they might have detected molecular gas closer to the central engine, with the closeness of the gas to the central engine being more than that in past observations. The 321\,GHz H$_2$O maser was also detected toward the nuclear continuum in the nearby AGN NGC\,4945 (Figure~\ref{fig:hagi}; Hagiwara et al., 2016). These results demonstrate that the submm  H$_2$O lines are potentially powerful tracers of molecular materials in the nuclear region closest to the central engine of AGNs. Higher-frequency submm and far-infrared (terahertz) maser emissions have been detected and/or predicted, and they could be useful for future observational studies \citep[e.g.][]{2016MNRAS.456..374G}. These lines fall in mm and submm bands, which are observable with ALMA. If bright maser lines are detected, they can serve as powerful probes for high-redshift galaxies \citep[e.g.][]{2008Natur.456..927I}. The submm H$_2$O masers can be used as a cosmological tool for H$_0$ determination in both nearby and distant AGNs \citep{1999Natur.400..539H}.

The submm  H$_2$O masers could originate from a region under more restricted physical conditions (e.g., T$_k$ $>$ 1000\,K)  and close to the central engine, possibly at radii of $\leq$0.1\,pc, separated spatially from the gas up to  the excitation of the 22\,GHz  H$_2$O maser. Thus, in combination with submm VLBI observations at milliarcsecond angular resolution, extragalactic submm masers could be a promising tool for probing the dynamics and conditions close to the central engines of AGNs.

Moreover, the detection of extragalactic  H$_2$O masers in other transitions, such as those at 325\,GHz and 658\,GHz, is expected, and continuing the search for masers of other molecular species is imperative.

\subsubsection{Galactic structure and astrometry}

Another use of maser lines is for probing Galactic structures. Absolute astrometry of maser sources enables direct measurements of their distances through the measurement of trigonometric annual parallaxes, along with absolute positions, proper motions, and radial velocities of the maser sources. Because a considerable fraction of SFRs harbor YSOs associated with 22\,GHz H$_{2}$O and/or 6.7\,GHz CH$_{3}$OH masers, these maser sources trace the location of Galactic spiral arms in the (thin) disk. Such VLBI astrometry projects involving VERA \citep{2012PASJ...64..136H,2015PASJ...67...70H} and VLBA \citep[The Bar and Spiral Structure Legacy Survey or BeSSeL; ][]{2009ApJ...700..137R,2014ApJ...783..130R} are now underway. To date, more than 100 maser sources have been observed for Galactic astrometry, and fundamental Galactic constants have been estimated \citep{2014ARA&A..52..339R}. 

One of the frontiers of Galactic astrometry is the Galactic structure that is accessible only from the Southern Hemisphere. Although half of the Galactic structures can be accessed by VLBI arrays in the Northern Hemisphere (VERA and VLBA), rest of the regions host essential parts of the Galactic structure, including the GC region, Galactic bars, and bulge, except for the southern hemisphere Long Baseline Array (LBA) \citep[e.g.][]{2015ApJ...805..129K}.  A VLBI network including phased ALMA can have an advantage over VERA and the VLBA if good mm/submm probes are detected.

Along with the H$_{2}$O and CH$_{3}$OH masers, high-resolution and high-sensitivity SiO maser observations of AGB stars will be crucial for determining the 3D structure of the GC, bar, and bulge. A thousand SiO maser sources have been detected in these regions through single-dish observations \citep[][ and references therein]{2012IAUS..287..265D}.  Most of the SiO maser sources are not sufficiently strong to be detected by current VLBI arrays, and phased ALMA is expected to enhance array sensitivity to increase the number of detectable target SiO maser sources for VLBI astrometry. AGB stars associated with SiO masers are thought to behave differently from SFRs (gas), which are traced by the H$_{2}$O and CH$_{3}$OH masers in the course of the dynamical evolution of the Galaxy. Thus, VLBI astrometry of SiO masers will be complementary to previous H$_{2}$O and CH$_{3}$OH maser astrometry. SiO maser astrometry is expected to serve as the interface between radio and optical/infrared astrometry, similar to GAIA and the series of Japan Astrometry Satellite Mission for Infrared Exploration (JASMINE) projects, because a considerable fraction of AGB stars associated with SiO masers will become common targets.

A major challenge is to measure proper motions and trigonometric annual parallax of maser sources associated with the Large Magellanic Cloud (LMC) and Small Magellanic Cloud (SMC). Assuming the distance to the Magellanic Clouds to be 50\,kpc, the modulation resulting from trigonometric annual parallax is only 0.02\,mas. Thus, the positional accuracy of astrometry in mm/submm VLBI
observations must be better than approximately 2\,$\mu$as for this purpose. The brightest maser line in the LMC/SMC is that of the 22\,GHz H$_{2}$O masers with flux densities of 50--100\,Jy \citep{2010MNRAS.404..779E, 2013MNRAS.432L..16I}. If the search
for new masers in mm/submm bands in the Magellanic Clouds is successful, it will have a strong impact on various fields of astronomy because of the resulting sharp increase in the accuracy of VLBI astrometry.

\clearpage

\newpage

\section{East Asian View of mm/submm VLBI}

In this section, we present the EA view of mm/submm VLBI. First, we summarize the current status of VLBI arrays used for cm/mm wavelength observations in the EA region. In Section 3.1, EA VLBI capabilities and the vision of the EA VLBI community for the near future vision, mainly for VLBI observations at wavelengths of 3, 7, and 13 mm, are discussed. We include Australian stations, which are crucial partners in extending north–south baselines. In Section 3.2, a summary of mm/submm VLBI is presented. In particular, we present our views on EA submm VLBI at 1.3 and 0.7 mm. In Section 3.3, unique capabilities developed in EA VLBI community are highlighted.

\subsection{cm/mm VLBI stations}

\subsubsection{VLBI Exploration of Radio Astrometry}

The VERA array in Japan is dedicated to astrometry of Galactic radio sources. The main bands of the VERA array are K-band (22\,GHz) and Q-band (43\,GHz), and they correspond to the lower frequency tail of the synchrotron emission to be observed with mm/submm VLBI. Therefore, VERA provides excellent complementary data for AGN studies by monitoring flux and image variation and also through astrometry of AGN positions and their variations. For instance, Sgr A* is one of the most crucial sources for studying the Galactic structure with VERA, and a large amount of monitoring data of Sgr A* are available. These data are useful for investigating the physical properties of the accretion disk around Sgr A* and their variations  \citep{2013PASJ...65...91A}.

VERA has also been used for monitoring M87 \citep{2014ApJ...788..165H}, demonstrating its capability to trace the flux and structural changes in the vicinity of M87 in accordance with $\gamma$-ray flux increase observed with ground-based Cherenkov telescopes. 
Several $\gamma$-ray active blazars have also been monitored in intervals of a few weeks in the GENJI program \citep{2013PASJ...65...24N}. GENJI observations have shown that $\gamma$-ray flares are often accompanied by an increase in the radio flux, suggesting that VLBI observations can provide valuable information about the $\gamma$-ray flare mechanism. For future mm/submm VLBI studies, time-domain analysis will be crucial to prove the physical processes in the vicinity of SMBHs, and the combination of mm/submm VLBI and VERA for monitoring will provide unique data for investigating AGNs on the basis of their multiwavelength radio properties.

It is noteworthy that there are a few possible extensions of monitoring programs beyond VERA. For instance, the VLBI network consisting of VERA and other Japanese radio telescopes, called the Japanese VLBI Network (JVN), provides an imaging array with approximately 10 stations. The main bands of the JVN are C-band
(6.7\,GHz) and X-band (8\,GHz), and therefore, it can complement even lower frequencies compared with VERA.
Moreover, one of the large programs of KaVA will be the monitoring of Sgr A*, M87, and selected blazars, and this involves mm/submm VLBI observations. A major feature of KaVA is that its imaging capability with seven stations is considerably higher compared with VERA, which has with only four stations. As discussed in Section~2,  \citet{2014PASJ...66..103N} demonstrated that KaVA can achieve high dynamic range to trace the extended structure of jets in AGNs such as M87 and blazars. Accordingly, the KaVA monitoring program should provide a considerable amount of information that will be useful for interpreting the mm/submm VLBI observations of the same target sources.

%
%
%
%
%
%


\subsubsection{Korean VLBI Network}

\begin{figure}[htbp]
   \centering
	\includegraphics[width=12cm, angle=0]{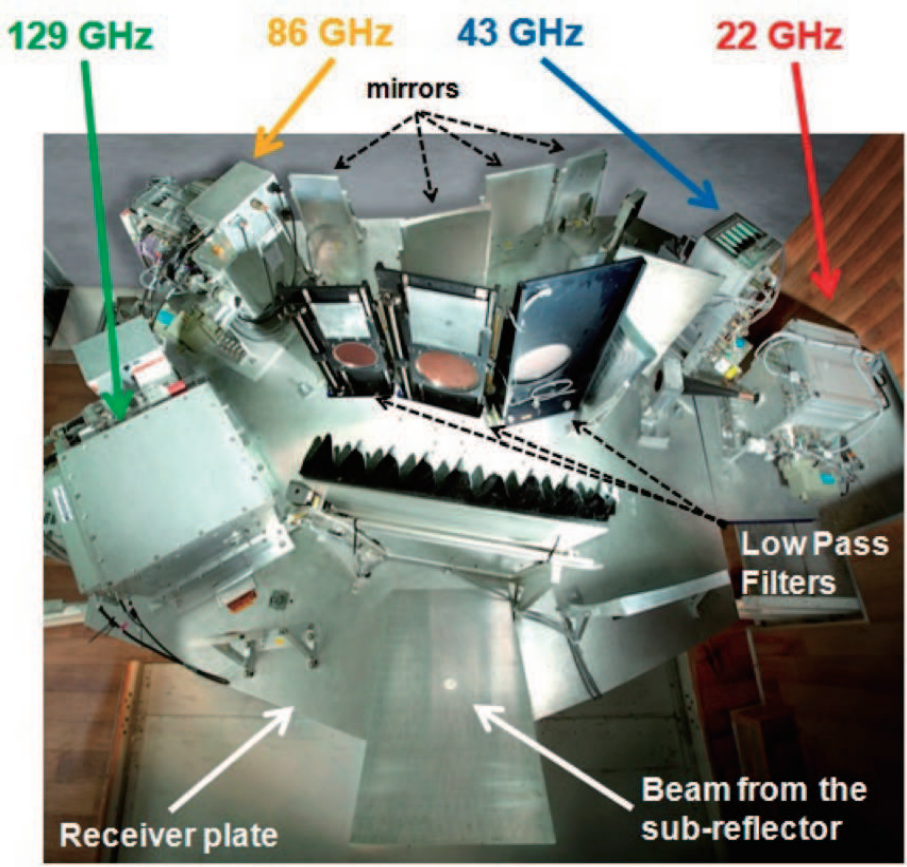}
   \caption{Top view of the simultaneous multifrequency (22, 43, 86, and 129\,GHz) receiver system of the KVN. The quasi-optical mirrors and three low-pass filters supply the signal from the subreflector to the appropriate receiver (Han et al., 2013).
   \label{fig:kvn1}
         }
   \end{figure}

\begin{figure}[htbp]
   \centering
	\includegraphics[width=12cm, angle=0]{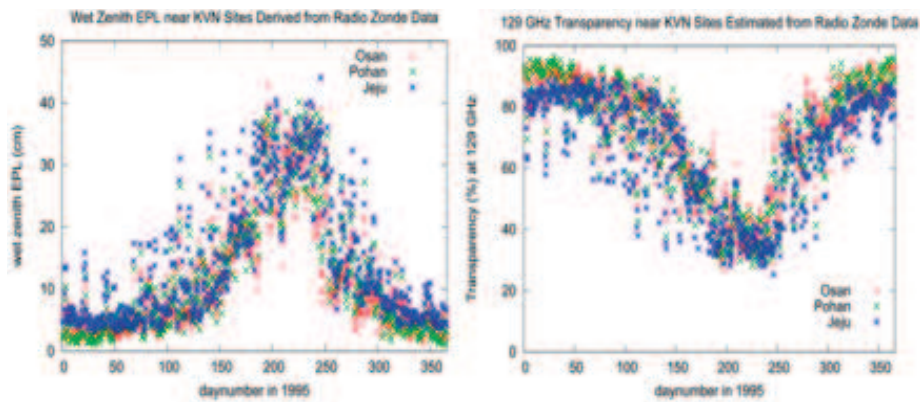}
   \caption{(Left) Wet zenith excess path length and (right) sky transparency at 129GHz derived from radiosonde data recorded near the KVN site in Korea (Sasao and Lee, 2006).
   \label{fig:kvn2}
         }
   \end{figure}

The KVN (http://radio.kasi.re.kr/kvn) consists of three 21-m radio telescopes located in Seoul, Ulsan, and Jeju Island, with a maximum baseline length of 476\,km. The network is dedicated to VLBI observations at mm wavelengths (currently up to 2\,mm, 142\,GHz), and its performance as a single dish and VLBI network was described by Lee et al.\ (2011, 2014). The KVN introduced a unique simultaneous multi-frequency receiver system that is designed for observing four different frequencies (22, 43, 86, and 129\,GHz) simultaneously, and quasi-optical mirrors and low-pass filters are used to supply the radio signal from the source to the appropriate receiver (Han et al., 2008, 2013; also see Figure \ref{fig:kvn1}). This receiver system has a major benefit on atmospheric calibration, which is a major diﬃculty in mm/submm VLBI observations (see details in Section 3.4.2). The KVN can therefore increase the coherence time effectively with the aid of the receiver system and provide new aspects of mm/submm VLBI science. The approximate antenna aperture eﬃciencies of the KVN, which has a shaped Cassegrain focusing system, are 50\% at 22 and 43\,GHz, 60\% at 86\,GHz, and  40\% at 129\,GHz. We are now considering conducting 230\,GHz VLBI experiments with the available telescopes in East Asia. On the basis of current specifications such as total surface accuracy ($\sim$124\,micrometer) and Tsys ($\sim$250\,K), the approximate values of the expected aperture eﬃciency and system equivalent flux density (SEFD) at 230\,GHz are 20\% and 8700\,Jy. The water vapor effects in the Korean peninsula for mm VLBI observations were determined on the basis of radiosonde measurements obtained in 1995 (Sasao and Lee, 2010; \ref{fig:kvn2}).
From November to April, the wet zenith excess path length near the KVN sites is less than 10\,cm and the sky transparency at 129\,GHz is above 70\%. Therefore, the water vapor conditions in Korea are appropriate for conducting mm VLBI observations, except in the summer season.

\subsubsection{East-Asian VLBI Network}

\begin{figure}[htbp]
\centering
\includegraphics[width=15cm]{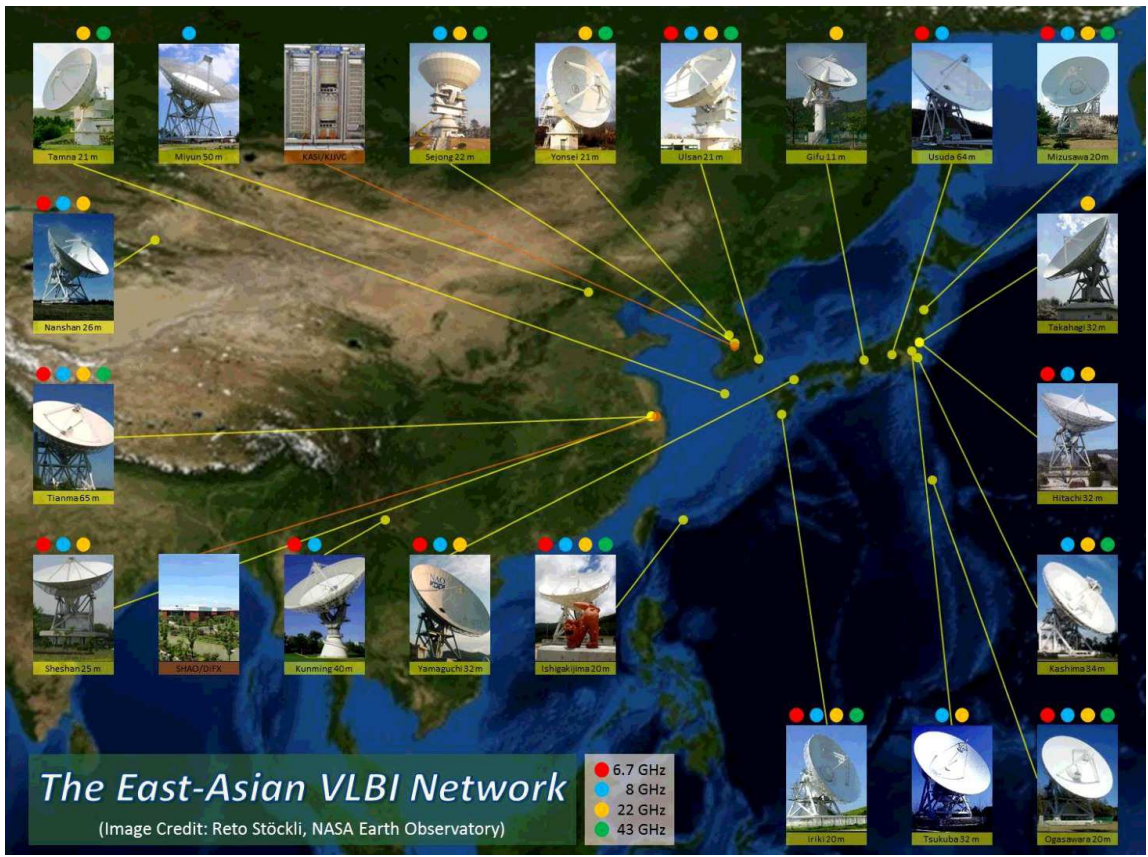}
\caption{
Overview of the EAVN. Photographs of potential EAVN telescopes (yellow points) and correlator sites (brown points) are overlaid on the ``Blue Marble" image (image credit: NASA's Earth Observatory). This figure is taken from \citet{2016ASPC..502...81W}  with a few minor revisions.
}
\label{fig:Sec_3.1.3_Fig-1}
\end{figure}

A new VLBI array in East Asia, the EAVN \citep[EAVN;][]{2016ASPC..502...81W}, is being planned under the auspices of the East Asia Core Observatory Association. The EAVN is the integration of VLBI facilities in EA countries: the Chinese VLBI Network \citep[CVN;][]{2008IAUS..248..182L} in China, the Korean VLBI Network
\citep[KVN;][]{2014AJ....147...77L} in Korea, the Japanese VLBI Network
\citep[JVN;][]{2006evn..confE..71D} and VLBI Exploration of Radio Astrometry
\citep[VERA;][]{2003ASPC..306P..48K} in Japan. Figure \ref{fig:Sec_3.1.3_Fig-1} shows an overall image of the EAVN, which consists of 20 potential telescopes (5 from China, 4 from Korea, and 11 from Japan) and two correlator sites, the Korea--Japan Correlation Center at the Korea Astronomy and Space Science Institute (KASI), and Shanghai Astronomical Observatory. Brief specifications of the EAVN are presented in Table~\ref{tbl:Sec_3.1.3_Tbl-1}. The EAVN will mainly be operated at 6.7, 8, 22, and 43\,GHz, although a part of it can perform observations at 1.6--129\, GHz. The EAVN has very high sensitivity at cm-wavelengths because of the large effective aperture and many large antennas, such as the TianMa 65-m telescope and Usuda 64-m telescope. Thus, the EAVN will be a complementary VLBI array to ALMA (with $\sim 100$-mas-scale angular resolution at submm wavelengths) and the Square Kilometer Array (with $\sim 10$-mas-scale angular
resolution in the 100\,MHz to 10\,GHz frequency range) in terms of angular resolution and frequency coverage.

We have performed eight test observations with the EAVN since 2013 at 8 and 22\,GHz, and fringes have been successfully detected between international baselines of the network. Furthermore, we conducted science commissioning observations of methanol masers in massive SFRs at 6.7\,GHz in 2010 and 2011 \citep{2014PASJ...66...31F} and also started regular operation of the Korea--Japan joint VLBI project and KaVA \citep[KaVA;][]{2014ApJ...789L...1M,2014PASJ...66..103N}.

On the basis of the results obtained with the aforementioned test observations, we conducted the first EAVN imaging test observation on 2015 December 13 at 8\,GHz to confirm the array performance and imaging capability of the EAVN. Nine telescopes participated in the observation, and the bright AGN 4C\,39.25 was the target source. We started imaging tests at both 22 and 43\,GHz and shared the observing time with the KaVA AGN Large Program. Notably, the observation at 43\,GHz was conducted as a joint program with the Australia Telescope Compact Array (ATCA), which is a good example of a future collaboration between the EAVN and Australia. We also plan to conduct imaging test observations with the EAVN at 6.7\,GHz by using a new 1\,Gbps back-end system in 2016, and in the current stage, we will invite proposals for observations with the EAVN in the second half of 2017.

\begin{table}[htbp]
\caption{Specifications of the EAVN.}
\label{tbl:Sec_3.1.3_Tbl-1}
\begin{center}
\begin{tabular}{ll}
\hline
Number of (potential) telescopes        & 20 \\
Frequency coverage                      & 6.7~GHz (12 stations), 8~GHz (16), 22~GHz (16), 43~GHz (9) \\
Angular resolution                      & 1.5~mas (8~GHz), 0.6~mas (22~GHz), 0.7~mas (43~GHz)        \\
7-$\sigma$ Fringe detection sensitivity & 1.6~mJy (8~GHz), 9.5~mJy (22~GHz)                          \\
                                        & (for a continuum source, $\tau = 60$~s)                    \\
Recording rate                          & $\ge 1$~Gbps ($B = 256$~MHz)                               \\
Correlator                              & Korea--Japan Joint VLBI Correlator (KASI), DiFX (KASI/SHAO) \\
\hline
\end{tabular}
\end{center}
\vspace{-0.3cm}
\end{table}

\newpage
\subsubsection{Nobeyama Radio Observatory 45-m telescope}

The Nobeyama Radio Observatory 45-m telescope (NRO45m) operated by NAOJ is one of the most powerful mm-wavelength radio telescopes in the world. Currently three main frequency bands --- 22, 43, and 86--112 GHz ($\lambda=$ 13mm, 7mm, and $\sim$3mm) --- are available, and the telescope is open to all astronomers during the winter season (December to May).

Historically, the NRO45m has participated in several VLBI sessions, including global mm VLBI observations. Additionally, 86\,GHz VLBI observations were performed with the Taeduk Radio Astronomy Observatory 14\,m telescope in Korea \citep{2004PASJ...56..475S}. The station position of the NRO45m has been regularly measured with high accuracy for VLBI observations. Currently, the NRO45m participates mainly in 13 and 7\,mm VLBI observations together with VERA, based on VERA open-use observations.

\begin{figure}[htbp]
	\begin{center}
	\begin{minipage}{0.45\hsize}
		\begin{center}
			\includegraphics[width=0.8\linewidth]{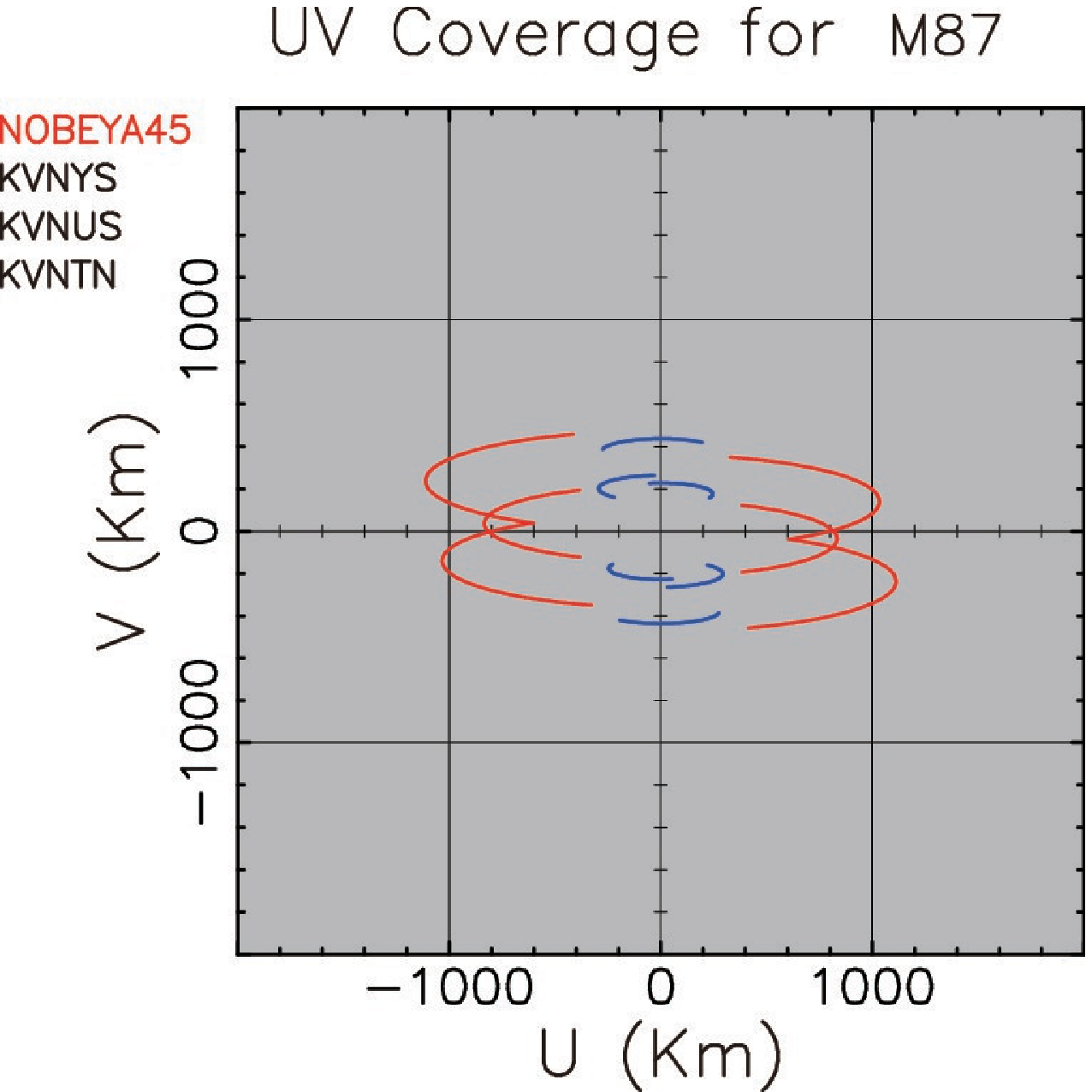}
		\end{center}
	\end{minipage}
	\begin{minipage}{0.45\hsize}
		\begin{center}
			\includegraphics[width=0.85\linewidth]{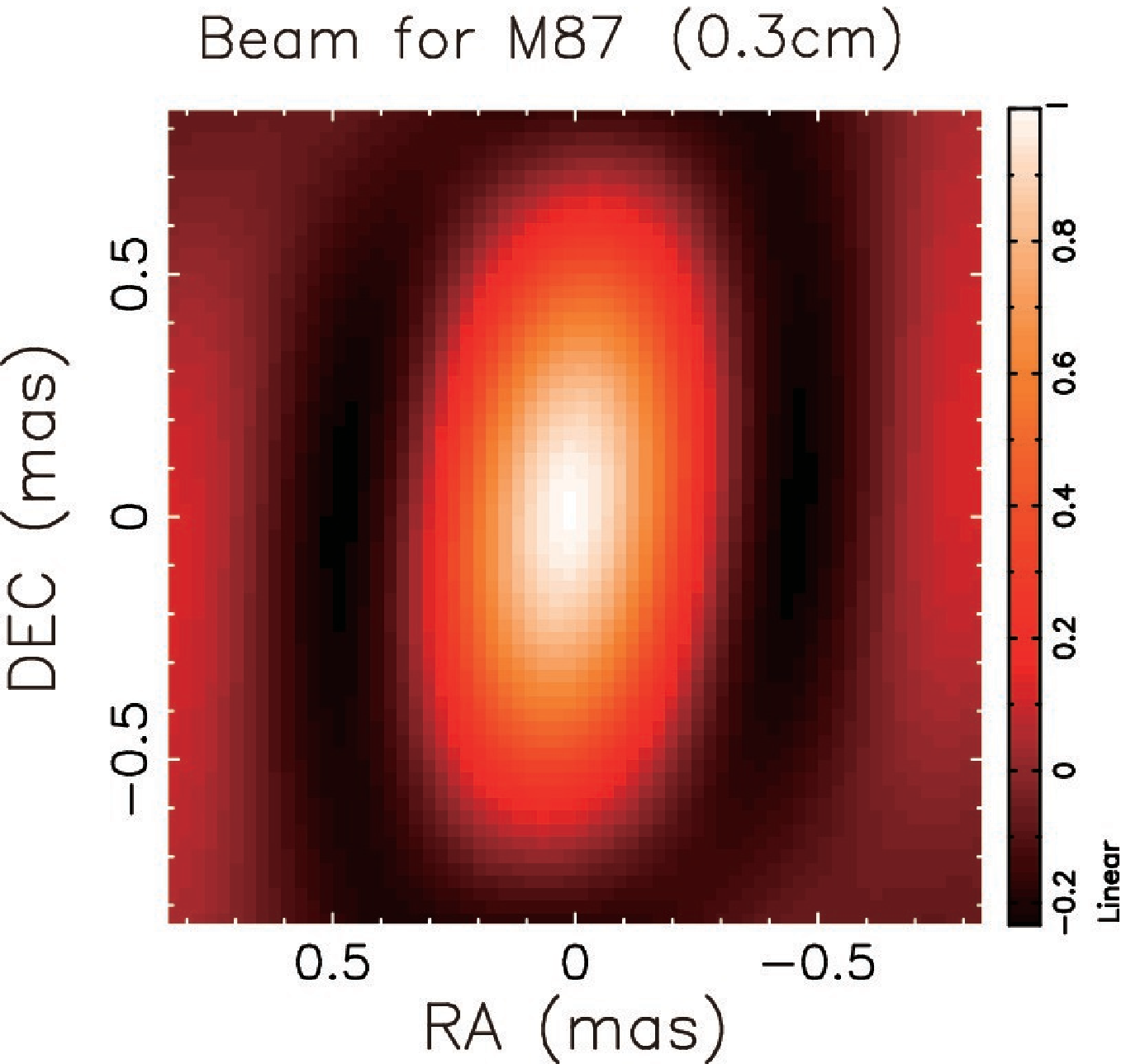}
		\end{center}
	\end{minipage}
	\caption{\textsl{Left}: (Left) The $uv$-coverage for M87 expected from the observation with 8-h tracks, obtained using the NRO45m and KVN baselines. The red solid line indicates the NRO45m baselines. (Right) The synthesized beam expected to be obtained with the NRO45m and KVN baselines for M87.\label{kvn+nob}}
	\end{center}
\end{figure}

In East Asia, the KVN has routinely performed VLBI observations at 3\,mm. However, the KVN has only short baselines ($< 500$\,km), and the accuracy of amplitude calibration is not high because it consists of three radio telescopes and three baselines. VLBI observation with the NRO45m and the KVN at 3\,mm will facilitate the achievement of a longer baseline greater than
1000\,km in the east–west direction (e.g., M87 in the left panel of Figure \ref{kvn+nob}) and improve the accuracy of amplitude calibration because of the use of four stations and six baselines. Therefore, 3\,mm VLBI observation in East Asia will be enhanced by the participation of the NRO45m.

The 3\,mm wavelength is crucial not only to achieve higher angular resolution but also to view the emission coming from upstream of the jet by avoiding the effect of the opacity structure seen in the M87 jet (e.g., Hada et al., 2011). Actually, the VLBI observation of M87 with the NRO45m and KVN at 3\,mm is expected to achieve an angular resolution of 0.4\,mas, corresponding to approximately 100\,$R_\mathrm{s}$ toward the direction of the de-projected jet
of M87 (see Figure \ref{kvn+nob} \textsl{right}). 
Here, we assumed that the $M_\mathrm{BH}=6\times10^9M_\mathrm{\odot}$, $D=16.7~\mathrm{Mpc}$ and viewing angle 
$\theta=35^{\circ}$ \citep{2015arXiv151203783H}.

\subsubsection{Shanghai 65\,m (TianMa) Telescope}

The Shanghai 65-m radio telescope is a joint project between the Chinese Academy of Sciences (CAS) and the city of Shanghai. Having been constructed for a range of scientific purposes, the telescope is a giant rotatable radio telescope with a 65-m-diameter dish, and was built with collaborative funding from the CAS, Shanghai Municipal Government, and Chinese Lunar Exploration Program. The telescope is located in Sheshan, at the Songjiang base of the Shanghai Astronomical Observatory, which is responsible for operating the telescope under the auspices of the CAS.

The TianMa radio telescope features a modified Cassegrain antenna comprising a 65-m primary reflector with a primary focal ratio of 0.32 and a 6.5-m subreflector. The primary reflector of the telescope is made up of 14 rings of 1008 high-precision solid panels, and the subreflector comprises 3 rings of 25 double-layered, aluminum honeycomb reflecting panels. The accuracies of the panels in the primary reflector and subreflector are better than 0.1\,mm rms and 0.05\,mm rms. respectively. A novel technology, known as active surface control, was adopted by the primary reflector. A total of 1104 actuators were installed at the locations where the panels join the antenna backbone structure, for compensating for the gravitational deformation of the reflecting surface during tracking. This compensating mechanism improves the receiving efficiency of the antenna in high-frequency observations. The TianMa telescope is characterized by high operating frequencies and wide receiving bandwidths. The telescope is fitted with receiving equipment for eight frequency bands that cover the range 1--50\,GHz. The telescope covers the entire cm band and part of the long mm band for astronomical observations, and it is China’s first radio telescope operating at a wavelength of 7\,mm.

There are two approaches to select the operating bands, namely changing the feedhorn through rotation and deflection of the incoming signals by the subreflector. Changing the feedhorn causes switching between different observing frequencies by first placing the feedhorns for all but the L-bands on a mechanical rotating disk, and then positioning and locking the selected feedhorn at the second focal point (or focus of the subreflector) through rapid rotation of the disk. In the other approach, the L-band feed is aligned solely by deflecting the incoming signals with the aid of the subreflector because the bulky L-band feed is fixed at a position that has deviated from the symmetric axis of the antenna. This frequency switching process can be completed within a minute.

The Shanghai 65-m radio telescope project was approved at the end of October 2008, its foundation was laid on 2009 December 29, on-site construction began on 2010 March 19, and the main work was completed on 2012 October 28. On 2012 October 26, the telescope was used for the first time to observe the massive SFR W3(OH) in the 18-cm band. On 2012 November 28, a single-baseline interferometry fringe was successfully detected in collaboration with a 25-m telescope at Sheshan, Shanghai. Subsequently, numerous test observations were performed. The telescope was given the name TianMa (Pegasus), on 2013 December 2. Because of its superior features such as high sensitivity and wide frequency coverage, the telescope, as a single dish, can play a crucial role in astronomical observations, research on molecular spectral lines, and the study of pulsars. The TianMa telescope will also be used for observing weak radio sources, and specific targets include radio blazars, microquasars and X-ray binaries. Studies on the fast time variation of AGNs and the transient phenomenon of X-ray binaries require high-sensitivity single-dish observations, which are often combined with other means of observation.

Following the successful installation and testing of the digital back-end system, which can support both spectral line and pulsar observing modes, calls for proposals of single-dish observations were made: the first call was made on 2014 September 15 for pulsar research, and the second call was made for spectroscopic observations on 2014 December 29. As a key component of the VLBI network in China and the world, the TianMa telescope considerably increases its resolution and capability for VLBI observations. This telescope is expected to open up new areas of mm VLBI observations. As a VLBI station, it has participated in observation programs involving most of the VLBI networks in the world, including the CVN, EVN, VLBA, LBA, International VLBI Service for Geodesy and Astrometry, EAVN, and space VLBI observations with RadioAstron. In particular, a test observation of TianMa along with the dedicated high-frequency VLBI network KaVA was successfully performed at 43\,GHz on 2015 December 7. The fringes to TianMa were well detected, revealing the potential of the telescope for 43\,GHz VLBI observations in the near future.

\subsubsection{Australia Telescope Compact Array  and  Long Baseline Array}

The ATCA consists of six 22-m antennas that operate in five bands: 16\,cm (1.1--3.1\,GHz), 4\,cm (4.5--12\,GHz), 15\,mm (16--25\,GHz), 7\,mm (30--50\,GHz), and 3\,mm (83--105\,GHz,
five antennas only). Observations at 7\,mm, and 3\,mm in particular, are typically conducted in the Southern Hemisphere winter months (roughly May to October).

As a stand-alone facility, the ATCA records data on two 2-GHz bandwidth dual polarization channels with spectral resolutions ranging from 64\,MHz to 0.5\,kHz \citep{2011MNRAS.416..832W}. The array is reconfigured every few weeks, cycling through 17 standard array configurations with maximum baselines ranging from 90\,m (between the innermost five antennas) to 6\,km. For higher phase stability, more compact array configurations are favored for VLBI observations, with baselines typically less than 350\,m. Only the inner five antennas are phased-up to form a ``tied array'' for VLBI observations. A channel bandwidth of 16 or 64 MHz is most commonly used for VLBI recording, and currently, it is possible to achieve a total sky bandwidth of up to 128\,MHz, dual polarization, and data rates up to 1\,Gbps.

The LBA, used for Southern Hemisphere VLBI \citep{2015PKAS...30..659E}, is used primarily in cm bands. One of the 34-m Tidbinbilla antennas has a 32-GHz receiver and has been used with the ATCA and Mopra 22\,m telescope for a three-station AGN survey at 32\,GHz. 
The LBA operates for a total of approximately 25 days per year, with the majority of days being concentrated in three or four observing sessions; additional out-of-session support is available
for RadioAstron observations and occasional time-critical observations. VLBI observations requiring only
the ATCA can be scheduled with more flexibility.

Disk-based recording systems are now used for all LBA telescopes. The Australian stations use the LBA Data Recorder \citep{2009evlb.confE..99P}. A bit rate of 512\,Mbps (2 polarizations, with 2-bit digitization and Nyquist sampling) can be sustained at all LBA telescopes and is the standard observing mode. Higher bit rates, up to 1\,Gbps, can be achieved but are not supported at all telescopes. Raw voltages are recorded in a local format, but tools exist to translate this into Mark5B or VDIF \citep{2009evlb.confE..99P} formats, improving the compatibility between Australian VLBI antennas and international antennas that involve other disk-based recording systems. At most stations, data are transferred to the correlator by using fast network connections.

\subsection{mm/submm VLBI stations}
\subsubsection{Greenland Telescope }

Submillimeter VLBI is aiming at proving the existence of SMBHs by imaging their shadow. The ASIAA VLBI group has been working to prepare a 12m antenna for operations at submm wavelengths.  The antenna was built by Vertex as a prototype for ALMA-NA.  In April 2011, the US National Science Foundation (NSF) awarded this antenna to the VLBI group led by ASIAA, collaborating with the Smithsonian Astrophysical Observatory (SAO), MIT Haystack Observatory, and the National Radio Astronomy Observatory (NRAO), based on a competitive proposal to redeploy this antenna for submm VLBI.  Since the award, ASIAA has successfully taken the lead on preparing the antenna for operations. 

During 2010 and 2011, we surveyed sites that were suitable for the new telescope for conducting submm VLBI observations of SMBH shadows, and we found a site, near the summit of the Greenland ice sheet, satisfying our requirements. An alternative site is near the high site of the South Pole, but the 10-m South Pole Telescope is already in place at the location. Summit Station, located near the summit of Greenland and within the Arctic Circle, is supported by the NSF for polar research with limited infrastructure. In the summer of 2011, we deployed a radiometer at Summit Station to monitor sky conditions and confirmed very good weather conditions at the site. Since then, the GLT project has been operating in close collaboration with the NSF.

Summit Station and (the planned ISI Station at the summit) has an average temperature between $-$30$^{\circ}$\,C and $-$40$^{\circ}$\,C, and extreme temperatures (below $-$70$^{\circ}$\,C). Materials and facilities have to be specifically designed for operations in such a low-temperature environment. Although ISI Station is at an altitude of only 3,200\,m, the atmosphere at the station has very low water contents because of its cold ambient temperature. The Greenland station will provide
high angular resolution information with ALMA in Chile, and JCMT and SMA in Hawaii, and therefore, commencement of the operations of the GLT is eagerly awaited.

During 2012--2015, the antenna has been retrofitted to operate in the cold climate of ISI Station.In 2016, more than 180 metric tons of antenna components and equipment were shipped to the Thule Air Base, Greenland.  Thule, located 1,200km inside the Arctic Circle, will serve as the initial antenna location and allow VLBI observations at 230 GHz until the Summit Station is prepared.  The project team completed the cold assembly of the reflector dish composite backup structure as well as assembly of the telescope base, azimuth and elevation structure.  The plan for 2017 is to integrate the reflector dish onto the base structure and complete integration, assembly and test of all necessary drive electronics by mid-summer.  Receiver components, electronics and Hydrogen Maser will be shipped for integration during the summer of 2017. 
 
In addition to the antenna work, we have been designing and fabricating receiver systems for the GLT in collaboration with the NAOJ and Osaka Prefecture University. We are also developing terahertz receivers for single-dish science studies.

Parallel to the engineering works above, we have been investigating physics on the M87 and relevant issues to dominate the VLBI related science on M87, which is our primary target to identify the shadow image and other sciences.  These are described elsewhere in this paper.  So, we have coordinated an international symposium on M87 and related topics in May 2016 at ASIAA in Taipei.

\subsubsection{James Clerk Maxwell Telescope}

The JCMT is a 15-m submm single-dish telescope located at the summit of Mauna Kea, Hawaii, USA. It started operating in 1987, and it was managed by the Joint Astronomy Center, which was funded by the United Kingdom, Canada, and the Netherlands. In 2015, the operation center moved to the EA Observatory (EAO), which is funded by institutes in China, Japan, Korea, and Taiwan.

The history of submm VLBI studies involving the JCMT is long; this telescope was one of the first submm telescopes to be used in submm VLBI experiments in the EHT project, together with a single antenna from CARMA in California, USA, and the Submillimeter Telescope (SMT) in Arizona, USA. In particular, the first observation made with these three telescopes was toward Sgr A*, the SMBH at the center of our Galaxy, in 2007, and it was successful. These telescopes helped detect the 230-GHz continuum emission of the SMBH, and measure the intrinsic size of the SMBH as approximately  40\,$\mu$as, which is roughly 5\,Rs (Doeleman et al., 2008). Since then, the JCMT has almost always been used for
EHT observations, and many results have been obtained using the telescope. The JCMT is therefore the most solid submm telescope for submm VLBI observations in the world.

After the JCMT moved to the EAO, a few submm VLBI runs were performed. Because of lack of experience of the EA VLBI community in submm VLBI observations, they faced many difficulties. However, fringes were successfully detected with other telescopes. The EA VLBI community requires more experience in JCMT VLBI operations.

For future EA submm VLBI observations, the JCMT must be included in the network, since it is the most used telescope in submm VLBI observations in this region. A cause of concern regarding the JCMT is that there is only an old single-polarization 230\,GHz receiver for VLBI. Currently, the installation of a new SIS mixer for improving the sensitivity is under consideration. Because there is space in the cabin for additional receivers, it would be useful to have more receivers at different frequencies. Another concern is that the VLBI reference signal and the back-ends are all located in the SMA building, which is located next to the JCMT. Some of the cables are using common path for the SMA normal operation, and therefore, performing VLBI observations by using the JCMT will affect the daily SMA operation. Separating the JCMT VLBI system from the SMA normal operation system is vital for future operations of the JCMT for VLBI.

\subsubsection{The Submillimeter Array}

The SMA consists of eight 6\,m diameter dishes located on Mauna Kea, Hawaii, USA \citep{2004ApJ...616L...1H}.   The array is equipped with receivers that span the frequency from 190 to 420\,GHz. An upgrade is underway to provide dual-polarization receivers with wideband back-ends over the entire frequency range. Current dual-polarization observations depend on time multiplexing through the insertion and rotation of quarter-wave plates. In the future, a birefringent optical element will separate incoming polarizations into two separate receivers.

The existing application-specific integrated circuit correlator will be eventually replaced by an FPGA-based correlator known as SMA Wideband Astronomical ROACH2 Machine \citep[SWARM;][]{2014AAS...22441405W}. SWARM will provide 8\,GHz bandwidth, but can be upgraded to provide higher bandwidths.

The SMA has been used extensively for VLBI observations at a wavelength of 1.3\,mm. Currently, it supports JCMT VLBI observations by providing a maser reference signal, a local oscillator signal for locking the JCMT receiver, and back-end recording capabilities. The SMA and JCMT can be operated jointly as a local interferometer in the eSMA mode, which is an important demonstration of system performance. The SWARM correlator includes a phased array back-end that adds in-phase the eight antenna signals delivered to the recording system. It is possible to include the JCMT as an element of the SMA phased array signal, and this would nearly double the sensitivity of the phased SMA alone. The SMA and JCMT have been used to provide orthogonal circular polarization.

\subsubsection{Solar Planetary Atmosphere Research Telescope}

A 10\,m mm-wave radio telescope, named SPART, is located at the Nobeyama Radio Observatory in Nagano prefecture, Japan. This telescope was an element antenna of the Nobeyama Millimeter Array, and it is owned and operated by Osaka Prefecture University. The telescope is used as single-dish with 100- and 200-GHz dual-band receiver. Prof.\ Maezawa, who is the leader of this telescope experiment, and his group investigate the planetary atmosphere by using molecular line emission/absorptions, such as the CO (J=2-–1) line at 230\,GHz.

\begin{figure}[htbp]
   \centering
	\includegraphics[width=14 cm, angle=0]{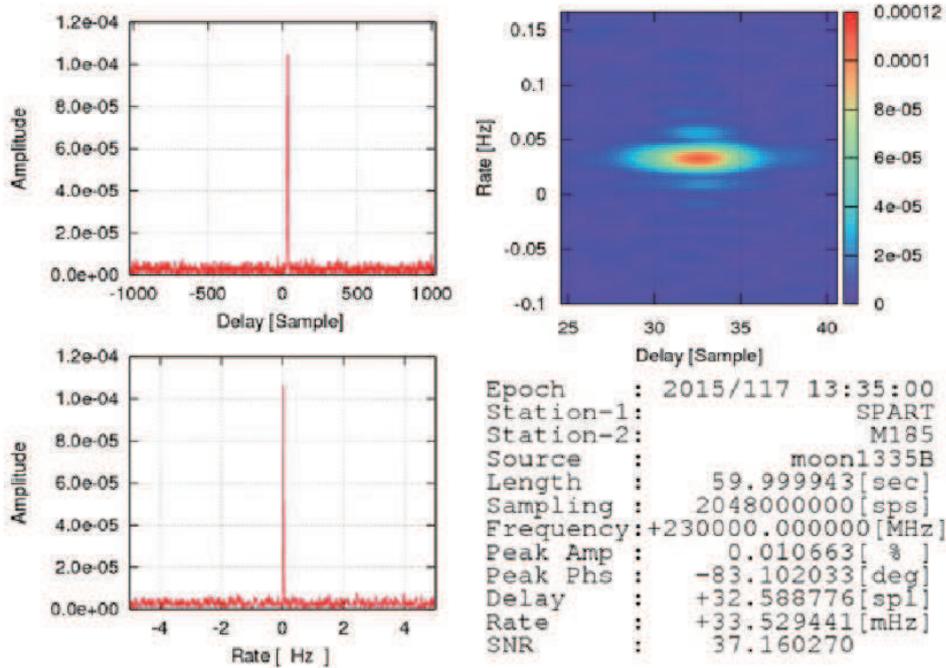}
   \caption{The detected fringe on the SPART 10\,m -- OPU 1.85\,m baseline. The integration time was 60 seconds and the signal-to-noise ratio was 37.
   \label{fig:SPART}
         }
   \end{figure}

At the site of the Nobeyama Radio Observatory, there is another small telescope operated at 230\,GHz. This Osaka Prefecture University (OPU) 1.85-m telescope (led by Prof.\ Ohnishi) is also used as a single-dish telescope for CO line observation. On 2015 April 27, a VLBI observation was carried out with SPART and the OPU 1.85\,m telescope at 230\,GHz for an engineering experiment; the experiment was a collaboration among four universities and three institutes. The baseline length was 150\,m, the observing frequency was 230\,GHz, the sampling rate was 2048 Msample/s with 1 bit/sample. The VLBI terminal was the K6 system, developed by the National Institute of Information and Communications Technology (NICT). An H-maser and Oven-Controlled Crystal Oscillator (OCXO) were used as independent frequency standards. The OCXO was a highly stable and relatively low-cost frequency standard. The target source was the edge of the Moon. Correlation was performed using the GICO3 software correlator, which was also developed by the NICT. Consequently, fringes with the expected amplitude and phase stability were successfully obtained (Figure~\ref{fig:SPART}).


Scientific VLBI observations with SPART at 230\,GHz and with longer baselines is under consideration. As a first step, a Korea--Japan experiment is being discussed by groups from both countries. A 6-m telescope at Seoul National University (SRAO) is operated at 230\,GHz as single-dish. This telescope is located 1,000\,km from SPART and it is a possible candidate for use in combination with SPART in future experiments. Another candidate is the JCMT at Hawaii, operated by the EAO.

\subsubsection{Seoul Radio Astronomy Observatory }

The SRAO is located on the Gwanak campus of Seoul National University (37.454$^{\circ}$\,N, 126.955$^{\circ}$\,E), to the south of Seoul, Korea. Its construction was completed in October 2001, and it features an aperture of six meters \citep{2003JKAS...36...43K} and a 1.3\,mm band dual linear polarization receiver covering the
frequency range 210--250\,GHz with a bandwidth of 1\,GHz and a spectral resolution of 61\,kHz \citep{2013JKAS...46..225L}. Across the frequency range, the aperture and beam efficiencies, derived from test observations of planets, are in the ranges 44\%--59\% and 47\%--61\%, respectively. Under good weather conditions, system temperatures are approximately 150\,K. With water vapor columns ranging from approximately 5\,mm in January to approximately 40\,mm in July, SRAO observations are typically limited to the time interval from November to April.

SRAO has been operated only as a stand-alone single-dish telescope. Integrating SRAO into VLBI networks is tempting, especially because of its ability to cover the 1.3\,mm band. Thus, SRAO is a candidate for future EHT VLBI networks. Integrating SRAO would require dedicated hardware upgrades: (a) various interventions in the signal chain, (b) conversion of linear to circular polarization through installation of quarter-wave plates in the receiver, (c) installation of a sampler, (d) installation of a clock, and (e) installation of a (or connection to a remote) VLBI recorder. If sufficient community support is provided, SRAO could join the EAVN at relatively short notice.

\subsection{Unique capabilities}
\subsubsection{Imaging with Sparse Modeling}

Since the size of a BH shadow is extremely small, very high angular resolution is essential for BH imaging with mm/submm VLBI. 
Basically, the resolution of any interferometric array is given by $\theta\sim \lambda/D$, where $\theta$ is the angular resolution, $\lambda$ is the observing wavelength, and $D$ is the diameter of the array (i.e., the maximum baseline length). 
However, improvement of the angular resolution by a factor of a few is possible by introducing a novel interferometric imaging technique, and recently, imaging with sparse modeling was proposed as an approach to obtain super-resolution images.

\begin{figure}[h]
\centering
\includegraphics[width=142mm]{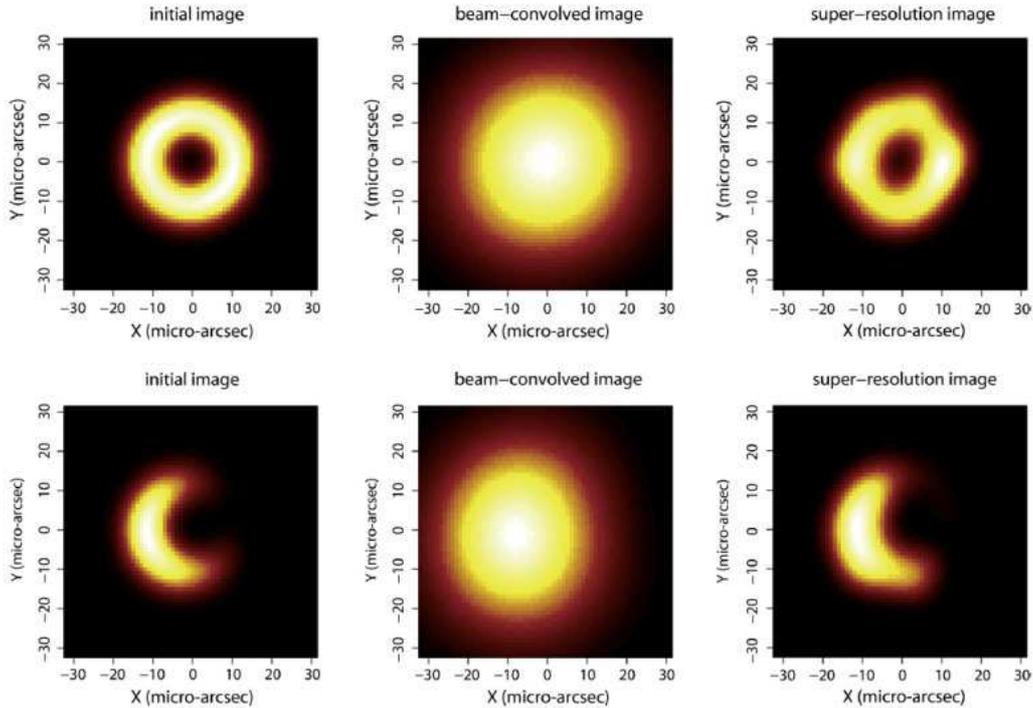}
\caption{
Imaging results for simulated EHT observations of M87 shadow. The upper panels correspond to the ring case, and the lower panels to the crescent case. From the left to the right, the panels show an initial image, convolution with the standard synthesized beam, and the solution with sparse modeling. Each image has 64\,$\times$\,64 grids, with a grid size of 1\,$\mu$as. The initial images as well as super-resolution reconstructed images are convolved with a restoring beam that is finer than the standard synthesized beam by a factor of four.
\label{fig:honma14}}
\end{figure}

\citet{2014PASJ...66...95H} demonstrated that a super-resolution image can be obtained by directly solving an underdetermined Fourier equation for interferometric imaging by using a class of sparse modeling techniques called least absolute shrinkage and selection operator (LASSO). They showed that sparse modeling can trace the BH shadow of M87 even in the smaller mass case of $M=3\times 10^9 M_\odot$, whereas
conventional imaging with CLEAN cannot (see Figure~\ref{fig:honma14}). To perform sparse modeling for real
mm VLBI data, further researches have been conducted from a more practical viewpoint, and recently (Ikeda et al., submitted), it was found that imaging with a closure phase is also possible by using a phase-retrieval technique called phase retrieval from closure phase (PRECL). Because directly-observable phases are likely to be closure phases, particularly in the early phase of mm/submm VLBI observations  \cite[e.g.,][]{2015ApJ...807..150A}, methods such as PRECL should be useful for imaging of BH shadows in the initial phase of EHT observations.

\subsubsection{Frequency Phase Transfer}

Rapid fluctuations of the troposphere at high frequencies cause large errors in visibility phases, and therefore, the sensitivity of mm/submm VLBI is considerably limited. For instance, the coherence time of the fringe phase at 100\,GHz in VLBI is only about a quarter of a minute. However, as mentioned in Section~3.1.2, a simultaneous multi-frequency receiver system offers a novel solution for calibrating the tropospheric phase fluctuations by using the phase solutions obtained from a lower frequency that is more stable than the higher frequency; the solution is based on the non-dispersive nature of troposphere to a radio system (Middelberg et al., 2005, Jung et al., 2011). We call this technique FPT technique.

\begin{figure}[htbp]
   \centering
	\includegraphics[width=12cm, angle=0]{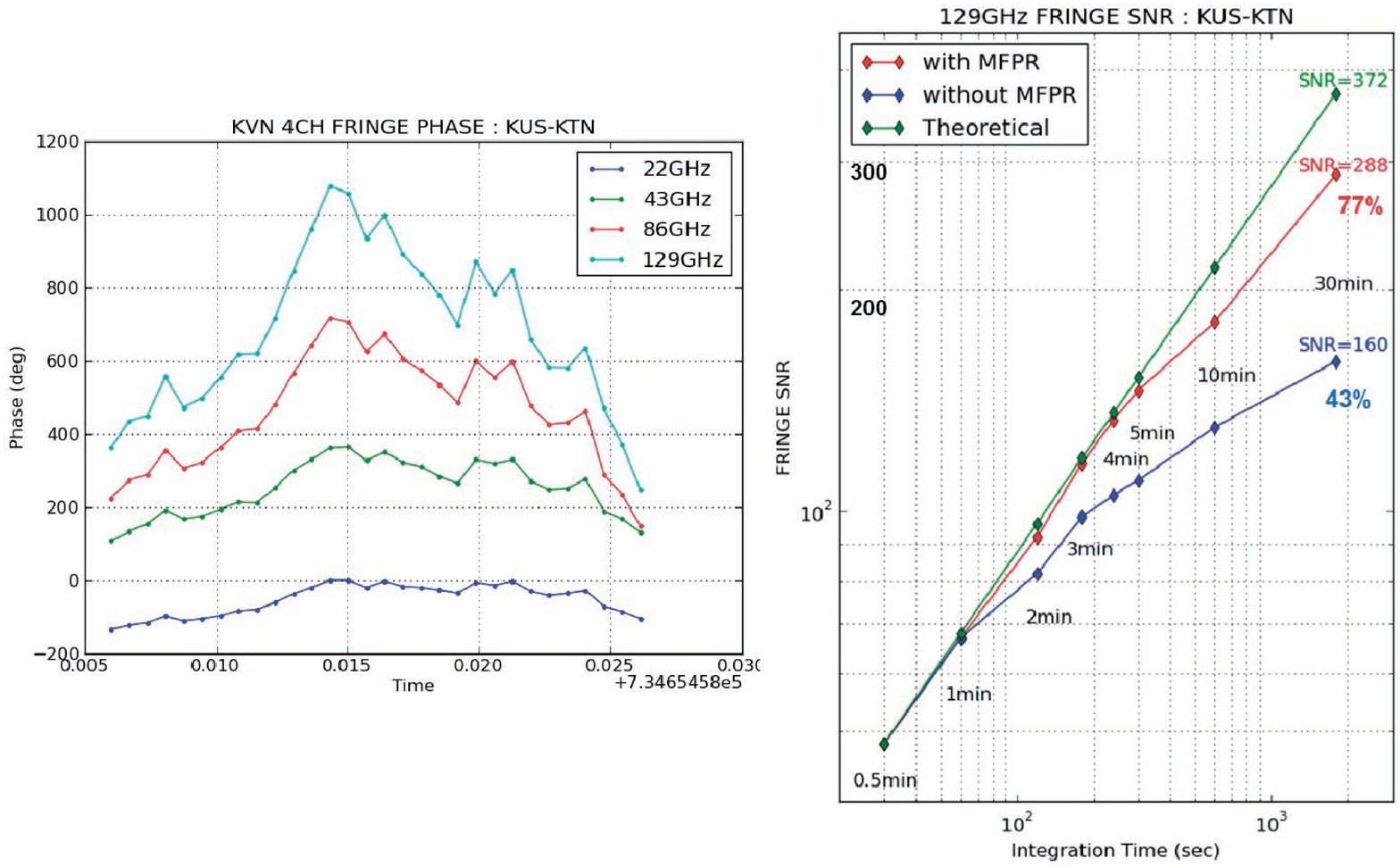}
   \caption{(Left) Four-frequency fringe phase variations for 3C\,279 on the Ulsan--Tamna baseline. (Right) 129\,GHz fringe SNRs of 3C\,279 for integration times of 0.5, 1, 2, 3, 4, 5, 10, and 30\,min. The red and blue lines represent SNR evolution when FPT was applied and when FPT was not applied, respectively. The green line shows the theoretical SNR evolution (Jung et al., 2012).
   \label{fig:fpr1}
         }
   \end{figure}

\begin{figure}[htbp]
   \centering
	\includegraphics[width=16cm, angle=0]{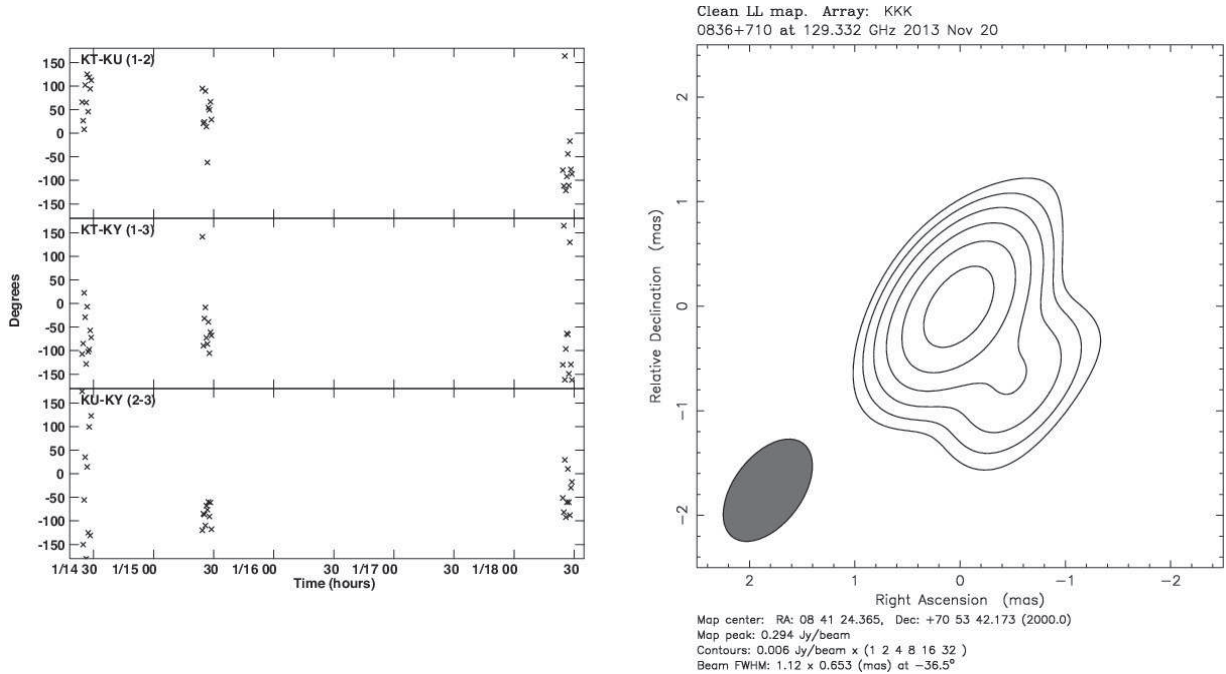}
   \caption{
   Application of the FPT technique to an iMOGABA observation. (Left) Phases of 0836+710 at 129\,GHz after FPT from 86\,GHz; no fringe fitting is applied. Clearly, the scatter is approximately 90$^{\circ}$ ($\pi$/2 rad) or lower, whereas originally no solutions were found for this source. (Right) Cleaned map of 0836+710 at 129\,GHz.
   \label{fig:fpr2} }
   \end{figure}

Jung et al.\ (2011, 2012), Dodson et al.\ (2014), and Rioja et al.\ (2014, 2015) have demonstrated the effectiveness of the FPT when it is applied to VERA and the KVN by showing an effective increase in the coherence time from few tens of seconds up to hours at the frequency range of 20--130\,GHz in dedicated observations. In Figure~\ref{fig:fpr1}, the FPT performance is clearly demonstrated, implying that the technique can effectively compensate for tropospheric phase fluctuations and offer new opportunities for mm/submm VLBI observations. Furthermore, Algaba et al.\ (2015) studied the viability of FPT for a non-dedicated KVN observation, with a sample of 30 all-sky sources, sparse sampling, and a limited number of scans. It was demonstrated that even for non-optimal weather conditions and very restricted ($u,v$)-coverage, FPT allowed phase solutions to be obtained at up to 129\,GHz and several previously undetected sources to be imaged (see Figure~\ref{fig:fpr2}). Clearly, with this technique, we can for the first time systematically monitor a large sample of relatively weak sources at 130\,GHz. We expect that this technique will work at even higher frequencies ($>$130\,GHz) if the non-dispersive characteristics of the troposphere is preserved.

The EA VLBI community has made efforts to extend the high-frequency analysis beyond FPT techniques. For example, attempts are currently being made to extend the FPT domain to calibrate not only tropospheric but also ionospheric phase solutions by using two lower frequencies. Furthermore, source frequency phase referencing can also be performed on mm/submm data. In this technique, once phase solutions have been obtained for high frequencies with FPT, the strongest sources can be used as calibrators for the weaker ones. Combining both these techniques for the KVN, Rioja et al.\ (2015) obtained a coherence time of several hours and the first astrometric results at 130\,GHz with a sample of sources with an angular separation of up to 11$^{\circ}$.

Recently, the KVN-style multi-frequency system was extended to global baselines to maximize the observing capability at mm wavelengths in the extended-KVN project. Successful demonstrations have been presented with VERA Mizusawa and Yebes 40\,m radio telescopes in the K/Q band (Jung et al., 2015), and more radio telescopes in East Asia and Europe will introduce this system over the next couple of years. The FPT technique along with a simultaneous multifrequency receiver system is certainly a powerful tool that will be applicable to mm/submm VLBI.


\subsubsection{Balloon-borne VLBI experiment}

Scientific balloons reach a level flight at an altitude up to approximately 40\,km above sea level in the stratosphere, where atmosphere-free conditions similar to space, in terms of attenuation and phase fluctuation, are available for radio astronomy. The Institute of Space and Astronautical Science (ISAS), a branch of the Japan Aerospace Exploration Agency (JAXA), has decided to launch a dedicated gondola system from the Taiki Aerospace Research Field for a balloon-borne VLBI experiment in the stratosphere.
The experiment aims to conduct a feasibility study and demonstration from technical points of view, for the purpose of determining the prospects for a future high-frequency VLBI project. The first balloon-borne VLBI gondola system \citep{Doi:2016} has been designed to obtain VLBI fringes at 19--23\,GHz toward ground-based radio telescopes of the JVN with a very high signal-to-noise ratio. This allows us to virtually evaluate the stability of the VLBI system as a submm VLBI element by observing a strong radio signal from a satellite; celestial objects such as 3C\,454.3, BL~Lac, and Orion~KL are to be observed as well. The duration of the level flight is planned to be at least 1.5\,h. The gondola will be picked up from the sea by boat.

The on-board observing equipment consists of a 1.5\,m diameter radio telescope, with non-cooled receivers for both left-hand circular polarization and right-hand circular polarization, an 8-Gbps A/D converter, a recorder developed at Mizusawa VLBI observatory \citep{2012ivs..conf...91O}, an 128-Mbps A/D converter, a recording system developed at the NICT as a redundant VLBI system, an oven-controlled crystal oscillator as a frequency standard, precise accelerometers, inclinometers, and gyroscopes for position determination of the antenna under a pendulum fluctuation between the balloon and the gondola with a flight train of approximately 100\,m. Most of the electronics are contained in pressurized vessels. The bus system consists of a GPS-compass and a geomagnetic sensor for coarse attitude determination, gyroscopes and day-time star trackers for precise attitude determination, actuators of a pivot and a reaction wheel for azimuth control, elevation motors based on the mechanical concept of the Wallops ArcSecond Pointing (WASP) system, lithium-ion batteries, and truss structures. The total dry mass is 500\,kg, and power consumption is expected to increase to 500\,Wh. Ground-based VLBI test observations were successfully performed by obtaining very stable fringes between a ground-based radio telescope and the VLBI gondola system that was hung with a pendulum fluctuation, and the activation of the attitude control system and position determination were verified, using the engineering model \citep{Kono:2016}.

\begin{figure*}
\centering \includegraphics[width=1.0\columnwidth]{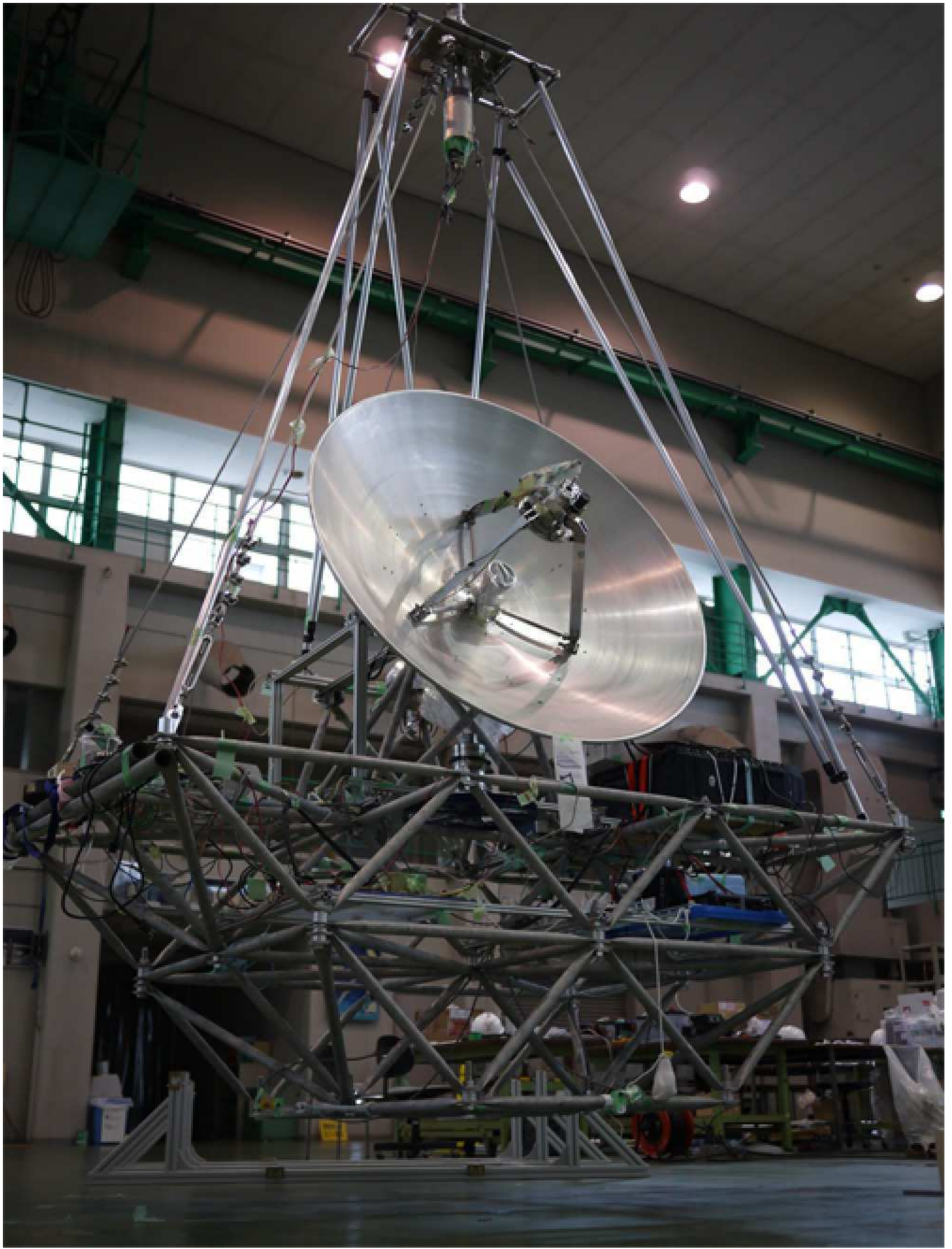}
\caption{Picture of VLBI gondola system being developed at ISAS/JAXA.}
\label{figure:BVLBIpicture}
\end{figure*}



\newpage

\section{Future Vision of East Asia VLBI}

\subsection{Extension of  Arrays}

For the future extension of the EAVN, close collaboration with other VLBI arrays is necessary to produce outstanding scientific results with higher angular resolution and better sensitivity at various observing frequencies. Here, we discuss possible collaborations of our VLBI network with neighboring countries/regions. Such collaborations will enable us to obtain unique data and enrich the overall scientific outcome.
\bigskip

\noindent
\textbf{Extension of arrays at 1.3 mm} 

\noindent
For the VLBI community, mm/submm VLBI is one of the ultimate goals. At 1.3\,m, the EHT network consists of the APEX, CARMA (closed), the JCMT, the Large Millimeter Telescope (LMT), the SMA, the Submillimeter Telescope Observatory (SMTO), the South Pole Telescope (SPT), Pico Veleta, and the Plateau de Bure Interferometer. Observations with the EHT network have been conducted in 2007, 2009, 2011, 2012, 2013, and 2015. From the EA region, the JCMT has been used for VLBI observations at 1.3\,mm (since 2015). In the near future, as summarized in the previous section, we expect to have more EA VLBI stations, such as the GLT, SPART, SRAO as part of the VLBI array at 1.3\,mm. In addition, the ALMA phasing-up capabilityi at 1.3 and 3\,mm was opened to the community from Cycle~4 and observations were conducted for 2017. The ALMA Phasing Project (APP) is one of the ALMA Development Projects that will provide a means to coherently sum all the individual ALMA antennas. This capability will allow ALMA to participate in global VLBI networks operating at mm/submm wavelengths, which will considerably improve the sensitivity of the networks and provide unique baselines. The MIT Haystack Observatory led this project, with ASIAA, NRAO, NAOJ, and other institutions also participating. In Table~\ref{tab:1mm-arry}, we show the sensitivities of each station together with the planned stations in EA regions for 1.3\,mm VLBI.

\begin{figure}[htbp]
\center \scalebox{0.6}{\includegraphics{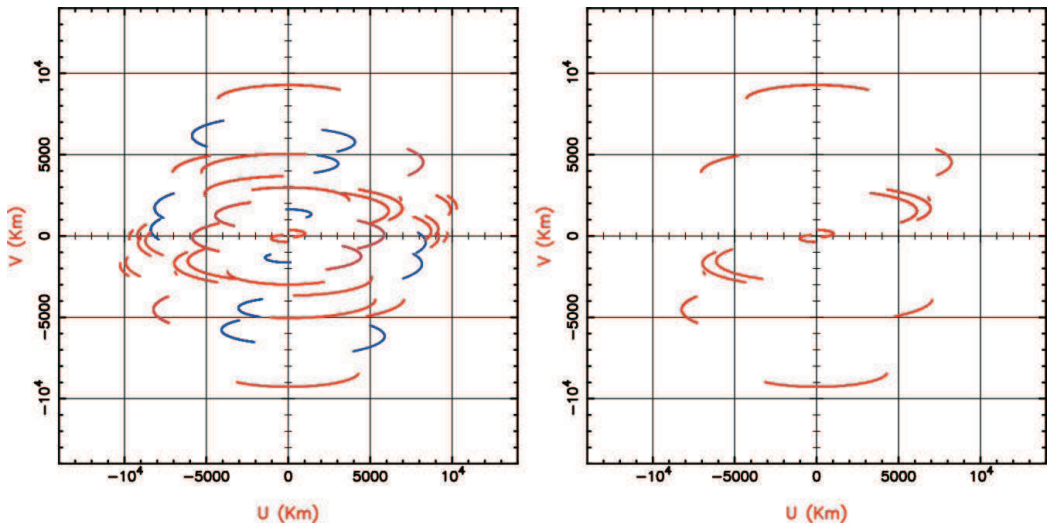}} \caption{
(Left) Expected ($u,v$)-coverage of the current EHT array (JCMT/SMA, SMTO, ALMA/APEX, LMT, IRAM30m) together with new EA stations (SPART, GLT, SRAO) toward M87. Baselines with EA stations (JCMT, SPART, GLT, SRAO) are indicated by the red lines. (Right) Expected ($u,v$)-coverage toward M87 with only EA stations (JCMT, SPART, GLT, SRAO) and ALMA.
\label{fig:M87-uv}}
\end{figure}

\begin{figure}[htbp]
\center \scalebox{0.6}{\includegraphics{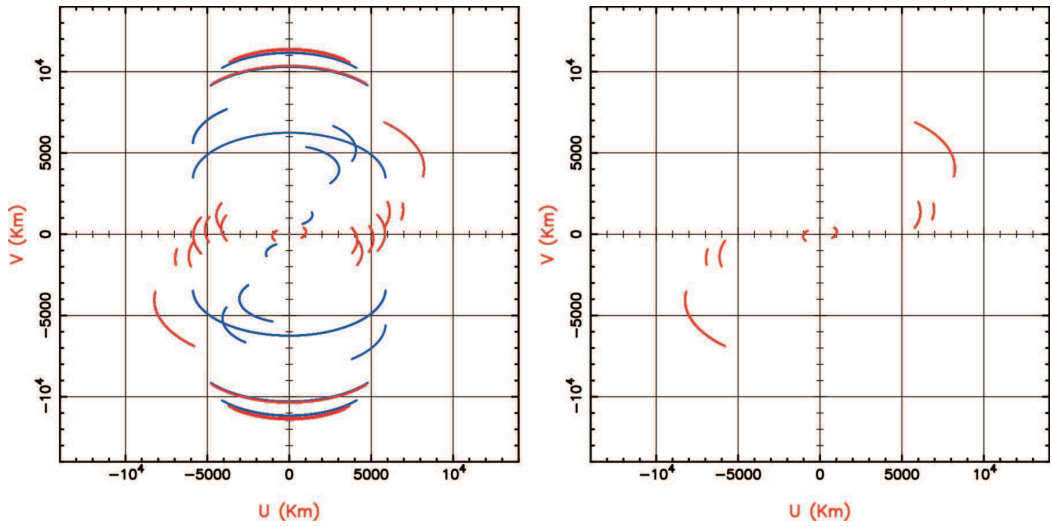}} \caption{
Same as Figure \ref{fig:M87-uv}, but toward Sgr A*. In this case, baselines with SPT are included, but not GLT.
\label{fig:SgrA-uv}}
\end{figure}

The primary scientific goal of 1.3\,mm observations would be imaging/detecting the shadow of the SMBHs Sgr A* and M87, as described in Sections 2.1.1, 2.1.2, and 2.1.3. We also expect the array to play a crucial role in probing blazar jet science as well. In Figures~\ref{fig:M87-uv} and \ref{fig:SgrA-uv}, we show the ($u,v$)-coverage of the entire EHT network (including planned EA stations) and that of only EA stations plus ALMA toward M87 and Sgr A*. For M87, baselines with EA stations provide complementary ($u,v$)-coverage, which will be helpful for improving the dynamic range. The baseline between GLT and phase ALMA is the longest baseline in north-south direction, which provides the smallest scale information on its structure.  The baseline between SPART and SRAO is the shortest, and these two stations can have mutual visibility with the JCMT(/SMA) for five hours. For Sgr A*, again, the baseline between SPART and SRAO is the shortest. More interestingly, baselines between SPART and SRAO to the SPT are very similar to those between the IRAM 30-m telescope and the SMTO to the SPT, but in different Greenwich sidereal time range. Because rapid time variation within a timescale of one day \citep{2014MNRAS.442.2797D} is expected, continuous monitoring of Sgr A* with the event horizon scale fringe spacing would be crucial. For both SMBHs, unfortunately, because of the geography, the SPART and SRAO stations in East Asia cannot provide mutual visibility with ALMA.

We show the ($u,v$)-coverage of the combination of EHT and EA stations and that of the combination of only EA stations and ALMA toward 3C\,84 in Figure~\ref{fig:3C84-uv}. Additional baselines with EA stations render the ($u,v$)-coverage much denser, although the baselines sample similar parameter spaces in the ($u,v$)-distance. This is similar to other blazar sources, and therefore, the coordinated network with EA VLBI stations can be used for studies of blazars as described above. In addition, similar to the case of Sgr A*, this array would provide unique opportunities for time-domain VLBI studies on nearby LLAGNs and blazars.

Therefore, we expect that EA stations will make a significant contribution to the existing 1.3-mm network. Furthermore, joint efforts with EA institutes toward establishing more mm/submm VLBI stations will provide us very unique and valuable opportunities to form our own array, which will be a very powerful array appropriate for time variation studies on Sgr A*, M87, and other AGNs.

\begin{table}[htbp]
  \caption{Current and planned stations for 1.3\,mm VLBI. EA contributions are shown in bold.}
\begin{center}
\begin{tabular}{lcccc}
\hline
Stations & Location & Diameter [m] & SEFD [Jy] & Status \\
\hline
\hline
ALMA 37     & Chile      & 37 $\times$ 12 & 100 & 2017 - \\
APEX        & Chile      & 12 & 3600 & operational \\
{\bf GLT}   & Greenland  & 12 & 7800 & 2018 (planned) \\
IRAM 30m    & Spain      & 30 & 1400 & operational \\
{\bf JCMT}  & Hawaii     & 15 & 4700 & operational \\
LMT         & Mexico     & 32 & 1400 & operational \\
NOEMA       & France     & 15 & 5200 &operational \\
SMA         & Hawaii     & 8 $\times$ 6 & 4000 & operational \\
SMT         & Arizona    & 10 & 11000 & operational \\
{\bf SPART} & Japan      & 10 & 10000 & 2018? (planned) \\
SPT         & South Pole & 10 & 9000 & operational \\
{\bf SRAO}  & Korea      & 6 & 40000 & 2018? (planned) \\
\hline
\label{tab:1mm-arry}
\end{tabular}
\end{center}
\vspace{-0.3cm}
\end{table}

\begin{figure}[htbp]
\center \scalebox{0.6}{\includegraphics{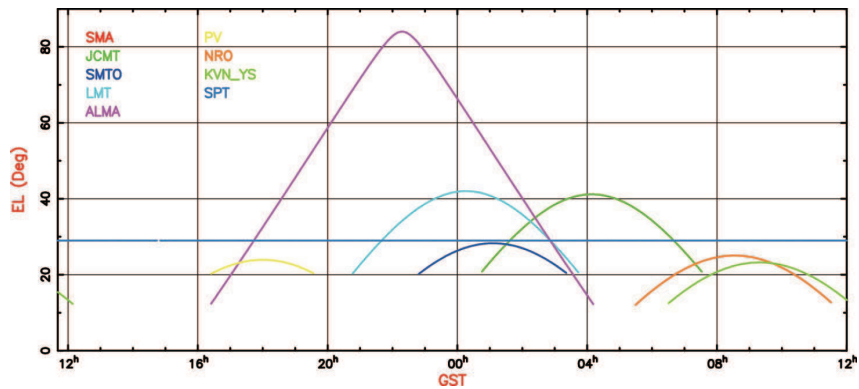}} \caption{
Elevation of each telescope toward Sgr A* versus time. EA stations (SPART and SRAO) will provide a unique window to continuously monitor Sgr A*, which would be very unique and crucial for detecting rapid time variability with changes in the event horizon scale on Sgr A*.
\label{fig:SgrA-EL}}
\end{figure}

\begin{figure}[htbp]
\center \scalebox{0.6}{\includegraphics{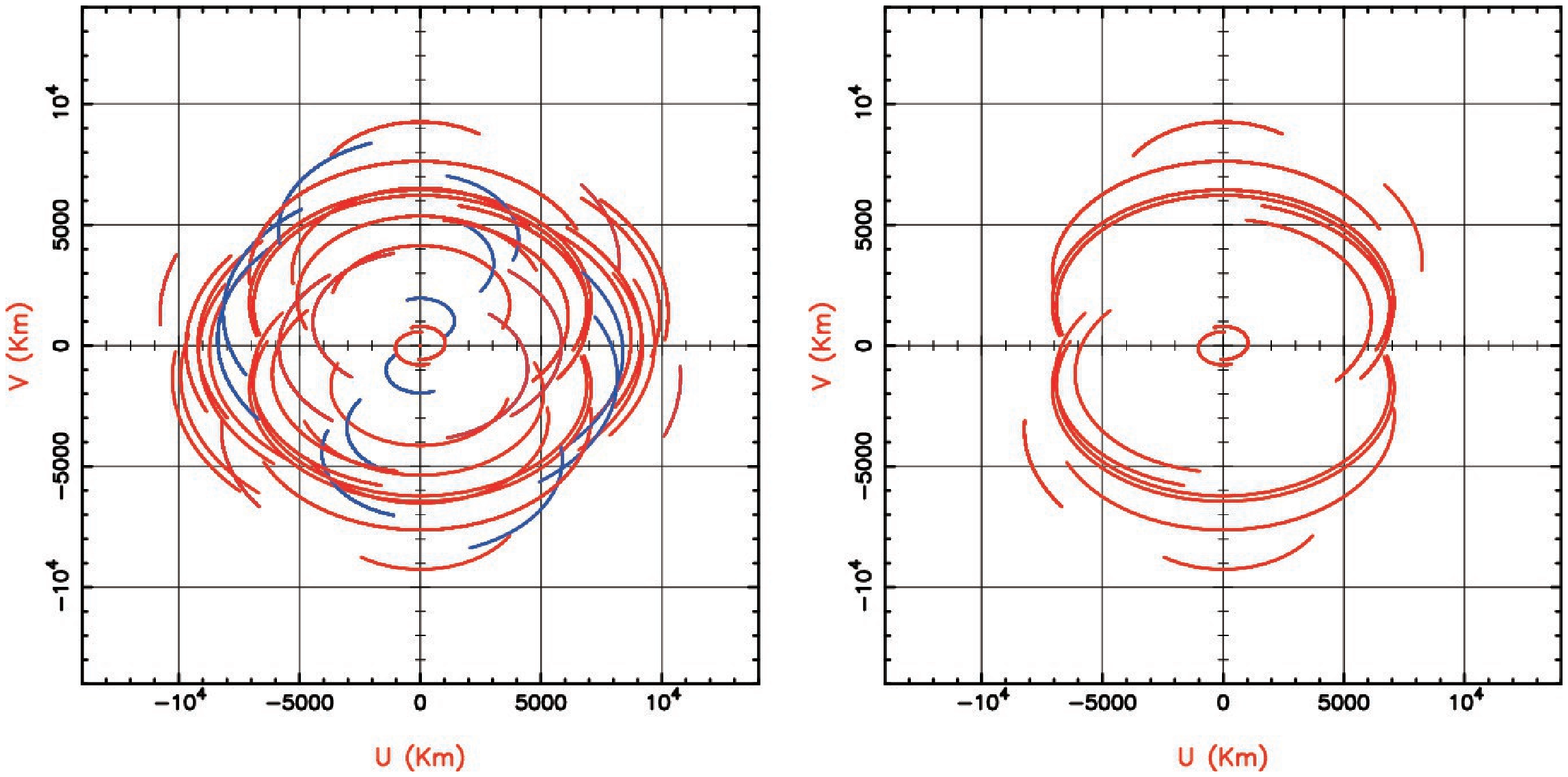}} \caption{
Same as Fig. \ref{fig:M87-uv}, but towards 3C\,84.
\label{fig:3C84-uv}}
\end{figure}

\vspace{3mm}
\noindent
\textbf{Extension of arrays at 0.85 mm and beyond} 

VLBI is a technique to pursue higher angular resolution. Current VLBI at 3 and 1.3\,mm has demonstrated the feasibility of VLBI observations at a few times the event horizon scale resolution. In the next couple of years, mm/submm VLBI at 1.3\,mm (and possibly 3\,mm) is expected to be used to image the shadow of the SMBHs Sgr A* and M87 cast on hot plasma emission, such as accretion flow and/or relativistic jet as described in Sections 2.1, 2.2, and 2.3, despite angular resolution being limited to only 3\,R$_{s}$. Detection of the shadow indicates the existence of the BH and will surely open up a new area of BH astrophysics. Along with pioneering efforts, spontaneous quests for much higher angular resolution are expected, and true submm VLBI observations at 0.85\,mm or shorter wavelengths would be a natural extension of 3 and 1.3\,mm VLBI observations.

To achieve further progress and to push forward the observing frequency, a careful selection of telescope sites is necessary; the number of telescope sites is very limited. On the basis of statistics obtained from the 225\,GHz tipping radiometer at the Caltech Submillimeter Observatory and atmospheric modeling at Mauna Kea, it is expected the quartiles of opacity at 0.85\,mm would be 0.23 and 0.36 for 25\% and 50\%, respectively (Radford \& Camberlin, 2000). For Chajnantor, where ALMA is located, these statistics are expected to improve to 0.14 and 0.24 on the basis of monitoring measurements of a tipping radiometer (Radford \& Camberlin, 2000). Together with those sites, the GLT at the ISI site of Greenland is an additional promising site for VLBI at 0.85\,mm. ASIAA had continuously monitored the opacity at 225\,GHz again by using a tipping radiometer, and the quartiles of opacity at 0.85 mm for the winter season were 0.18 and 0.24 for 25\% and 50\% on the basis of three years of monitoring (Martin-Cocher et al., 2014). Moreover, the planned Balloon-borne VLBI experiment will be free from atmospheric limitations, and it will be expected to add mobile stations and improve the sampled baseline coverage together with ground-based stations.

Therefore, by using four extremely good weather and/or atmospheric-limitation-free sites, such as the JCMT/SMA, ALMA, the GLT, and balloon-borne VLBI stations, as the anchor of the 0.85\,mm VLBI array, EA coordinated submm VLBI array will be able to advance the current submm VLBI efforts to much higher angular resolution. The expected angular resolution would be 12 and 10\,$\mu$as at 0.85 and 0.7\,mm (350 and 420\,GHz), respectively, and corresponds to 1.5 and 1.25\,R$_{s}$ for M87. By pushing the observing frequency up to 0.4\,mm (690\,GHz), we can achieve an apparent angular resolution of 6\,$\mu$as, which corresponds to 0.75\,R$_s$ for M87. Therefore, in the future, we will be able to conduct observations at sub-event horizon angular resolution, and this will impel BH astrophysics toward being an exact science.

\vspace{3mm}
\noindent
\textbf{Extension of arrays at 3 mm}



\noindent
Currently, global VLBI observations at 3 mm are operated in coordination in the framework of the GMVA (http://www3.mpifr-bonn.mpg.de/div/vlbi/globalmm/), whose host institute is the Max-Planck-Institut für Radioastronomie. The sensitivities of each station together with those of the planned stations in EA regions are presented in Table~3. The KVN joined the GMVA in 2012 May on a test basis, and now the KVN array officially participates in observation programs involving the GMVA. The inclusion of the KVN in the GMVA improves the ($u,v$)-coverage and imaging
capabilities. Although the European (EU)--Korea baseline length, which can extend up to approximately 9000\,km, is comparable to the EU--USA baseline length, the increase in the number of longest baselines improves the imaging capability of a 40\,$\mu$as scale structure. Another unique feature of the GMVA including the KVN is the short inner KVN baseline spacings, which facilitate
a more accurate amplitude calibration of the GMVA and further improves detectability of extended resolved structures.

The tera-electronvolt $\gamma$-blazar Mrk 501 ($z$=0.0337) is one of the best targets for exploring the innermost part of blazars. Mrk 501 has a newly born jet component in the northeast direction, approximately 200\,$\mu$as from the radio core, as opposed to the persistent southeast jet (Koyama et al., 2016). The GMVA incorporating the KVN will be necessary for resolving the newly born jet component and studying its relationship to the extended persistent jet. It is evident that the superior imaging capability of GMVA + KVN would be beneficial for resolving the innermost part of other blazars as well.

\begin{figure}[h]
\centering
\includegraphics[width=120mm]{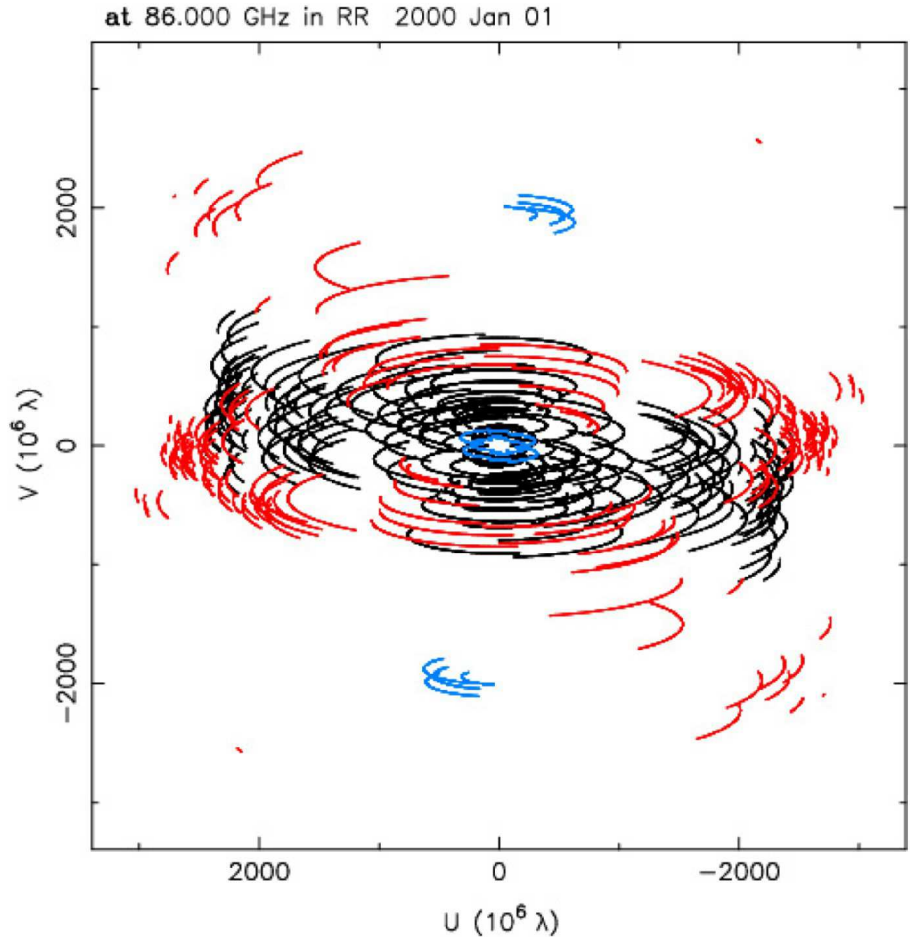}
\caption{
Expected ($u,v$)-coverage with the extended EAVN array at 3\,mm. Here, we present the case of 24-h track observation toward M87. Baselines within EA--Oceania stations (KVN, NRO45m, GLT, and ATCA) are shown by the blue lines, and the baselines between EA--Oceania and GMVA stations are marked by the red lines. The lower limit of each antenna elevation is set as 20$^{\circ}$.
\label{fig:M87-KVNATCA}}
\end{figure}

Extension of arrays at 3\,mm will lead to new advances in M87 science. The jet collimation profile of the M87 jet that is extensively down to the $10~R_{s}$ scale from the central BH has been investigated by using the VLBA between 15 and 3\,mm (Hada et al., 2013). Interestingly, a possible break of the jet collimation profile around $30~R_{s}$ was detected, and it was interpreted as reflecting a sudden change in the jet flow direction near the jet base. Nevertheless, this image is not conclusive because the VLBA-synthesized beam for M87 is elongated in the NS direction. With the objective of extending NS baselines, a new trial between KVN and ATCA has been started (Dodson et al. 2014, also see Figure \ref{fig:M87-KVNATCA}).

A new mode has been made available at the ATCA, and it enables observers to split the array and simultaneously observe different frequencies with different receivers. The change was specifically made to facilitate simultaneous 7 and 3\,mm VLBI observations between the ATCA and the KVN, and it makes the ATCA the sole facility in the Southern Hemisphere to simultaneously observe at these frequencies. Lower frequency measurements can be used to correct the phase at higher frequencies through a technique called source-frequency phase referencing. Because the Compact Array Broadband Backend has two intermediate frequencies (IFs), the array can be split into only two parts, with one set of antennas using one receiver in IF1 and the other set of antennas using another receiver in IF2. There are restrictions on the bands and frequencies that can be chosen together (because of constraints on the choice of local oscillator frequencies).

Finally, we expect that the NRO45m and GLT (see Section~4) will participate in coordinated 3\,mm VLBI observations in the near future. The inclusion of the NRO45m will considerably improve the fringe detection rate of transcontinental baselines between EU stations and Japan/Korea stations at 3\,mm. The GLT will also be able to link the stations between EU and US stations.




\begin{table}[htbp]
\caption{Current and planned stations for 3\,mm VLBI. EA contributions are shown in bold.}
\begin{center}
\begin{tabular}{lcccc}
\hline
Stations & Location & Diameter [m] & SEFD [Jy] & Status \\
\hline
\hline
ALMA           & Chile & 37 - 50 $\times$ 12 & 75  & operational  \\
ATCA           & Australia & 1--5 $\times$ 22 & 7200 - 1440  & operational  \\
Effelsberg     & Germany   & 80 & 1000 & operational \\
GBT            & Virginia  & 100 & 140 & operational \\
{\bf GLT}      & Greenland & 12 & 4900 & 2018 (planned) \\
IRAM 30m       & Spain     & 30 & 650 & operational \\
{\bf KVN}      & Korea     & 3 $\times$ 21 & 3200 & operational \\
LMT            & Mexico    & 32 & 1700 & operational \\
Metsahovi      & Finland   & 14 & 18000 & operational \\
NOEMA          & France    & 33 & 820 &operational \\
{\bf NRO 45 m} & Japan     & 10 & 10000 & 2018? (planned) \\
Onsala         & Sweden    & 20  & 5100 & operational \\
VLBA           & US        & 8 $\times$ 25 & 2500 & operational \\
Yebes          & Spain     & 40  & 1700 & operational \\
\hline
{\footnotesize write footnote here.
\label{tab:3mm-arry}}
\end{tabular}
\end{center}
\vspace{-0.3cm}
\end{table}

\bigskip

\noindent
\textbf{Extension of arrays up to at 7 mm (43 GHz)}


\noindent
In terms of overall scientific merits, the joint EAVN + Thai VLBI Network (TVN) + LBA network is a unique array because of its location across the equator, and it will be valuable for observations toward notable radio sources, such as Sgr A*, Cen A, and TANAMI blazars, with higher angular resolution in the north–south (NS) direction.

A new radio astronomy project, the TVN, is underway under the initiative of the National Astronomical Research Institute of Thailand (NARIT). The TVN will be the first VLBI array in southeastern Asia, and it consists of two phases. Two new radio telescopes will be constructed in the phase-I term (2016--2020). One of them will be dedicated to radio-astronomical research, and it will have an antenna diameter of 40--45 m and
 a frequency coverage of 1--50 GHz; it will be constructed in the northern part of Thailand. The other radio telescope will have a diameter of 13\,m and a frequency coverage of 2--14 GHz, and it will be mainly used for geodetic observations in accordance with specifications of the VLBI Global Observing System. NARIT has already started a detailed site survey, and it has submitted the TVN phase-I project to the government on the basis of the survey results. The project will be launched in 2016, provided the project proposal is approved by the government. In the TVN phase-II term (2020--), two additional telescopes will be constructed in the central and southern parts of Thailand, in parallel with the commissioning of a 40-m telescope. The TVN will have four telescopes, with the maximum baseline length being approximately 1,300\,km at cm and mm wavelengths after completion of construction.

In the Asia-Pacific region, the southern hemisphere LBA has been operational for many years. Collaborative work between the EAVN and the LBA has been made as part of the framework in the Asia-Pacific Telescope (APT), and we will start joint EAVN-LBA VLBI test observations from 2016. A maximum baseline length of above 10,000\,km can be obtained with the array in the north–south direction. This can facilitate much higher angular resolution and improved ($u,v$) coverage toward sources at low declinations and southern sources.



\newpage

{\bf \Large Acronyms and Abbreviations}

\begin{table}[htbp]
\begin{center}
\begin{tabular}{ll}
  \hline
ADAF   & advection-dominated accretion flow \\
ADIOS  & adiabatic inflow-outflow solution \\
AGB    & asymptotic giant branch  \\
AGN    & Active Galactic Nucleus \\
ALMA   & Atacama Large Millimeter Array  \\
APEX   & Atacama Pathfinder Experiment  \\
APP    & ALMA Phasing Project \\
APT    & Asia-Pacific Telescope \\
ASIAA  & Academia Sinica Institute of Astronomy and Astrophysics  \\
ATCA   & Australia Telescope Compact Array \\
CARMA  & Combined Array for Research in Millimeter-wave Astronomy \\
CDAF   & convection-dominated accretion flow \\
CSE    & circumstellar envelope \\
CSO    & Caltech Submillimeter Observatory \\
CSO    & Compact Symmetric Object \\
CSS    & Compact Steep Spectrum \\
DPNL   & double-peaked narrow-line \\
EAO    & East Asia Observatory \\
EAVN   & East Asia VLBI Network  \\
EHT    & Event Horizon Telescope \\
EVN    & European VLBI Network \\
FIRST  & Faint Images of the Radio Sky at Twenty Centimeters \\
GC     & Galactic Center \\
GLT    & Greenland Telescope \\
GMVA   & Global Millimeter VLBI Array  \\
GRMHD  & general relativistic magnetohydrodynamic \\
IRAM   & Institut de Radioastronomie Millim{\'e}trique \\
ISCO   & innermost stable circular orbit \\
JCMT   & James Clerk Maxwell Telescope   \\
KASI   & Korea Astronomy and Space Science Institute \\
KaVA   & KVN and VERA Array \\ 
KVN    & Korean VLBI Network  \\
LLAGN  & low-luminosity AGN \\
LMT    & Large Millimeter Telescope \\
\hline
\label{tab:acronyms}
\end{tabular}
\end{center}
\end{table}

\begin{table}[htbp]
\begin{center}
\begin{tabular}{ll}
  \hline

NAOJ   & National Astronomical Observatory of Japan  \\
NARIT  & National Astronomical Research Institute of Thailand \\
NICT   & National Institute of Information and Communications Technology \\
NLR    & Narrow Line Region \\
NLS1   & narrow-line Seyfert 1 (galaxy) \\
NOEMA  & Northern Extended Millimeter Array   \\
NRAO   & National Radio Astronomy Observatory  \\
OCXO   &  Oven-Controlled Crystal Oscillator \\
PdBI   & Plateau de Bure Interferometer \\
RIAF   & radiatively inefficient accretion flow \\
RLNLS1 & tadio-loud narrow-line Seyfert 1 (galaxy) \\
RM     & rotation measure \\
RRL    & radio recombination line \\
RRO    & rapid response observation \\
RSG    & red supergiant \\
SD     & single-dish \\
  SFR    & star-forming region \\
SGI    & sphere of gravitational influence \\
SMA    & Submillimeter Array \\
SMBH   & Super-massive Black Hole \\
SMT    & Sub-Millimeter Telescope  \\
SMTO   & Sub-Millimeter Telescope Observatory  \\
SNR    & supernova remnant \\
SPART  & Solar Planetary Atmosphere Research Telescope   \\
SPT    & South Pole Telescope \\
SRAO   & Seoul Radio Astronomy Observatory  \\
SRON   &   \\
SSA    & synchrotron self-absorption \\
SWARM  & SMA Wideband Astronomical ROACH2 Machine \\
TVN    & Thailand VLBI Network \\
VERA   & VLBI Exploration of Radio Astrometry  \\
VLA    & Very Large Array \\
VLBA   & Very Long Baseline Array \\
VLBI   & Very Long Baseline Interferometry \\
YSO   & young stellar object \\
\hline
\label{tab:acronyms}
\end{tabular}
\end{center}
\end{table}

\newpage

\newpage

\section*{Acknowledgements}

This work is partly based on observations made with the KaVA, 
which is operated by KASI and NAOJ.
M.~Kino (KASI) and B-W.~Sohn (KASI)
wish to thank all of the members for their 
intensive research activities in KaVA AGN Science Working Group.
The complete list of the members is as follows;
Kazunori Akiyama    (MIT, NAOJ),
Juan-Carlos Algaba  (KASI),
Ilje Cho            (KASI),          
Richard Dodson     (ICRAR),
Akihiro Doi        (ISAS/JAXA),
Tomoya Hirota      (NAOJ),   
Kazuhiro Hada       (NAOJ),
Yoshiaki Hagiwara   (Toyo Univ.),
Mareki Honma        (NAOJ),            
Jeffrey Hodgson    (KASI),
Hiroshi Imai       (Kagoshima Univ.),    
Wu JIANG           (SHAO),
Taehyun Jung        (KASI),
Sincheol Kang      (KASI),       
Noriyuki Kawaguchi  (SHAO),
Dae-Won Kim         (SNU),
Minsun Kim         (KASI),
Shoko Koyama        (MPIfR),
Cheulhong Min      (GUAS/NAOJ),      
Atsushi Miyazaki   (Hosei Univ.),      
Kotaro Niinuma      (Yamaguchi Univ.),
Se-Jin Oh          (KASI),
Junghwan Oh         (SNU),      
Gabor Orosz        (Kagoshima Univ.),    
Jong-Ho Park        (SNU),            
Taeseok Lee         (SNU),            
Jeong Ae Lee        (KASI),          
Jee Won Lee        (KASI),          
Sang-Sung Lee      (KASI),
Satoko Sawada-Satoh (Ibaraki Univ.),
Sascha Trippe       (SNU),
Fumie Tazaki        (NAOJ),
Jan Wagner         (KASI),
Kiyoaki Wajima      (KASI),
Yuanwei Wu         (NAOJ),
Maria Rioja        (ICRAR),
Hyunwook Ro         (Yonsei Univ.),       
Hyemin Yoo          (Yonsei Univ.), and       
Guang-Yao Zhao      (KASI).
As for maser science,
T.~Hirota (NAOJ) acknowledges support from
Hiroshi Imai (Kagoshima Univ.), 
Hiroko Shinnaga (NAOJ), and
Kazuhito Motogi (NAOJ).

\bigskip

The EAVN is a collaborative project between core institutes of the
East Asian Core Observatories Association (EACOA) and universities
in East Asia, and is operated as part of activities of East Asia
VLBI Working Group under EACOA. The authors should like to express
our thanks to the members of the EAVN Tiger Team; Tao An (SHAO),
Willem A. Baan (SHAO), Kenta Fujisawa (Yamaguchi Univ.), Kazuhiro Hada
(NAOJ), Yoshiaki Hagiwara (Toyo Univ.), Longfei Hao (Yunnan Astronomical
Observatory (YAO)), Wu Jiang (SHAO), Taehyun Jung (KASI), Noriyuki
Kawaguchi (NAOJ), Jongsoo Kim (KASI), Hideyuki Kobayashi (NAOJ),
Se-Jin Oh (KASI), Duk-Gyoo Roh (KASI), Min Wang (YAO), Yuanwei Wu
(NAOJ), Bo Xia (SHAO), Kiyoaki Wajima (KASI), and Ming Zhang (Xinjiang
Astronomical Observatory) The authors acknowledge Do-Young Byun (KASI),
Mareki Honma (NAOJ), Zhi-Qiang Shen (SHAO), and Yoshinori Yonekura
(Ibaraki Univ.) for their encouragement and support for observations.

\bigskip

M.I. would like to thank all of engineers of the GLT group.  He also would like to thank T. Norton of SAO/CfA for his great support, particularly arrangements for shipping the GLT components at Norfolk, USA, and Thule, Greenland, and Ching-Tang Liu's team of ASRD/NCSIST for their great support in GLT antenna structure modifications and component designs, and particularly help for antenna assembling and test at Norfolk, USA, and Thule, Greenland.

\bigskip

Akihiro Doi (JAXA)
should like to express our thanks to the developing team members of the balloon-borne VLBI project; 
Yusuke Kono (NAOJ), Mitsuhisa Baba (JAXA),
Kimihiro Kimura, Nozomi Okada (Osaka Prefecture Univ.), 
Naoko Matsumoto (Yamaguchi Univ./NAOJ), 
Tomoaki Oyama, Syunsaku Suzuki, Kazuyoshi Yamashita (NAOJ), 
Yasuhiro Murata (JAXA), 
Satomi Nakahara (SOKENDAI), and other minor partners.  
The balloon-borne VLBI experiment is supported by MEXT Grant-in-Aid for Scientific Research on Innovative Areas (AD; No. 26120537), JSPS Grant-in-Aid for Scientific Research (B) (YK; 16K05305), CASIO science promotion foundation (AD; 201421), Inamori Foundation (YK; 201515), JAXA strategic fundamental development and research (AD, 2014-2015), JAXA fundamental development and experiment for on-board devices (YK; 2014-2015).

\bigskip

We are grateful to Yong-Sun Park (SNU) and Bon-Chul Koo (SNU)
for establishing the SRAO and many years of hard work with the observatory. 

\bigskip

The VLBA
is operated by the US National Radio Astronomy Observatory 
(NRAO), a facility of the National Science Foundation operated
under cooperative agreement by Associated Universities, Inc.
This White Paper has made use of data from the MOJAVE 
database that is maintained by the MOJAVE team (Lister et al., 2009, AJ, 137, 3718).

\bigskip

T.H. is supported by MEXT/JSPS KAKENHI grant Nos. 24684011,
25108005, 15H03646, and 15H03646 and the ALMA Japan Research
Grant of NAOJ Chile Observatory, NAOJ-ALMA-0006 and NAOJ-ALMA-0028.
S.T.
acknowledges financial support from the National Research Foundation
of Korea (NRF) via Basic Research Grant NRF-2015-R1D1A1A01056807".
H.N. is supported by MEXT KAKENHI Grant Number 15K17619. 
K.N. is supported by JSPS KAKENHI Grant Number 15H00784 and the grant of
Visiting Scholar Program by the Research Coordination Committee, NAOJ.
M.F.G. is supported by the National Science Foundation
of China (grants 11473054 and U1531245) and by the Science and Technology Commission
of Shanghai Municipality (14ZR1447100).
K.H. is supported by MEXT/JSPS KAKENHI grant No. 26800109. 
M.H. is supported by MEXT Grant-in-Aid for Scientific Research 
on Innovative Areas No. 25120007.
Y.H. is supported by Japan Society for the Promotion of Science (JSPS) 
Grant-in-Aid for Scientific Research (B) (Grant Number: 15H03644)
T.J. and G.Z. are supported by Korea Research Fellowship Program 
through the National Research Foundation of Korea (NRF) funded 
by the Ministry of Science, ICT and Future Planning (NRF-2015H1D3A1066561).
AD is supported by Japan Society for the Promotion of Science (JSPS) Grant-in-Aid for Scientific Research (B) (Grant Number: 24340042).




\bibliography{EAwhitepaper}

\end{document}